\documentclass[aos, preprint]{imsart}

\RequirePackage{amsthm,amsmath,amsfonts,amssymb}
\RequirePackage{graphicx,bm}
\usepackage[numbers,sort&compress]{natbib}
\usepackage{lipsum,bm}
\usepackage{algorithm}
\usepackage{algpseudocode}
\usepackage[colorlinks = true]{hyperref}
\usepackage{titlesec}
\usepackage{xcolor}
\usepackage{subcaption}
\algrenewcommand\algorithmicindent{2mm}
\usepackage{lineno}

\usepackage{xcolor}
\usepackage{soul}
\usepackage{rotating}
 \usepackage{multirow}
 \usepackage{tabularx}
 \usepackage{bigints}
 \usepackage[english]{babel}
 \usepackage{blindtext}
 \usepackage{float}

\usepackage{enumitem}

\usepackage{titlesec}

\usepackage{amsthm}



\allowdisplaybreaks
\newtheorem{theorem}{Theorem}
\newtheorem{lemma}{Lemma}
\newtheorem{remark}{Remark}
\newtheorem{corollary}{Corollary}

\newtheorem{example}{\quad Example}
\theoremstyle{definition}

 \DeclareMathOperator{\spn}{span}

\newcommand{\comment}[1]{}

\def\tr{\mathop{\rm tr}}

\def\P{\mathbb P}
\def\bP{\mathbb P}
\def\bE{\mathbb E}
\def\E{\mathbb E}
\def\R{\mathbb R}

\def\*#1{\mathbf{#1}}

\def\argmin{\mathop{\rm arg\, min\,}}

\def\arg{\mathop{\rm arg}}

\def\btheta{\bm\theta}
\def\bnu{\bm\nu}
\def\bxi{\bm \xi}
\def\bbeta{\bm\beta}

\newcommand\numberthis{\addtocounter{equation}{1}\tag{\theequation}}
\begin{document}

\begin{frontmatter}

\title{
Finite- and Large-Sample Inference for Model and Coefficients in High-dimensional Linear Regression with Repro Samples}
\runtitle{High-dimensional inference for model and coefficients with repro samples}

\begin{aug}
\author[A]{\fnms{Peng} \snm{Wang}\ead[label=e1]{wangp9@ucmail.uc.edu}}
\and
\author[B]{\fnms{Min-ge} \snm{Xie}\ead[label=e2,mark]{mxie@stat.rutgers.edu}}
\and
\author[B]{\fnms{Linjun} \snm{Zhang}\ead[label=e3,mark]{linjun.zhang@rutgers.edu}}
\address[A]{Department of Operations, Business Analytics and Information Systems,University of Cincinnati, Cincinnati, OH 45221, USA.
\printead{e1}}
\address[B]{Department of Statistics, Rutgers University, New Brunswick, NJ 08854, USA.
\printead{e2}; \printead{e3}}
\end{aug}

\begin{abstract}
\, In this paper, we present a novel and effective inference
approach
to conduct both finite- and large-sample inference
for high-dimensional linear regression models.
This approach is developed under the so-called {\it repro samples} framework, in which we conduct statistical inference by creating and studying the behavior of artificial samples that are obtained by mimicking the sampling mechanism of the data. We constr confidence sets for (a) the true model corresponding to the nonzero coefficients, (b) a single or any collection of regression coefficients, and (c) both the model and regression coefficients jointly.
To facilitate the constructions of these confidence sets and overcome computational difficulties of searching all possible models, we use an innovative Fisher inversion technique to construct a model candidate set that includes the true sparse model with the probability close to 1 for models with both Gaussian and non-Gaussian errors.
The proposed approach fills in two major gaps in the high-dimensional regression literature: (1) lack of effective approaches to addressing model selection uncertainty and providing valid inference for the underlying true model; (2) lack of effective inference approaches to guaranteeing finite-sample performance.   We provide both finite-sample and asymptotic results to theoretically guarantee the performance of the proposed methods.
In addition, our numerical results demonstrate that the proposed methods are valid and achieve better coverage with smaller confidence sets than the current state-of-the-art approaches, such as debiasing and bootstrap approaches.

\end{abstract}



\end{frontmatter}

\footnotetext{
The authors acknowledge support of research grants: NSF-DMS2015373,
2027855,
2015378,
2319260,
2311064, 2340241,
2515766 and
NIH-1R01GM157610.}

\section{Introduction}

High-dimensional linear regression plays an important role in modern statistics, with applications ranging from signal processing \cite{wang2020beamforming} to econometrics \cite{belloni2018high,korobilis2021high} to bioinformatics \cite{wang2019precision}.
There has been a large amount of literature on this topic in the past 30 years.
The earlier research
focused more on estimation/detection problems such as coefficients estimation \cite{candes2007dantzig,tibshirani1996regression} and support recovery \cite{zhao2006model,zou_adaptive_2006}.
Starting with the
work of debiased Lasso \cite{zhang_confidence_2014,nickl_confidence_2013,javanmard_confidence_2014},
the more difficult task of inference comes to the central stage. Some recent works on inference include confidence interval construction \cite{zhang_confidence_2014,nickl_confidence_2013,javanmard_confidence_2014}, multiple testing of regression coefficients \cite{liu2013gaussian}, and post-selection inference \cite{taylor2015statistical,tibshirani2016exact,leiner2021data}.

Despite many works on
the topic,
several important open problems remain.
First,  most existing works focus on
the inference for regression coefficients, while the inference for true model (including uncertainty quantification for model selection) in the high-dimensional regression
model is mostly absent.
This is partly due to the challenges arising from the discrete nature of the model space,
which makes the conventional inference tools
built for continuous parameters, such as the central limit theorem and bootstrap theorems, inapplicable. Furthermore, all the results in the literature on high-dimensional inference are asymptotic,
assuming
the sample size goes to infinity, and
there are no theories concerning the performance of these procedures under finite-sample settings. The performance of these asymptotic procedures is frequently  empirically unsatisfactory, especially when sample size is limited. Thus
a  procedure with guaranteed finite-sample performance is desirable. Finally,
the post-selection inference framework attempts to sidestep the problem of model uncertainty by only making conditional inferences for regression coefficients of the predictors selected by a model selection procedure. If some predictor variables are significant but not selected, no inference results are available on these predictors and
we may miss
some important signals.

To solve the above problems,  we develop a repro samples method that quantifies both the uncertainty in model selection and  that in estimation of regression coefficients and their functions. Specifically, we provide a comprehensive inferential approach with which we can construct confidence sets for (a) the true model, (b) a single or any collection of regression coefficients, and (c) both the model and regression coefficients jointly.
Moreover,
the proposed repro samples approach enjoys finite-sample performance guarantees without requiring a large sample size for all of (a)-(c).
Although our work focuses primarily on finite-sample performances, we also provide related large-sample results.

Consider the high-dimensional linear regression problem where we observe an $n \times 1$ response vector ${\bf y}_{obs}$ with an $n \times p$ design matrix ${\bf X}$, where $p \gg n.$
Suppose that
the observed ${\bf y}_{obs}$  relates only to a subset of predictor variables indexed by
$\tau_0 \subset \{1, \dots, p\}$ with
\begin{equation}
\label{eq:model-rel}
{\bf y}_{obs}
= \*X \bbeta_0^{full}  + \sigma_0 \*u^{rel} = \*X_{\tau_0} \bbeta_0 + \sigma_0 \*u^{rel}.
\end{equation}
Here, $\*u^{rel}$ is the realization of the error term, $\btheta_0= (\tau_0, {\bbeta}_0, \sigma_0^2)$ are unknown model parameters, $\*X= (\*X_{\tau_0}, \*X_{\tau^C_0})$, $\bbeta^{full}_{0}=(\bbeta^\top_{0}, \*0^\top_{\tau_0^C})^\top$ and $\tau_0^C = \{1, \ldots, p \} \setminus \tau_0$.
Corresponding to model (\ref{eq:model-rel}),
there is a random sample (or population) version of data generation model
\begin{equation}
\label{eq:model-random}
{\bf Y} =
\*X \bbeta_0^{full}  + \sigma_0 \*U = \*X_{\tau_0} \bbeta_0 + \sigma_0 \*U,
\end{equation}
of which model (\ref{eq:model-rel}) is a realization. In general, we assume $\mathbb{E}({\*U}) = \*0$, although
further
conditions may be required for different inference tasks.

To carry out the inference tasks, two inversion
techniques are devised in the proposed repro samples method to handle the discrete parameter $\tau_0$.
The first inversion technique, developed in Section~\ref{sec:finding_cand} and referred to as {\it Fisher inversion}, aims to obtain a model candidate set that includes the true $\tau_0$ with a high (close to 1) probability using the observed data and a large set of {\it repro}duced (simulated)
artificial error terms.
This model candidate set,
 typically of a reasonable size,
 is then used to facilitate the constructions of
level-$\alpha$ confidence sets and intervals later in  Sections~\ref{Sec:CS} and~\ref{sec:coef}.
To  obtain a level-$\alpha$ model confidence set for $\tau_0$
in Section~\ref{Sec:CS}, we use the second inversion technique, referred to as {\it Fisher-Dempster inversion} for distinction, to  overcome the difficulty that the central limit theory does not apply on a discrete parameter space.
The Fisher-Dempster inversion technique inverts a level-$\alpha$ Borel set of possible error realizations to get a level-$\alpha$ confidence set for  $\tau_0$.
Unlike the conventional Wald-test type of methods, the proposed repro samples method directly provides the desired confidence sets without having to estimate $(\tau_0, \bbeta_0)$ or any other model parameters.

\subsection{Contributions}

To summarize, this paper has the following
contributions.
\begin{enumerate}
    \item We propose a repro samples method to effectively construct model confidence set and quantify model selection uncertainty for the high-dimensional linear regression model.
    To the best of our knowledge, it is the only computationally efficient approach that provides a performance-guaranteed model confidence set for $\tau_0$ without data splitting or a prior assumed
    candidate model set.
    \item We develop a novel inference procedure for regression coefficients $\bbeta_0$. Contrary to other existing methods, our approach does not rely on covariance matrix estimation or a consistent model selection procedure. Therefore, it sidesteps potential issues caused by any inaccurate estimation of the covariance matrix  or mis-selection of the model.

\item { We propose a novel and efficient way to find candidate models
using synthetically generated residuals. Theoretically, we show that this set of candidate models have a high probability to include the true model $\tau_0$ for both Gaussian and non-Gaussian errors, and we also provide an upper bound for the expected size of the set. Numerically, we have shown in our simulation studies that the proposed model candidate set is of reasonable size, and covers $\tau_0$ with a probability close to 1 even in challenging settings. The model candidate set facilitates the inference for both the true model and the regression coefficients. It can also be used for
 variable screening,
providing
superior performances to the
traditional screening approaches relying on only marginal relationship.}
    \item Theoretically, we show that the proposed inference procedures for both the model and the regression coefficients achieve finite-sample coverage guarantees,
    while most literature on high-dimensional models focuses only on asymptotic properties.
To our knowledge, the proposed method is the first approach that guarantees coverage for finite samples. Additionally, our theory suggests a complementary effect between computational power and sample size: one can achieve valid coverage as long as either is sufficiently large.
    \item Theoretically, we do not need to impose the standard conditions that high-dimensional statistics literature typically requires to obtain a consistent estimation, such as the restricted isometry property
    or restricted eigenvalue conditions  \cite{buhlmann2011statistics, zhang_confidence_2014}. Neither do we need to require  signal strength conditions \cite{zhao2006model,buhlmann2011statistics}, which is usually necessary for consistent model selection.
    { We also provide a discussion on conditions required for different implementations of the proposed procedure under computational~considerations. }
    \item Finally, through extensive numerical studies, we show that the proposed  method
    produces better performed confidence sets than those of the state-of-the-art debiased Lasso estimators \citep{zhang_confidence_2014,  javanmard_confidence_2014}.
    Because of the finite-sample validity guarantee, our method achieves the desired coverage even in small-sample regimes, while the existing~methods~can~not.
\end{enumerate}
Overall, we provide a comprehensive framework that subsumes existing inference approaches by two means: we consider a broader set of marginal and joint inference problems to account for uncertainties of estimating both the model and the regression coefficients; and we provide supporting theories to guarantee both finite- and large-sample performances.

\subsection{Related works} \label{sec:related_work}
 There has been much effort in recent years to develop inference procedures for regression coefficients $\bbeta_0^{full}$ or functions of $\bbeta_0^{full}$ in high-dimensional linear regression models. On the inference for a single coefficient parameter,
\cite{zhang_confidence_2014,van_de_geer_asymptotically_2014,javanmard_confidence_2014} propose the debiased Lasso estimator and develop its asymptotic distribution. Other works along this line include  \cite{chatterjee_bootstrapping_2011, das_perturbation_2019_aos, das_distributional_2019_bio,chernozhukov_doubledebiasedneyman_2017, chernozhukov_post-selection_2015}. Moreover,  \cite{zhu_high-dimensional_2020, zhang_simultaneous_2017, dezeure_high-dimensional_2017,   cai_confidence_2017, nickl_confidence_2013, zhou_honest_2019, zhu_linear_2018} investigate simultaneous inference on a subset or all of $\bbeta_0^{full}.$
Additionally, quadratic and more general functions of $\bbeta_0^{full}$ have been studied in \citep{athey_approximate_2018,  guo_optimal_2019,  javanmard_flexible_2019,  zhu_projection_2017, zhu_linear_2018,  meinshausen_p-values_2009}. However, all the existing approaches are developed using
large-sample theories and do not have
any finite-sample performance guarantees.   The inference for $\tau_0$, on the other hand,
is almost entirely absent in high-dimensional statistics literature, although there are a few works in low-dimensional setting.  \cite{hansen_model_2011-1, ferrari_confidence_2015} construct a model confidence set utilizing sequential testing procedures against a pre-specified finite-dimensional full model, which cannot be well-defined in
the high-dimensional setting. 
\cite{li_model_2019} proposes a novel concept of model confidence bounds
to confine $\tau_0$ within
a pair of nested
models.
However, the method relies on selection consistency and bootstrap validity, and is computationally expensive for high-dimensional data.

A recent work by the authors \cite{xie_repro_2022} provides a repro samples framework for statistical inference under a general setup, in which the number of parameters $p$ is less than the number of observations $n$.
The current paper focuses on the high-dimensional $p \gg n$ case that was not discussed in \cite{xie_repro_2022}. New procedures and theoretical results with conditions tailored to high-dimensional models that guarantee the performance of the proposed method in both finite and large-sample cases are developed.
Finally,
as discussed in \cite{xie_repro_2022}, the repro samples approach is related to other modern simulation-based procedures, such as the bootstrap \cite{efron_bootstrap_1992, chatterjee_bootstrapping_2011},  the approximate Bayesian computation
\cite{beaumont_approximate_2002, craiu_approximate_2023}, the inferential models \cite{Martin2015} and the generalized fiducial inference \cite{hannig_generalized_2016}, where artificial data are used to address inference problems.

\subsection{Notation}
For any $p \in \mathbb N^+$, we let $[p] = \{1,\dots,p\}$. For a vector $\*v$, we let $v_i$ be the $i$-th entry. For a set $S$, let $|S|$ be the cardinality of $S$. For two positive sequences $\{a_k\}$ and $\{b_k\}$, write $a_k =O(b_k)$,
if $\lim_{k\rightarrow\infty}(a_k/b_k) < \infty$;  write $a_k = o(b_k)$, if $\lim_{k\rightarrow\infty}(a_k/b_k) =0$.
We use $\P$ for probability and $\E$ for expectation and add subscripts (eg., $\P_{\*U}$ and $\E_{\*U}$) to indicate source of randomness. We use $\hat \bP$ and $\hat \bE$ for empirical probability and expectation. For a $\bbeta\in\R^p$ and  $\tau\subset
 [p]$, we use $\bbeta_{\tau}$ to denote the sub-vector of $\bbeta$, containing the entries of $\bbeta$ that are associated with the indices in $\tau$. The model space ${\cal M} = 2^{[p]}.$
For a matrix $\*M\in\mathbb R^{m\times n}$, let $\spn(\*M)$ be the vector space spanned by the columns of $\*M$: $\spn(\*M)=\{\*M\*v:\*v\in\R^n\}$. We also call $\*M(\*M^\top \*M)^{-1}\*M^\top$ the projection matrix of $\*M$, and $\*I$ is the identity matrix.  Lastly, we use $\Gamma_{\alpha}^{\btheta}(\*y_{obs}),$ $\Gamma_{\alpha}^{\tau}(\*y_{obs})$ and $\Gamma_{\alpha}^{\bbeta_\Lambda}(\*y_{obs})$ to denote the level-$\alpha$ confidence set for $\btheta_0$, $\tau_0$ and $\bbeta_{0, \Lambda}$,  respectively, where $\Lambda$ is any subset of $[p]$.
Here, the superscript $\btheta$ of $\Gamma_{\alpha}^{\btheta}(\*y_{obs})$ simply indicates the target parameter of the confidence set is $\btheta$, and the set $\Gamma_{\alpha}^{\btheta}(\*y_{obs})$ does not depend on any particular value of~$\btheta$. Notations for other confidence sets, such as $\Gamma_{\alpha}^{\tau}(\*y_{obs})$, $\Gamma_{\alpha}^{\bbeta_\Lambda}(\*y_{obs})$, etc.,
are defined similarly.

Finally, we refer to  a simulated copy of artificial $\*u^* \sim \*U$ as a {\it repro} copy of the realized $\*u^{rel}$ and the artificial data $\*y^* = X_\tau  \bbeta_\tau + \sigma \*u^*$ as a {\it repro sample} of $\*y^{obs}$ for a potential set of values $(\tau, \beta_\tau, \sigma^2)$. The key of our approach is to study and relate this $\*u^*$ with $\*u^{rel}$ and the $\*y^*$ with $\*y^{obs}$. We generally refer to our method, developed by using the copies of $\*u^*$ and $\*y^*,$ as a {\it repro samples method}. We will provide more details in each of the sections.

\subsection{Organization}
The paper is organized as follows.
Section~\ref{sec:finding_cand} provides a data-driven approach
to effectively construct a set of candidate models that will include the true model $\tau_0$ with a high (close to 1) probability.  Section~\ref{Sec:CS} utilizes the candidate set to construct a level-$\alpha$ confidence set for $\tau_0$, and provides  both finite-sample and large-sample guarantees. Section~\ref{sec:coef} studies the inference problems of regression coefficients, including inference for $\bbeta_0^{full},$  linear transformations of $\bbeta_0^{full}$ and functions of $\bbeta_0^{full}$. Section~\ref{sec:simulation} provides numerical illustrations of the proposed methods and compares the coverage and size of the constructed confidence sets with the bootstrap and state-of-the-art debiased Lasso methods.  In Section~\ref{sec:real_data}, we perform a real data analysis.
Section~\ref{sec:discussion} concludes the paper with a discussion of our results and future research directions. Theoretical proofs, technical lemmas, as well as additional discussions and numerical results are deferred to Appendices~\ref{sec:constraint}--\ref{sec:joint_beta_full} in the supplementary materials.

\section{Finding candidate models for $\tau_0$}
\label{sec:finding_cand} In this section, we propose a novel procedure to efficiently find possible candidate models for $\tau_0.$ In Section~\ref{sec:identifiablity}, to rigorously set up our problem and eliminate possible non-identifiability issues, we formally define the target true sparse model $\tau_0$ as the smallest model that generates the data. Section~\ref{sec_cand} introduces an effective computing algorithm and Fisher inversion method to obtain a set of candidate models for $\tau_0$. Section~\ref{sec:cand_theory_normal} proves that the model candidate set obtained in Section~\ref{sec_cand}  is guaranteed to cover $\tau_0$ with a probability close to 1 under the Gaussian error model assumption, in both finite-sample and asymptotic settings.  Moreover, we also provide a theoretical upper bound for the size of the model candidate set.  In Section~\ref{sec:cand_non_normal}, we show that the finite-sample
coverage results also hold
when the error term follows a number of other distributions, such as when $\*U$ is heterogeneous, Cauchy, $t$-distributed, contaminated Gaussian or sub-Gaussian distributed.
In addition, we show that the large-sample result continues to hold under sub-Gaussian errors.

\subsection{Identifiability and definition of $\tau_0$}  \label{sec:identifiablity}
\label{paragraph:identify}
Under the high-dimensional setting with $p\gg n$,
there might be another model $\tilde\tau_0$ and corresponding coefficients $\bbeta_{\tilde \tau_0}$ such that  $\*X_{\tau_0}\bbeta_0 = \*X_{\tilde \tau_0} \bbeta_{\tilde \tau_0}$.
Even when we know both ${\bf y}_{obs}$ and the realized noise $\*u^{rel}$, it is not possible to tell apart $\tau_0$ and $\tilde \tau_0$, since ${\bf y}_{obs} = \*X_{\tau_0} \bbeta_0 + \sigma_0 \*u^{rel} =  \*X_{\tilde \tau_0}  \bbeta_{\tilde \tau_0} + \sigma_0  \*u^{rel}.$ We refer to this as an {\it identifiability issue}.
To address this issue and uniquely define $\tau_0$, the conventional practice in the high-dimensional regression literature \citep[e.g.,][]{tibshirani1996regression, buhlmann2011statistics} is to favor the smaller model, since as stated in \citep{tibshirani1996regression}, in real applications researchers would often prefer and be interested in the simplest (smallest) model that generates the observed data for better prediction performance and model interpretation. Commonly used penalized regression approaches, such as Lasso \citep{tibshirani1996regression}, SCAD \citep{fan_variable_2001}, and MCP \citep{zhang_nearly_2010}, all employ penalty terms designed to favor smaller models. In this paper, to address this identifiability issue, we follow the same practice
to rigorously re-define $\tau_0$ as the smallest model among the set $\{ \tau\in\mathcal M| \*X_{\tau} \bbeta_{\tau} = \*X_{\tau_0} \bbeta_{0}, \hbox{for some } \bbeta_{\tau}\}$:
\begin{align}
\label{eq:def_true_model_0}
\tau_0'  =
\underset{\small \big\{  \tau|  \*X_{\tau} \bbeta_{\tau}  = \*X_{\tau_0} \bbeta_{0} \big\}}{\rm argmin}  |\tau|.
\end{align}
 Throughout the paper, we assume that the true model $\tau_0'$ defined in \eqref{eq:def_true_model_0} is unique, which we refer to as the {\it identifiability condition}.
For notational simplicity, we will still refer $\tau_0'$ as $\tau_0$ and $\bbeta_{\tau_0'}$ as $\bbeta_0$ in the remainder of the paper. Our inference target is this set of $(\tau_0, \bbeta_0)$~just~defined.

{
Furthermore, we follow \citep{shen_constrained_2013} to define the degree of separation between model $\tau_0$ and other models of equal or smaller model sizes as
\begin{align}
    C_{\min} = \min_{\{\tau: \tau \neq \tau_0, |\tau| \leq |\tau_0|\} } \frac{1}{n \max(|\tau_0 \setminus \tau|,1)} \|\*X_{\tau_0}\bbeta_0 - \*X_\tau \bbeta_\tau \|_2^2.  \nonumber
 \end{align}
Under the identifiability condition mentioned above, we have
$C_{\min} > 0$.
The notion $C_{\min}$ is related to $\bbeta_{\min}$ of the
$\bbeta$-min condition in the literature \cite[e.g.,][]{tibshirani1996regression}. However, unlike the existing literature,
we do not impose any
assumption on $C_{\min}$ other than that $C_{\min} >0$.}

\subsection{Algorithm for finding candidate models}
\label{sec_cand}

Here,  we
use an inversion method to
construct a set of candidate models for $\tau_0$.
To illustrate the basis of this inversion
method, we
first show that we can
recover the true model
$\tau_0$ in an ideal (unrealistic) case assuming that the realization of the error term $\*u^{rel}$ is known.
 In particular, Lemma~\ref{lemma_def1} below states that $\tau_0$ defined in \eqref{eq:def_true_model_0} can be expressed in terms of the given realization~$({\bf y}_{obs}, \*u^{rel})$~using~an~optimization~statement.
\begin{lemma}\label{lemma_def1}
Let $\*H_\tau$ be the projection matrix of $\*X_\tau$ and $\*H_{\tau,\*u^{rel}}$ be the projection matrix of $(\*X_\tau, \*u^{rel}).$  Let $\gamma^2_{(\*u^{rel}, \tau_0)} = 1- \min_{\{\tau: |\tau| < |\tau_0|\}}\frac{\|(I -  \*H_{\tau, \*u^{rel}})\*X_{\tau_0}\bbeta_0\|^2}{\|(\bm I -  \*H_{\tau})\*X_{\tau_0}\bbeta_0\|^2} < 1.$
Then, given $\*u^{rel}$, $\tau_0$ defined in \eqref{eq:def_true_model_0} satisfies
\begin{eqnarray} \label{eq:lemma1a}
\tau_0 &  = &  \underset{\tau}{\rm argmin}\left\{ \min_{\bbeta_{\tau}, \sigma} \left\{    \|\*y_{obs}- \*X_{\tau}\bbeta_{\tau}-\sigma \*u^{rel}\|^2_2 + \lambda |\tau| \right\}\right\},
\end{eqnarray}
and moreover $(\tau_0,\bbeta_0, \sigma_0)   =   \underset{\tau, \bbeta_{\tau}, \sigma}{\rm argmin} \left\{  \|\*y_{obs}- \*X_{\tau}\bbeta_{\tau}-\sigma \*u^{rel}\|^2_2 + \lambda |\tau|\right\}$
for any $0<\lambda < n [ 1-{\gamma^2_{(\*u^{rel}, \tau_0)}}]C_{\min}.$
\end{lemma}

In practice, however, we do not know the realized errors $\*u^{rel}$ so we cannot directly apply Lemma~\ref{lemma_def1}. Nonetheless, equation \eqref{eq:lemma1a} offers  guidance on constructing a set of candidate models for $\tau_0.$
More specifically,
we generate  a large number of, say $d$, copies of Monte Carlo $\*u_{1}^*,\dots, \*u^*_{d} \stackrel{i.i.d. }{\sim} \*U$.
Then instead of solving \eqref{eq:lemma1a} with the realized $\*u^{rel},$ we compute
\begin{align}
\label{eq::obj}
 \hat\tau_{b}= \underset{\tau}{\rm argmin}\left\{ \min_{\bbeta_{\tau}, \sigma} \left\{  \|\*y_{obs}- \*X_\tau\bbeta_{\tau}-\sigma \*u^*_{b}\|^2_2 + \lambda|\tau|\right\}\right\},
\end{align}
for each $\*u^*_{b}, b=1, \dots, d.$ After that, we  collect all  $\hat\tau_{b}$'s to form a candidate set for $\tau_0$:
\begin{align}
\label{eq:S_d}
    S^{(d)} = \left\{\hat\tau_{b}: b=1, \dots, d\right\}.
\end{align}
Since the mapping function from $\*u^*_{b} \in \R^n$ to $\hat\tau_{b} \in {\cal M} = 2^{[p]}$ in (\ref{eq::obj}) is a many-to-one mapping,
many of the $d$ copies of $\hat\tau_{b}$'s obtained by \eqref{eq::obj} are identical.
The size $|S^{(d)}|$ is often much smaller than $d$. See Theorem~\ref{the:size} of Section~\ref{sec:cand_theory_normal} for a theoretical result~ on~the~size~of~the~candidate~set.

The only difference between \eqref{eq::obj} and \eqref{eq:lemma1a} is that we replace $\*u^{rel}$ with $\*u^*_{b}.$ Since the mapping function from $\*u^*_{b}$ to $\hat \tau_b$ in \eqref{eq::obj} is many-to-one, many $\*u^*_{b}$'s that are close to each other map to an identical $\hat\tau_{b}$.
One could imagine that if some $\*u^*_{b}$ is in a neighborhood of $\*u^{rel}$,
then for such $\*u^*_{b}$'s, the event $\{\hat\tau_{b} = \tau_0\}$ is very likely to happen.
The size of such a neighborhood depends on the separation metric $C_{\min}$ and the sample size, yet its probability measure is always positive under the identifiability condition described in Section~\ref{sec:identifiablity}. As a result, as long as $d$, the number of repro copies, is sufficiently large, eventually some $\*u_{b}^*$ will fall in this neighborhood, leading to $\hat\tau_{b} = \tau_0$ and
hence the candidate set $S^{(d)}$ contains the true model $\tau_0$.
Formal theorems that support this method for different error distributions are presented~in~Section~\ref{sec:cand_theory_normal}--\ref{sec:cand_non_normal}.

To put it succinctly, we summarize the aforementioned procedure in Algorithm~\ref{alg:candidate} below.
\begin{algorithm}[H]
\caption{Search of Candidate Models}\label{alg:candidate}
\begin{algorithmic}
\State{\bf Input:} Design matrix $\*X$, response vector $\bm y_{obs}$, the number of repro samples $d$.
\State{\bf Output:} Candidate Models $S^{(d)}$.
\State{\bf Step 1:} Simulate a large number $d$ copies of ${\bf u}^* \sim \*U \sim N(0, \*I_n)$. Denote the $d$ copies by ${\bf u}^*_{b}$, $b = 1, \ldots, d.$
\State{\bf Step 2:} Compute $\hat\tau_{b, \lambda} =     \underset{\tau}{\rm argmin}\, \underset{(\bbeta_{\tau}, \sigma)}{\min} \left[  \lambda |\tau|+  \|\*y_{obs}- \*X_{\tau}\bbeta_{\tau}-\sigma {\bf u}^*_{b}\|^2_2\right]$ for $b = 1, \ldots, d$ and a grid of $\lambda$ values.  For each $b$, use certain selection criteria to pick a subset of all values of $\lambda$, denoted as $\Lambda_b$.
\State{\bf Step 3:} Construct $S^{(d)}= \left\{\hat\tau_{b, \lambda}: \lambda \in \Lambda_b, b=1, \dots,  d \right\}$.
\end{algorithmic}
\end{algorithm}

\begin{remark} \label{remark:implementation}
[Practical implementation of Algorithm~\ref{alg:candidate}] When we implement Algorithm~\ref{alg:candidate}, we need to consider two practical issues: a) how to handle the tuning parameter $\lambda$ in the penalty term, and b) solving an optimization problem with a $L_0$ penalty $\lambda|\tau|$ is often computationally difficult for high-dimensional data.
In our implementation in the numerical study Sections~\ref{sec:simulation} and \ref{sec:real_data}, we follow common practices in the literature to handle these two issues.
First,
it is common to use a selection criterion to determine the value of the tuning parameter $\lambda$ \cite{chen_extended_2008-1, fan_variable_2001, tibshirani1996regression}.
We use the extended BIC (EBIC) \citep{chen_extended_2008-1} to determine $\lambda$, due to its good empirical performance and asymptotic model selection consistency in high-dimensional settings.
Second, solving an optimization problem with the $L_0$ penalty is computationally expensive and yields unstable results \citep{liu2007variable}.
In practice,
researchers often use
a surrogate to replace the $L_0$ penalty.
In our numerical studies, we adopt the adaptive Lasso \citep{zou_adaptive_2006} as a surrogate for the $L_0$ penalty in  \eqref{eq::obj} because of its simplicity and convexity. One may also use other surrogates like the truncated Lasso penalty \citep{shen_likelihood-based_2012-1}, smoothly clipped absolute deviation penalty (SCAD) \citep{fan_variable_2001}, or the minimax concave penalty (MCP) \citep{zhang_nearly_2010}, among others. Although computationally more efficient,
using some of these penalties may require us to make additional assumptions on the design matrix. See Remark~\ref{remark:condition} for further discussions.
In this paper, we develop our general theories using the $L_0$ penalty and constraint rather than a specific surrogate penalty function, since we would like to understand the fundamental properties and allow for the flexibility of using any penalty or constraint within the proposed repro samples framework.
\end{remark}

Equations (\ref{eq:lemma1a}) and (\ref{eq::obj}) are  inversion operations that solve for $\tau$ when given $\*y_{obs}$ and the error term $\*u$. The difference is that (\ref{eq:lemma1a}) assumes the realized
$\*u^{rel}$ is known while (\ref{eq::obj}) uses a simulated $\*u^*_b$.
This technique of using a random $\*u^*_b$ to replace $\*u^{rel}$ in an inversion can be traced back to Fisher's  fiducial inference \cite{hannig_generalized_2016,thornton2024bridging}. Therefore, we refer to the inversion method used in Algorithm~\ref{alg:candidate} as
{\it Fisher inversion}. Here, we use it to assemble potential candidate models for $\tau_0$, which reduces the size of the effective model space from $2^p$ to $|S^{(d)}|$. In Section~\ref{Sec:CS}, we will develop a different inversion technique to construct the level-$\alpha$ confidence set for $\tau_0$.

{
\subsection{Theoretical results for models with  Gaussian errors}
\label{sec:cand_theory_normal}

In this subsection, we present theoretical guarantees of our method under Gaussian error $\*U \sim N(\*0, \*I_{n})$, and extend the results beyond the Gaussian error model in the next subsection.
Here, we show in Theorems \ref{the:finite_pen}-\ref{the:aymp_pen}
that
$ \P_{({\cal U}^d, \*Y)}(\tau_0 \notin S^{(d)} ) \rightarrow 0,$
in the following two cases: 1) the sample size $n$ is finite and  $d \to \infty$, 2) $d$ is finite and  $n \to \infty$, respectively.
The probability $\P_{({\cal U}^d, \*Y)}(\cdot)$ refers to the joint distribution of $\*U$ and ${\cal U}^d = (\*U^*_1,\dots, \*U^*_d),$ where $\*U^*_b$ is a Monte Carlo copy of $N(\*0, \*I_n).$

\begin{theorem}
 \label{the:finite_pen}
Suppose $n-|\tau_0|>4.$ For any $\delta>0$, there exists a constant $\gamma_{\delta}\in(0,1)$  such that when $ \lambda \in \big[4n\gamma_\delta^{1/2}\big\{2 + 2(|\tau_0|+1)\frac{\log(p/2)}{n} \big\}\sigma_0^2,
     \frac{n\gamma_\delta^{1/4}}{6}C_{\min} \big],$
 the finite-sample probability bound that the true model is not included in the model candidates set $S^{(d)}$, obtained by \eqref{eq:S_d} with the objective function \eqref{eq::obj}, is as follows,
\begin{align}
\label{eq:bound_cs_finite}
    \P_{({\cal U}^d, \*Y)}(\tau_0 \notin S^{(d)} ) \leq \left(1- \frac{\gamma_{\delta}^{n-1}}{n-1}\right)^d + \delta.
\end{align}
Therefore as $d \rightarrow \infty,$ $ \P_{({\cal U}^d, \*Y)}(\tau_0 \notin S^{(d)} ) \rightarrow \delta, $ where $\delta>0$ is arbitrarily small.
\end{theorem}

\begin{theorem}
    \label{the:aymp_pen}
Suppose
$\frac{\lambda}{n} \in \big[6 \sigma_0^2\frac{(|\tau_0|+1)\log(p/2)}{n}+t, 0.05C_{\min} \big]$ for a positive constant  $t>0$. Then the finite-sample probability bound that the true model is not included in the confidence set $ S^{(d)}$, obtained by \eqref{eq:S_d} with the objective function \eqref{eq::obj} for any finite $d$ is as follows,
\begin{align}
\label{eq:prob_asympto_pen}
 \P_{({\cal U}^d, \*Y)} (\tau_0 \notin  S^{(d)} ) \nonumber
   \leq &6\exp\left[- \frac{n}{18\sigma_0^2}\{0.3C_{\min}- 36 \frac{\log (p+1)}{n}\sigma_0^2\}\right] + 3\exp\left(-\frac{nt}{3\sigma_0^2}\right) \nonumber\\
   \qquad & + \exp\left\{-nd\left(0.23- \frac{|\tau_0|\log(p)+2}{n}\right) \right\},
\end{align}
Therefore  $\P_{({\cal U}^d, \*Y)} (\tau_0 \notin  S^{(d)} ) \rightarrow 0$ for any $d$ as $n \rightarrow \infty$, if $\frac{|\tau_0|\log(p)}{n} <0.23$ and $C_{\min}> 120 \frac{(|\tau_0|+2)\log (p+1)}{n}\sigma_0^2$ when $n$ is large enough.
\end{theorem}

The two theorems above suggest two complementary driving forces of the coverage validity: the sample size and the computation time measured by $d$. In cases when the sample is limited, Theorem~\ref{the:finite_pen} implies that we can recover the signal with a valid coverage as long as the computation time (linearly scaled with $d$) goes to infinity; in cases when the computational resources are limited, Theorem~\ref{the:aymp_pen} then indicates collecting sufficient samples will result in a valid coverage guarantee. { In Theorem 1, for any finite $n,p$, the lower bound for $\lambda$ is of the same order as $\gamma^{1/2}_{\delta},$ and the upper bound is of the same order as $\gamma_\delta^{1/4}.$ Therefore the range of $\lambda$ always exists for a  $\gamma_\delta$ that is small enough. In Theorem~\ref{the:aymp_pen}, the existence of the range of $\lambda$ follows from $C_{\min}> 120 \frac{(|\tau_0|+2)\log (p+1)}{n}\sigma_0^2$ when $n$ is large enough. Therefore, in both theorems, the required range for $\lambda$ is a non-empty interval of positive length, although this interval is smaller for a smaller $C_{\min}.$ } \label{par:lambda_values}
\begin{remark}
\label{remark:condition}
In this paper, we develop our general theorems using the $L_0$ penalty or constraints, rather than any specific version of surrogates, to keep the theory general and allow researchers to select the surrogate that best suits their needs.
In our implementation in the paper, we used the adaptive Lasso penalty as a surrogate for the $L_0$ penalty since it is computationally efficient and performs comparably to commonly used non-convex penalties.
 To obtain similar result of Theorem~\ref{the:finite_pen} tailored specifically for the adaptive Lasso penalty,
we would need impose an additional condition called the minimum adaptive restrictive eigenvalue condition \citep[][Ch 6\& 7]{buhlmann2011statistics}. This condition is similar, but slightly weaker than the restricted strong convexity and it is also weaker than the irrepresentable condition \citep[][Ch 6\& 7]{buhlmann2011statistics}.
Moreover, the simulation results of Model (M5) in Section~\ref{sec:simulation} suggest that our current implementation of the repro samples approach still performs well empirically even when the minimum adaptive restricted eigenvalue condition required for the adaptive Lasso does not hold. Besides the adaptive Lasso,
there are other possible surrogates (e.g., Lasso, adaptive Lasso, SCAD, MCP, etc.) for the $L_0$ penalty.
Whether we need additional conditions on the design matrix and what these conditions are depend on the specific $L_0$ surrogate we use in our implementation.
In general, there is a trade-off between additional conditions required and computational cost.
For example, if we choose to adopt the truncated Lasso penalty (TLP) proposed by \cite{shen_constrained_2013}, then no additional condition is required on the design matrix.  Alternatively, we can choose to use a constrained least squares approach as opposed to the penalized approach (see Appendix~\ref{sec:constraint} for the formulation and theories regarding the constraint approach). In this case, if we choose to use the constrained $L_0$ regression to estimate the models, which we can achieve with the modern mixed integer optimization approach \citep{bertsimas_best_2016}, we would not need any condition on the design matrix either.
However, both of these approaches demand substantially higher computational cost compared to a convex penalty function like the adaptive~Lasso.
\end{remark}

Besides the coverage results above,
another important aspect
is the size  of the candidate set $|S^{(d)}|.$
Theorem~\ref{the:size} below provides a theoretical bound for
 the expectation of the size of the model candidate set $ \mathbb{E}(|S^{d}|)$.
In the theorem,
for any model $\tau$ with $|\tau| \leq |\tau_0|,$ we define the model distance between $\tau$ and the truth $\tau_0$ as  $  C_{\tau} =  \frac{1}{n \max(|\tau_0 \setminus \tau|,1)} \|\*X_{\tau_0}\bbeta_0 - \*X_\tau \bbeta_\tau \|_2^2. $
\begin{theorem}
\label{the:size}
Let $\Xi(c) = \{\tau: |\tau| \leq |\tau_0|, C_\tau \leq c\}$ be the set of $\tau$ smaller than $\tau_0$ that are close to $\tau_0.$
Then for $\frac{\lambda}{n} \geq \frac{3\sigma_0^2(|\tau_0|+1)\{\log(p-|\tau_0|)  +\log(|\tau_0|)+ \frac{2}{3}\}}{n}+t,$ where $t>0,$
\begin{align*}
    \E(|S^{(d)}|) \leq |\Xi(\bar c)| +  \sum_{\left\{\tau: |\tau| \leq |\tau_0|, C_{\tau} > \bar c\right\}} \exp\left\{  - \frac{n}{20 \sigma^2} (C_\tau - \bar c)\right\}  + 3\exp\left(-\frac{nt}{3\sigma_0^2}\right),
\end{align*}
where $\bar c =  \{7\lambda+ 14\sigma_0^2( 1+ 1.5\log d )\}/n = O(\log(d)/n),$ where $d$ is the number of repro samples used in Algorithm~\ref{alg:candidate}.
\end{theorem}

Intuitively, if an alternative model $\tau$ is closer to $\tau_0$ in that $C_\tau$ is small, it should be more likely to be included in the model candidate set $S^{(d)}.$ Therefore the candidate set would include models that are close to $\tau_0$,
and models that are farther away from $\tau_0$
would be included with a smaller probability. As a result, the size of the candidate set depends on (a) how many models are close to $\tau_0$, (b) the probability of other models included. This intuition is verified explicitly by the result in Theorem~\ref{the:size}. Specifically, if $\log(p)/n = o(1)$, then with high probability, the candidate set will  include all the models with $C_\tau = o(\log(d)/n),$ where $d$ is the number of repro copies in Algorithm~\ref{alg:candidate},  and it will include those with $C_\tau = O(\log(d)/n)$ with a positive probability. The larger the $C_\tau$ is, the smaller the probability $\tau$ being included. Moreover, a larger sample size leads to smaller model selection uncertainty, typically resulting in a smaller candidate set, aligning with our expectations. Additionally, the impact of the repro sample size $d$ on the size of $S^{(d)}$ is logarithmic. 
Finally, we see from Theorem~\ref{the:size} that the contribution to the cardinality of the model candidate set from all models larger than $\tau_0$ is bounded by $3\exp\left(-\frac{nt}{3\sigma_0^2}\right)$, therefore the candidate models include only models of size similar or smaller than $\tau_0$ with probability close to 1  due to the regularization in Step~2~of~Algorithm~\ref{alg:candidate}.
}

\subsection{Heterogeneous, non-Gaussian and sub-Gaussian error models}
\label{sec:cand_non_normal}
{
In this subsection, we show that even when the model error assumption $\*U \sim N(0, \*I_{n})$ is violated, the model candidate set $S^{(d)}$ obtained using Algorithm~\ref{alg:candidate} with $\*u^*_b \sim N(0, \*I_{n})$ can still cover the true model $\tau_0$ with a high probability.

We first show in Theorem~\ref{the:uw} below that  the results in Theorem~\ref{the:finite_pen} still hold when
 the linear model in \eqref{eq:model-random} is now generalized to the following,
\begin{equation}
\label{eq:model-random_generalized}
{\bf Y} =
\*X \bbeta_0^{full}  + \sigma_0 \*U_{\Omega} = \*X_{\tau_0} \bbeta_0 + \sigma_0 \*U_{\Omega}.
\end{equation}
Here,  the error term
$\*U_{\*\Omega} = diag(\*\Omega) \*U$,
$\*U\sim N(0, I_n)$ and
$\*\Omega = (\Omega_1, \dots, \Omega_n)$ is an $n \times 1$ fixed vector with each $\Omega_i = O(1), 1 \leq i \leq n,$ or a  random vector independent of $\*U$, with each $\Omega_i = O_p(1), 1\leq i \leq n.$

\begin{theorem}
\label{the:uw}
    Suppose $\*Y$ is generated by \eqref{eq:model-random_generalized}, $n-|\tau_0|>4,$ $|\{\Omega_i: \Omega_i > 0\}| > |\tau_0|$.
    Then for any $\delta>0$, there exists a constant $\gamma_{\delta}\in(0,1)$ such that when $ \lambda \in \big[n\gamma_{\delta}^{1/2}\big\{2 + 2(|\tau_0|+1)\frac{\log(p/2)}{n} \big\}, \allowbreak
     n\gamma_{\delta}^{1/4}\frac{C_{\min}}{6}\big],$
 the finite-sample probability bound that the true model is not included in the model candidates set $S^{(d)}$, obtained by \eqref{eq:S_d} with the objective function \eqref{eq::obj}, is as follows,
\begin{align}
    \P_{({\cal U}^d, \*Y)}(\tau_0 \notin S^{(d)} ) \leq \left(1- \frac{\gamma_{\delta}^{n-1}}{n-1}\right)^d + \delta. \nonumber
\end{align}
Therefore as $d \rightarrow \infty,$ $ \P_{({\cal U}^d, \*Y)}(\tau_0 \notin S^{(d)} ) \rightarrow \delta, $ where $\delta>0$ is arbitrarily small.
\end{theorem}

The above theorem shows that when the error term is generalized from $\*U$ to $\*U_{\Omega},$ the candidate set $S^{(d)}$  still cover the truth $\tau_0$ with an arbitrarily high probability, as long as $d$ is large enough. The generalized error term $\*U_{\Omega}$ covers a wide range of  non-Gaussian error models, including heterogeneous variances, Cauchy distribution, $t$-distribution, contaminated Gaussian distribution, etc.
Below, we discuss how these non-Gaussian errors are connected~with~$\*U_{\Omega}.$

\begin{enumerate}
    \item[] {\bf Heterogeneous Variance.} In this case, $\*U_\Omega \sim N(0, \Sigma),$ where $\Sigma = diag(\sigma^2_1, \dots, \sigma_2^2).$ Therefore we can just make $\Omega$ fixed such that $P(\Omega_i = \sigma_i) =1$.
  \item[] {\bf Cauchy Distribution.} When the error is Cauchy distribution, we can simply make $\Omega_i = 1/|Z_i|,$ where $Z_i$ are i.i.d. $N(0,1).$
  \item[]{\bf T-distribution with degree of freedoms $\nu$.} Since random variables with $t$-distribution can be formulated by a ratio of a normal random variable and the square root of a Chi-square random variable divided by its degree of freedom, we make $\Omega_i \sim \frac{1}{\sqrt{\chi^2_{\nu}/\nu}}.$
  \item[]{\bf Contaminated Gaussian Distribution.} In cases where some observations are contaminated, leading to a larger variance for the contaminated samples, we can make $\Omega_i = (1-Z_i) + wZ_i,$
   where $w$ is a constant, usually larger than 1, and $Z_i, i = 1, \dots, n$ are i.i.d $Bernoulli(p).$ Here $p$ represents the proportions of contaminated distributions.
\item[]{\bf Combination of the above.} This is when $\*\Omega_i$ follows a mixture of the above distributions.
\enlargethispage{\baselineskip}

\end{enumerate}

Sub-Gaussian error models are another set of non-Gaussian error models used in high-dimensional settings due to their flexibility and robustness.
We show that the results of Theorem~\ref{the:finite_pen} and Theorem~\ref{the:aymp_pen} still hold for the sub-Gaussian error models, and thus candidate set $S^{(d)}$ obtained in Algorithm~\ref{alg:candidate} can still cover the true $\tau_0$ with a high probability.

\begin{theorem}
\label{the:sub_finite_pen}
Suppose  $\*Y$ is generated by \eqref{eq:model-random} with $\*U$ being a sub-Gaussian vector with sub-Gaussian norms bounded by a universal constant, and $n-|\tau_0|>4.$ For any $\delta>0$, there exists a constant $\gamma_{\delta}\in(0,1)$ and $\zeta_\delta > 0 $ such that when $ \lambda \in \big[n\gamma_{\delta}^{1/2}\big\{2 + 2(|\tau_0|+1)\frac{\log(p/2)}{n} \big\}, \allowbreak
     n\gamma_{\delta}^{1/4}\frac{C_{\min}}{6}\big],$
 the finite-sample probability bound that the true model is not included in the model candidates set $S^{(d)}$, obtained by \eqref{eq:S_d} with the objective function \eqref{eq::obj}, is as follows,
\begin{align}
\label{eq:bound_cs_finite_sub}
    \P_{({\cal U}^d, \*Y)}(\tau_0 \notin S^{(d)} ) \leq e^{-\zeta_\delta d} + \delta.
\end{align}
Therefore as $d \rightarrow \infty,$ $ \P_{({\cal U}^d, \*Y)}(\tau_0 \notin S^{(d)} ) \rightarrow \delta, $ where $\delta>0$ is arbitrarily small.
\end{theorem}

For the model candidate set $S^{(d)}$ in both theorems,
the repro errors used in Algorithm~\ref{alg:candidate} are still sampled from $\*u^*_b \sim N(0, \*I_n)$, even though the true underlying  $\*u^{rel}$ is not. This is possible because $\*u^{rel}$ is a vector in $\R^n$, as long as $\*u^{rel}$ is not too extreme, we often can find some $\*u^*_b \sim N(0, \*I_n)$ in its neighborhood when $d \to \infty$. Under the non-Gaussian settings considered in Theorems~\ref{the:uw} and \ref{the:sub_finite_pen}, we are able to quantify such a neighborhood of $\*u^{rel}$ that also maps $\*u^*_b$ to $\tau_0$. A nice implication of these results is that we do not need to know exactly the error distribution of the model, as long as it is one of those in Theorems~\ref{the:uw} and \ref{the:sub_finite_pen}, the model candidate set $S^{(d)}$ obtained by Algorithm~\ref{alg:candidate} contains the true $\tau_0$ with a high~probability.
{Furthermore, we later
extend the finite-sample result here to arbitrary error distributions with finite second moments; see the Discussion section.}

\begin{theorem}
    \label{the:sub_aymp_pen}
Suppose  $\*Y$ is generated by \eqref{eq:model-random} with $\*U$  being a sub-Gaussian vector with sub-Gaussian norms bounded by a universal constant and
$\frac{\lambda}{n} \in \big[6 \sigma_0^2\frac{(|\tau_0|+1)\log(p/2)}{n}+t, 0.05C_{\min} \big]$ for a positive constant  $t>0$. Then the probability bound that the true model is not included in the confidence set $ S^{(d)}$, obtained by \eqref{eq:S_d} with the objective function \eqref{eq::obj} for any finite $d$ is as follows,
\begin{align}
 \P_{({\cal U}^d, \*Y)} (\tau_0 \notin  S^{(d)} ) \nonumber
   \leq & \exp\left[- \frac{n}{18\sigma_0^2}\{0.3C_{\min}- 36 \frac{\log (p+1)}{n}\sigma_0^2\}\right] + 3\exp\left(-\frac{nt}{3\sigma_0^2}\right) \nonumber\\
   \qquad & + \exp\left\{-nd\left(0.23- \frac{|\tau_0|\log(p)+2}{n}\right) \right\}, \nonumber
\end{align}
Therefore  $\P_{({\cal U}^d, \*Y)} (\tau_0 \notin  S^{(d)} ) \rightarrow 0$ for any $d$ as $n \rightarrow \infty$, if $\frac{|\tau_0|\log(p)}{n} <0.23$ and $C_{\min}> 120 \frac{(|\tau_0|+2)\log (p+1)}{n}\sigma_0^2$ when $n$ is large enough.
\end{theorem}

The above theorem extends the result in Theorem~\ref{the:aymp_pen} to models with sub-Gaussian errors. It indicates that when sample size is large, we can recover the truth $\tau_0$ with Algorithm~\ref{alg:candidate} with a limited number of repro samples for models with sub-Gaussian errors. Similar to the implications of Theorem~\ref{the:uw} and \ref{the:sub_finite_pen}, here we do not need to know the distribution of $\*U$, and only require $\*U$ to be sub-Gaussian.

}

\section{Construction of a level-$\alpha$ Model Confidence Set}\label{Sec:CS}
In this section, we construct a level-$\alpha$ confidence set for model $\tau_0$ by developing a conditional repro samples method tailored to the problem. Here, we assume $\*U \sim N(0, \*I_n)$.

For the ease of presenting the general idea of the repro samples method as described in \citep{xie_repro_2022}, let us first assume that we are interested in making a joint inference about $\btheta_0= (\tau_0,\bbeta_0^\top, \sigma_0)^\top$ and describe how the method proceeds in this case. The idea is that given any possible value of the parameters  $\btheta = (\tau,\bbeta_\tau^\top, \sigma)^\top$, we create an artificial repro sample data $\*y^* = \*X_\tau\bbeta_\tau + \sigma\*u^*,$ where $\*u^* \sim N(0, \*I_n).$
If $\*u^*$ is close to $\*u^{rel}$ and $\btheta$ is equal or close to $\btheta_0$, then we expect $\*y^*$ and $\*y_{obs}$ to be equal or close. Inversely, for a given value  $\btheta$, if we can find a $\*u^*$ likely matching $\*u^{rel}$ such that $\*y^*$ matches  $\*y_{obs}$ (i.e., $\*y_{obs} \approx \*y^*$), then
we cannot rule out that this $\btheta$ is a potential value of $\btheta_0$.
Mathematically,
we define
\begin{equation}\label{eq:GammaTheta}
    \Gamma^{\theta}_{\alpha}(\*y_{obs}) = \left\{\btheta: \exists \*u^*\ \ \mbox{s.t. } \*y_{obs} = G(\btheta, \*u^*),  T(\*u^*, \btheta) \in B_{\alpha}(\btheta)
    \right\}.
\end{equation}
Here, the function $T(\cdot, \cdot)$ is referred to as a {\it nuclear mapping function} and the set $B_{\alpha}(\btheta)$ is a fixed level-$\alpha$ Borel set in $\R^d$ such that
\begin{equation}
\label{eq:Borel_p}
    {\P}_{\*U}\left(T(\*U, \btheta) \in B_{\alpha}(\btheta)
    \right) \ge \alpha.
\end{equation}
Again, we clarify that the $\btheta$ in the superscript of $\Gamma$ simply indicates the target parameter is $\btheta$ and  $\Gamma^{\theta}_{\alpha}(\*y_{obs})$ in (\ref{eq:GammaTheta}) does not rely on any particular value of $\btheta.$

The repro samples method uses ${\P}_{\*U}\left(T(\*U, \btheta) \in B_{\alpha}(\btheta) \right)$, for each given value $\btheta$, as a way to quantify the uncertainty of $\*U$ thus also the uncertainty of $\*Y$.
Moreover, for any nuclear mapping function $T(\*U, \btheta)$, as long as we have a set $B_\alpha(\btheta)$ such that (\ref{eq:Borel_p}) holds,
we can show that the set $\Gamma_{\alpha}(\*y_{obs})$ in (\ref{eq:GammaTheta}) is a level-$\alpha$ confidence set \citep{xie_repro_2022}.
Here, the role of $T(\*U, \btheta)$ under the repro samples framework is similar to that of a test statistic under the classical (Neyman-Pearson) hypothesis testing framework.
Besides, a good choice for $T(\*U, \btheta)$ is problem-specific. Effectively, the operation in (\ref{eq:GammaTheta}) can be considered as an inversion operation that maps a set of $\*u^* \in \R^n$ to a set of $\btheta \in \Theta$. Such a mapping is a key element of the Dempster-Shafer calculus \cite{shafer1976mathematical, Martin2015}. To distinguish the Fisher inversion method introduced in Section~\ref{sec:finding_cand} that produces a model candidate set,
we refer to the techniques used in this section to produce a level-$\alpha$ confidence set for $\tau_0$ as  {\it Fisher-Dempster inversion}.

Our goal in this section
to make inference only for the true model $\tau_0$ with $(\bbeta_\tau^\top, \sigma)$ being the unknown nuisance parameters.
First,
we write  $\*y_{\btheta}=\*X_{\tau} \*\bbeta_{\tau} + \sigma \*u$, for a $\*u \sim \*U$. This $\*y_{\btheta}$ is a copy of artificial
data generated from a given set of parameters $\btheta= (\tau, \bbeta_\tau^\top, \sigma)^\top$. The corresponding random version is
\begin{align}
\label{eq:fake_data}
\*Y_{\bm\theta} = \*X_{\tau} \*\bbeta_{\tau} + \sigma \*U.
\end{align}
Then based on the artificial repro sample data $(\*X, \*y_{\bm \theta}),$ one can obtain an estimate of $\tau$, denoted by $\hat \tau(\*y_{\bm\theta}).$  In this paper, we use
\begin{align}
\label{eq:tau_hat}
\hat \tau(\*y_{\bm\theta}) = \argmin_{\tilde \tau \in \mathcal M, \bbeta_{\tilde\tau} \in \mathbb R^{|\tilde \tau|}}
\| \*y_{\bm\theta} - \*X_{\tilde\tau}\bbeta_{\tilde\tau}\|^2  \mbox{ s.t. }  |\tilde\tau| \leq |\tau|,
\end{align}
although in principle we can choose to use another estimator of reasonable performance.
We thereafter use \eqref{eq:tau_hat} to define the nuclear mapping function  as $T(\*u, \bm\theta) = \hat\tau(\*y_{\btheta}) = \tilde T(\*y_{\btheta}, \tau).$
Then
we need to find a Borel set $B_{\alpha}(\btheta)$
that satisfies \eqref{eq:Borel_p}, i.e.,
    ${\P}_{\*U}\left(T(\*U, \btheta) \in B_{\alpha}(\btheta)
    \right)  = \P_{\*U}(\tilde T(\*Y_{\btheta}, \tau) \in B_{\alpha}(\btheta)) \geq \alpha,$
for the nuclear mapping $\tilde T(\*y_{\btheta}, \tau)$ defined above.
     However, the distribution of $\tilde T(\*y_{\btheta}, \tau)$ involves all of the parameters $\btheta = (\tau, \bbeta_\tau^\top, \sigma)^\top$, including the nuisance parameters $\bbeta_\tau$ and $\sigma^2$.
Therefore,  directly
obtaining $B_{\alpha}(\btheta)$ for all values of $\btheta$ is  computationally challenging, if not infeasible.

Since it is not feasible to search through the entire space of $(\bbeta_\tau^\top, \sigma)$,  we introduce below an effective conditional repro samples method to handle the nuisance parameters and construct a level-$\alpha$ confidence set for $\tau_0$. The idea is to first find a quantity $W(\*U, \btheta)$, such that the conditional distribution of the nuclear statistic $T(\*U, \btheta)$ given $W(\*U, \btheta) = \*w$ is free of the nuisance parameters $(\bbeta, \sigma)$.
For now, assume we have such $W(\*U, \btheta)$; we will discuss how to obtain $W(\*U, \btheta)$ for our purpose later in the section.
Then, based on the conditional distribution of $T(\*U, \btheta)|W(\*U, \btheta)$,
we construct a Borel set $B_\alpha(\tau, \*w)$ that depends on $\*w$, the value of the random quantity $\*W(\*U, \btheta)$,
but not on $(\bbeta, \sigma)$, such that
\begin{align}
\label{eq:cond_P_nuclear}
   & \P_{\*U|W}\{T(\*U, \btheta) \in B_{\alpha}(\tau, \*w)|W(\*U, \btheta) = \*w\} \nonumber \\   & \qquad = \P_{\*U|W}\{\tilde T(\*Y_{\btheta}, \tau) \in B_{\alpha}(\tau, \*w)|W(\*U, \btheta) = \*w\}\geq \alpha.
\end{align}
Accordingly, the marginal probability $\P_{\*U}(T(\*U, \btheta) \in B_{\alpha}(\tau, W(\*U, \btheta)) \geq \alpha.$

Now, instead of directly following \eqref{eq:GammaTheta}, we construct a subset in the model space ${\cal M}$:
\begin{align}
\label{eq:cond_cs}
    \Gamma^{\tau}_{\alpha}(\*y_{obs}) = \{\tau \in \mathcal M: \exists \*u^*\ \mbox{and }\  (\bbeta_\tau, \sigma)\  \mbox{s.t. }\;  {\*y}_{obs} = \*X_\tau\bbeta_\tau + \sigma\*u^*, \nonumber\\
    T(\*u^*, \btheta) \in B_{\alpha}\left(\tau, W(\*u^*, \btheta)\right), \btheta = (\tau, \bbeta_\tau, \sigma)   \}.
\end{align}
The following theorem suggests that $\Gamma^{\tau}_{\alpha}(\*y_{obs})$ constructed above is a level-$\alpha$ confidence set for the true model $\tau_0$.
\begin{theorem}
\label{thm:cond_rps}
    Suppose the conditional distribution of $T(\*U, \btheta)$ given $W(\*U, \btheta) = \*w$ is free of $(\bbeta_\tau, \sigma)$ and the Borel set $B_{\alpha}(\tau, \*w)$ satisfies \eqref{eq:cond_P_nuclear}, then  $\P(\bm\tau_0 \in \Gamma^{\tau}_{\alpha}(\*Y)) \geq \alpha,$
    where the confidence set $\Gamma^{\tau}_{\alpha}(\*Y)$ is defined by \eqref{eq:cond_cs}.
\end{theorem}

The remaining task is
to find the random quantity $\*W(\*U, \btheta)$
and a Borel set $B_\alpha(\tau, \*w)$~such that the conditional distribution of $T(\*U, \btheta)|W(\*U, \btheta) = \*w$ is free of the nuisance parameters $(\bbeta, \sigma)$
and the inequality
\eqref{eq:cond_P_nuclear} holds.
Note that we can rewrite (\ref{eq:fake_data}) as $\*Y_{\bm\theta} = \*H_\tau \*Y_{\bm\theta} +  (\*I-\*H_\tau) \*Y_{\bm\theta}$ $ =  \*H_\tau \*Y_{\bm\theta} +  \sigma (\*I-\*H_\tau) \*U$, where $\*H_\tau = \*X_{\tau} (\*X_{\tau}^\top \*X_{\tau})^{-1} \*X_{\tau}^\top$ is the projection matrix of $\*X_\tau$.
Write $ \*A_{\bm\theta}(\*U) = \*H_\tau \*Y_{\bm\theta} = \widetilde {\*A}_{\bm\theta}({\*Y}_{\bm\theta})$
and $b_{\bm\theta}(\*U) = \|(\*I-\*H_\tau) \*Y_{\bm\theta}\| = \widetilde b_{\bm\theta}(\*Y_{\bm\theta})$. We have
\begin{align*}
\*Y_{\bm\theta}
& = \*A_{\bm\theta}(\*U) +   b_{\bm\theta}(\*U)\frac{(\*I-\*H_\tau) \*U}{\|(\*I-\*H_\tau) \*U \|} = \widetilde{\*A}_{\bm\theta}(\*Y_{\bm \theta}) +   \widetilde{b}_{\bm\theta}(\*Y_{\bm\theta})\frac{(\*I-\*H_\tau) \*U}{\|(\*I-\*H_\tau) \*U \|}. \numberthis \label{eq:split}
\end{align*}
In this equation, the ``randomness'' of $\*U$ (and also $\*Y_{\bm\theta}$) are decomposed into three components, $\*A_{\bm\theta}(\*U),  b_{\bm\theta}(\*U)$ and
$(\*I-\*H_\tau) \*U/ \|(\*I-\*H_\tau) \*U \|$.
Under (\ref{eq:fake_data}), $(\*A_{\bm\theta}(\*U), \, b_{\bm\theta}(\*U))$ $= \big(\widetilde {\*A}_{\bm\theta}({\*Y}_{\bm\theta}),  \, \widetilde{b}_{\bm\theta}(\*Y_{\bm\theta})\big)$ is a sufficient statistic and the last piece $(\*I-\*H_\tau) \*U/\|(\*I-\*H_\tau) \*U \|$ is an ancillary statistic that is free of the nuisance parameters $(\bbeta_\tau, \sigma^2)$.
Based on this partition, we define $\*W(\*U, \btheta)= \left( \*A_{\bm\theta}(\*U),  b_{\bm\theta}(\*U)\right) = \big(\widetilde {\*A}_{\bm\theta}({\*Y}_{\bm\theta}),  \, \widetilde{b}_{\bm\theta}(\*Y_{\bm\theta})\big) = \widetilde{\*W}(\*Y_{\bm \theta}, \btheta)$. It then follows immediately that the conditional distribution of $\*Y_{\btheta} \mid \*W(\*U, \btheta) = \*w$ is free of $(\bbeta_\tau, \sigma^2),$ so is the conditional probability mass function of  $ \tilde T(\*Y_{\bm\theta}, \tau)$ given $\*W(\*U, \btheta) = \*w$,
\begin{equation}\label{eq:ptilde} \quad
  p_{(\*w, \tau)}(\tau') = {\P}_{\*U|\*W}\big\{ \tilde T(\*Y_{\bm\theta}, \tau) = \tau' \big| \*W(\*U, \btheta) = \*w\big\}, \quad
\end{equation}
for any $\tau' \in {\cal M}$.
Note that, when given $(\*w, \tau)$, we can use the model equation (\ref{eq:fake_data}) to generate many copies of $\*Y_{\btheta}$ by repeated draws from $\*U$. Therefore we can obtain the conditional probability mass function in (\ref{eq:ptilde}) through a Monte-Carlo method.

We therefore define the Borel set $B_\alpha(\tau, \*w)$ as
\begin{align}
\label{eq:borel_model_cs}
\small
    B_\alpha(\tau, \*w) = \left\{\tau^*\in\mathcal M: \sum_{\{\tau': \, p_{(\*w, \tau)}(\tau') \leq p_{(\*w, \tau)}(\tau^*) \}} p_{(\*w, \tau)}(\tau') \geq 1- \alpha \right\}.
\end{align}
In the proof of Theorem~\ref{thm:cond_rps} in Appendix~\ref{sec:app_proof}, we prove that the conditional probability
\begin{align}
    \label{eq:cond_prob_borel}
{\P}_{\*Y_{\btheta}| \*W}\big\{\tilde T(\*Y_{\theta}, \tau) \in B_{\alpha}(\tau, \*w) \big| \widetilde{\*W}(\*Y_{\btheta}, \tau) = \*w \big\} \geq \alpha.
\end{align}
It follows that, marginally,
${\P}_{\*Y_{\btheta}}\big\{\tilde T(\*Y_{\theta}, \tau) \in B_\alpha\big(\tau, \widetilde{\*W}(\*Y_{\btheta}, \tau)\big)  \big\} \geq \alpha.
$
Then by \eqref{eq:cond_cs} and using the candidate set $S^{(d)}$, we propose the following confidence set for $\tau_0,$
\begin{equation}\label{eq:bar-CS}
{\bar\Gamma}^\tau_\alpha(\*y_{obs}) =  \Gamma^\tau_\alpha(\*y_{obs})  \bigcap S^{(d)} =
\left\{\tau \in S^{(d)}: \tilde T(\*y_{obs}, \tau) \in B_\alpha\left(\tau,\widetilde{\*W}(\*y_{obs}, \tau)\right) \right\}.
\end{equation}
To obtain the above confidence set,  we use a Monte-Carlo method to compute the conditional probability in \eqref{eq:ptilde}. We summarize the procedure of constructing the above model confidence set in Algorithm~\ref{alg:tau},
with the size of the Monte-Carlo simulations $J.$

\begin{algorithm}
\caption{Confidence set construction for $\tau_0$}\label{alg:tau}
\begin{algorithmic}
\State{\bf Input:} Design matrix $\*X$, response vector $\bm y_{obs}$, candidate set $S^{(d)}$, simulation size $J$
\State{\bf Output:} Confidence set of $\tau_0$
\State{{\bf Step 1:} For each candidate model $\tau_b \in S^{(d)}$, calculate $\hat\tau_{jb}^*, j=1, \dots, J.$ }
 \For{$\tau_b\in S^{(d)}$ and $j \in {1, \dots, J}$}
 \State{\bf Step 1(a):} Calculate $\*w_{obs} = (\*a_{obs}, b_{obs})= (\*H_{\tau_b}\*y_{obs}, \|(\*I-\*H_{\tau_b}) \*y_{obs}\|).$
\State{\bf Step 1(b):}
Generate $\*u_j^* \sim N(0,\*I_n)$, and compute
    \begin{align}
    \*y_{jb}^* = \*a_{obs} + b_{obs}\frac{(\*I - \*H_{\tau_b}) \*u_j^*}{\|(\*I-\*H_{\tau_b}) \*u_j^*\|}. \nonumber
    \end{align}
    In addition, obtain the estimated model $\hat\tau^*_{jb} = \hat\tau(\*y_{jb}^*)$ by
    \begin{align}
    \label{eq:tau_hat_y_star}
\hat \tau(\*y^*_{jb}) = \argmin_{\tilde \tau \in {\cal M}, \bbeta_{\tilde\tau} \in R^{|\tilde \tau|}}
\| \*y_{jb}^* - X_{\tilde\tau}\bbeta_{\tilde\tau}\|^2  \mbox{ s.t. }  |\tilde\tau| \leq |\tau_b|.
\end{align}
\EndFor
\State{\bf Step 2:} Estimate $p_{(\*w_{obs}, \tau_b)}(\tilde \tau)$ for all $\tilde\tau \in \mathcal{M}$ by
    $\hat p_{(\*w_{obs}, \tau_b)}(\tilde \tau)=\frac{1}{J}\sum_{j=1}^{J} {\bf 1}_{\{\hat\tau_{jb}^*=\tilde\tau\}}, $
where ${\bf 1}_{\{\cdot\}}$ is the indicator function.
\State{\bf Step 3:} Calculate
$\tilde T(\*y_{obs}, \tau_b) = \hat\tau^{obs}_b $ by $\hat\tau^{obs}_b =  \argmin_{\tilde \tau \in {\cal M}, \bbeta_{\tilde\tau} \in R^{|\tilde \tau|}}
\| \*y_{obs} - X_{\tilde\tau}\bbeta_{\tilde\tau}\|^2  \mbox{ s.t. }  |\tilde\tau| \leq |\tau_b|.$
\State{\bf Step 4:} We then compute the estimated tail probability of $\tilde T(\*y_{obs}, \tau_b)$  as
\begin{align}
   \hat F_{(\*w_{obs}, \tau_b)}\left\{ \tilde T(\*y_{obs}, \tau_b)\right\} = \sum_{\{\tilde\tau:p_{(\*w_{obs}, \tau_b)}(\tilde \tau) \leq p_{(\*w_{obs}, \tau_b)}(\hat\tau^{obs}_b) \}}\hat p_{(\*w_{obs}, \tau_b)}(\tilde \tau). \nonumber
\end{align}
We therefore obtain the level-$\alpha$ confidence set for $\tau_0$
\begin{align*}
      \bar\Gamma^{\tau}_{\alpha} (\*y_{obs})=
      \left\{\tau_b \in S^{(d)}:   \hat F_{(\*w_{obs}, \tau_b)}\left\{ \tilde T(\*y_{obs}, \tau_b)\right\} \geq 1- \alpha \right\}.
\end{align*}
\end{algorithmic}
\end{algorithm}

Theorem \ref{the:confidence_set} below states that ${\bar\Gamma}^\tau_\alpha(\*y_{obs})$ in (\ref{eq:bar-CS})
is a level-$\alpha$ confidence set for $\tau_0$
with a guaranteed finite-sample coverage rate, as long as $d$, the number of repro samples used to construct the candidate set $S^{(d)}$ in Algorithm~\ref{alg:candidate}, is large.
Theorem \ref{the:confidence_set_asymptotic} states that even when  $d$ is limited,  ${\bar\Gamma}^\tau_\alpha(\*y_{obs})$
is still a level-$\alpha$ confidence set for $\tau_0$ if we have a large sample size $n.$

\begin{theorem}
    \label{the:confidence_set}
    Under the conditions in Theorem~\ref{the:finite_pen}, for any finite $n$ and $p$, and arbitrarily small $\delta>0$, the coverage probability of model confidence set  ${\bar\Gamma}^\tau_\alpha(\*y_{obs})$ constructed above
    is
    $\P_{({\cal U}^d, \*Y)}\left\{\tau_0 \in  {\bar\Gamma}^{\tau}_{\alpha}(\*Y) \right\}  \geq \alpha - \delta - o(e^{-c_1d})$ as $d \rightarrow \infty$ for some $c_1>0$. Further $\P_{\*Y|{\cal U}^d}\left\{\tau_0 \in  {\bar\Gamma}^{\tau}_{\alpha}(\*Y) \right\}  \geq \alpha - \delta - o_p(e^{-c_1d}).$
\end{theorem}

\begin{theorem}
    \label{the:confidence_set_asymptotic}
    Under the conditions in Theorem~\ref{the:aymp_pen}, for any finite $d$, the coverage probability of model confidence set  ${\bar\Gamma}^\tau_\alpha(\*y_{obs})$ constructed above
    is
    $\P_{({\cal U}^d, \*Y)}\left\{\tau_0 \in  {\bar\Gamma}^{\tau}_{\alpha}(\*Y) \right\}  \geq \alpha  - o(e^{-c_2 n})$ as $n \rightarrow \infty$ for some $c_2>0$. Further $\P_{\*Y|{\cal U}^d}\left\{\tau_0 \in  {\bar\Gamma}^{\tau}_{\alpha}(\*Y) \right\}  \geq \alpha -  o_p(e^{-c_2 n}).$
\end{theorem}

\begin{remark}
\label{remark:cs_nonnormal}
When the error $\*U$ is non-Gaussian, the theoretical results in Section~\ref{sec:cand_non_normal} show that the model candidate set $S^{(d)}$ itself, constructed in Algorithm~\ref{alg:candidate}, is a valid but conservative confidence set for $\tau_0$, even without knowing the distribution of $\*U.$
It is challenging to derive a tighter model confidence set with a closer-to-level-$\alpha$ coverage in cases when the distribution of $\*U$ is unknown, since we often need the knowledge of $\*U$'s distribution  to obtain $B_{\alpha}(\*\tau, \*w)$ in (\ref{eq:cond_P_nuclear}) to quantify the uncertainty of $\*u^{rel}$. This is because both $\tau$ and $\hat\tau(\*Y)$ are discrete, and typical concentration inequalities or central limit theorem do not apply.
When the distribution of $\*U$ is known but non-Gaussian,
the approach in Algorithm~\ref{alg:tau} is not valid anymore because the independence among the three components in the decomposition \eqref{eq:split} does not hold. In such cases,
a profiling approach as described in \cite{xie_repro_2022, hou2024repro} can be potentially applied instead, although it may pose computational challenges and yield less efficient results.
Alternatively, since we now have a  model candidate set with a high coverage probability, we may follow \cite{lei_cross-validation_2020} to construct a distribution-free confidence set for the best working models, that is, the model that provides the best predictive performance. The idea is to construct a confidence interval for the smallest cross-validation error, and use that to construct the confidence set for the best predictive model. Since the candidate models  are low-dimensional and covers the truth $\tau_0$ with a high probability, we believe that  $\tau_0$ is very likely to be the best predictive model among all the candidates. The theoretical justification
and the coverage of this confidence set constructed using the ``best predictive model''
is an intriguing future research topic.
\end{remark}

\section{Inference for regression coefficients accounting for model selection uncertainty}
\label{sec:coef}

 Section~\ref{sec:subset} proposes a confidence set for any subset of $\bbeta_0^{full} = (\beta_{0,1}, \dots, \beta_{0,p})^\top$ that accounts for model selection uncertainty, and extend the work to make inference for any linear transformation of $\bbeta^{full}_0$.
Section~\ref{sec:special} discusses two special cases of practical importance: (a) inference for a single regression coefficient $\beta_{0,i}, i = 1, \dots, p.$ and (b) joint inference for all regression coefficients $\bbeta_0^{full}$.  Note that, most existing methods focus only on one of the two special cases, and there are few effective approaches on making inference for any subset or linear transformation of $\bbeta^{full}_0$ in the literature. Moreover, our work guarantees both finite-sample and large-sample coverage, while existing methods provide only asymptotic inferences.  In Section~\ref{sec:subset}--~\ref{sec:special}, we assume that the error terms are Gaussian $\*U \sim N(0, \*I_n)$. In Section~\ref{sec:beta_nonGaussian}, we extend the confidence sets for the regression coefficients in Section~\ref{sec:subset} to cases where non-Gaussian errors are present. Due to space limits, we defer the joint inference for the model and regression coefficients to Appendix~\ref{sec:joint_inference_model_coefficients}.

\subsection{Inference for a subset of regression coefficients}\label{sec:subset}
Let  $\bbeta_{0, \Lambda}$ be a collection of $\beta_{0,i}$'s that are of interests, where the index set $\Lambda \subset [p]$. The remaining parameters $\beta_{0, i}, i \not\in \Lambda$, $\sigma_0$ and $\tau_0$ are nuisance parameters.
The subset $\Lambda$ is given based on the problem of interest,  and it may overlap with or separate from $\tau_0$.
Here our strategy is to first remove the influence of the nuisance parameters $\beta_{0, i}, i \not\in \Lambda$ and $\sigma_0$ by defining a nuclear mapping function that only involves $\bm \eta_{\Lambda} = (\bbeta_{\Lambda}, \tau)$, where $\tau$ is a potential value of $\tau_0$. The role of the nuclear mapping is similar to test statistics in the classical hypothesis testing framework, but in general, the definition of the nuclear mapping is broader and more flexible than the definition of test statistics. See \cite{xie_repro_2022} for a detailed discussion.
We then utilize the model candidate set $S^{(d)}$ constructed in Section~\ref{sec:finding_cand} to
handle the impact of $\tau$, leading to a valid confidence set for $\bbeta_{0,\Lambda}$.

For a given $\bm \eta_{\Lambda} =(\bbeta_{\Lambda}, \tau)$, we define the nuclear mapping as follows
\begin{align}
\label{eq:nulcear_beta_lambda_0}
  T(\*u, \bm\eta_\Lambda)    = \begin{cases}
    \frac{\*u^\top \*O_{\tau, \Lambda} \*u}{\*u^\top (\*I-\*H_\tau) \*u/(n-|\tau|)} & \text{if $\Lambda \cap \tau \neq \emptyset$ \& $\beta_i=0$ for  any $i \in \Lambda \setminus \tau$} \\
    \infty &  \text{if $\beta_i \neq 0$ for any $i \in \Lambda \setminus \tau$}\\
    0 & \text{if $\Lambda \cap \tau = \emptyset$ \& $\beta_i = 0$ for any $i \in \Lambda$
    }
  \end{cases},
 \end{align}
 where $\*O_{\tau, \Lambda}$ is the projection matrix of $(\*I-\*H_{\tau\setminus \Lambda}) \*X_{\Lambda \cap \tau},$ and $\*H_{\tau\setminus \Lambda}$ is the projection matrix of  $\*X_{\tau\setminus \Lambda}.$
 We can rewrite the above nuclear mapping as a function of $\*y_{\bm \theta} = \*X_{\tau} \bbeta_{\tau} + \sigma \*u$:
 \begin{align}
 \label{eq:nulcear_beta_lambda}
{\small \tilde T(\*y_{\bm \theta}, \bm \eta_\Lambda)  =  \begin{cases}
    \frac{(\*y_{\bm \theta}- \*X_\Lambda\bbeta_\Lambda)^\top \*O_{\tau, \Lambda} (\*y_{\bm \theta}- \*X_\Lambda\bbeta_\Lambda)}{(\*y_{\bm \theta}- \*X_\Lambda\bbeta_\Lambda)^\top (\*I-\*H_\tau) (\*y_{\bm \theta}- \*X_\Lambda\bbeta_\Lambda)/(n-|\tau|)} & \text{if $\Lambda \cap \tau \neq \emptyset$ \& $\beta_i=0$ for  any $i \in \Lambda \setminus \tau$} \\
    \infty & \text{if $\beta_i \neq 0$ for any $i \in \Lambda \setminus \tau$} \\
    0 & \text{if $\Lambda \cap \tau = \emptyset$ \& $\beta_i = 0$ for any $i \in \Lambda$
    }
  \end{cases}. }
\end{align}
 Since when $\Lambda \cap \tau \neq \emptyset$ and $\beta_i=0$ for  any $i \in \Lambda \setminus \tau,$ the distribution of the nuclear mapping defined above is $T(\*U, \bm\eta_\Lambda) = \tilde T(\*Y_{\bm \theta}, \bm \eta_\Lambda) \sim F_{|\tau\cap\Lambda|, n-|\tau\cap \Lambda|},$ we let the Borel set
 be $B_{\bm \eta_\Lambda}(\alpha) =  [0, F_{|\tau\cap\Lambda|, n-|\tau\cap \Lambda|}^{-1}(\alpha)],$ such that $\P(\tilde T(\*Y_{\bm \theta}, \bm \eta_\Lambda)  \in B_{\bm \eta_\Lambda}(\alpha)) \geq \alpha.$
 We can show that a valid level-$\alpha$ repro samples confidence set for $\bm\eta_{0,\Lambda}  = (\bbeta_{0, \Lambda}, \tau_0)$ is
 \begin{align}
\label{eq:cs_eta}
    \Gamma^{\bm\eta_\Lambda}_{\alpha}(\*y_{obs})   =\left \{\bm\eta_\Lambda: \tilde T(\*y_{obs}, \bm \eta_\Lambda) \in B_\alpha({\bm \eta_\Lambda})\right\}.
\end{align}

 Now with both $\beta_{0,i}, i \not\in \Lambda$ and $\sigma_0$ out of the picture,  we need to deal with the only remaining nuisance parameter $\tau.$ To handle the impact of $\tau,$ we utilize the model candidate set $S^{(d)}$ constructed in Section~\ref{sec_cand} and take a union approach. That is, for certain $\beta_{\Lambda}$, if $(\beta_{\Lambda}, \tau)$ is defined above for any $\tau$ in the candidate set $S^{(d)},$ we then retain the $\beta_{\Lambda}$ in the confidence set for $\beta_{0,\Lambda}.$ Specifically,
\begin{align}
    \label{eq:conf_beta_Lambda}
    \Gamma^{\bbeta_\Lambda}_{\alpha}(\*y_{obs}) &  =\left \{\bbeta_\Lambda: \tilde T(\*y_{obs}, \bm \eta_\Lambda) \in B_\alpha({\bm \eta_\Lambda}), \bm\eta_\Lambda=(\bbeta_\Lambda, \tau) \text{ for some } \tau \in S^{(d)}\right\} \nonumber \\
    & =  \bigcup_{\tau \in S^{(d)}} \left\{\bbeta_\Lambda: \tilde T(\*y_{obs}, \bm \eta_\Lambda) \in B_\alpha({\bm \eta_\Lambda}), \bm\eta_\Lambda=(\bbeta_\Lambda, \tau)\right\}.
\end{align}
 Note that this confidence set is a union of multiple smaller sets. This is different than confidence sets produced by existing methods which are often single intervals or ellipsoid sets.
An illustration of such a confidence set is provided in the special cases that $\Lambda = [p]$, { please see Figure~\ref{fig:B2C} of Appendix~\ref{sec:joint_beta_full}}.

We observe that inside the union in \eqref{eq:conf_beta_Lambda}, each set is a confidence set based on certain low-dimensional model $\tau$ in the candidate set $S^{(d)}.$
Although we do not know the true underlying model $\tau_0$, with Algorithm~\ref{alg:candidate},  we are able to construct a candidate set of reasonable size that would include $\tau_0$ with a high probability. This enables us to guarantee the coverage rate, both in finite samples and asymptotically, as indicated in the following theorems.

\begin{theorem}
\label{cor:coverage_beta_Lambda}
 Under the conditions in Theorem~\ref{the:finite_pen}, for any finite  $n$ and $p$ and an arbitrarily small $\delta>0,$ the coverage probability of the confidence interval $\Gamma^{\bbeta_\Lambda}_{\alpha}(\*Y)$ defined in \eqref{eq:conf_beta_Lambda} is
$\P_{({\cal U}^d, \*Y)} \{\bbeta_{0, \Lambda} \in \Gamma^{\bbeta_\Lambda}_{\alpha}(\*Y)\} \geq \alpha - \delta - o(e^{-c_1d})$ for some $c_1>0$. Further $\P_{\*Y| {\cal U}^d} \{\bbeta_{0, \Lambda} \in \Gamma^{\bbeta_\Lambda}_{\alpha}(\*Y)\} \geq \alpha - \delta - o_p(e^{-c_1d}).$
\end{theorem}

\begin{theorem}
\label{cor:coverage_beta_Lambda_asymp}
 Under the conditions in Theorem~\ref{the:aymp_pen}, for any finite $d$, the coverage probability of  $\Gamma^{\bbeta_\Lambda}_{\alpha}(\*Y)$ defined in \eqref{eq:conf_beta_Lambda} is
$\P_{({\cal U}^d, \*Y)} \{\bbeta_{0,\Lambda} \in \Gamma^{\bbeta_\Lambda}_{\alpha}(\*Y)\} \geq \alpha  - o(e^{-c_2n})$ for some $c_2>0$. Further $\P_{ \*Y| {\cal U}^d} \{\bbeta_{0,\Lambda} \in \Gamma^{\bbeta_\Lambda}_{\alpha}(\*Y)\} \geq \alpha  - o_p(e^{-c_2n}).$
\end{theorem}

\begin{remark}
[Extension to make inference for any linear transformation of $\bbeta^{full}_{0}$]
Let $\*L\bbeta^{full}_0$ be a linear transformation of $\bbeta^{full}_0,$ where $\*L$ is a $l \times p$ transformation matrix. Let
\begin{align*}
    \widetilde{\*L}  = \left[\begin{array}{c}
         \*L \\
         \begin{array}{cc}
             \*I_{(p-l)\times (p-l)},  & \*0_{(p-l) \times l}
         \end{array}
    \end{array}\right],
\end{align*}
$\tilde\bbeta_0^{full} = \widetilde{\*L} \bbeta_0^{full},$ and $\widetilde{\*X} = \*X  \widetilde{\*L}^{-1}.$ The inference for  $\*L\bbeta^{full}_0$ based on the data $(\*y_{obs}, \*X)$ is now equivalently transformed to the inference for a subset of $\tilde\bbeta_0^{full}$ based on the transformed data $(\*y_{obs}, \widetilde{\*X}).$ Therefore we are able to construct the confidence set for  $\*L\bbeta^{full}_0$ by applying \eqref{eq:conf_beta_Lambda} on $(\*y_{obs}, \widetilde{\*X}).$ Note that one should also derive the candidate set $S^{(d)}$ from the transformed data $(\*y_{obs}, \widetilde{\*X})$ using Algorithm~\ref{alg:candidate}.

\end{remark}

\subsection{Two special cases of interest}\label{sec:special}

As stated in \cite{zhang_confidence_2014, javanmard_confidence_2014}, we are often interested in marginal inference for a single regression coefficient $\beta_{0,i}$ in practice. Another interesting inference problem that has been studied in the literature is to jointly infer all regression coefficients $\bbeta_0^{full}$ \cite{zhang_simultaneous_2017, dezeure_high-dimensional_2017}. In this subsection, we consider these interesting special cases and provide a new solution using the repro samples method.

{\bf Inference for a single regression coefficient.}
To obtain the repro samples confidence set for $\beta_{0,i},$ we simplify the nuclear mapping function defined in \eqref{eq:nulcear_beta_lambda_0} and \eqref{eq:nulcear_beta_lambda} by making $\Lambda = \{i\}$. For a given $\bm \eta_i=(\beta_i, \tau)$, it is
\begin{align*}
\small
\resizebox{\hsize}{!}{$
    T(\*u, \bm\eta_i)   = \tilde T(\*y_{\bm \theta}, \bm \eta_i) = \begin{cases}
    \frac{\*u^\top \*O_{\tau, i} \*u}{\*u^\top (\*I-\*H_\tau) \*u/(n-|\tau|)} =  \frac{(\*y_{\bm \theta}- \*X_i\beta_i)^\top \*O_{\tau, i} (\*y_{\bm \theta}- \*X_i\beta_i)}{(\*y_{\bm \theta}- \*X_i\beta_i)^\top (\*I-\*H_\tau) (\*y_{\bm \theta}- \*X_i\beta_i)/(n-|\tau|)}  & \text{if $i  \in \tau$} \\
    \infty & \text{if $i \notin \tau$, $\beta_i \neq 0$} \\
    0 & \text{if $i \notin \tau$, $\beta_i = 0$
    }
   \end{cases}$, }
\end{align*}
where  $\*O_{\tau, i}$ is the projection matrix of $(\*I-\*H_{\tau_{-i}})\*X_i$, and $\tau_{-i} = \tau \setminus \{i\}.$ Note that for $i \in \tau,$ the nuclear statistics $\tilde T(\*y_{\bm \theta}, \bm \eta_i)$ is equivalent to the square of $t$-statistics for testing $H_0: \beta_{0,i} = \beta_i$,  obtained by fitting a linear regression of $\*y_{\bm \theta}$ on $\*X_{\tau}.$

Then following \eqref{eq:conf_beta_Lambda}, we obtain the confidence set for $\beta_{0,i},$
\begin{align}
   \label{eq:conf_beta_i}
    \Gamma^{\beta_i}_{\alpha}(\*y_{obs}) &  =\left \{\beta_i: \tilde T(\*y_{obs}, \bm \eta_i) \in B_\alpha({\bm \eta_i}), \bm\eta_i=( \beta_i, \tau) \text{ for some } \tau \in S^{(d)}\right\} \nonumber \\
    & =  \bigcup_{\tau \in S^{(d)}} \left\{\beta_i: \tilde T(\*y_{obs}, ( \beta_i, \tau)) \leq F_{1, n-|\tau|}^{-1}(\alpha)
    \right\},
\end{align}
 where we let $B_\alpha({\bm \eta_i}) = B_\alpha({\tau}) =  [0, F_{1, n-|\tau|}^{-1}(\alpha)].$ Then by Theorems~\ref{cor:coverage_beta_Lambda} and \ref{cor:coverage_beta_Lambda_asymp},  $\Gamma^{\beta_i}_{\alpha}(y_{obs})$ is a level-$\alpha$ confidence set for $\beta_{0,i}$.

\begin{remark}[Comparison with the debiased method]
We discuss the difference between our method and the debiased Lasso.
First of all, our method offers the finite-sample coverage guarantee, while the debiased Lasso method can only achieve the asymptotic coverage rate.
More specifically, the debiased Lasso method needs the sample size $n\to\infty$ to make sure the bias, which comes from the regularized estimation and is of order $O(|\tau_0|\sqrt{\log p/n})$, goes to 0. In contrast, our method bypasses the estimation step and constructs the confidence sets directly via the repro sampling framework, and is therefore unbiased in nature. Second, the debiased Lasso method \citep{javanmard_confidence_2014} is designed to make inferences for an individual regression coefficient. The idea of the debiased Lasso method was later generalized to make inferences for functions of the regression coefficients, such as co-heritability\citep{guo_optimal_2019} and group inference statistics \citep{guo_group_2020}. However, such a generalization relies on specific forms of the functions and does not allow arbitrary functions. Our method, however, as we will show in Remark~\ref{rem:functional}, can be used to construct the confidence sets for arbitrary functions. Third, we will show in Section~\ref{sec:simu_coef} that, when the sample size is small,  the debiased Lasso method may have either coverage issues or overly large intervals for large regression coefficients. In contrast, our method achieves the nominal coverage for both large and zero regression coefficients in the small sample setting with preferable interval lengths.
See Section~\ref{sec:simu_coef}~for more details.
\end{remark}

\enlargethispage{\baselineskip}

{\bf Joint inference for all regression coefficients.}
Let $\Lambda = [p]$ and we make joint inference for all regression coefficients $\bbeta_0^{full}$ here. Note that $\tau \subset \Lambda$, so $\tau \setminus \Lambda = \emptyset.$
Following \eqref{eq:nulcear_beta_lambda_0} and \eqref{eq:nulcear_beta_lambda},  the nuclear mapping function for  $\bm\eta= (\bbeta, \tau) = \left(\left(\bbeta_\tau, \bm 0_{\tau^C}\right), \tau\right)$ is
\begin{align}
\label{eq:nuclea_joint}
 T(\*u, \bm\eta)   & = \frac{ \*u^\top \*H_\tau \*u/|\tau|}{ \*u^\top (I -  \*H_\tau) \*u/(n-|\tau|)}   \nonumber \\
 & =  \frac{ {(\*y_{\bm\theta}  -  \*X_{\tau} \bbeta_{\tau})}^\top \*H_\tau  (\*y_{\bm\theta}  -  \*X_{\tau} \bbeta_{\tau})/|\tau|}{ {(\*y_{\bm\theta}  -  \*X_{\tau} \bbeta_{\tau})}^\top (I -  \*H_\tau) (\*y_{\bm\theta} -  \*X_{\tau} \bbeta_{\tau})/(n-|\tau|)} := \tilde T(\*y_{\bm\theta} , \bm\eta).
\end{align}
We then let the Borel set be  $B_\alpha({\bm \eta}) = B_\alpha(\tau) = [0,  F^{-1}_{|\tau|, n - |\tau|}(\alpha) ],$ and construct the joint confidence set for $\bbeta_0^{full}$ following
from \eqref{eq:conf_beta_Lambda},
 \begin{align}
 \label{eq:conf_joint}
\Gamma_{\alpha}^{\bbeta}(\*y_{obs}) & =    \left\{\bbeta:  \tilde T\left(\*y_{obs} , \bm\eta\right)   \leq F^{-1}_{|\tau|, n - |\tau|}(\alpha), \bbeta=\left(\bbeta_\tau, \bm 0_{\tau^C}\right),  \bm\eta=(\bbeta, \tau)  \text{ for some }  \tau \in S^{(d)}\right\} \nonumber \\
& =   \bigcup_{\tau \in S^{(d)}} \left\{\bbeta:  \tilde T\left(\*y_{obs} , \bm\eta\right)   \leq F^{-1}_{|\tau|, n - |\tau|}(\alpha), \bbeta=\left(\bbeta_\tau, \bm 0_{\tau^C}\right),  \bm\eta=(\bbeta, \tau) \right\}.
\end{align}
Again, following Theorems~\ref{cor:coverage_beta_Lambda} and \ref{cor:coverage_beta_Lambda_asymp}, $\Gamma_{\alpha}^{\bbeta}(\*y_{obs})$ is a level-$\alpha$ confidence set for the entire coefficients vector $\bbeta_0^{full}$. {In addition, to better understand the confidence set in \eqref{eq:conf_joint}, we have presented a visualization of \eqref{eq:conf_joint} for an example in Appendix~\ref{sec:joint_beta_full}.}

\begin{remark}[Extension to inference for any function of $\bbeta^{full}_{0}$]
\label{rem:functional}
We can extend the joint confidence set in \eqref{eq:conf_joint} to obtain a repro samples confidence set for any function of $\bbeta^{full}_{0},$ say $h(\bbeta^{full}_{0})$.
To put it more clearly, for each $(\tau, \bbeta_\tau)$ in the confidence set \eqref{eq:conf_joint}, we collect the  function value $h(\bbeta_\tau^{full}),$ where $\bbeta_\tau^{full} = (\bbeta_\tau, \bm 0_{\tau^C})$ to form the confidence set for $h(\bbeta^{full}_{0}),$ i.e.
$\Gamma^{h}_{\alpha}(\*y_{obs}) =  \big\{h\left((\bbeta_\tau, \bm 0_{\tau^C})\right):  \tilde T\left(\*y_{obs} , \bm\eta\right)   \leq F^{-1}_{|\tau|, n - |\tau|}(\alpha), \bm\eta=(\tau,  \bbeta_\tau)$ for $\tau \in S^{(d)}\big\}.$
\end{remark}
 {
\subsection{Extension to models with non-Gaussian errors}
\label{sec:beta_nonGaussian}
In this section, we extend the confidence set for the regression coefficients in \eqref{eq:conf_beta_Lambda} to models with non-Gaussian errors. We first introduce the following corollary.
\begin{corollary}
\label{cor:nonGaussian}
    Let   $\widetilde \Gamma^{\bm\beta_\Lambda|\tau}(\*y_{obs})$ be a data-dependent set of $\bm\beta_\Lambda$ that also depends on a given model $\tau,$
    and let $\tilde\alpha = \P\big(\bm\beta_{0, \Lambda } \in \widetilde \Gamma^{\bm\beta_\Lambda|\tau_0}(\*Y)\big)$ be the coverage probability for $\bm\beta_{0, \Lambda}$  given the true model $\tau_0,$  then the confidence set for $\bbeta_{0,\Lambda}$ constructed by $\widetilde \Gamma^{\bbeta_{\Lambda}}(\*y_{obs}) = \bigcup_{\tau \in S^{(d)}}\widetilde \Gamma^{\bm\beta_\Lambda|\tau}(\*y_{obs})$  has a coverage rate bounded by $\P(\bbeta_{0,\Lambda} \in \widetilde \Gamma^{\bbeta_{\Lambda}}(\*Y)) \geq \tilde\alpha - \P(\tau_0 \not\in S^{(d)}).$
\end{corollary}

The above corollary implies that models with non-Gaussian errors can still follow the idea in \eqref{eq:conf_beta_Lambda} to construct confidence sets for the regression coefficients. This is because the model candidate set proposed in Section~\ref{sec_cand} still covers the true model $\tau_0$ with a high probability according to the results in Section~\ref{sec:cand_non_normal} for common non-Gaussian errors.
Once we have a candidate set $S^{(d)}$ that guarantees to cover the true model such that $\P(\tau_0 \not\in S^{(d)})$ is close to 0, all we need is a valid inference approach in low-dimensional settings that can achieve the desired coverage given $\tau_0.$ In other words, we only need a data-dependent set given a low-dimensional model $\tau,$ namely  $\widetilde \Gamma^{\bm\beta_\Lambda|\tau}(\*y_{obs}),$ such that given the true model, $\widetilde \Gamma^{\bm\beta_\Lambda|\tau_0}(\*y_{obs})$ is a valid confidence set for $\bm\beta_{0, \Lambda }$ with a desired coverage rate of (approximately) $\alpha.$ It then follows immediately from Corollary~\ref{cor:nonGaussian} that $\widetilde \Gamma^{\bbeta_{\Lambda}}(\*y_{obs}) = \bigcup_{\tau \in S^{(d)}}\widetilde \Gamma^{\bm\beta_\Lambda|\tau}(\*y_{obs})$ is a valid level-$\alpha$ confidence set for $\bm\beta_{0, \Lambda }.$

     Fortunately, such a $\Gamma^{\bm\beta_\Lambda|\tau}(\*y_{obs})$ is not difficult to find, since the
     inference for linear models with non-Gaussian errors in low-dimensional settings have been extensively studied. For example, \cite{lange_robust_1989} proposed a likelihood-based approach for $t$ and $Cauchy$ errors, and \cite{pek_how_2018} provided a thorough review on different approaches to deal with non-Gaussianity in the error terms of linear models. To make inference for a model with contaminated errors, see \cite{lemdani_likelihood_1999, dai_inferences_2007} and the reference therein. For sub-Gaussian errors, it is well known that most of the results for Gaussian errors hold asymptotically due to the central limit theorem \citep{williams2019assumptions}.

}

{

\section{Simulation studies}\label{sec:simulation}
In this section, we conduct simulation studies to evaluate the numerical performance of the proposed repro samples methods.
 The synthetic data are generated from the following five models:

\begin{itemize}
    \item[(M1)]
     (Extremely high dimension)
	 {Let $\bbeta_0^{full}=(3, 2, 1.5, 0, \dots, 0)$.}  For $j_1,j_2\in[p]$, the correlation between $x_{j_1}$
	and $x_{j_2}$ is set to $0.5^{|j_1 - j_2 |}$. We set $n = 50,
	p = 1000$ and $\sigma = 1$.

\item[(M2)] (Decaying signal) Let $\bbeta_0^{full}=(2, 1.5, 1, 0.8, 0.6, 0, \dots, 0).$ For $j_1,j_2\in[p]$, the correlation between $x_{j_1}$
and $x_{j_2}$ is set to $0.1^{|j_1 - j_2 |}$. We let $n = 80,
p = 150$ and $\sigma = 1$.
\item[(M3)] (High-dimensional, decaying signal) Let $\bbeta_0^{full}=(3,2, 1.5, 1, 0.8, 0.6, 0, \dots, 0).$ The correlation between $x_{j_1}$
and $x_{j_2}$ is $0.1^{|j_1 - j_2 |}, j_1,j_2\in[p]$. Let $n = 100,
p = 500,$ and $\sigma = 1$.
{\item[(M4)] (From \cite{li_model_2019} with many signals, but changed $n$ to create a high-dimensional setting with $p>n$) Let the true values $\bbeta_0^{full}=
(1, \ldots, 1, 0, \dots, 0)$, where the first $12$ coefficients $\beta_{0,j} = 1$, for $1 \leq j \leq 12$, and remaining $\beta_{0,j} = 0$, for $j \ge 12$. For $j_1,j_2\in[p]$, the correlation between $x_{j_1}$
and $x_{j_2}$ is set to $0.5^{|j_1 - j_2 |}$. We let $n = 150,
p = 200$ and $\sigma = 1$.
\item[(M5)]
     (Perfect colinearity, restricted eigenvalue conditions do not hold)
	 {Let the true values $\bbeta_0^{full}=(3, 2, 1.5, 0, \dots, 0)$.}  For $j_1,j_2\in[p]$, the correlation between $x_{j_1}$ and $x_{j_2}$ is set to $0.5^{|j_1 - j_2 |}$. To create perfect collinearity, we set $2x_{10} + 3x_{11} + x_{10} = 3x_{1} + 2x_{2} + 1.5x_{3},$
     such that the restricted eigenvalue condition does not hold, and $C_{\min} = 0.$
 We set $n = 50,p = 1000$ and $\sigma = 1$.
}
\end{itemize}
The first model (M1) represents an extremely high-dimensional setting with $p \gg n.$ The second model (M2) represents a challenging case of decaying signals with the weakest one just 0.6. We set (M3) by increasing the dimension of  (M2) to study the performance of the proposed approach when both a high-dimensional design matrix and weak signals are present in the data.
We also add a strong signal of $\beta_{1,0} =3$ so the range of signal strength is from $0.6$ to $3$.
Model (M4), a high-dimensional adaptation of the scenario (f) in \cite{li_model_2019}, features a setting that involves many true signals.
Finally, we present an extremely challenging case in (M5), where typical conditions in the literature such as the restricted eigenvalue condition do not hold anymore.  We replicate the simulation for 200 times for each model. }

\subsection{Model candidates and inference for the true model $\tau_0$}
\label{sec:simu_model}

{  \label{para:point_estimation}

}

We first study the numerical performance of the data-driven model candidate set $S^{(d)}$ in \eqref{eq:S_d}, produced by Algorithm~\ref{alg:candidate}, and the 95\% repro samples model confidence set in \eqref{eq:bar-CS}, constructed by Algorithm~\ref{alg:tau}.

To motivate the use of
a set of models to quantify model selection uncertainty, we first examine the performance of a single selected (estimated) model based on  common model selection criteria, AIC, BIC, cross-validation (CV), and the knockoff \cite{candes_panning_2018}.
Table~\ref{tab:point_estimation} reports the proportions of times a single selected model correctly identifies the true model under (M1)--(M5).
 We observe that the single selected model frequently differs from the true model: for Model (M2), the majority of the times the estimated model is wrong; for the other three settings, AIC, BIC, and CV methods are correct 60\%--80\% of the times. The knockoff approach performs poorly in estimating the true model, since it is developed mainly to control the false discovery rate. Moreover, there is no finite-sample theory to ensure that the estimated models are accurate, and even asymptotic results depend on assumptions that are often hard to verify. Therefore, incorporating data uncertainty into model selection or estimation is essential, underscoring the importance of using a valid and efficient model confidence set, such as the one we have proposed.

\begin{table}[!t]

\centering
\caption{Percentage of times when point estimation is the true model among the 200 simulations. }
\label{tab:point_estimation}
\begin{tabular}{c|cccc}
\hline
 Model & AIC &  BIC & CV & Knockoff \\ \hline
M1 & 0.695  & 0.790  & 0.645  & 0.000 \\
 M2 & 0.250  & 0.510  & 0.205  & 0.000 \\
 M3 & 0.780  & 0.810   & 0.750   & 0.000 \\
 M4 & 0.730  & 0.860  & 0.655  & 0.600\\
 M5 &  0.845  & 0.880 &0.830 &0.000\\ \hline
\end{tabular}
\end{table}

{ In our implementation of Algorithm~\ref{alg:candidate}, we use the EBIC  \citep{chen_extended_2008-1} to select the tuning parameter when constructing the model candidate set. Further implementation details are provided in Appendix~\ref{sec:joint_beta_full}.} When applying Algorithm~\ref{alg:tau} to obtain our model confidence set, we calculate the   $\hat\tau(\*y_j^*)$ in \eqref{eq:tau_hat_y_star} by obtaining the largest estimated model that is not larger than $|\tau_b|$ in the adaptive Lasso solution path. Also, in our analysis, we set the number of repro samples for the candidate set in Algorithm \ref{alg:candidate} to be $d = 1000$ for Models (M1), {(M4), and (M5)}.
For Models (M2)-(M3) with
weak signals, identifying the true model is a known challenging problem. In this case, we set the number of repro samples to be a large $d = 10,000$ for (M2) and $d = 100,000$ for (M3). Regarding the number of repro samples in Algorithm \ref{alg:tau} for calculating the distribution of the nuclear statistics, we set $J=200$ for all five models.

We compare our repro samples approach with the residual bootstrap approach in the literature (e.g., \citep{chatterjee_bootstrapping_2011}).
The numbers of bootstrap samples are $1000 $ {for (M1), (M4) and (M5)}, and $10,000$, and $100,000$ for (M2) and (M3) respectively, matching the numbers of repro samples for used for finding the candidate models.
{In each setting, the collection of all models obtained using the bootstrap samples forms a bootstrap model candidate set.
Here, to implement the bootstrap approach, we use  AIC, BIC, and CV to choose the tuning parameter.}
The bootstrap model ``confidence'' sets are obtained by removing the least frequent model estimations from the bootstrap {candidate model set}, with the total (cumulative) frequency of the removed models not larger than 5\%.
We note that the bootstrap method here is an ad hoc method commonly
used in current practice. Due to the discreteness of the model space and estimated models, there is no theoretical support for the ``confidence'' claim that such a bootstrap method can get a valid level-$95\%$  model confidence set for the true model~$\tau_0$.

\label{paragraph:simu_candidate_set}
Table~\ref{tab:combined_model}, columns 3--4, compares the model candidate sets from the proposed repro samples approach and the  residual bootstrap approaches with different tuning criteria. We report the average cardinality of the model candidate sets (Cardinality) and the percentage of simulation cases where the true model $\tau_0$ is included (Inclusion of $\tau_0$). From Table~\ref{tab:combined_model}, we see that the proposed repro samples approach provides much smaller model candidate sets.
For (M1) where $n=50, p=1000$, the repro samples candidate sets achieve 100\% coverage for $\tau_0$ with a size of only 2--3 on average out of the 1000 repro samples. Even for (M2) and (M3), where weak signals make identifying the true model notoriously challenging, the proposed procedure produces a candidate set of around 30 and 5 models on average, respectively, while covering the true model 98\% and 99.5\% of the times.   The bootstrap method, on the other hand, yields {50--600 models out of  1000 replicates for (M1), (M4) and (M5)}, around 1700--4000 models out of 10,000 bootstrap runs for (M2), and around 3000--6500 models out of 100,000 bootstrap samples for (M3), proportionate to a random search.
In summary, Table~\ref{tab:combined_model} clearly demonstrates the advantage of our proposed
method in finding candidate models. The size of the candidate model set by our repro samples method is small and manageable, while those by the corresponding bootstrap methods are all unreasonably large, making them inapplicable in practice. 
Furthermore, following a reviewer's suggestion,
we have also tried constructing an alternative model candidate set by including all models on the solution path of EBIC using just the original data without the repro samples procedure.
The empirical coverage of such an EBIC candidate set is not satisfactory except for (M1), as the coverage for (M2)--(M5) ranges only from 81.5\% to 94.5\%, which is not comparable to our repro samples approach in general.

\begin{table}[!t]
\centering
\caption{Performances of Model Candidate Sets and $95\%$ Model Confidence Sets for the  True Model $\tau_0$}
\label{tab:combined_model}
\scriptsize
\begin{tabular}{ll|ll|ll}
\hline
&  & \multicolumn{2}{c|}{\it  Model Candidate Sets} & \multicolumn{2}{c}{\it Level-$95\%$ Model Confidence Sets} \\
  Model & {Method} & Cardinality of $S^{(d)}$ & Inclusion of  $\tau_0$ & Cardinality of $\Gamma^{\tau}_{0.95}$ & Coverage of $\Gamma^{\tau}_{0.95}$ \\
 \hline
\multirow{4}{*}{ M1}
 & Repro samples & 2.605 (0.191) & 1.000 (0.000) & 2.180 (0.102) & 1.000 (0.000) \\
 & Bootstrap AIC & 215.425 (10.855) & 1.000 (0.000) & 165.960 (7.201) & 1.000 (0.000) \\
 & Bootstrap BIC & 146.100 (7.423) & 1.000 (0.000) & 110.440 (5.423) & 1.000 (0.000) \\
 & Bootstrap CV & 259.535 (11.891) & 1.000 (0.000) & 207.500 (7.890) & 1.000 (0.000) \\
\hline
\multirow{4}{*}{ M2}
 & Repro samples & 29.455 (3.080) & 0.980 (0.010) & 12.050 (0.708) & 0.955 (0.015) \\
 & Bootstrap AIC & 4350.850 (134.000) & 1.000 (0.000) & 4350.850 (134.000) & 0.995 (0.005) \\
 & Bootstrap BIC & 2303.190 (75.708) & 1.000 (0.000) & 2286.015 (77.898) & 0.995 (0.005) \\
 & Bootstrap CV & 5033.700 (134.233) & 1.000 (0.000) & 5033.700 (134.233) & 0.995 (0.005) \\
\hline
\multirow{4}{*}{ M3}
 & Repro samples & 4.710 (0.558) & 0.995 (0.005) & 3.810 (0.316) & 0.985 (0.009) \\
 & Bootstrap AIC & 5088.030 (456.021) & 1.000 (0.000) & 3481.215 (497.552) & 0.935 (0.017) \\
 & Bootstrap BIC & 2944.325 (245.670) & 1.000 (0.000) & 1425.125 (265.383) & 0.905 (0.021) \\
 & Bootstrap CV & 6458.345 (570.104) & 1.000 (0.000) & 4989.195 (618.298) & 0.955 (0.015) \\
\hline
\multirow{4}{*}{ M4}
 & Repro samples & 1.875 (0.144) & 0.995 (0.005) & 1.745 (0.087) & 0.995 (0.005) \\
 & Bootstrap AIC & 165.410 (6.178) & 1.000 (0.000) & 162.250 (6.495) & 1.000 (0.000) \\
 & Bootstrap BIC & 96.240 (3.210) & 1.000 (0.000) & 85.415 (4.107) & 1.000 (0.000) \\
 & Bootstrap CV & 204.175 (7.291) & 1.000 (0.000) & 203.090 (7.414) & 1.000 (0.000) \\
\hline
\multirow{4}{*}{ M5}
 & Repro samples & 2.775 (0.199) & 1.000 (0.000) & 2.640 (0.164) & 0.985 (0.009) \\
 & Bootstrap AIC & 72.240 (6.341) & 0.990 (0.007) & 57.875 (6.820) & 0.990 (0.007) \\
 & Bootstrap BIC & 55.285 (4.794) & 0.990 (0.007) & 41.410 (5.191) & 0.990 (0.007) \\
 & Bootstrap CV & 87.515 (7.406) & 0.990 (0.007) & 74.065 (7.925) & 0.990 (0.007) \\
\hline
\end{tabular}
\end{table}

 Table~\ref{tab:combined_model}, columns 5--6, reports the average cardinality of the confidence sets
 obtained using Algorithm~\ref{alg:tau} and the bootstrap approaches, along with their coverage of the true model $\tau_0$ out of the 200 repetitions.
From Table~\ref{tab:combined_model}, we see that, for  (M1), {(M4) and (M5),  the model confidence set based on the repro samples approach only contains 1.7--2.6 models on average, while the ``confidence" sets by the bootstrap methods have sizes between 40--600}.  For  (M2) and (M3), the model confidence sets generated by the bootstrap are impractically large, containing between 1400--5000 models on average. Even with those many models, the bootstrap confidence sets with AIC and BIC slightly undercover the true model $\tau_0$ for (M3). In contrast, for (M2) and (M3), the repro samples approach achieves much more efficient results: its confidence sets contain only about 12 and 4 models on average, respectively, yet still maintain coverage above 95\%.
This strong performance holds even under the challenging decaying-signal scenarios, where the smallest nonzero coefficient is as low as 0.6.

{ \label{para:coverage}
The empirical coverage rate of our model confidence set is higher than 95\% except for (M2). This is mostly due to the discrete nature of the inference target here, the true model $\tau_0.$ Unlike a typical confidence interval for a parameter whose value is a real number, the coverage rate for $\tau_0$ is by no means a continuous or linear function of the size of the confidence set. Therefore, reducing the size of the confidence set will reduce the coverage rates below the desired level of 95\%. However, there would be less overcoverage when the candidate set contains more models, as in (M2), where the coverage rate of the confidence set is 95.5\%.
}

To sum up, the results in Table~\ref{tab:combined_model} demonstrate that the proposed repro samples method constructs valid and efficient model confidence sets for the true model $\tau_0,$
even under the challenging settings among (M1)--(M5). In contrast, the bootstrap method exaggerates the uncertainty of model selection by producing extremely large number of models in its ``confidence'' sets,
rendering results that are not very useful in practice.

\subsection{Inference for regression coefficients accounting for model selection uncertainty}
\label{sec:simu_coef}

Here, we examine the performance of the proposed repro samples confidence set for a  single regression coefficient in \eqref{eq:conf_beta_i} and compare it with the state-of-the-art debiased methods.
Due to space limits, the results of the joint confidence set for $\bbeta_0^{full}$ in \eqref{eq:conf_joint} are placed in Appendix~\ref{sec:joint_beta_full}.

\begin{table}[!t]
\caption{Repro Confidence Sets for (Scalar) Regression Parameter $\beta_{0,i}$ with Comparison to Debiased Lasso}
\label{tab:confidence_interval}
\scriptsize
\begin{tabular}{lr|cc|cc|cc}
\hline
 & & \multicolumn{2}{c|}{Repro Samples}    & \multicolumn{2}{c|}{Debiased Lasso (JM) } & \multicolumn{2}{c}{Debiased Lasso (ZZ)}    \\
Model &  $\beta_{0,i}$ & Coverage & Width& Coverage & Width &  Coverage & Width   \\ \hline
\multirow{3}{*}{M1} & All $\beta_{0,i} $ & 1.000(0.000) & 0.003(0.000)         & 0.984(0.000)  & 0.247(0.000) & 0.961(0.000)  & 1.335(0.000)\\
 & $\beta_{0,i} \neq 0$    & 0.952(0.009)     & 0.748(0.006)  & 0.357(0.020)   & 0.247(0.005)  & 0.982(0.005)     & 1.321(0.009)  \\
 & $\beta_{0,i} = 0$      & 1.000(0.000)   & 0.001(0.000)   & 0.986(0.000)    & 0.247(0.000)  & 0.961(0.000)   & 1.335(0.000)    \\
\hline
\multirow{3}{*}{M2}  & All $\beta_{0,i} $ &   0.999(0.000)&  0.038(0.000) & 0.982(0.001) & 0.498(0.000) & 0.983(0.001)  & 0.942(0.001)  \\
 & $\beta_{0,i} \neq 0$    & 0.975(0.005)     & 0.539(0.003)  & 0.938(0.008)     & 0.497(0.002) & 0.989(0.003)     & 0.939(0.004)  \\
 & $\beta_{0,i} = 0$       & 1.000(0.000)   & 0.021(0.000)   & 0.984(0.001)   & 0.498(0.000)  & 0.983(0.001)   & 0.942(0.001) \\
\hline
\multirow{3}{*}{M3}  & All $\beta_{0,i} $ &   1.000(0.000)&  0.007(0.000) &  0.989(0.000)  & 0.429(0.000)  & 0.974(0.001)  & 0.943(0.000) \\
 & $\beta_{0,i} \neq 0$    & 0.959(0.006)     & 0.459(0.002)   & 0.868(0.010)     & 0.428(0.001)  & 0.977(0.004)   & 0.943(0.004) \\
 & $\beta_{0,i} = 0$       & 1.000(0.000)   & 0.002(0.000)   & 0.991(0.000)   & 0.429(0.000)  & 0.974(0.001)   & 0.943(0.000) \\
\hline

\multirow{3}{*}{M4}  & All $\beta_{0,i} $ &   0.997(0.000)  & 0.027(0.001) & 0.985(0.001)  & 0.441(0.000) & 0.972(0.001)  & 0.513(0.000)  \\
 & $\beta_{0,i} \neq 0$    & 0.957(0.004)  & 0.437(0.001)  & 0.942(0.005)  & 0.439(0.001) & 0.953(0.004)    & 0.511(0.001) \\
 & $\beta_{0,i} = 0$       & 1.000(0.000)   & 0.001(0.000)  & 0.988(0.001)   & 0.441(0.000) &  0.974(0.001)   & 0.513(0.000) \\
\hline

\multirow{3}{*}{M5}  & All $\beta_{0,i} $ &  1.000(0.000)  &  0.003(0.000) & 0.984(0.000) &  0.244(0.000) & NA & NA\\
 & $\beta_{0,i} \neq 0$    &  0.953(0.009)  &  0.755(0.006) & 0.337(0.019) &  0.246(0.005) & NA & NA\\
 & $\beta_{0,i} = 0$       & 1.000(0.000)  &  0.001(0.000) & 0.986(0.000)  &  0.244(0.000)  & NA & NA\\
\hline
\end{tabular}
\end{table}

Table~\ref{tab:confidence_interval} compares the proposed  $95\%$ repro samples confidence sets for single regression coefficients with the two state-of-the-art debiased approaches from \cite{javanmard_confidence_2014} (debiased Lasso (JM),  authors' code at \cite{mont_code_nodate}) and \cite{zhang_confidence_2014} (debiased Lasso (ZZ), authors' R code). Although the debiased Lasso (JM) can achieve the desired coverage rate when averaging over all $\beta_j$'s.
it significantly undercovers the non-zero coefficients (signals) for all models {except for (M4), particularly for (M1) where the correlations among the covariates are large, and for (M5) where the restricted eigenvalue condition does not hold.}   This undercoverage, noted by both \cite{zhang_confidence_2014} and \cite{javanmard_confidence_2014}, arises because large correlations inflate the estimation error $|\hat\beta_i-\beta_{0,i}|$ for nonzero signals when sample sizes are limited.
\cite{zhang_confidence_2014} provides an enhanced method (ZZ) to overcome this issue by including an independent set of highly correlated variables when debiasing to enforce small correlations between the score vector and covariates. This approach improves the coverage rates in the three simulation settings, but at the expense of larger interval widths.
In contrast, the repro samples method consistently achieves the desired coverage for all the signals in all five models, with confidence sets at least 40\% shorter than those from the debiased Lasso~(ZZ). {Besides, the debiased Lasso (ZZ) fails to produce results for (M5) because the required condition on the design matrix is not satisfied in this case.  Additionally, we also compare the confidence intervals of each non-zero coefficients, confirming the aforementioned advantages of our repro samples method over the debiased approaches. See Appendix~\ref{sec:joint_beta_full} for the details. }

Moreover, compared to both debiased approaches, the repro samples confidence sets for zero coefficients appear to be much narrower, making the average width of the proposed confidence sets much smaller.  In addition, the computing code of~the~enhanced~debiased~Lasso~(ZZ)~approach requires arbitrarily  pre-selecting the number of highly correlated variables to calculate the score vector. Conversely, the repro samples approach is a data-dependent procedure that avoids such an ad hoc decision.
In summary, the repro samples approach achieves better coverage with data-dependent smaller confidence set for a single regression coefficient.

{
\section{Real Data Analysis}
\label{sec:real_data}

In this section, we apply our repro samples method along with the bootstrap approaches to analyze the glioblastoma gene expression data from The Cancer Genome Atlas (TCGA) consortium. The goal of the analysis is to identify the highly informative genes regarding the survival time of glioblastoma.  The gene expression data set we obtained, which includes 428 samples,  were measured experimentally by the University of North Carolina TCGA genomic characterization center.  In our analysis, the logarithm of the survival time is the response variable.  We first screen the $17814$ genes using sure independence screening \citep{fan_sure_2008} to identify $1000$ genes that are most correlated with the response~\citep{wang_random_2011}.

Here we set the number of the repro samples $d=5000$ when constructing the model candidate set with Algorithm~\ref{alg:candidate}. Similar to the simulation studies, we compare the proposed repro samples procedure with the bootstrap approaches for inference on the true models.  Out of the 5000 iterations, bootstrap yields 4935 candidate models with BIC, 5000 with AIC or CV. This means almost every bootstrap sample will generate a different model,  and the candidate models exhaust all the 1000 covariates, confirming earlier findings in Section~\ref{sec:simu_model} that bootstrap
includes too many models with spurious variables.
In contrast, the repro samples method identifies six candidate models out of 5000 iterations, as shown in Table~\ref{tab:combined_table_data}(a). The results are also consistent with the previous findings from the simulations, demonstrating the efficient construction of candidate sets by the repro samples method. Moreover,
based on the six candidate models, we follow Algorithm~\ref{alg:tau} to construct a model confidence set with $J=200$. We summarize the tail probability $\hat F_{(\*w_{obs}, \tau_b)}\big\{ \tilde T(\*y_{obs}, \tau_b)\big\}$ in Algorithm~\ref{alg:tau}, also interpreted as a p-value,  in Table~\ref{tab:combined_table_data}(a). According to the tail probabilities, the 95\% model confidence set consists of four models: $\phi$, (ZNF208), (ZNF208, TOP1), and~(ZNF208, NETO2).

Comparing  with the top covariates obtained using the bootstrap, listed in Table~\ref{tab:combined_table_data}(b), the repro samples confidence set highlights  two key genes: TOP1 and NETO2. The significance of TOP1 gene in glabliostoma diagnosis and treatment has been proven by numerous studies \citep[eg.,][]{wang_exth-09_2017, butler_exth-58_2019, sarcar_vorinostat_2010}. Furthermore, the link between the NETO2 gene and glabliostoma is well documented by a recent study in \cite{li2023glioma}.
 Meanwhile, CCDC19---the second most frequent gene under the bootstrap method---does not appear in our candidate models, and there is no established evidence supporting its association with glioblastoma. Overall, existing scientific findings suggest that the proposed repro samples procedure is more reliable when inferring the true underlying models.

\begin{table}[!t]
    \centering
    \caption{ Repro Samples Candidate Models and Top 10 Variables from the Bootstrap}
    \label{tab:combined_table_data}
    \resizebox{\textwidth}{!}{
\begin{tabular}{ccc cll|ll|ll}
        \multicolumn{3}{c}{\it \small (a) Repro samples candidate models} & &  \multicolumn{5}{c}{\it \small   (b) Top 10 most often selected variables by the bootstrap} \\
        \cline{1-3} \cline{5-10}
        Candidate Model & Tail Probability &  & & \multicolumn{2}{c|}{AIC} & \multicolumn{2}{c|}{BIC} & \multicolumn{2}{c}{CV} \\
        \cline{1-3} \cline{5-10}
        $\phi$ & 1.000 &  & & ZNF208 & 4877 & ZNF208 & 4350 & ZNF208 & 4727 \\
        (ZNF208) & 1.000 &  & & CCDC19 & 4751 & CCDC19 & 2259 & CCDC19 & 3946 \\
        (ZNF208, NETO2) & 1.000 &  & & SAA4 & 4319 & GRM8 & 2215 & GRM8 & 3643 \\
        (ZNF208, TOP1) & 0.265 &  & & GRM8 & 4233 & TOP1 & 2192 & SAA4 & 3517 \\
        (ZNF208, GRM8) & 0.000 &  & & SLC25A23 & 3975 & NETO2 & 1932 & CETP & 3436 \\
        (ZNF208, RPS28, TOP1) & 0.000 &  & & ATP5G3 & 3903 & CETP & 1915 & NETO2 & 3134 \\ \cline{1-3}
        & & & & NETO2 & 3820 & SAA4 & 1901 & TOP1 & 3036 \\
        & & & & CLCNKB & 3769 & SLC25A23 & 1405 & SLC25A23 & 2868 \\
        & & & & CETP & 3754 & HCRTR2 & 1210 & PAX3 & 2300 \\
        & & & & CPNE1 & 3718 & PAX3 & 979 & ATP5G3 & 2182 \\ \cline{5-10}
    \end{tabular}}
\end{table}

}

\section{Discussion}\label{sec:discussion}
We have developed a repro samples approach to address inference problems concerning high-dimensional linear models.
The paper contains
three technical innovations.
\begin{enumerate}
    \item
We develop a data-driven approach
to obtain an efficient model candidate set, which  covers the true model with high probability by including just a reasonable number of model candidates. Using this model candidate set effectively addresses the computational issue since it avoids searching the entire model space.
The approach is based on the matching attempt of repro samples with the observed data, leading to the many-to-one mapping function in (\ref{eq::obj}). Specifically, this many-to-one mapping tells us that there always exists a neighborhood of $\*U,$ within which a repro copy $\*U^*$ can help recover the true model with a high probability.
With this insight, we propose a formal procedure and provide supporting theories and numerical evidence, both of which also help to outline trade-offs among sample size, the signal strength, and the performance of the model candidate set. {Additionally, the number of repro samples $d$ in Algorithm~\ref{alg:candidate} is analog to the bootstrap sample size in the bootstrap approach, and our simulation study shows empirically that $d$ does not need to be exceptionally large. In theory,  we have derived a theorem, which is placed in Appendix~\ref{sec:size_d} along with its discussions due to space constraints, to study the scale of $d$ required to theoretically guarantee the model candidate set includes the true model $\tau_0.$  }

{

{Moreover, we have developed supporting theoretical results that accommodate both Gaussian and common non-Gaussian error distributions. We further extend the finite-sample result for sub-Gaussian errors in Theorem~\ref{the:sub_finite_pen} to any continuous error distribution with a finite second moment, as stated in the following corollary. The proof is provided in Appendix~\ref{sec:non_normal}.
\begin{corollary}
\label{cor:second_moment_u}
    Suppose  $\*Y$ is generated by \eqref{eq:model-random} with $\*U$ being a continuous random vector with a finite second moment $\E(\|\*U\|^2) < \infty,$ then when $ \lambda \in \big[n\gamma_{\delta}^{3/4} , n\gamma_{\delta}^{1/4}\frac{C_{\min}}{6}\big],$ the probability bound \eqref{eq:bound_cs_finite_sub} in Theorem~\ref{the:sub_finite_pen} still holds.
\end{corollary}}

 In addition, our model candidate set can be used to achieve model selection consistency under non-Gaussian errors. Specifically, Section~4.2 of \cite{lei_cross-validation_2020} shows that the smallest model in their cross-validation confidence set is consistent in model selection under the assumption of a finite sixth moment if the candidate models include the true model. By Corollary~\ref{cor:second_moment_u}, the proposed model candidate set $S^{(d)}$ contains $\tau_0$ with high probability for sufficiently large $d$ under a finite second-moment assumption; hence, under the stronger sixth-moment condition required by \cite{lei_cross-validation_2020}, the inclusion still holds, and applying their result to our repro-samples candidate set $S^{(d)}$ delivers selection consistency. Alternatively, Theorem~\ref{the:sub_aymp_pen} implies the same consistency under sub-Gaussian errors for suitably small $d$.

}

\item
When making inference for the true model $\tau_0$, we develop a conditional repro samples approach to remove the impact of the nuisance parameters $(\bbeta_\tau, \sigma^2)$. This conditional approach  works in general for inference problems beyond the scope of this paper.  In particular, let $\btheta_0 = (\bnu_0, \bxi_0)$, where $\bnu_0$ and $\bxi_0$ are the target and nuisance parameters, respectively. If we have a nuclear mapping $T(\*U, \btheta)$ and a quantity $\*W(\*U, \btheta),$ such that the conditional distribution of $T(\*U, \btheta)$ given $W(\*U, \btheta) = \*w$ is free of $\bxi_0,$ then there exists a Borel set $B_{\alpha}(\bm\nu, \*w)$ free of the nuisance $\bxi$ such that
   $\P(T(\*U, \btheta) \in B_{\alpha}(\bm\nu, w)|W(\*U, \btheta) = w) \geq \alpha.
$
Consequently, similar to \eqref{eq:cond_cs}, we obtain a valid level-$\alpha$ confidence set for $\bnu_0$.
\item
We propose confidence sets both for a single and for any subset of regression coefficients.  In contrast, existing literature only focuses on one aspect of these inference problems. This is because, unlike existing approaches, we take a union of intervals or multi-dimensional ellipsoids based on each low-dimensional model in the model candidate set. Therefore, our approach takes into account the uncertainty in model estimation. Not only does it provide the desired coverage, it also produces confidence sets that are sparse and generally smaller than the existing methods, including the debiased approach. { \label{para:computation} We also would like to note that the computational complexity of the proposed repro samples procedure is $O(dp^3),$ while for debiased Lasso it is $O(p^4).$ Therefore the proposed procedure is likely to be more efficient computationally in high-dimensional settings when $p\gg n.$ }
\end{enumerate}

Finally, there are several potential directions for extensions of the work.
First,  it is possible to extend the proposed approach to a generalized linear model,
{\label{para:glm}where two challenges arise: (a) the geometry between the underlying linear function, error term and the response is much more complicated;
  (b) the conditional approach to handle nuisance regression parameters does not directly apply. We have reported on how to overcome these challenges for binary regression in separate papers \citep{hou2024repro, hou2025repro}.}
{\label{para:general_T}Second,  an interesting extension is to
robust models such as the median regression or more generally the quantile regression. Without loss of generality, consider inference on a median regression, we have    $\sum_{i=1}^n I(Y_i - \*X_i\bbeta < 0)  -\sum_{i=1}^n U_i  = 0,$
where $U_i \sim Bernoulli(0.5).$ { And a natural choice of the nuclear mapping is  $T(\*U, \theta) = \sum_{i=1}^n U_i$ if joint inference for $\bbeta$ is our target.}  We refer the readers to the discussions in \cite{xie_repro_2022} for more examples and details on the nuclear mapping and repro samples approach.}
Another research direction concerns weak signals.
Although we do not impose any conditions on the signal strength, the proposed approach may demand a high computational cost to recover weak signals.
Therefore a natural question is, under limited computational resources, how to adjust the proposed approach for weak signals.
Additionally,
the identifiability condition or $C_{\min} >0$ ensures that there is no perfect co-linearity between the true model and an alternative model of equal size. When there is, then multiple equivalent ``true'' models exist. Our procedure is still valid to cover one of these ``true'' models.  However, constructing a confident set to cover all of these equivalent ``true'' models is an open problem for future research.

\section{Acknowledgements} The authors wish to thank the editors and reviewers for their constructive suggestions that have helped significantly improve the paper.
They would also like to thank Professor Cun-Hui Zhang
for his insightful knowledge and in-depth discussions, and for sharing the R code used in his seminal paper \cite{zhang_confidence_2014}. The results shown here are in part based upon data generated by the TCGA Research Network: https://www.cancer.gov/tcga.

\bibliographystyle{imsart-number}
\bibliography{Library_Final}

@book{shafer1976mathematical,
  title={A Mathematical Theory of Evidence},
  author={Shafer, G},
  year={1976},
  publisher={Princeton University Press}
}

@book{buhlmann2011statistics,
  title={Statistics for High-dimensional Data: Methods, Theory and Applications},
  author={B{\"u}hlmann, Peter and Van De Geer, Sara},
  year={2011},
  publisher={Springer Science \& Business Media}
}

@article{wang2020beamforming,
  title={Beamforming with small-spacing microphone arrays using constrained/generalized LASSO},
  author={Wang, Xianghui and Benesty, Jacob and Chen, Jingdong and Cohen, Israel},
  journal={IEEE Signal Processing Letters},
  volume={27},
  pages={356--360},
  year={2020},
  publisher={IEEE}
}

@article{leiner2021data,
  title={Data fission: splitting a single data point},
  author={Leiner, James and Duan, Boyan and Wasserman, Larry and Ramdas, Aaditya},
  journal={arXiv preprint arXiv:2112.11079},
  year={2021}
}

@article{tibshirani2016exact,
  title={Exact post-selection inference for sequential regression procedures},
  author={Tibshirani, Ryan J and Taylor, Jonathan and Lockhart, Richard and Tibshirani, Robert},
  journal={Journal of the American Statistical Association},
  volume={111},
  number={514},
  pages={600--620},
  year={2016},
  publisher={Taylor \& Francis}
}

@article{taylor2015statistical,
  title={Statistical learning and selective inference},
  author={Taylor, Jonathan and Tibshirani, Robert J},
  journal={Proceedings of the National Academy of Sciences},
  volume={112},
  number={25},
  pages={7629--7634},
  year={2015},
  publisher={National Acad Sciences}
}

@article{tibshirani1996regression,
  title={Regression shrinkage and selection via the lasso},
  author={Tibshirani, Robert},
  journal={Journal of the Royal Statistical Society: Series B (Methodological)},
  volume={58},
  number={1},
  pages={267--288},
  year={1996},
  publisher={Wiley Online Library}
}

@article{zhao2006model,
  title={On model selection consistency of Lasso},
  author={Zhao, Peng and Yu, Bin},
  journal={The Journal of Machine Learning Research},
  volume={7},
  pages={2541--2563},
  year={2006},
  publisher={JMLR. org}
}

@article{candes2007dantzig,
  title={The Dantzig selector: Statistical estimation when p is much larger than n},
  author={Candes, Emmanuel and Tao, Terence},
  journal={The annals of Statistics},
  volume={35},
  number={6},
  pages={2313--2351},
  year={2007},
  publisher={Institute of Mathematical Statistics}
}

@article{wang2019precision,
  title={Precision Lasso: accounting for correlations and linear dependencies in high-dimensional genomic data},
  author={Wang, Haohan and Lengerich, Benjamin J and Aragam, Bryon and Xing, Eric P},
  journal={Bioinformatics},
  volume={35},
  number={7},
  pages={1181--1187},
  year={2019},
  publisher={Oxford Academic}
}

@article{korobilis2021high,
  title={High-dimensional macroeconomic forecasting using message passing algorithms},
  author={Korobilis, Dimitris},
  journal={Journal of Business \& Economic Statistics},
  volume={39},
  number={2},
  pages={493--504},
  year={2021},
  publisher={Taylor \& Francis}
}

@article{belloni2018high,
  title={High-dimensional econometrics and regularized GMM},
  author={Belloni, Alexandre and Chernozhukov, Victor and Chetverikov, Denis and Hansen, Christian and Kato, Kengo},
  journal={arXiv preprint arXiv:1806.01888},
  year={2018}
}

@article{athey_approximate_2018,
  title = {Approximate {{Residual Balancing}}: {{De-Biased Inference}} of {{Average Treatment Effects}} in {{High Dimensions}}},
  shorttitle = {Approximate {{Residual Balancing}}},
  author = {Athey, Susan and Imbens, Guido W. and Wager},
  year = {2018},
  month = jan,
  journal = {arXiv:1604.07125 [econ, math, stat]},
  eprint = {1604.07125},
  eprinttype = {arxiv},
  primaryclass = {econ, math, stat},
  archiveprefix = {arXiv}
}

@article{beaumont_approximate_2002,
  title = {Approximate {{Bayesian}} Computation in Population Genetics},
  author = {Beaumont, Mark A. and Zhang, Wenyang and Balding, David J.},
  year = {2002},
  journal = {Genetics},
  volume = {162},
  number = {4},
  pages = {2025--2035},
  publisher = {{Genetics Soc America}}
}

@article{bertsimas_best_2016,
  title = {Best {{Subset Selection}} via a {{Modern Optimization Lens}}},
  author = {Bertsimas, Dimitris and King, Angela and Mazumder, Rahul},
  year = {2016},
  month = apr,
  journal = {The Annals of Statistics},
  volume = {44},
  number = {2},
  pages = {813--852},
  issn = {0090-5364},
  doi = {10.1214/15-AOS1388},
  langid = {english}
}

@article{butler_exth-58_2019,
  title = {Exth-58. {{Inhibition Of Dna Topoisomerase}} 1 {{And Poly}} ({{Adp-ribose}}) {{Polymerase Synergistically Induces Cell Death In Glioblastoma With Pten Loss}}},
  author = {Butler, Madison and Su, Yu-Ting and Hwang, Lee and Marzi, Laetitia and Gilbert, Mark and Pommier, Yves and Wu, Jing},
  year = {2019},
  journal = {Neuro-Oncology},
  volume = {21},
  number = {Supplement\_6},
  pages = {vi94--vi95},
  publisher = {{Oxford University Press US}}
}

@article{cai_confidence_2017,
  title = {Confidence Intervals for High-Dimensional Linear Regression: {{Minimax}} Rates and Adaptivity},
  shorttitle = {Confidence Intervals for High-Dimensional Linear Regression},
  author = {Cai, T. Tony and Guo, Zijian},
  year = {2017},
  journal = {The Annals of statistics},
  volume = {45},
  number = {2},
  pages = {615--646},
  publisher = {{Institute of Mathematical Statistics}}
}

@article{liu2013gaussian,
  title={Gaussian graphical model estimation with false discovery rate control},
  author={Liu, Weidong},
  journal={The Annals of Statistics},
  volume={41},
  number={6},
  pages={2948--2978},
  year={2013},
  publisher={Institute of Mathematical Statistics}
}

@article{candes_panning_2018,
  title = {Panning for Gold: `Model-{{X}}' Knockoffs for High Dimensional Controlled Variable Selection},
  shorttitle = {Panning for Gold},
  author = {Cand{\`e}s, Emmanuel and Fan, Yingying and Janson, Lucas and Lv, Jinchi},
  year = {2018},
  month = jun,
  journal = {Journal of the Royal Statistical Society: Series B (Statistical Methodology)},
  volume = {80},
  number = {3},
  pages = {551--577},
  issn = {1467-9868},
  doi = {10.1111/rssb.12265},
  langid = {english}
}

@article{chatterjee_bootstrapping_2011,
  title = {Bootstrapping {{Lasso Estimators}}},
  author = {Chatterjee, A. and Lahiri, S. N.},
  year = {2011},
  month = jun,
  journal = {Journal of the American Statistical Association},
  volume = {106},
  number = {494},
  pages = {608--625},
  issn = {0162-1459},
  doi = {10.1198/jasa.2011.tm10159}
}

@article{chen_extended_2008-1,
  title = {Extended {{Bayesian Information Criteria}} for {{Model Selection}} with {{Large Model Spaces}}},
  author = {Chen, Jiahua and Chen, Zehua},
  year = {2008},
  journal = {Biometrika},
  volume = {95},
  number = {3},
  pages = {759--771},
  issn = {0006-3444}
}

@article{chernozhukov_doubledebiasedneyman_2017,
  title = {Double/Debiased/Neyman Machine Learning of Treatment Effects},
  author = {Chernozhukov, Victor and Chetverikov, Denis and Demirer, Mert and Duflo, Esther and Hansen, Christian and Newey, Whitney},
  year = {2017},
  journal = {American Economic Review},
  volume = {107},
  number = {5},
  pages = {261--65}
}

@article{chernozhukov_post-selection_2015,
  title = {Post-Selection and Post-Regularization Inference in Linear Models with Many Controls and Instruments},
  author = {Chernozhukov, Victor and Hansen, Christian and Spindler, Martin},
  year = {2015},
  journal = {American Economic Review},
  volume = {105},
  number = {5},
  pages = {486--90}
}

@article{dezeure_high-dimensional_2017,
  title = {High-Dimensional Simultaneous Inference with the Bootstrap},
  author = {Dezeure, Ruben and B{\"u}hlmann, Peter and Zhang, Cun-Hui},
  year = {2017},
  month = dec,
  journal = {TEST},
  volume = {26},
  number = {4},
  pages = {685--719},
  issn = {1133-0686, 1863-8260},
  doi = {10.1007/s11749-017-0554-2},
  langid = {english}
}

@incollection{efron_bootstrap_1992,
  title = {Bootstrap Methods: Another Look at the Jackknife},
  shorttitle = {Bootstrap Methods},
  booktitle = {Breakthroughs in Statistics},
  author = {Efron, Bradley},
  year = {1992},
  pages = {569--593},
  publisher = {{Springer}}
}

@article{fan_sure_2008,
  title = {Sure Independence Screening for Ultrahigh Dimensional Feature Space},
  author = {Fan, Jianqing and Lv, Jinchi},
  year = {2008},
  month = nov,
  journal = {Journal of the Royal Statistical Society: Series B (Statistical Methodology)},
  volume = {70},
  number = {5},
  pages = {849--911},
  issn = {1467-9868},
  doi = {10.1111/j.1467-9868.2008.00674.x},
  copyright = {\textcopyright{} 2008 Royal Statistical Society},
  langid = {english}
}

@article{fan_variable_2001,
  title = {Variable {{Selection}} via {{Nonconcave Penalized Likelihood}} and Its {{Oracle Properties}}},
  author = {Fan, Jianqing and Li, Runze},
  year = {2001},
  month = dec,
  journal = {Journal of the American Statistical Association},
  volume = {96},
  number = {456},
  pages = {1348--1360},
  issn = {0162-1459, 1537-274X},
  doi = {10.1198/016214501753382273},
  langid = {english}
}

@article{ferrari_confidence_2015,
  title = {Confidence Sets for Model Selection by {{F-testing}}},
  author = {Ferrari, Davide and Yang, Yuhong},
  year = {2015},
  journal = {Statistica Sinica},
  pages = {1637--1658}
}

@article{guo_group_2020,
  title = {Group {{Inference}} in {{High Dimensions}} with {{Applications}} to {{Hierarchical Testing}}},
  author = {Guo, Zijian and Renaux, Claude and B{\"u}hlmann, Peter and Cai, T. Tony},
  year = {2020},
  month = mar,
  journal = {arXiv:1909.01503 [stat]},
  eprint = {1909.01503},
  eprinttype = {arxiv},
  primaryclass = {stat},
  archiveprefix = {arXiv}
}

@article{guo_optimal_2019,
  title = {Optimal Estimation of Genetic Relatedness in High-Dimensional Linear Models},
  author = {Guo, Zijian and Wang, Wanjie and Cai, T. Tony and Li, Hongzhe},
  year = {2019},
  journal = {Journal of the American Statistical Association},
  volume = {114},
  number = {525},
  pages = {358--369},
  publisher = {{Taylor \& Francis}}
}

@article{hannig_generalized_2016,
  title = {Generalized {{Fiducial Inference}}: {{A Review}} and {{New Results}}},
  shorttitle = {Generalized {{Fiducial Inference}}},
  author = {Hannig, Jan and Iyer, Hari and Lai, Randy C. S. and Lee, Thomas C. M.},
  year = {2016},
  month = jul,
  journal = {Journal of the American Statistical Association},
  volume = {111},
  number = {515},
  pages = {1346--1361},
  issn = {0162-1459, 1537-274X},
  doi = {10.1080/01621459.2016.1165102},
  langid = {english}
}

@article{hansen_model_2011-1,
  title = {The {{Model Confidence Set}}},
  author = {Hansen, Peter R. and Lunde, Asger and Nason, James M.},
  year = {2011},
  journal = {Econometrica},
  volume = {79},
  number = {2},
  pages = {453--497},
  issn = {1468-0262},
  doi = {10.3982/ECTA5771},
  langid = {english}
}

@article{javanmard_confidence_2014,
  title = {Confidence Intervals and Hypothesis Testing for High-Dimensional Regression},
  author = {Javanmard, Adel and Montanari, Andrea},
  year = {2014},
  journal = {The Journal of Machine Learning Research},
  volume = {15},
  number = {1},
  pages = {2869--2909},
  publisher = {{JMLR. org}}
}

@article{javanmard_flexible_2019,
  title = {A {{Flexible Framework}} for {{Hypothesis Testing}} in {{High-dimensions}}},
  author = {Javanmard, Adel and Lee, Jason D.},
  year = {2019},
  month = sep,
  journal = {arXiv:1704.07971 [cs, math, stat]},
  eprint = {1704.07971},
  eprinttype = {arxiv},
  primaryclass = {cs, math, stat},
  archiveprefix = {arXiv}
}

@article{meinshausen_p-values_2009,
  title = {P-Values for High-Dimensional Regression},
  author = {Meinshausen, Nicolai and Meier, Lukas and B{\"u}hlmann, Peter},
  year = {2009},
  journal = {Journal of the American Statistical Association},
  volume = {104},
  number = {488},
  pages = {1671--1681},
  publisher = {{Taylor \& Francis}}
}

@article{nickl_confidence_2013,
  title = {Confidence Sets in Sparse Regression},
  author = {Nickl, Richard and Van De Geer, Sara},
  year = {2013},
  journal = {The Annals of Statistics},
  volume = {41},
  number = {6},
  pages = {2852--2876},
  publisher = {{Institute of Mathematical Statistics}}
}

@article{sarcar_vorinostat_2010,
  title = {Vorinostat Enhances the Cytotoxic Effects of the Topoisomerase {{I}} Inhibitor {{SN38}} in Glioblastoma Cell Lines},
  author = {Sarcar, Bhaswati and Kahali, Soumen and Chinnaiyan, Prakash},
  year = {2010},
  journal = {Journal of Neuro-oncology},
  volume = {99},
  number = {2},
  pages = {201--207},
  publisher = {{Springer}}
}

@article{shen_constrained_2013,
  title = {On Constrained and Regularized High-Dimensional Regression},
  author = {Shen, Xiaotong and Pan, Wei and Zhu, Yunzhang and Zhou, Hui},
  year = {2013},
  month = oct,
  journal = {Annals of the Institute of Statistical Mathematics},
  volume = {65},
  number = {5},
  pages = {807--832},
  issn = {0020-3157, 1572-9052},
  doi = {10.1007/s10463-012-0396-3},
  langid = {english}
}

@article{shen_likelihood-based_2012-1,
  title = {Likelihood-{{Based Selection}} and {{Sharp Parameter Estimation}}},
  author = {Shen, Xiaotong and Pan, Wei and Zhu, Yunzhang},
  year = {2012},
  month = mar,
  journal = {Journal of the American Statistical Association},
  volume = {107},
  number = {497},
  pages = {223--232},
  issn = {0162-1459},
  doi = {10.1080/01621459.2011.645783}
}

@article{van_de_geer_asymptotically_2014,
  title = {On Asymptotically Optimal Confidence Regions and Tests for High-Dimensional Models},
  author = {{Van de Geer}, Sara and B{\"u}hlmann, Peter and Ritov, Ya'acov and Dezeure, Ruben},
  year = {2014},
  journal = {The Annals of Statistics},
  volume = {42},
  number = {3},
  pages = {1166--1202},
  publisher = {{Institute of Mathematical Statistics}}
}

@article{wang_exth-09_2017,
  title = {Exth-09. {{Tdp1}}/Top1 {{Ratio As A Predictive Indicator For The Response Of Glioblastoma Cancer Cells To Irinotecan Treatment}}},
  author = {Wang, Wenjie and Silva, Monica Rodriguez and Chambers, Jeremy and {Tse-Dinh}, Yuk-Ching},
  year = {2017},
  journal = {Neuro-oncology},
  volume = {19},
  number = {Suppl 6},
  pages = {vi74},
  publisher = {{Oxford University Press}}
}

@article{wang_random_2011,
  title = {Random Lasso},
  author = {Wang, Sijian and Nan, Bin and Rosset, Saharon and Zhu, Ji},
  year = {2011},
  month = mar,
  journal = {The Annals of Applied Statistics},
  volume = {5},
  number = {1},
  pages = {468--485},
  issn = {1932-6157},
  doi = {10.1214/10-AOAS377},
  langid = {english}
}

@article{xie_confidence_2013,
  title = {Confidence Distribution, the Frequentist Distribution Estimator of a Parameter: {{A}} Review},
  shorttitle = {Confidence Distribution, the Frequentist Distribution Estimator of a Parameter},
  author = {Xie, Minge and Singh, Kesar},
  year = {2013},
  journal = {International Statistical Review},
  volume = {81},
  number = {1},
  pages = {3--39}
}

@article{xie_repro_2022,
  title = {Repro {{Samples Method}} for {{Finite-}} and {{Large-Sample Inferences}}},
  author = {Xie, Minge and Wang, Peng},
  year = {2022},
  month = jun,
  journal = {arXiv e-prints},
  eprint = {2206.06421},
  eprinttype = {arxiv},
  primaryclass = {math, stat},
  pages = {arXiv.2206.06421 (Invited revision for The Journal of the American Statistical Association)},
  doi = {10.48550/arXiv.2206.06421},
  archiveprefix = {arXiv}
}

@article{zhang_confidence_2014,
  title = {Confidence Intervals for Low Dimensional Parameters in High Dimensional Linear Models},
  author = {Zhang, Cun-Hui and Zhang, Stephanie S.},
  year = {2014},
  month = jan,
  journal = {Journal of the Royal Statistical Society: Series B (Statistical Methodology)},
  volume = {76},
  number = {1},
  pages = {217--242},
  issn = {13697412},
  doi = {10.1111/rssb.12026},
  langid = {english}
}

@article{zhang_nearly_2010,
  title = {Nearly Unbiased Variable Selection under Minimax Concave Penalty},
  author = {Zhang, Cun-Hui},
  year = {2010},
  journal = {The Annals of statistics},
  volume = {38},
  number = {2},
  pages = {894--942}
}

@article{zhang_simultaneous_2017,
  title = {Simultaneous {{Inference}} for {{High-Dimensional Linear Models}}},
  author = {Zhang, Xianyang and Cheng, Guang},
  year = {2017},
  month = apr,
  journal = {Journal of the American Statistical Association},
  volume = {112},
  number = {518},
  pages = {757--768},
  issn = {0162-1459, 1537-274X},
  doi = {10.1080/01621459.2016.1166114},
  langid = {english}
}

@article{zhou_honest_2019,
  title = {Honest Confidence Sets for High-Dimensional Regression by Projection and Shrinkage},
  author = {Zhou, Kun and Li, Ker-Chau and Zhou, Qing},
  year = {2019},
  journal = {arXiv preprint arXiv:1902.00535},
  eprint = {1902.00535},
  eprinttype = {arxiv},
  archiveprefix = {arXiv}
}

@article{zhu_high-dimensional_2020,
  title = {On High-Dimensional Constrained Maximum Likelihood Inference},
  author = {Zhu, Yunzhang and Shen, Xiaotong and Pan, Wei},
  year = {2020},
  journal = {Journal of the American Statistical Association},
  volume = {115},
  number = {529},
  pages = {217--230},
  publisher = {{Taylor \& Francis}}
}

@article{zhu_linear_2018,
  title = {Linear Hypothesis Testing in Dense High-Dimensional Linear Models},
  author = {Zhu, Yinchu and Bradic, Jelena},
  year = {2018},
  journal = {Journal of the American Statistical Association},
  volume = {113},
  number = {524},
  pages = {1583--1600},
  publisher = {{Taylor \& Francis}}
}

@article{zhu_projection_2017,
  title = {A Projection Pursuit Framework for Testing General High-Dimensional Hypothesis},
  author = {Zhu, Yinchu and Bradic, Jelena},
  year = {2017},
  month = may,
  journal = {arXiv:1705.01024 [math, stat]},
  eprint = {1705.01024},
  eprinttype = {arxiv},
  primaryclass = {math, stat},
  archiveprefix = {arXiv}
}

@article{zou_adaptive_2006,
  title = {The {{Adaptive Lasso}} and {{Its Oracle Properties}}},
  author = {Zou, Hui},
  year = {2006},
  month = dec,
  journal = {Journal of the American Statistical Association},
  volume = {101},
  number = {476},
  pages = {1418--1429},
  issn = {0162-1459, 1537-274X},
  doi = {10.1198/016214506000000735},
  langid = {english}
}

@book{Martin2015,
	Author = {Martin, Ryan  and  Liu, Chuanhai},
	Publisher = {Chapman \& Hall/CRC},
	Title = {Inferential Models: Reasoning with Uncertainty},
	Year = {2015}}

@misc{mont_code_nodate,
  author = {Javanmard, Adel and Montanari, Andrea},
	title = {Confidence Intervals and Hypothesis Testing for High-Dimensional Regression},
	howpublished = {\url{https://web.stanford.edu/~montanar/sslasso/code.html}},
	urldate = {2022-09-29},
}

@article{das_perturbation_2019_aos,
  title = {Perturbation Bootstrap in Adaptive Lasso},
  author = {Das, Debraj and Gregory, Karl and Lahiri, S. N.},
  year = {2019},
  journal = {The Annals of Statistics},
  volume = {47},
  number = {4},
  pages = {2080--2116},
  publisher = {{Institute of Mathematical Statistics}}
}

@article{das_distributional_2019_bio,
  title = {Distributional Consistency of the Lasso by Perturbation Bootstrap},
  author = {Das, Debraj and Lahiri, S N},
  year = {2019},
  month = dec,
  journal = {Biometrika},
  volume = {106},
  number = {4},
  pages = {957--964},
  issn = {0006-3444, 1464-3510},
  doi = {10.1093/biomet/asz029},
  langid = {english}
}

@article{craiu_approximate_2023,
	title = {Approximate {Methods} for {Bayesian} {Computation}},
	volume = {10},
	issn = {2326-8298, 2326-831X},
	url = {https://www.annualreviews.org/doi/10.1146/annurev-statistics-033121-110254},
	doi = {10.1146/annurev-statistics-033121-110254},
	language = {en},
	number = {1},
	urldate = {2022-12-21},
	journal = {Annual Review of Statistics and Its Application},
	author = {Craiu, Radu V. and Levi, Evgeny},
	month = mar,
	year = {2023},
	pages = null,
}

@article{lange_robust_1989,
	title = {Robust {Statistical} {Modeling} {Using} the t {Distribution}},
	volume = {84},
	issn = {0162-1459},
	url = {https://www.jstor.org/stable/2290063},
	doi = {10.2307/2290063},
	number = {408},
	urldate = {2024-07-17},
	journal = {Journal of the American Statistical Association},
	author = {Lange, Kenneth L. and Little, Roderick J. A. and Taylor, Jeremy M. G.},
	year = {1989},
	pages = {881--896},
}

@article{pek_how_2018,
	title = {How to {Address} {Non}-normality: {A} {Taxonomy} of {Approaches}, {Reviewed}, and {Illustrated}},
	volume = {9},
	issn = {1664-1078},
	shorttitle = {How to {Address} {Non}-normality},
	url = {https://www.frontiersin.org/journals/psychology/articles/10.3389/fpsyg.2018.02104/full},
	doi = {10.3389/fpsyg.2018.02104},
	language = {English},
	urldate = {2024-07-17},
	journal = {Frontiers in Psychology},
	author = {Pek, Jolynn and Wong, Octavia and Wong, Augustine C. M.},
	month = nov,
	year = {2018},
	note = {Publisher: Frontiers},
}

@article{lemdani_likelihood_1999,
	title = {Likelihood {Ratio} {Tests} in {Contamination} {Models}},
	volume = {5},
	issn = {13507265},
	url = {https://www.jstor.org/stable/3318698?origin=crossref},
	doi = {10.2307/3318698},
	language = {en},
	number = {4},
	urldate = {2024-07-17},
	journal = {Bernoulli},
	author = {Lemdani, Mohamed and Pons, Odile},
	month = aug,
	year = {1999},
	pages = {705},
}

@article{dai_inferences_2007,
	title = {Inferences in {Contaminated} {Regression} and {Density} {Models}},
	volume = {69},
	issn = {0972-7671},
	url = {https://www.jstor.org/stable/25664592},
	number = {4},
	urldate = {2024-07-17},
	journal = {Sankhyā: The Indian Journal of Statistics (2003-2007)},
	author = {Dai, Hongying and Charnigo, Richard},
	year = {2007},
	note = {Publisher: Springer},
	pages = {842--869},
}

@article{lei_cross-validation_2020,
	title = {Cross-{Validation} {With} {Confidence}},
	volume = {115},
	issn = {0162-1459, 1537-274X},
	url = {https://www.tandfonline.com/doi/full/10.1080/01621459.2019.1672556},
	doi = {10.1080/01621459.2019.1672556},
	language = {en},
	number = {532},
	urldate = {2024-07-17},
	journal = {Journal of the American Statistical Association},
	author = {Lei, Jing},
	month = oct,
	year = {2020},
	pages = {1978--1997},
}

@article{hou2024repro,
  title={Repro Samples Method for High-dimensional Logistic Model},
  author={Hou, Xiaotian and Zhang, Linjun and Wang, Peng and Xie, Minge},
  journal={arXiv preprint arXiv:2403.09984},
  year={2024}
}

@article{li_model_2019,
  title = {Model Confidence Bounds for Variable Selection},
  author = {Li, Yang and Luo, Yuetian and Ferrari, Davide and Hu, Xiaonan and Qin, Yichen},
  year = {2019},
  journal = {Biometrics},
  volume = {75},
  number = {2},
  pages = {392--403},
  issn = {1541-0420},
  doi = {10.1111/biom.13024},
  urldate = {2024-09-06},
  langid = {english}
}

@article{segura_sharp_2016,
	title = {Sharp bounds for cumulative distribution functions},
	volume = {436},
	issn = {0022-247X},
	url = {https://www.sciencedirect.com/science/article/pii/S0022247X15011579},
	doi = {10.1016/j.jmaa.2015.12.024},
	number = {2},
	urldate = {2024-10-08},
	journal = {Journal of Mathematical Analysis and Applications},
	author = {Segura, Javier},
	month = apr,
	year = {2016},
	pages = {748--763},
}

@article{williams2019assumptions,
  title={Assumptions of multiple regression: Correcting two misconceptions},
  author={Williams, Matt N and Grajales, Carlos Alberto G{\'o}mez and Kurkiewicz, Dason},
  journal={Practical Assessment, Research, and Evaluation},
  volume={18},
  number={1},
  pages={11},
  year={2019}
}

@incollection{thornton2024bridging,
  title={Bridging Bayesian, frequentist and fiducial inferences using confidence distributions},
  author={Thornton, Suzanne and Xie, Minge},
  booktitle={Handbook of Bayesian, Fiducial, and Frequentist Inference},
  pages={106--131},
  year={2024},
  publisher={Chapman and Hall/CRC}
}

@article{liu2007variable,
  title={Variable selection via a combination of the l 0 and l 1 penalties},
  author={Liu, Yufeng and Wu, Yichao},
  journal={Journal of Computational and Graphical Statistics},
  volume={16},
  number={4},
  pages={782--798},
  year={2007},
  publisher={Taylor \& Francis}
}

@article{hsu2012tail,
  title={A tail inequality for quadratic forms of subgaussian random vectors},
  author={Hsu, Daniel and Kakade, Sham and Zhang, Tong},
  year={2012},
 journal={Electronic Communications in Probability},
  volume={17},
  pages={1--6},
}

@article{li2023glioma,
  title={Glioma-derived LRIG3 interacts with NETO2 in tumor-associated macrophages to modulate microenvironment and suppress tumor growth},
  author={Li, Youwei and Wang, Wei and Hou, Xiaoshuang and Huang, Wenda and Zhang, Po and He, Yue and Wang, Baofeng and Duan, Qiuhong and Mao, Feng and Guo, Dongsheng},
  journal={Cell Death \& Disease},
  volume={14},
  number={1},
  pages={28},
  year={2023},
  publisher={Nature Publishing Group UK London}
}

@article{hou2025repro,
    title={Repro Samples Method for Model-Free Inference in High-Dimensional Binary Classification},
    author={Xiaotian Hou and Peng Wang and Minge Xie and Linjun Zhang},
    year={2025},
    journal={arXiv preprint arXiv:2510.01468},
}

 \normalsize
\newpage
\begin{appendix}
\begin{center}
\LARGE
Supplemental Materials
\end{center}

In these supplementary materials, in Appendix~\ref{sec:constraint}, we present an alternative formulation of using the constrained regression approach to find candidate models described in Section~\ref{sec_cand} and the relevant theoretical results. In Appendix~\ref{sec:joint_inference_model_coefficients}, we present a method for joint inference for model and regression coefficients. This complements the method we developed in Section~\ref{sec:coef}. In Appendix~\ref{sec:app_proof}, we present the technical proofs for Lemma~\ref{lemma_def1} and  the results with assumptions of Gaussian errors, including  Lemmas~\ref{lem::finite_penalize}, \ref{lem::finite}, \ref{lem::finite_penalize_complete}, \ref{lem::finite_complete}, \ref{lem::symptotic_bound_penalize0} and Theorems \ref{the:finite_pen}, \ref{the:aymp_pen}, \ref{the:finite_con},  \ref{the:aymp_con}, \ref{the:size}, \ref{thm:cond_rps}- \ref{the:confidence_set_asymptotic}, \ref{cor:coverage_beta_Lambda}, \ref{cor:coverage_beta_Lambda_asymp}, \ref{the:coverage_eta} and \ref{the:coverage_eta_asym}. In Appendix~\ref{sec:non_normal}, we provide the technical proofs for results developed for the non-Gaussian errors, including Theorems~\ref{the:uw}, \ref{the:sub_finite_pen}, \ref{the:sub_aymp_pen} and Corollary~\ref{cor:nonGaussian}--\ref{cor:second_moment_u}. In Appendix~\ref{sec:size_d}, we explore the number of the repro samples in Algorithm~\ref{alg:candidate}
to ensure a high probability that the model candidate set includes the true model. Appendix~\ref{sec:joint_beta_full} contain details in  choosing the tuning parameter $\lambda$ in Algorithm~\ref{alg:candidate}, additional simulation results and visualizations.

\section{An alternative formulation in Section~\ref{sec_cand}}
\label{sec:constraint}

In Section~\ref{sec:finding_cand}, we obtain the candidate set by solving the objective function \eqref{eq::obj}.
In addition to \eqref{eq::obj}, there is also an almost equivalent form that imposes a constraint on $|\tau| = \|\bbeta_\tau\|_0$ other than adding a regularization term, i.e.
\begin{align}
    \label{eq::obj_constraint}
    \min \|\*y_{obs}- X_\tau\bbeta_{\tau}-\sigma \*U^*_{b}\|^2_2, \quad
    \hbox{s.t.  }  |\tau| \leq k,
\end{align}
where $k$ is a constraint on the model size, playing a similar role as the $\lambda$ in \eqref{eq::obj}. One can opt to use \eqref{eq::obj_constraint} in Step 2 of Algorithm~\ref{alg:candidate}. Similarly, We can obtain the following results, where Theorem~\ref{the:finite_con}, Lemma~\ref{lem::finite}, Lemma~\ref{lem::finite_complete}, Theorem~\ref{the:aymp_con} and Lemma~\ref{lem::asymptotic_constraint} are  counterparts of Theorem~\ref{the:finite_pen}, Lemma~\ref{lem::finite_penalize}, Lemma~\ref{lem::finite_penalize_complete}, Theorem~\ref{the:aymp_pen}, and Lemma~\ref{lem::symptotic_bound_penalize0} respectively. We provide the proofs of the following theorems and lemmas in Appendix~\ref{sec:app_proof}.

\begin{theorem}
\label{the:finite_con}
For any $\delta>0$, there exists a  constant $ \gamma_\delta$ such that under the constraint $|\tau| \leq |\tau_0|,$ the finite-sample probability bound that the true model is not covered by the model candidates set $S^{(d)}$, obtained by Algorithm~\ref{alg:candidate} with the objective function \eqref{eq::obj_constraint}, is as follows,
\begin{align*}
    \P_{({\cal U}^d, \*Y)}(\tau_0 \notin S^{(d)} ) \leq \left\{1- \frac{(\gamma_{\delta})^{n-1}}{n-1}\right\}^d + \delta.
\end{align*}
Therefore as $d \rightarrow \infty,$ $ \P_{({\cal U}^d, \*Y)}(\tau_0 \notin S^{(d)} ) \rightarrow 0. $
\end{theorem}

\begin{lemma}
    \label{lem::finite}
 Suppose $n-|\tau_0| > 4.$ Under the constraint $|\tau| \leq |\tau_0|,$ let $\*U^*$ be a random repro sample of $\*U$, such that $\*U^*, \*U \sim N(0, \*I_n),$ and
$$\hat \tau_{\*U^*} = \argmin_{\{\tau| |\tau| \leq |\tau_0|\}}\left\{\min_{\bbeta_\tau, \sigma} \|\*Y - \*X_{\tau} \bbeta_{\tau} -\sigma \*U^* \|^2\right\}.$$
Then for any $0<\gamma_2< 1/64$ such that $C_{\min} > 24\sqrt\gamma_2\left( \frac{ \log(p/2)}{n}+ \gamma_2\right) \sigma_0^2,$
{
\begin{align*}
    & \P_{{(\*U, \*U^*)}}\left\{\hat \tau_{\*U^*} \neq \tau_0| \rho(\*U^*, \*U) > 1- \gamma_2^2\right\} \\
    &  \leq 3\exp\left\{- \frac{n}{12\sigma_0^2}\left[ \frac{ C_{\min}}{\sqrt\gamma_2} - 24\left( \frac{  \log(p/2)}{n}+ \gamma_2\right) \sigma_0^2\right]
    \right\} +  4(64\gamma_2)^{\frac{n-|\tau_0|-1}{6}} p^{|\tau_0|}.
\end{align*}
}
\end{lemma}

\begin{lemma}
    \label{lem::finite_complete}
Suppose $n-|\tau_0| > 4.$ Then for any $0<\gamma_2< 1/64$ such that $C_{\min} >
24\sqrt\gamma_2\left( \frac{ \log(p/2)}{n}+ \gamma_2\right) \sigma_0^2,$ the finite-sample probability bound that the true model is not covered by the model candidates set $S^{(d)}$, obtained by Algorithm~\ref{alg:candidate} with the objective function \eqref{eq::obj_constraint} under the constraint  $|\tau| \leq |\tau_0|$, is as follows,
\begin{align}
\label{eq:prob_bound_constraint}
   \P_{({\cal U}^d, \*Y)} (\tau_0 \notin S^{(d)} )  & \leq 3\exp\left\{- \frac{n}{12\sigma_0^2}\left[ \frac{ C_{\min}}{\sqrt\gamma_2} - 24\left( \frac{  \log(p/2)}{n}+ \gamma_2\right) \sigma_0^2\right]
    \right\} \nonumber
    \\& +   4(64\gamma_2)^{\frac{n-|\tau_0|-1}{6}} p^{|\tau_0|}  + \left(1-\frac{\gamma_2^{n-1}}{n-1}\right)^d.
\end{align}
\end{lemma}

\begin{theorem}
    \label{the:aymp_con}
Under the constraint $|\tau| \leq |\tau_0|$, the probability bound that the true model is not covered by the model candidates set $S^{(d)}$, obtained by Algorithm~\ref{alg:candidate} with the objective function \eqref{eq::obj_constraint} for any finite $d$ is as follows,
\begin{align}
\label{eq:prob_asympto_con}
   \P_{({\cal U}^d, \*Y)} (\tau_0 \notin S^{(d)} ) \leq 6\exp\left[- \frac{n}{18\sigma_0^2}\{0.3C_{\min}- 36 \frac{\log (p)}{n}\sigma_0^2\}\right] \nonumber\\
   + \exp\left\{-nd\left(0.23-\frac{|\tau_0|\log(p)+2}{n}\right) \right\}.
\end{align}
Therefore  $\P_{({\cal U}^d, \*Y)} (\tau_0 \notin S^{(d)} ) \rightarrow 0$ for any $d$ as $n \rightarrow \infty$ , if $\frac{|\tau_0|\log(p)}{n} <0.23$ and $C_{\min}> 120 \frac{\log (p+1)}{n}\sigma_0^2$ when $n$ is large enough.
\end{theorem}

\begin{lemma}
    \label{lem::asymptotic_constraint}
Under the constraint $|\tau| \leq |\tau_0|$, the finite-sample probability bound that the true model is not covered by the model candidate  set $S^{(d)}$, obtained by Algorithm~\ref{alg:candidate} with the objective function \eqref{eq::obj_constraint}, is as follows,
\begin{align}
\label{eq:prob_bound_asymp_constraint}
   \P_{({\cal U}^d, \*Y)} (\tau_0 \notin S^{(d)} )  \leq L(\gamma_1) +\left[2\{\arccos (\gamma_1)\}^{n-|\tau_0|-1 } p^{|\tau_0|}\right]^d
\end{align}
where
\begin{align*}
L(\gamma_1)=6\exp\left[-\frac{n}{18 \sigma_0^2}\left\{(1-\gamma_1^2)C_{\min} -36\frac{\log p}{n}\sigma_0^2\right\}\right],
\end{align*}
and $\cos(0.3\pi) < \gamma_1<1$ is any real number.
\end{lemma}

{
\section{Joint inference for model and regression coefficients}
\label{sec:joint_inference_model_coefficients}

Besides constructing confidence sets for the true model $\tau_0$ and certain regression coefficients $\beta_{\Lambda, 0}$ respectively, we are also able to construct joint confidence set for the model and coefficients $\bm \eta_0 =  (\tau_0, \bbeta_0).$  Specifically, let $\bm \eta_\tau = (\tau, \bbeta_\tau), $ we then follow \eqref{eq:nuclea_joint} to define the nuclear mapping as
\begin{align*}
 T(\*u, \bm\eta_\tau) & = \frac{ \*u^\top \*H_\tau \*u/|\tau|}{ \*u^\top (I -  \*H_\tau) \*u/(n-|\tau|)} \nonumber \\
  & =  \frac{ {(\*y_{\bm\theta}  -  \*X_{\tau} \bbeta_{\tau})}^\top \*H_\tau  (\*y_{\bm\theta}  -  \*X_{\tau} \bbeta_{\tau})/|\tau|}{ {(\*y_{\bm\theta}  -  \*X_{\tau} \bbeta_{\tau})}^\top (I -  \*H_\tau) (\*y_{\bm\theta} -  \*X_{\tau} \bbeta_{\tau})/(n-|\tau|)} = \tilde T(\*y_{\bm\theta} , \bm\eta_\tau).
\end{align*}
Then it follows immediately that  $\P_{\*U}\big\{\tilde T\big(\*Y_{\bm\theta}, \bm\eta_\tau\big)  \in  B_{\bm \eta_\tau}(\alpha) \big\} = \alpha$ if we let $B_{\bm \eta_\tau}(\alpha) =  \big[0,  F^{-1}_{|\tau|, n - |\tau|}(\alpha) \big].$

If we use the above nuclear mapping and follow a similar approach to \eqref{eq:cs_eta} to construct the joint confidence set for $\eta_0=(\tau_0, \bbeta_0),$ the resulting confidence set is not tight for the true model $\tau_0$ since it includes all models in the model candidate set. To make the joint confidence set informative about $\tau_0$,  we can limit $\tau$ in a level-$\alpha_1$ model confidence set $\Gamma_{\alpha_1}^\tau(\*y_{obs})$
 obtained in Section 2 using
 (\ref{eq:bar-CS}). Here, $\alpha_1 \in (\frac 12, 1)$ and close to $1$.
 Similarly, take another $\alpha_2 \in (\frac 12, 1)$, and let $\alpha =\alpha_1 + \alpha_2 - 1$. We then use a modified version of \eqref{eq:conf_joint} to construct the confidence set for $\eta_0 = (\tau_0, \bbeta_0)$:
 \begin{align}
 \label{eq:joint_tau_beta}
     {\Gamma}^{\bm \eta_\tau}_{\alpha}(y_{obs}) =  \hbox{$\bigcup_{\tau \in {\bar \Gamma}_{\alpha_1}^\tau(\*y_{obs})}$} \big\{\bm\eta_\tau: \tilde T(\*y_{obs}, \bm \eta_\tau) \in B_{\alpha_2}({\bm \eta_\tau}) \big\}.
 \end{align}

 The following Theorems \eqref{the:coverage_eta} and \eqref{the:coverage_eta_asym}
 guarantee  that  ${\Gamma}^{\bm \eta}_{\alpha}(y_{obs})$ is
 a level-$\alpha$ joint confidence set for $\bm\eta_0$.
 If, for instance, we take $\alpha_1 = \alpha_2 = 0.975$,  then $\alpha = \alpha_1 + \alpha_2 -1 = 0.95$ and the above $\Gamma^{\bm \eta}_{\alpha}(y_{obs})$ has at least $95\%$ guaranteed coverage. This scheme also applies to the confidence set \eqref{eq:conf_beta_Lambda} discussed in the previous subsection, including the two special cases of $\bbeta_{\Lambda, 0}.$   Proofs  of the theorems are in { Appendix~\ref{sec:app_proof} }.

 \begin{theorem}
\label{the:coverage_eta}
Under the conditions in Theorem~\ref{the:finite_pen}, for any finite sample size $n$ and an arbitrarily small $\delta>0,$ the coverage probability of the confidence interval $\Gamma^{\bm\eta_\tau}_{\alpha}(\*Y)$ defined in \eqref{eq:joint_tau_beta} is
$ P_{({\cal U}^d, \*Y)}\{(\tau_0,\bbeta_0) \in \Gamma^{\bm\eta_\tau}_{\alpha}(\*Y)\} \geq \alpha - \delta - o(e^{-cd})$ for some $c_1>0$, provided that $\alpha_1 + \alpha_2 -1 = \alpha.$ Further $ P_{\*Y | {\cal U}^d, }\{(\tau_0,\bbeta_0) \in \Gamma^{\bm\eta_\tau}_{\alpha}(\*Y)\} \geq \alpha - \delta - o_p(e^{-c_1d}).$
\end{theorem}

\begin{theorem}
\label{the:coverage_eta_asym}
 Under the conditions in Theorem~\ref{the:aymp_pen}, for any finite $d$, the coverage probability of the confidence interval $\Gamma^{\bm\eta_\tau}_{\alpha}(\*Y)$ defined in \eqref{eq:joint_tau_beta} is $\P_{\*Y} \{(\tau_0,\bbeta_0) \in \Gamma^{\bm\eta_\tau}_{\alpha}(\*Y)\} \geq \alpha  - o(e^{-c_2n})$ for some $c_2>0$, provided that $\alpha_1 + \alpha_2 -1 = \alpha.$ Further $ P_{\*Y | {\cal U}^d, }\{(\tau_0,\bbeta_0) \in \Gamma^{\bm\eta_\tau}_{\alpha}(\*Y)\} \geq \alpha - o_p(e^{-c_2 n}).$
\end{theorem}

}
\section{Technical Proofs: Lemma~\ref{lemma_def1} and Results for Gaussian Errors}
\label{sec:app_proof}

\subsection{Proof of Lemma~\ref{lemma_def1}}
\begin{proof}[Proof of Lemma~\ref{lemma_def1}]
 By the definition \eqref{eq:def_true_model_0}, there exist a $\beta_{0}$ and a $\sigma_0$ such that $ X_{\tau_0} \beta_{0} = {\bf y}_{obs} -  \sigma_0  \*u^{rel}$. Since
\begin{eqnarray*}
0 &\leq& \underset{\tau, \bbeta_{\tau}, \sigma} \min \|\*y_{obs}- \*X_{\tau}\bbeta_{\tau}-\sigma \*u^{rel}\|^{2}_2 \leq   \|\*y_{obs}- X_{\tau_0}\beta_{0}-\sigma _{0}\*u^{rel}\|^{2}= 0,
\end{eqnarray*}
it follows that
\begin{align}
\underset{\tau, \bbeta_{\tau}, \sigma} \min \|\*y_{obs}- \*X_{\tau}\bbeta_{\tau}-\sigma \*u^{rel}\|^{2}_2= 0. \nonumber
\end{align}

Now, let
\begin{equation} \label{lemma.tau}
(\tilde{\tau}, \beta_{\tilde \tau}, \sigma_{\tilde\tau}) = \underset{\tau, \bbeta_\tau, \sigma}{\rm argmin}  \big\{\lambda |\tau| +  \|\*y_{obs}- \*X_{\tau}\bbeta_{\tau}-\sigma \*u^{rel}\|^{2}_2\big\}.
\end{equation}
We show below that $\|\*y_{obs}- X_{\tilde \tau}\beta_{\tilde \tau}-\sigma_{\tilde\tau} \*u^{rel}\|^{2}_2 = 0$ using the ``proof by contradiction'' method.

First, we show that, if $\|\*y_{obs}- X_{\tilde \tau}\beta_{\tilde \tau}-\sigma_{\tilde\tau} \*u^{rel}\|^{2}_2 \neq 0,$ then size of $\tilde \tau$ must be smaller than $\tau_0$, i.e. $|\tilde\tau| < |\tau_0|.$ This is because otherwise if $|\tilde\tau| \geq |\tau_0|,$ then $\lambda |\tilde\tau| +  \|\*y_{obs}- X_{\tilde \tau}\beta_{\tilde \tau}-\sigma_{\tilde \tau} \*u^{rel}\|^{2}_2 > \lambda |\tilde\tau| \geq \lambda |\tau_0| = \lambda |\tau_0| +
\|\*y_{obs}- X_{ \tau_0}\beta_{\tau_0}-\sigma_0 \*u^{rel}\|^{2}_2, $ which contradicts with \eqref{lemma.tau}.

Now, with the triplet $(\tilde{\tau}, \beta_{\tilde \tau}, \sigma_{\tilde\tau})$ defined in (\ref{lemma.tau}) and $|\tilde\tau| < |\tau_0| < n$,
we have
for the given $\tilde \tau$,
\begin{align*}
& \|
\*y_{obs}- \*X_{\tilde \tau} \bbeta_{\tilde\tau} -\sigma_{\tilde\tau} \*u^{rel}
\| \geq \| (\*I - \*H_{\tilde\tau, \*u^{rel}})\*y_{obs}\|,
\end{align*}
where $\*H_{\tilde\tau, \*u^{rel}} = \*X_{\tilde\tau, \*u^{rel}} (\*X_{\tilde\tau, \*u^{rel}}^\top \*X_{\tilde\tau, \*u^{rel}})^{-1} \*X_{\tilde\tau, \*u^{rel}}^\top$ with $\*X_{\tilde\tau, \*u^{rel}} = (\*X_{\tilde\tau},  \*u^{rel})$ is the projection matrix to the space expanded by $\*X_{\tilde\tau}$ and $\*u^{rel}$. It follows that
\begin{align}
\label{eq:aaa}
\|
\*y_{obs}- \*X_{\tilde \tau} \bbeta_{\tilde\tau} -\sigma_{\tilde\tau} \*u^{rel}
\| \geq \| (\*I - \*H_{\tilde\tau, \*u^{rel}})
\*y_{obs}
\| = \|(\*I - \*H_{\tilde\tau, \*u^{rel}}) \*X_{\tau_0}\bbeta_0 \|,
\end{align}
where the equality holds because $\*u^{rel}$ is orthogonal to $(\*I - \*H_{\tilde\tau, \*u^{rel}})$.

By \eqref{eq:aaa}, the definitions of $\gamma^2_{(\*u^{rel}, \tau_0)}$ and $C_{\min}$,  and under the condition that
$0<\lambda \leq n\{1-\gamma^2_{(\*u^{rel},\tau_0)}\}C_{\min},$
\begin{align*}
    \|\*y_{obs}- \*X_{\tilde \tau} \bbeta_{\tilde\tau} & -\sigma_{\tilde\tau} \*u^{rel}\|^2  + \lambda |\tilde\tau|
    \geq  \| (\*I - \*H_{\tilde\tau, \*u^{rel}})\*X_{\tau_0}\bbeta_0\|^2  + \lambda |\tilde\tau|\\
    &\geq \left\{1-\gamma^2_{(\*u^{rel},\tau_0)}\right\} \|(\*I - \*H_{\tilde\tau})\*X_{\tau_0}\bbeta_0\|^2 + \lambda |\tilde\tau|\\
    &\geq \left\{1-\gamma^2_{(\*u^{rel},\tau_0)}\right\} \, n |\tau_0 \setminus \tilde\tau| C_{\min} + \lambda |\tau_0| - \lambda |\tau_0 \setminus \tilde\tau|\\
    & \geq \lambda |\tau_0| = \lambda |\tau_0| +
\|\*y_{obs}- \*X_{ \tau_0}\bbeta_{0}-\sigma_0 \*u^{rel}\|^{2}_2,
\end{align*}
which contradicts with \eqref{lemma.tau}. Thus, $\|\*y_{obs}- \*X_{\tilde \tau}\bbeta_{\tilde \tau}-\sigma_{\tilde\tau} \*u^{rel}\|^{2}_2 \neq 0$ does not hold and we only have  $\|\*y_{obs}- \*X_{\tilde \tau}\bbeta_{\tilde \tau}-\sigma_{\tilde\tau} \*u^{rel}\|^{2}_2=0.$
Because $\*u^{rel} \not\in \spn(\*X_{\tau_0}, \*X_\tau)$ for any $\tau$ with $|\tau|\leq |\tau_0|,$  by definition \eqref{eq:def_true_model_0},  we have $\tilde \tau= \tau_0$ and  thus the conclusion of the lemma follows.

\end{proof}

\subsection{Proofs of Theorems \ref{the:finite_pen}-\ref{the:aymp_pen} and Theorems \ref{the:finite_con}-\ref{the:aymp_con}}
In this section, we prove our results in Theorems \ref{the:finite_pen}--\ref{the:aymp_pen},  and their counter parts Theorems \ref{the:finite_con}-\ref{the:aymp_con} for the constrained regression formulation in Appendix~\ref{sec:constraint}.
We would like to point out that Theorems \ref{the:finite_pen} and \ref{the:aymp_pen} in Section~\ref{sec:cand_theory_normal} are particularly challenging. In both cases, we have to control the behavior of the repro samples $\*U^*,$ not only in relation to the error term $\*U,$ but also in relation to $(\*I - \*H_\tau)\*X_{\tau_0}\beta_0$ for any $|\tau| \leq |\tau_0|,$ within the proximity of which $\*U^*$ could possibly lead to $\tau$ instead of $\tau_0.$  We also would like to note that there have not been any finite-sample theories like Theorem~\ref{the:finite_pen} in the literature.
{
 Together, Theorems \ref{the:finite_pen} and \ref{the:aymp_pen} imply that when either the number of repro simulations $d$ or the sample size $n$ is large enough, our candidate model sets $S^{(d)}$ will contain the true model $\tau_0$ with high probability.
\subsubsection{Proofs of Theorem~\ref{the:finite_pen} and Theorem~\ref{the:finite_con}}
First we define a similarity measure between two vectors $\*v_1, \*v_2$ as the square of cosine of the angles between $\*v_1$ and $\*v_2,$ i.e. $\rho(\*v_1, \*v_2) =\|\*H_{\*v_1}\*v_2\|^2/\|\*v_2\|^2 =(\*v_1^\top\*v_2)^2/(\|\*v_1\|^2\|\*v_2\|^2).$
We therefore use $\rho(\*u^*, \*u^{rel})$ to measure the similarity between a single repro sample $\*u^*$ and the realization $\*u^{rel}$. Apparently, the closer $\rho(\*u, \*u^{rel})$ is to 1, the smaller the angle between $\*u$ and $\*u^{rel}.$ Hence we use $\rho(\*u,\*u^{rel})$ to measure the similarity between $\*u$ and $\*u^{rel}$.

We then present a technical lemma that derives the probability bound of obtaining the true model $\tau_0$ when the repro sample $\*u^*$ falls within close proximity of $\*u^{rel}$ in that $\rho(\*u^*, \*u^{rel}) > 1- \gamma^2_2$ for a small $\gamma_2>0. $
We provide the proof of Lemma~\ref{lem::finite_penalize}
in Appendix~\ref{sec:proof_lem24}.

\begin{lemma}
    \label{lem::finite_penalize}
Suppose $n-|\tau_0|>4.$ Let $\*U^*$ be a random repro sample of $\*U$, such that $\*U^*, \*U \sim N(0, \*I_n),$ and
$\hat \tau_{\*U^*} = \argmin_{\tau}\left\{\min_{\bbeta_\tau, \sigma} \|\*Y - \*X_{\tau} \bbeta_{\tau} -\sigma \*U^* \|^2 + \lambda |\tau|\right\}.$
Then for any $0<\gamma^{1/4}_2 < \min\big\{  \frac{C_{\min}}{24 \{2 + 2(|\tau_0|+1)\log(p/2)/n\}\sigma_0^2}, 0.35\big\}$ that is small enough such that $C_{\min} >52{\sqrt{\gamma_2}}\big( \frac{  \log(p/2)}{n}+ \gamma_2\big) \sigma_0^2$ and  for  $ \lambda \in \big[4n\gamma_2^{1/2}\big\{2 + 2(|\tau_0|+1)\frac{\log(p/2)}{n} \big\}\sigma_0^2$, $
     \frac{n\gamma_2^{1/4}}{6}C_{\min} \big],$

\begin{align}\label{eq:lemma2}
    & \P_{(\*U, \*U^*)}\left\{\hat \tau_{\*U^*} \neq \tau_0| \rho(\*U^*, \*U) > 1- \gamma_2^2\right\}\notag \\
    & \leq   3\exp\left\{- \frac{n}{26\sigma_0^2}\left[ \frac{C_{\min} }{\sqrt{\gamma_2}} - 52\left( \frac{  \log(p/2)}{n}+ \gamma_2\right) \sigma_0^2\right]
    \right\} + 3\exp\left(-\frac{n}{4\gamma_2^{1/2}}   \right) \\
     & \hspace{4cm} +  4(64\gamma_2)^{\frac{n-|\tau_0|-1}{6}} p^{|\tau_0|}. \notag
\end{align}
\end{lemma}

Unlike existing literature in the high-dimensional regime, the results in Lemma~\ref{lem::finite_penalize} do not require any conditions on $C_{\min},$ nor do it even depend on any conditions necessary for achieving consistent regression parameter estimation. This is because the probability bound on the right-hand side of \eqref{eq:lemma2} depends on $C_{\min}$ only through ${C_{\min} }/{\sqrt{\gamma_2}}$. When the quantity ${C_{\min} }/{\sqrt{\gamma_2}}$ becomes larger,
the probability bound becomes smaller. Therefore no matter how small ${C_{\min} }$ is, as long as ${C_{\min} >0},$  the quantity ${C_{\min} }/{\sqrt{\gamma_2}}$ can be arbitrarily large when $\gamma_2$ is small enough. Consequently, however small the separation between the true model $\tau_0$ and the alternative models is, we can always recover $\tau_0$ with high probability with a repro sample $\*U^*$ that is close to $\*U.$

By the finite-sample probability bound obtained in the above lemma, when $\gamma_2$ goes to 0, that is, $\*U^*$ proximate $\*U$ more closely, the probability of $\hat\tau_{\*U^*} \neq \tau_0$ goes to 0 for any finite $n$ and $p$. This indicates that we do not need $\*U^*$ to hit $\*U$ exactly, rather we would only need $\*U^*$ to be in a neighborhood of $\*U$ in order to recover $\tau_0$ with high probability.
Additionally we observe that as the sample size $n$ increases, the probability bounds in Lemma~\ref{lem::finite_penalize}
decay exponentially. Therefore, for a larger sample, the estimation $\hat\tau_{\*U^*}=\tau_0$ with large probability even for a large $\gamma_2.$ As a result, the neighborhood of $\*u^{rel}$, within which $\*U$ yields $\hat\tau_{\*U^*}=\tau_0$ with high probability, will expand as the sample size $n$ grows larger.

As Lemma~\ref{lem::finite_penalize} shows the probability bound given a single repro sample $\*U^*$ being close to $\*U$, in the following
Lemma~\ref{lem::finite_penalize_complete}, we develop the probability bound for at least one of the $d$ independent samples of $\*U^*$ being close to $\*U$. This probability bound, together with the bound in \eqref{eq:lemma2},  then implies a finite-sample probability bound of $\tau_0$ not included in the candidate set $S^{(d)}$ constructed by Algorithm~\ref{alg:candidate}. The proof of Lemma~\ref{lem::finite_penalize_complete} is deferred to Appendix~\ref{sec:proof_lem35}.
\begin{lemma}\label{lem::finite_penalize_complete}
Suppose $n-|\tau_0|>4.$  Then for any $0<\gamma^{1/4}_2 < \min\big\{  \frac{C_{\min}}{24 \{2 + 2(|\tau_0|+1)\log(p/2)/n\}\sigma_0^2}$, $0.35\big\},$
such that $C_{\min} >52{\sqrt{\gamma_2}}\big( \frac{  \log(p/2)}{n}+ \gamma_2\big) \sigma_0^2,$ and $ \lambda \in \big[4n\gamma_2^{1/2}\big\{2 + 2(|\tau_0|+1)\frac{\log(p/2)}{n} \big\}\sigma_0^2$, $
     \frac{n\gamma_2^{1/4}}{6} C_{\min}\big],$
the finite-sample probability bound that the true model is not covered by the model candidates set $S^{(d)}$, obtained by Algorithm~\ref{alg:candidate} with the objective function \eqref{eq::obj}, is
{\small \begin{align}
    \label{eq:prob_bound_penalize_finite}
      & \P_{({\cal U}^d, \*Y)}(\tau_0 \notin S^{(d)} )   \leq   3\exp\left\{- \frac{n}{26\sigma_0^2}\left[ \frac{C_{\min} }{\sqrt{\gamma_2}} - 52\left( \frac{  \log(p/2)}{n}+ \gamma_2\right) \sigma_0^2\right]
    \right\} \nonumber \\ & \quad + 3\exp\left(-\frac{n}{4\gamma_2^{1/2}}   \right) +
     4(64\gamma_2)^{\frac{n-|\tau_0|-1}{6}} p^{|\tau_0|}  + \left(1-\frac{\gamma_2^{n-1}}{n-1}\right)^d.
\end{align}}
\end{lemma}

We are now to present the proof of Theorem~\ref{the:finite_pen}

\begin{proof}[Proof of Theorem~\ref{the:finite_pen} and Theorem~\ref{the:finite_con}:]
The first four terms of $\eqref{eq:prob_bound_penalize_finite}$ go to 0 as $\gamma_2$ goes to $0.$  Therefore for any $\delta >0, $ there exists a $\gamma_\delta>0,$ such that  when $\gamma_2=\gamma_\delta,$ sum of the first { three} terms of $\eqref{eq:prob_bound_penalize_finite}$ is smaller than $\delta,$ which implies the probability bound in \eqref{eq:bound_cs_finite} of Theorem~\ref{the:finite_pen}. Similarly Theorem~\ref{the:finite_con} follows from Lemma~\ref{lem::finite_complete} by making $\gamma_2=\gamma_\delta.$
\end{proof}
}

\subsubsection{Proofs of Theorem~\ref{the:aymp_pen} and Theorem~\ref{the:aymp_con}}
Similar to the last section, we first introduce a key lemma. The proof of Lemma~\ref{lem::symptotic_bound_penalize0} is in Appendix~\ref{sec:proof_lem6}.

\begin{lemma}
    \label{lem::symptotic_bound_penalize0}
For any finite $n$ and $p$, if $\frac{\lambda}{n} \in \big[\frac{3\sigma_0^2(|\tau_0|+1)(\log(p-|\tau_0|)  +\log(|\tau_0|)+ \frac{2}{3})}{n}+t,$ $ \frac{(1-\gamma_1^2)C_{\min}}{6}\big],$
a finite-sample probability bound that the true model is not covered by the model candidates set $S^{(d)}$, obtained by Algorithm~\ref{alg:candidate} with the objective function \eqref{eq::obj}, is,

\begin{align}
\label{eq:prob_bound_penalize_asymptotic}
     \P_{({\cal U}^d, \*Y)}(\tau_0 \notin S^{(d)} ) \leq L(\gamma_1) + 3\exp\left(-\frac{nt}{3\sigma_0^2}\right) + \left[2\{\arccos (\gamma_1)\}^{n-|\tau_0|-1 } p^{|\tau_0|}\right]^d,
\end{align}
where
$L(\gamma_1)=6 \exp\left[-\frac{n}{18 \sigma_0^2}\left\{(1-\gamma_1^2)C_{\min} -36\frac{\log p}{n}\sigma_0^2\right\}\right],
$
and $\cos(0.3\pi) < \gamma_1<1$ is any real number.
\end{lemma}

Lemma~\ref{lem::symptotic_bound_penalize0} aims to offer insights on the asymptotic property of the candidate set $S^{(d)},$, therefore, it gives a different probability bound than Lemma~\ref{lem::finite_penalize_complete}.  The interpretation is that for any fixed $d,$
the probability of $\tau_0 \notin S^{(d)}$ is $O(e^{-n})$ under the conditions in Theorem~\ref{the:aymp_pen}. This provides us the insight that for large samples, we actually do not need an extremely large number of repro samples in order to recover the true model in the candidate set $S^{(d)}.$

To explain the intuition behind the probability bound in Lemma~\ref{lem::symptotic_bound_penalize0}, we denote the angle between the repro sample $\*U^*$ and $(\*I- \*H_\tau) \*X_{\tau_0}\bbeta_0$ as $\gamma_1^\tau.$ If $\gamma_1^\tau \geq \gamma_1$ for all $|\tau| \leq |\tau_0|,$ then  the probability of $\hat\tau_{\*U^*} \neq \tau_0$ is bounded by the first two terms of \eqref{eq:prob_bound_penalize_asymptotic}. The reason that we want to bound  $\*U^*$  away from  $(\*I- \*H_\tau) \*X_{\tau_0}\bbeta_0$ is that when $\*U^* \approx (\*I- \*H_\tau) \*X_{\tau_0}\bbeta_0,$ $\*X_\tau$ will explain $\*Y - \*U^*$ as well as $\*X_{\tau_0},$ possibly leading to $\hat\tau_{\*U^*} = \tau \neq \tau_0.$ The last term of  \eqref{eq:prob_bound_penalize_asymptotic} is derived from the probability bound that $\gamma_1^\tau \leq \gamma_1$ for some $|
\tau|
\leq |\tau_0|$ for all the $d$ copies of repro samples $\*U^*.$ Therefore, all the three terms together give a probability bound for $\tau_0 \notin S^{(d)}.$

We now present the proof of Theorem~\ref{the:aymp_pen}.
\begin{proof}[Proof of Theorem~\ref{the:aymp_pen} and Theorem~\ref{the:aymp_con}]
By Lemma \ref{lem::symptotic_bound_penalize0}, we obtain \eqref{eq:prob_asympto_pen} in  Theorem~\ref{the:aymp_pen} by making $\gamma^2_1=0.7$. The lower bound for $\frac{\lambda}{n}$ is simplified by  applying  $\log(|\tau_0|)+ \log(p-|\tau_0|)\leq 2\log(p/2).$ Similarly, by Lemma \ref{lem::asymptotic_constraint}, we make $\gamma^2_1=0.7$, then the probability bound \eqref{eq:prob_asympto_con} in Theorem~\ref{the:aymp_con} follows from \eqref{eq:prob_bound_asymp_constraint}.
\end{proof}

\subsection{Proofs of  Lemma~\ref{lem::finite} and Lemma~\ref{lem::finite_penalize}} \label{sec:proof_lem24}
Before we proceed to the proofs of Lemma~\ref{lem::finite_penalize} and Lemma~\ref{lem::finite}, we first provide two technical lemmas that facilitate the proofs.

\begin{lemma}
\label{lem_proj}
For any $\tau$ and $\*u^*$,
\begin{align*}
    \*I - \*H_{\tau,\*u^*} =  I - \*H_\tau - \*O_{\tau^{\bot} \*u^*},
\end{align*}
where $H_{\tau,\*U^*}=\begin{pmatrix} X_\tau & \*u^* \end{pmatrix} \begin{pmatrix} X_\tau^\top X_\tau & X_\tau^\top \*u^* \\
(\*u^*)^\top  X_\tau & (\*u^*)^\top \*u^* \end{pmatrix}^{-1} \begin{pmatrix} X_\tau^\top \\ (\*u^*)^\top \end{pmatrix}$ is the projection matrix on the space spanned by $(X_\tau, \*u^*)$ and  $ \*O_{\tau^{\bot} \*u^*}= \frac{(I - \*H_\tau) \*u^*  (\*u^*)^\top  (I - \*H_\tau)}
{(\*u^*)^\top (I - \*H_\tau)\*u^* }$ is the projection matrix on the space spanned by $(I - \*H_\tau) \*u^*.$
\end{lemma}
\begin{proof} By a direct calculation, we have
{\small \begin{align*}
 & \*I - \*H_{\tau,\*U^*}  =
I - \begin{pmatrix} X_\tau & \*u^* \end{pmatrix} \begin{pmatrix} X_\tau^\top X_\tau & X_\tau^\top \*u^* \\
(\*u^*)^\top  X_\tau & (\*u^*)^\top \*u^* \end{pmatrix}^{-1} \begin{pmatrix} X_\tau^\top \\ (\*u^*)^\top \end{pmatrix}  \\
& = I - \begin{pmatrix} X_\tau & \*u^* \end{pmatrix} \begin{pmatrix} (X_\tau^\top X_\tau)^{-1} +  \frac{(X_\tau^\top X_\tau)^{-1}X_\tau^\top \*u^*  (\*u^*)^\top  X_\tau (X_\tau^\top X_\tau)^{-1}}{(\*u^*)^\top (I - \*H_\tau)\*u^* } & - \frac{(X_\tau^\top X_\tau)^{-1}X_\tau^\top \*u^*} {(\*u^*)^\top (I - \*H_\tau)\*u^* } \\
- \frac{(\*u^*)^\top  X_\tau (X_\tau^\top X_\tau)^{-1}} {(\*u^*)^\top (I - \*H_\tau)\*u^* } & \frac{1}{(\*u^*)^\top (I - \*H_\tau)\*u^* } \end{pmatrix} \begin{pmatrix} X_\tau^\top \\ (\*u^*)^\top \end{pmatrix} \\a
 & = I - \begin{pmatrix} X_\tau & \*u^* \end{pmatrix} \begin{pmatrix} (X_\tau^\top X_\tau)^{-1}X_\tau^\top +  \frac{(X_\tau^\top X_\tau)^{-1}X_\tau^\top \*u^*  (\*u^*)^\top  X_\tau (X_\tau^\top X_\tau)^{-1}X_\tau^\top}{(\*u^*)^\top (I - \*H_\tau)\*u^* } - \frac{(X_\tau^\top X_\tau)^{-1}X_\tau^\top \*u^*(\*u^*)^\top} {(\*u^*)^\top (I - \*H_\tau)\*u^* } \\
- \frac{(\*u^*)^\top  X_\tau (X_\tau^\top X_\tau)^{-1}X_\tau^\top} {(\*u^*)^\top (I - \*H_\tau)\*u^* } + \frac{(\*u^*)^\top}{(\*u^*)^\top (I - \*H_\tau)\*u^*  } \end{pmatrix}
\\
 & = I -  X_\tau (X_\tau^\top X_\tau)^{-1}X_\tau^\top - \frac{X_\tau (X_\tau^\top X_\tau)^{-1}X_\tau^\top \*u^*  (\*u^*)^\top  X_\tau (X_\tau^\top X_\tau)^{-1}X_\tau^\top}{(\*u^*)^\top (I - \*H_\tau)\*u^* }  \\
 &+ \frac{X_\tau(X_\tau^\top X_\tau)^{-1}X_\tau^\top \*u^*(\*u^*)^\top} {(\*u^*)^\top (I - \*H_\tau)\*u^* }   + \frac{\*u^*(\*u^*)^\top  X_\tau (X_\tau^\top X_\tau)^{-1}X_\tau^\top} {(\*u^*)^\top (I - \*H_\tau)\*u^* } - \frac{\*u^*(\*u^*)^\top}{(\*u^*)^\top (I - \*H_\tau)\*u^*  } \\
& = I -  \*H_\tau -  \frac{\*H_\tau \*u^*  (\*u^*)^\top  \*H_\tau}{(\*u^*)^\top (I - \*H_\tau)\*u^* } + \frac{\*H_\tau \*u^*(\*u^*)^\top} {(\*u^*)^\top (I - \*H_\tau)\*u^* }
+ \frac{\*u^*(\*u^*)^\top  \*H_\tau} {(\*u^*)^\top (I - \*H_\tau)\*u^* } - \frac{\*u^*(\*u^*)^\top}{(\*u^*)^\top (I - \*H_\tau)\*u^*  } \\
& = I -  \*H_\tau -   \frac{(I - \*H_\tau) \*u^*  (\*u^*)^\top  (I - \*H_\tau)}
{(\*u^*)^\top (I - \*H_\tau)\*u^* } = I - \*H_\tau - \*O_{\tau^{\bot} \*u^*}.
\end{align*}
}
\end{proof}

Let $\rho(\*v_1, \*v_2) = \cos^2(\*v_1, \*v_2) = \frac{\|\*H_{\*v_2}\*v_1\|^2}{\|\*v_1\|^2}$ be the square of the cosine of the angle between any two $n \times 1$ vectors $\*v_1$ and $\*v_2.$ Further, for any given $\tau$, let $\rho_{\tau^\bot}(\*v_1, \*v_2) = \rho\{(\*I-\*H_\tau)\*v_1, (\*I-\*H_\tau) \*v_2\}$ be the cosine of the angle between $(\*I-\*H_\tau)\*v_1$ and $(\*I-\*H_\tau)\*v_2.$

\begin{lemma}
\label{lem::angle} Suppose $|\tau| < n$. For any $-1 \leq \gamma_1, \gamma_2 \leq 1,$ if $\*U^* \sim N(0, \*I),$
\begin{align*}
   \P_{\*U^*}\left\{\rho_{\tau^\bot}(\*U^*, \*X_{\tau_0}\bbeta_0)< \gamma_1^2\right\} =  \P_{\*U}\left\{\rho_{\tau^\bot}(\*U, \*X_{\tau_0}\bbeta_0)< \gamma_1^2\right\}>  1-   2 \{\arccos (\gamma_1)\}^{n-|\tau|-1},
\end{align*}
and
\begin{align*}
    \P_{(\*U^*, \*U)}\{\rho(\*U^*, \*U) > 1-\gamma_2^2\} > \frac{\gamma_2^{n-2}\arcsin (\gamma_2)}{n-1}.
\end{align*}
 Moreover, $\rho(\*U^*,\*U)$ and $\*U$ are independent. Further, if both $\*U$ and $\*U^*$ are Gaussian, i.e. $\*U \sim \*U^* \sim N(0, \*I),$ $\rho(\*U^*,\*U)$ and $\*U^*$ are also independent, $\rho_{\tau^\bot}(\*U^*, \*X_{\tau_0}\bbeta_0)$ and $\rho(\*U^*, \*U)$ are independent, and $(\rho_{\tau^\bot}(\*U^*, \*X_{\tau_0}\bbeta_0) ,\rho(\*U^*, \*U))$ are independent of $\|\*U\|.$

\begin{proof}

Let $(\*I-\*H_\tau) = \sum_{i=1}^{n-|\tau|} D_i D_i^\top$ be the eigen decomposition of $(\*I-\*H_\tau)$. Denote by $Z_i = D_i^\top \*U$ and $w_i = D_i^\top \*X_{\tau_0}\bbeta_0$, for $i = 1, \ldots, n-|\tau|$. It follows that $Z_1, \dots, Z_{n-|\tau|}$ are i.i.d $N(0,1)$
and
\begin{align*}
     & \P_{\*U^*}\left\{\rho_{\tau^\bot}(\*U^*, \*X_{\tau_0}\bbeta_0)< \gamma_1^2\right\} = \P_{\*U}\left\{\rho_{\tau^\bot}(\*U, \*X_{\tau_0}\bbeta_0)< \gamma_1^2\right\}\\ & =
    \P_{\*U}\left\{\frac{\sum_{i = 1}^{n - |\tau|} w_i Z_i}{\sqrt{\sum_{i = 1}^{n - |\tau|} w_i^2} \sqrt{\sum_{i = 1}^{n - |\tau|} Z_i^2}} <\gamma_1 \right\} \\
    & =
    \P_{\*U}\left\{|\cos(\varphi)| <\gamma_1 \right\},
\end{align*}
where $\varphi = \varphi(\*U)$ (or $\pi - \varphi$) is the angle between $(Z_1, \dots, Z_{n-|\tau|})$ and $(w_1, \dots, w_{n - |\tau|})$ for $0 \leq \varphi \leq \pi$.

We transform the co-ordinates of $Z_1, \dots, Z_{n-|\tau|}$ into sphere co-ordinates, with $\varphi$ as the first angle coordinate.
It follows from the Jacobian of the spherical transformation the density function of $\varphi$ is
\begin{align}
\label{eq:density_angle}
    f(\varphi)= \sin^{n-|\tau|-2}(\varphi)/c, \quad 0 \leq \varphi \leq \pi,
\end{align}
where $c = \int_0^{\pi} \sin^{n-|\tau|-2}(\varphi) d\varphi = 2 \int_0^{\frac \pi2} \sin^{n-|\tau|-2}(\varphi) d\varphi $ is the normalizing constant.

Note that,  for $0<\varphi<\pi/2$,  we have
\begin{equation}
\frac{2}{\pi} \varphi <
\sin(\varphi) <  \min\{\varphi, 1 \} = \varphi 1_{(0 < \varphi <1)} + 1_{(1\leq \varphi <\pi/2)}, \nonumber
\end{equation}
where $1_{(\cdot)}$ is an indicator function. It follows that
\begin{equation}
\frac{\pi}{2(n-|\tau|-1)} < c <\frac{1}{n-|\tau|-1} + (\frac{\pi}{2} - 1) < 2. \nonumber
\end{equation}
Therefore, we have
\begin{align*}
    & \P_{\*U}\left\{|\cos \varphi| <\gamma_1 \right\}  =\frac{2}{c}\int_{\arccos (\gamma_1)}^{\pi/2} \sin^{n-|\tau|-2}(s) ds  =1 - \frac{2}{c} \int_0^{{\arccos (\gamma_1)}}\sin^{n-|\tau|-2}(s) d s\\
     & \quad > 1 - \frac{ 2(n-|\tau|-1)\int_0^{{\arccos (\gamma_1)}}s^{n-|\tau|-2} ds}{\pi} = 1-  2\{\arccos (\gamma_1)\}^{n-|\tau|-1}.
\end{align*}

Next conditioning on $\*U^* = \*u^*$, with similar procedure as above but replacing $n-|\tau|$ with $n$, we can show that
\begin{align}\label{eq:u_epsi}
    &
    \P_{ \*U}\left\{{\|(\*u^*)^\top \*U\|}\big/{(\|\*u^*\|\| \*U\|)} > \sqrt{1 - \gamma_2^2} \bigg| \*u^* \right\}
    =\P_{ \*U}\left\{  |\cos(\psi)|  > \sqrt{1 - \gamma_2^2}
    \bigg| \*u^* \right\} \nonumber \\
    & \qquad = \frac{2}{c_1}\int_0^{\arcsin \gamma_2} \sin^{n-2}(s) ds   \\
    & \qquad >  \frac{2}{c_1}\int_0^{\arcsin \gamma_2} (\frac{s \gamma_2 }{\arcsin \gamma_2})^{n-2} ds> \frac{\gamma_2^{n-2}\arcsin \gamma_2}{n-1}, \nonumber
\end{align}
where $\psi = \psi(\*u,^* \*u)$ (or $\pi - \psi$) is the angle between $\*u$ and $\*u^*$ and the normalizing constant $c_1 = \int_0^{\pi} \sin^{n-2} (\psi) d\psi = 2 \int_0^{\frac \pi 2} \sin^{n-2} (\psi) d\psi \leq 2$. The first inequality follows from the fact that $\sin(s)$ is a concave function for $0 \leq s \leq \pi/2.$ The same derivation works when the conditional is on $\*U=\*u$:
\begin{align}\label{eq:u_epsi-1}
    &
    \P_{ \*U^*}\left\{{\|\*U^*{}^\top \*u\|}\big/{(\|\*U^*\|\| \*u\|)} > \sqrt{1 - \gamma_2^2} \bigg| \*u \right\}
    =\P_{ \*U^*}\left\{  |\cos(\psi)| > \sqrt{1 - \gamma_2^2}   \bigg| \*u \right\} \\
    & \qquad = \frac{2}{c_1}\int_0^{\arcsin \gamma_2} \sin^{n-2}(s) ds > \frac{\gamma_2^{n-2}\arcsin \gamma_2}{n-1}, \nonumber
\end{align}

Because \eqref{eq:u_epsi} and \eqref{eq:u_epsi-1} do not involve $\*u^*$ or $\*u$, we have
\begin{align*}
  \P_{(\*U^*, \*U)}\big\{ \rho(\*U^*, \*U)> 1- \gamma_2^2 \big\} & =    \P_{\*U}\left\{ \rho(\*U^*, \*U)> 1- \gamma_2^2\bigg| \*U^* \right\} \\ & = \P_{ \*U^*}\left\{ \rho(\*U^*, \*U)> 1- \gamma_2^2 \bigg| \*U \right\}  > \frac{\gamma_2^{n-2}\arcsin \gamma_2}{n-1}. \numberthis \label{eq:u_ustar}
\end{align*}
The above statement also suggests that $\rho(\*U^*,\*U)$ and $\*U$ are independent. Similarly,  $\rho(\*U^*,\*U)$ and $\*U^*$ are independent, therefore $\rho(\*U^*,\*U)$ and   $\rho_{\tau^\bot}(\*U^*, \*X_{\tau_0}\bbeta_0)$ are independent.

Furthermore, since the distribution of $\rho(\*U^*, \*U) = \rho(\*U^*, \*U/\|\*U\|)$ is free of $\|\*U\|,$ it then follows immediately from the above that the joint distribution of $(\rho_{\tau^\bot}(\*U^*, \*X_{\tau_0}\bbeta_0) ,\rho(\*U^*, \*U))$ is free of $\|\*U\|,$ therefore $(\rho_{\tau^\bot}(\*U^*, \*X_{\tau_0}\bbeta_0) ,\rho(\*U^*, \*U))$ are independent of $\|\*U\|.$

\end{proof}

\end{lemma}

\begin{proof}[Proof of Lemma~\ref{lem::finite}]
For a fixed $\tau$, let $$D(\tau, \*u^*) = \min_{\bbeta_\tau, \sigma} \|\*Y - \*X_{\tau} \bbeta_{\tau} -\sigma \*u^* \|^2 = \|(\*I- \*H_{\tau, \*u^*}) \*Y\|^2,$$
where $\*Y = \*X_{\tau_0}\bbeta_0 + \sigma_0 \*U$ is a random sample from the true model (\ref{eq:model-random}) with the error term $\*U \sim N(0, I_n),$ and $\*H_{\tau, \*u^*}$ is the projection matrix for $(\*X_{\tau}, \*u^*).$

Define $$\hat \tau_{\*u^*} = \argmin_{\{\tau| |\tau| \leq |\tau_0|\}} D(\tau, \*u^*).$$ By (\ref{eq::obj_constraint}) with constraint $ |\tau| = \|\bbeta_\tau\|_0 \leq |\tau_0|$, if there exists a $\tau$, $ |\tau| \leq |\tau_0|$, such that $\{D(\tau, \*u^*) - D(\tau_0,\*u^*) < 0\}$, then $\{\hat \tau_{\*u^*} \neq \tau_0\}$. On the other hand, if $\{\hat \tau_{\*u^*} \neq \tau_0\}$, then $D(\hat \tau_{\*u^*}, \*u^*) - D(\tau_0,\*u^*) < 0$.
Thus,  $\bigcup_{\{\tau||\tau| \leq |\tau_0| \}} \{D(\tau, \*u^*) - D(\tau_0,\*u^*) < 0\} = \{\hat \tau_{\*u^*} \neq \tau_0\}$.

For each $\*Y$,
\begin{align}
\label{eq:dist}
   & D(\tau, \*u^*) -D(\tau_0,\*u^*)  =\|(\*I - \*H_{ \tau, \*u^*})\*Y\|^2 - \|(\*I - \*H_{\tau_0, \*u^*})\*Y\|^2 \nonumber\\
  & \qquad = \|(\*I - \*H_{\tau, \*u^*})(\*X_{\tau_0}\bbeta_0 + \sigma_0 \*U)\|^2 - \sigma_0^2\|(\*I - \*H_{\tau_0, \*u^*})\*U\|^2 \nonumber\\
   & \qquad = \|(\*I - \*H_{\tau, \*u^*}) \*X_{\tau_0}\bbeta_0\|^2 + 2\sigma_0\*U^\top(\*I - \*H_{ \tau, \*u^*})\*X_{\tau_0}\bbeta_0  - \sigma_0^2\*U^\top( \*H_{\tau, \*u^*} -  \*H_{ \tau_0, \*u^*} )\*U.
\end{align}

Now, define an event set
\begin{align}
\label{eq:def_E}
E(\gamma_1, \gamma_2)= \left\{ (\*u^*, \*u): \max_{\tau \neq \tau_0, |\tau|\leq |\tau_0|} \rho_{\tau^\bot}(\*u^*, \*X_{\tau_0}\bbeta_0) < \gamma_1^2, \rho(\*u^*, \*u) > 1- \gamma_2^2  \right\},
\end{align}
we have, for any $\delta \in (0,1)$, conditional on the event $E(\gamma_1, \gamma_2),$
 \begin{align*}
   & \P_{(\*U^*, \*U |\cdot)}\left\{ D(\tau,\*U^*) -D(\tau_0,\*U^*)< 0 \middle | (\*U^*, \*U) \in
E(\gamma_1, \gamma_2)\right\}\\
    & \leq \P_{(\*U^*, \*U| \cdot)}\bigg\{(1- \gamma^2_1) \|(\*I-\*H_\tau)\*X_{\tau_0}\bbeta_0\|^2 -\sigma_0^2\*U^\top( H_{\tau,\*U^*} - H_{\tau_0,\*U^*})\*U \\
    &\qquad + 2\sigma_0\*U^\top(\*I-\*H_{\tau,\*U^*})\*X_{\tau_0}\bbeta_0 <0 \bigg |  (\*U^*, \*U) \in
E(\gamma_1, \gamma_2)\bigg\}\\
    & \leq \P_{(\*U^*, \*U | \cdot)}\bigg\{(1-\gamma^2_1)(1-\delta)\|(\*I-\*H_\tau)\*X_{\tau_0}\bbeta_0\|^2 \nonumber \\ & \qquad \qquad - \sigma_0^2\*U^\top( \*H_{\tau,\*U^*} - \*H_{\tau_0,\*U^*})\*U <0 \bigg | (\*U,^* \*U) \in
E(\gamma_1, \gamma_2)\bigg\} \\
    & \qquad + \P_{(\*U^*, \*U | \cdot)}\bigg\{(1-\gamma^2_1)\delta\|(\*I-\*H_\tau)\*X_{\tau_0}\bbeta_0\|^2 \nonumber \\
    & \qquad \qquad + 2\sigma_0 \*U^\top(\*I- \*H_{\tau,\*U^*})\*X_{\tau_0}\bbeta_0 < 0  \bigg | (\*U,^* \*U) \in
E(\gamma_1, \gamma_2)\bigg\} \\ & = (I_1) + (I_2).
\end{align*}

To derive an upper bound for $(I_1)$, we note that, by Lemma~\ref{lem_proj}, for any $(\*U,^* \*U)$ that satisfies  $\rho(\*U,^* \*U) > 1- \gamma_2^2,$
\begin{align*}
    & \*U^\top(
    \*H_{\tau,\*U^*} - \*H_{\tau_0,\*U^*})\*U
    = \*U^\top( \*I - \*H_{\tau_0,\*U^*})\*U -
    \*U^\top( \*I - \*H_{\tau,\*U^*} )\*U \leq \*U^\top( \*I - \*H_{\tau_0,\*U^*})\*U
    \\
    & \qquad =  \|(\*I -  \*H_{\tau_0}-\*O_{\tau_0^\bot\*U^* })\*U\|^2 = \|(\*I- \*H_{\tau_0})(\*I - \*O_{\tau_0^\bot\*U^* }) \*U \|^2 \\
&  \qquad   \leq \|(\*I- \*H_{\tau_0})(\*I - \*H_{\*U^*}) \*U\|^2 \leq \|(\*I - \*H_{\*U^*}) \*U\|^2 \leq \gamma_2\|\*U\|^2,
\end{align*}
where $\*O_{\tau_0^\bot\*U^* }$ is the projection matrix of $(\*I- \*H_{\tau_0})\*U^*$ and the first inequality follows from the definition of projection.

To bound $I_1$, it follows from  Lemma~\ref{lem::angle} and the definition of $C_{\min}$ that,
\begin{align*}
    (I_1) & <\P_{(\*U^*, \*U | \cdot)}\left\{\|\*U\|^2 > \frac{(1-\gamma^2_1)(1-\delta)}{\gamma^2_2}\frac{\|(\*I-\*H_\tau)\*X_{\tau_0}\bbeta_0\|^2 }{\sigma_0^2}  \middle | (\*U^*, \*U) \in
E(\gamma_1, \gamma_2)\right\}\\
   & < \P_{(\*U^*, \*U )}\left\{\|\*U\|^2 > \frac{(1-\gamma^2_1)(1-\delta)}{\gamma^2_2}\frac{\|(\*I-\*H_\tau)\*X_{\tau_0}\bbeta_0\|^2 }{\sigma_0^2}\right\}
   \\
   & \leq  \P_{\chi^2_n}\left\{\chi^2_n> \frac{(1-\gamma^2_1)(1-\delta)}{\gamma^2_2}\frac{n |\tau_0 \setminus \tau| C_{\min}}{\sigma_0^2} \right\}\\
   & < \exp\left\{-\frac{n}{2}\log(1-2t_1) - t_1\frac{(1-\gamma^2_1)(1-\delta)}{\gamma^2_2}\frac{n |\tau_0 \setminus \tau| C_{\min}}{\sigma_0^2}\right\},
\end{align*}
for any $0<t_1<1/2$, where $\chi^2_n$ is a random variable that follows $\chi^2_n$ distribution.
The last inequality is derived from Markov inequality and moment-generating function of Chi-square distribution.

For $(I_2)$, we note that, for any $(\*U^*, \*U) $ such that $ \rho(\*U^*, \*U) > 1- \gamma_2^2$, $\| (\*I - \*H_{\tau}-\*O_{ \tau^{\bot}  \*U^*}) \*U \|^2
    = \|(\*I - \*H_{\tau})(\*I - \*O_{ \tau^{\bot}  \*U^*}) \*U\|^2 \leq  \|(\*I - \*H_{\tau})(\*I - \*H_{\*U^*}) \*U\|^2 \leq \|(\*I - \*H_{\*U^*}) \*U\|^2 \leq \gamma_2^2 \|\*U\|^2.$

Thus, by Cauchy-Schwartz inequality,
\begin{align*}
    & |\*U^\top(\*I - \*H_{\tau}-\*O_{ \tau^{\bot}  \*U^*})\*X_{\tau_0}\bbeta_0|
     =  |\*U^\top(\*I - \*H_{\tau}-\*O_{ \tau^{\bot}  \*U^*})(\*I - \*H_{\tau})\*X_{\tau_0}\bbeta_0|
    \\ & \qquad \leq \|(\*I - \*H_{\tau}-\*O_{ \tau^{\bot}  \*U^*})\*U\| \|(\*I - \*H_{\tau})\*X_{\tau_0}\bbeta_0\|
    \leq \gamma_2 \|\*U\| \, \|(\*I - \*H_{\tau})\*X_{\tau_0}\bbeta_0\|.
\end{align*}
Therefore it follows from Lemma~\ref{lem::angle},
\begin{align*}
    (I_2) & \leq \P_{(\*U^*, \*U | \cdot)}\big\{(1- \gamma^2_1) \delta \|(\*I-\*H_\tau)\*X_{\tau_0}\bbeta_0\|^2  <  2 \sigma_0 |\*U^\top(\*I - \*H_{\tau}-\*O_{ \tau^{\bot}  \*U^*})\*X_{\tau_0}\bbeta_0| \big | \\
    & \qquad \qquad \, (\*U^*, \*U) \in
E(\gamma_1, \gamma_2) \big\} \\
   & \leq \P_{(\*U^*, \*U | \cdot )}\big\{(1- \gamma^2_1) \delta \|(\*I-\*H_\tau)\*X_{\tau_0}\bbeta_0\|^2 <  2\sigma_0 \gamma_2 \|\*U\| \, \|(\*I - \*H_{\tau})\*X_{\tau_0}\bbeta_0\| \big | \\
    & \qquad \qquad \, (\*U^*, \*U) \in
E(\gamma_1, \gamma_2) \big\} \\
    & = \P_{(\*U^*, \*U | \cdot)}\left\{\|\*U\|^2> \frac{(1-\gamma^2_1)^2\delta^2}{4\gamma^2_2}\frac{\|(\*I-\*H_\tau)\*X_{\tau_0}\bbeta_0\|^2}{\sigma_0^2} \middle | (\*U^*, \*U) \in
E(\gamma_1, \gamma_2)\right\}\\
    & \leq \P_{\*U}\left\{\|\*U\|^2> \frac{(1-\gamma^2_1)^2\delta^2}{4\gamma^2_2}\frac{\|(\*I-\*H_\tau)\*X_{\tau_0}\bbeta_0\|^2}{\sigma_0^2}\right\}\\
    & = \P_{\chi_n^2}\left\{\chi^2_n > \frac{(1-\gamma^2_1)^2\delta^2}{4\gamma^2_2}\frac{n |\tau_0 \setminus \tau| C_{\min}}{\sigma_0^2}\right\} \\
   &  \leq \exp\left\{-\frac{n}{2}\log(1-2t_2)- t_2\frac{(1-\gamma^2_1)^2\delta^2}{4\gamma^2_2}\frac{n |\tau_0 \setminus \tau| C_{\min}}{\sigma_0^2}\right\},
\end{align*}
for any $0<t_2<1/2.$

Now, by making of $(1-\gamma^2_1)(1 - \delta) =  (1-\gamma^2_1)^2\delta^2/4$, we obtain $\delta = \frac{2}{1- \gamma^2_1}(\sqrt{2 - \gamma_1^2} - 1)$.
Further
we make $t_1=t_2= \frac{\gamma_2}{2.04}$, so we have $-\frac{n}{2}\log(1-2t_1)=-\frac{n}{2}\log(1-2t_2)=-\frac{n}{2}\log(1-\frac{\gamma_2}{1.02}) \leq 2n\gamma_2$.
Then, intersect with the event $\{(\*U^*, \*U) \in E(\gamma_1, \gamma_2)\}$, we have
\begin{align*}
    & \P_{(\*U^*, \*U| \cdot)}\big\{\hat\tau_{\*U^*} \neq \tau_0 \big | (\*U^*, \*U) \in E(\gamma_1, \gamma_2)\big\} \\
    & \qquad <  \sum_{i=1}^{|\tau_0|} \sum_{j=0}^{i} \genfrac(){0pt}{0}{p-|\tau_0|}{j}\genfrac(){0pt}{0}{|\tau_0|}{i} \exp\left\{-\left(\sqrt{2 - \gamma_1^2} - 1\right)^2\frac{niC_{\min}}{2.04 \gamma_2\sigma_0^2} + 2n\gamma_2
    \right\} \\
    & \qquad <
    \sum_{i=1}^{|\tau_0|} \sum_{j=0}^{i} \genfrac(){0pt}{0}{p-|\tau_0|}{j}\genfrac(){0pt}{0}{|\tau_0|}{i} \exp\left\{- \frac{niC_{\min}}{12\sigma_0^2} \frac{\left(1 - \gamma_1^2\right)^2}{\gamma_2} + 2n\gamma_2
    \right\}.\end{align*}
The last inequality holds since $(\sqrt{2 - \gamma_1^2} - 1)^2 \geq 1.02 (1 - \gamma_1^2)^2/6$ for $\gamma_1^2 \in (0,1)$.
Since $\genfrac(){0pt}{0}{a}{b}\leq a^b$ and $\log(p-|\tau_0|)+\log(|\tau_0|) \leq \log(p^2/4) = 2 \log(p/2)$,
it follows
\begin{align}
\label{eq:U12a}    & \P_{(\*U^*, \*U | \cdot)}\{\hat\tau_{\*U^*} \neq \tau_0 | (\*U^*, \*U) \in E(\gamma_1, \gamma_2)\}\nonumber \\
      & \qquad \leq  \sum_{i=1}^{|\tau_0|} |\tau_0|^i
      \exp\left\{- \frac{niC_{\min}}{12\sigma_0^2} \frac{\left(1 - \gamma_1^2\right)^2}{\gamma_2} + 2n\gamma_2
    \right\} \sum_{j=0}^i(p-|\tau_0|)^j\nonumber\\
      & \qquad < 2 \sum_{i=1}^{|\tau_0|} \exp\left\{- n\left[i \frac{ C_{\min}}{12\sigma_0^2} \frac{\left(1 - \gamma_1^2\right)^2}{\gamma_2} - i \frac{2 \log(p/2)}{n}- 2\gamma_2\right]
    \right\}
    \nonumber\\
      & \qquad < \frac{2\exp\left\{- n\left[ \frac{ C_{\min}}{12\sigma_0^2} \frac{\left(1 - \gamma_1^2\right)^2}{\gamma_2}- \frac{2 \log(p/2)}{n}- 2\gamma_2\right]
    \right\}}{1-\exp\left\{- n\left[ \frac{ C_{\min}}{12\sigma_0^2} \frac{\left(1 - \gamma_1^2\right)^2}{\gamma_2}- \frac{2 \log(p/2)}{n}- 2\gamma_2\right]
    \right\}} \nonumber  \\
    & \qquad < 3\exp\left\{- \frac{n}{12\sigma_0^2}\left[ \frac{\left(1 - \gamma_1^2\right)^2}{\gamma_2} C_{\min} - 24\left( \frac{  \log(p/2)}{n}+ \gamma_2\right) \sigma_0^2\right]
    \right\}= L_c(\gamma_1, \gamma_2),
    \end{align}
since $ \frac{\left(1 - \gamma_1^2\right)^2}{\gamma_2} C_{\min} - 24\left( \frac{  \log(p/2)}{n}+ \gamma_2\right) \sigma_0^2 >0. $ The last inequality holds because
\begin{align*}
    &  \P_{(\*U^*, \*U| \cdot)}\{\hat\tau_{\*U^*} \neq \tau_0 | (\*U^*, \*U) \in E(\gamma_1, \gamma_2)\} \\
    &  \qquad \leq [2+  \P_{(\*U^*, \*U)}\{\hat\tau_{\*U^*} \neq \tau_0 | (\*U^*, \*U) \in E(\gamma_1, \gamma_2)\}]\\
    &  \hspace{2cm} \exp\left\{- n\left[ \frac{ C_{\min}}{12\sigma_0^2} \frac{\left(1 - \gamma_1^2\right)^2}{\gamma_2}- \frac{2 \log(p/2)}{n}- 2\gamma_2\right] \right\}\\
    & \qquad \leq  3\exp\left\{- n\left[ \frac{ C_{\min}}{12\sigma_0^2} \frac{\left(1 - \gamma_1^2\right)^2}{\gamma_2}- \frac{2 \log(p/2)}{n}- 2\gamma_2\right] \right\}.
\end{align*}

{

Then for any events $A, B$ and $C,$ we have
\begin{align}
\label{eq:ABC_indequality}
   \frac{\P(A \cap B)}{\P(B)} \leq \frac{\P(A \cap B \cap C)}{\P(B \cap C)}\frac{\P(B \cap C)}{\P(B)}  + \frac{\P(A \cap B \cap C^c)}{\P(B)} \leq  \P(A | B \cap C ) + \P(B \cap C^C)/\P(B).
\end{align}
 Make $A = \{\hat\tau_{\*U^*} \neq \tau_0\},$ $B = \{\rho(\*U^*, \*U) > 1- \gamma_2^2\}$ and $C =\left\{ \max_{\tau \neq \tau_0, |\tau|\leq |\tau_0|} \rho_{\tau^\bot}(\*U^*, \*X_{\tau_0}\bbeta_{0}) < \gamma_1^2 \right\}.$
 By Lemma~\ref{lem::angle}, we know $\*U^*$ is independent of  $\rho(\*U^*, \*U)$, therefore it follows from the above that $B$ and $C$ are independent and
\begin{align}
     & \P_{(\*U^*, \*U | \cdot)}(\hat \tau_{\*U^*} \neq \tau_0| \rho(\*U^*, \*U) > 1- \gamma_2^2) \nonumber\\
     &\leq  \P_{(\*U^*, \*U | \cdot)}\{\hat\tau_{\*U^*} \neq \tau_0 | (\*U^*, \*U) \in E(\gamma_1, \gamma_2)\} + \P\left(\max_{\tau \neq \tau_0, |\tau|\leq |\tau_0|} \rho_{\tau^\bot}(\*U^*, \*X_{\tau_0} \bbeta_0) \geq \gamma_1^2\right) \nonumber\\
     & \leq 3\exp\left\{- n\left[ \frac{ C_{\min}}{12\sigma_0^2} \frac{\left(1 - \gamma_1^2\right)^2}{\gamma_2}- \frac{2 \log(p/2)}{n}- 2\gamma_2\right]\right\} +  4(\arccos \gamma_1)^{n-|\tau_0|-1} p^{|\tau_0|} \nonumber
\end{align}

We then make $\gamma_1 = \sqrt{1-\gamma_2^{1/4}} \geq 1-1.6\gamma_2^{1/3} > 0$ for $\gamma_2 \in [0,0.24].$ Therefore $\arccos\gamma_1 \leq \arccos (1-1.6\gamma_2^{1/3})
\leq 2\gamma_2^{1/6} <1. $ Hence the above probability bound reduces to

\begin{align*}
     & \P_{(\*U^*, \*U | \cdot)}(\hat \tau_{\*U^*} \neq \tau_0 | \rho(\*U^*, \*U) > 1- \gamma_2^2) \\
     & \leq 3\exp\left\{- n\left[ \frac{ C_{\min}}{12\sigma_0^2\sqrt\gamma_2}- \frac{2 \log(p/2)}{n}- 2\gamma_2\right]\right\} +  4(64\gamma_2)^{\frac{n-|\tau_0|-1}{6}} p^{|\tau_0|}
\end{align*}
}\end{proof}

\begin{proof}[Proof of Lemma~\ref{lem::finite_penalize}]
By Lemma~\ref{lem_proj}, we let
{
$D(\tau, \*u^*)=  \frac{1}{2}\|(\*I - \*H_{\tau,\*u^*})\*Y\|^2 + \lambda |\tau| =  \frac{1}{2}\|(\*I-\*H_\tau-  \*O_{ \tau_0^{\bot} \*u^*})\*Y\|^2 + \lambda |\tau|$
for any $\tau \neq \tau_0,$ then
\begin{align}
\label{eq:dist_penalize}
& D(\tau, \*u^*) -D(\tau_0,\*u^*) \nonumber \\
& =\|(\*I - \*H_{\tau}-  \*O_{ \tau^{\bot}  \*u^*})\*Y\|^2 - \|(\*I - \*H_{\tau_0}- \*O_{ \tau_0^{\bot} \*u^*})\*Y\|^2 + \lambda(|\tau| - |\tau_0| ) \nonumber\\
&  = \|(\*I - \*H_{\tau}- \*O_{ \tau^{\bot}  \*u^*}) \*X_{\tau_0}\bbeta_0\|^2 - \*U^\top( \*H_{\tau}+ \*O_{ \tau^{\bot}  \*u^*} -  \*H_{\tau_0}-\*O_{ \tau_0^{\bot} \*u^*} )\*U \nonumber\\
& \quad   + 2\*U^\top(\*I - \*H_{\tau}-\*O_{ \tau^{\bot}  \*u^*})\*X_{\tau_0}\bbeta_0 + 2\lambda(|\tau|-|\tau_0|).\end{align}
}

Let $E(\gamma_1, \gamma_2)$ be the event defined in \eqref{eq:def_E}, then conditional on $E(\gamma_1, \gamma_2),$
\begin{align*}
   & \P_{(\*U^*, \*U| \cdot)}\left\{ D(\tau, \*U^*) -D(\tau_0,\*U^*)< 0|  (\*U^*, \*U) \in E(\gamma_1, \gamma_2) \right\}\\
      & \leq \P_{(\*U^*, \*U|\cdot)}\big\{\|(\*I - \*H_{\tau}- \*O_{ \tau^{\bot}  \*u^*}) \*X_{\tau_0}\bbeta_0\|^2- \sigma_0^2\*U^\top( \*H_{\tau}+ \*O_{ \tau^{\bot}\*U^*}-  \*H_{\tau_0}- \*O_{ \tau_0^{\bot} \*U^*} )\*U \\
       & \qquad + 2\sigma_0\*U^\top(\*I - \*H_{\tau}-\*O_{ \tau^{\bot}  \*U^*})\*X_{\tau_0}\bbeta_0 +  2\lambda(|\tau|-|\tau_0|) <0\big| (\*U^*, \*U) \in E(\gamma_1, \gamma_2) \big\}\\
        & \leq  \P_{(\*U^*, \*U| \cdot)}\big\{(1-\delta)\|(\*I - \*H_{\tau}- \*O_{ \tau^{\bot}  \*u^*}) \*X_{\tau_0}\bbeta_0\|^2 \\
    & \qquad{} - \sigma_0^2 \*U^\top(\*H_{\tau}+\*O_{ \tau^{\bot}  \*U^*} -  \*H_{\tau_0}-\*O_{ \tau_0^{\bot} \*U^*} )\*U  + \lambda(|\tau|-|\tau_0|) <0\big| (\*U^*, \*U) \in E(\gamma_1, \gamma_2)\big\} \\
    & \quad + \P_{(\*U^*, \*U |\cdot)}\big\{\delta\|(\*I - \*H_{\tau}- \*O_{ \tau^{\bot}  \*u^*}) \*X_{\tau_0}\bbeta_0\|^2\\
    & \qquad{} + 2\sigma_0\*U^\top(\*I - \*H_{\tau}-\*O_{ \tau^{\bot}  \*U^*})\*X_{\tau_0}\bbeta_0 +\lambda(|\tau|-|\tau_0|) ) < 0\big| (\*U^*, \*U) \in E(\gamma_1, \gamma_2)\big\} \\
    & \leq  \P_{(\*U^*, \*U| \cdot)}\big\{(1-\gamma^2_1)(1-\delta)\|(\*I-\*H_\tau)\*X_{\tau_0}\bbeta_0\|^2 \\
    & \qquad{} - \sigma_0^2 \*U^\top(\*H_{\tau}+\*O_{ \tau^{\bot}  \*U^*} -  \*H_{\tau_0}-\*O_{ \tau_0^{\bot} \*U^*} )\*U  + \lambda(|\tau|-|\tau_0|) <0\big| (\*U^*, \*U) \in E(\gamma_1, \gamma_2)\big\} \\
    & \quad + \P_{(\*U^*, \*U |\cdot)}\big\{\delta\|(\*I - \*H_{\tau}- \*O_{ \tau^{\bot}  \*u^*}) \*X_{\tau_0}\bbeta_0\|^2 \\
    & \qquad{} + 2\sigma_0\*U^\top(\*I - \*H_{\tau}-\*O_{ \tau^{\bot}  \*U^*})\*X_{\tau_0}\bbeta_0 +\lambda(|\tau|-|\tau_0|) ) < 0\big| (\*U^*, \*U) \in E(\gamma_1, \gamma_2)\big\} \\
    &= (I_1) + (I_2), \numberthis \label{eq:decom_prob}
\end{align*}
for any $\delta \in (0,1).$

To derive an upper bound for $I_1$, we have
\begin{align*}
  &\| (\*I - \*H_{\tau}- \*O_{ \tau^{\bot}  \*U^*}) \*U \|^2
   = \|(\*I - \*H_{\tau})(\*I - \*O_{ \tau^{\bot}  \*U^*}) \*U \|^2\\
    & \leq  \|(\*I - \*H_{\tau})(\*I - \*O_{ \tau^{\bot}  \*U^*}) \*U\|^2 \leq \|(\*I -\*O_{ \tau^{\bot}  \*U^*}) \*U\|^2 \leq \gamma_2^2 \|\*U\|^2.
\end{align*}
First,
\begin{align*}
    \*U^\top(\*I - \*H_{\tau}-\*O_{ \tau^{\bot}  \*U^*})\*X_{\tau_0}\bbeta_0 &  =  \*U^\top(\*I - \*H_{\tau}-\*O_{ \tau^{\bot}  \*U^*})(\*I - \*H_{\tau})\*X_{\tau_0}\bbeta_0
    \\ & \leq \|\*U^\top(\*I - \*H_{\tau}-\*O_{ \tau^{\bot}  \*U^*})\| \|(\*I - \*H_{\tau})\*X_{\tau_0}\bbeta_0\|.
\end{align*}
Then
\begin{align*}
    \*U^\top( \*H_{\tau}+\*O_{ \tau^{\bot}  \*U^*} -  \*H_{\tau_0}-\*O_{ \tau_0^{\bot} \*U^*} )\*U\leq 2\gamma_2\|\*U\|^2.
\end{align*}
Because
\begin{align*}
     \|\*U^\top(\*I - \*H_{\tau}-O_{ \tau^{\bot}  \*U^*})\|^2 < \gamma^2_2  \|\*U\|^2,
\end{align*}
it then follows from Lemma~\ref{lem::angle} that if $|\tau| \leq  |\tau_0|$ and if $\frac{\lambda}{n} < \frac{1}{6}(1-\gamma_1^2)C_{\min} \leq \frac{ (1- \gamma_1^2) \|(\*I-\*H_\tau)\*X_{\tau_0}\bbeta_0\|^2}{6|\tau_0 \setminus \tau|},$ and then from $|\tau_0 \setminus\tau| \geq |\tau_0| - |\tau|,$ we have
\begin{align*}
    (I_1) & \leq \P_{(\*U^*, \*U|\cdot)}\left\{\|\*U\|^2 > \frac{(1-\gamma^2_1)(1-\delta)}{\gamma^2_2}\frac{\|(\*I-\*H_\tau)\*X_{\tau_0}\bbeta_0\|^2 }{\sigma_0^2} + \frac{\lambda(|\tau|-|\tau_0|)}{\gamma_2^2\sigma_0^2} \middle | (\*U^*,
    \*U) \in E(\gamma_1, \gamma_2)\right\} \numberthis \label{eq: I_bound}\\
    & \leq \P_{\*U}\left\{\|\*U\|^2 > \frac{(1-\gamma^2_1)(1-\delta)}{\gamma^2_2}\frac{\|(\*I-\*H_\tau)\*X_{\tau_0}\bbeta_0\|^2 }{\sigma_0^2} + \frac{\lambda(|\tau|-|\tau_0|)}{\gamma_2^2\sigma_0^2}\right\}\\
    &\leq  \P\left\{\chi^2_n> \frac{(1-\gamma^2_1)(1-\delta-1/6)}{\gamma^2_2}\frac{niC_{\min}}{\sigma_0^2}\right\}\\
    &\leq  \exp\left\{-\frac{n}{2}\log(1-2t_1) - t_1\frac{(1-\gamma^2_1)(1-\delta-1/6  )}{\gamma^2_2}\frac{niC_{\min}}{\sigma_0^2}  \right\},
\end{align*}
for any $0<t_1<1/2.$ Otherwise when $|\tau| > |\tau_0|,$ we would have
\begin{align*}
    (I_1) \leq  \exp\left\{-\frac{n}{2}\log(1-2t_1) -  t_1 \frac{\lambda(|\tau|-|\tau_0|)}{\gamma_2^2\sigma_0^2} \right\}.
\end{align*}
The above inequalities are derived from Markov inequality and moment generating function of chi-square distribution.
For $(I_2)$,  if $\frac{\lambda}{n} < \frac{1}{6}(1-\gamma_1^2)C_{\min} \leq \frac{(1- \gamma_1^2) \|(\*I-\*H_\tau)\*X_{\tau_0}\bbeta_0\|^2}{6 |\tau_0 \setminus \tau| },$
by Cauchy-Schwartz inequality and Lemma~\ref{lem::angle}, when $|\tau| \leq |\tau_0|,$  we have
\begin{align*}
    & (I_2) \leq
     \P_{(\*U^*, \*U |\cdot)}\big\{(1- \gamma^2_1) \delta \|(\*I-\*H_\tau)\*X_{\tau_0}\bbeta_0\|^2 \\
    & \qquad{} + 2\sigma_0\*U^\top(\*I - \*H_{\tau}-\*O_{ \tau^{\bot}  \*U^*})\*X_{\tau_0}\bbeta_0 +\lambda(|\tau|-|\tau_0|) ) < 0\big| (\*U^*, \*U) \in E(\gamma_1, \gamma_2)\big\} \\
    &\leq \P_{(\*U^*, \*U|\cdot)}\big\{(1- \gamma^2_1) \delta \|(\*I-\*H_\tau)\*X_{\tau_0}\bbeta_0\|^2  <  2\sigma_0\|\*U^\top(\*I - \*H_{\tau}-\*O_{ \tau^{\bot}  \*U^*})\|\|(\*I-\*H_\tau)\*X_{\tau_0}\bbeta_0\|\\
     & \qquad  \hspace{4cm}  - \lambda(|\tau|-|\tau_0|) \big| (\*U^*, \*U) \in E(\gamma_1, \gamma_2)\big\} \\
    &= \P_{(\*U^*, \*U|\cdot)}\big\{2\sigma_0\|\*U^\top(\*I - \*H_{\tau}-\*O_{ \tau^{\bot}  \*U^*})\| >(1-\gamma_1^2 ) \delta \|(\*I-\*H_\tau)\*X_{\tau_0}\bbeta_0\|\\
     & \qquad \hspace{4cm} + \lambda(|\tau|-|\tau_0|)/\|(\*I-\*H_\tau)\*X_{\tau_0}\bbeta_0\|\big| (\*U^*, \*U) \in E(\gamma_1, \gamma_2)\big\} \\
         &\leq \P_{(\*U^*, \*U |\cdot)}\left\{\|\*U\|^2 > \frac{(1-\gamma^2_1)^2(\delta-1/6)^2}{4\gamma^2_2}\frac{\|(\*I-\*H_\tau)\*X_{\tau_0}\bbeta_0\|^2}{\sigma_0^2}\middle | (\*U^*, \*U) \in E(\gamma_1, \gamma_2)\right\}\\
    &\leq \P_{\*U}\left\{\|\*U\|^2 > \frac{(1-\gamma^2_1)^2(\delta-1/6)^2}{4\gamma^2_2}\frac{\|(\*I-\*H_\tau)\*X_{\tau_0}\bbeta_0\|^2}{\sigma_0^2}\right\}\\
    &\leq \P\left\{\chi^2_n > \frac{(1-\gamma^2_1)^2(\delta-1/6)^2}{4\gamma^2_2}\frac{niC_{\min}}{\sigma_0^2}\right\} \\
    &\leq \exp\left\{-\frac{n}{2}\log(1-2t_2)- t_2\frac{(1-\gamma^2_1)^2(\delta-1/6)^2}{4\gamma^2_2}\frac{niC_{\min}}{\sigma_0^2}\right\},
\end{align*}
for any $0<t_2<1/2$ and $\delta > 1/6.$

When $|\tau|  >  |\tau_0|$, from the fact that { $\delta\|(\*I - \*H_{\tau}- \*O_{ \tau^{\bot}  \*u^*}) \*X_{\tau_0}\bbeta_0\|^2
    + 2\sigma_0\*U^\top(\*I - \*H_{\tau}-\*O_{ \tau^{\bot}  \*U^*})\*X_{\tau_0}\bbeta_0  \geq \frac{-\sigma_0^2 \|(\*I - \*H_{\tau}-\*O_{ \tau^{\bot}  \*U^*}) \*U\|^2 }{\delta}   \geq -\frac{\sigma_0^2\gamma_2^2 \|\*U\|^2}{\delta}$, we have
\begin{align*}
   (I_2)
     &\leq \P_{(\*U^*, \*U|\cdot)}\left\{\sigma_0^2\gamma_2^2\|\*U\|^2\|\*X_{\tau_0}\bbeta_0\|^2 >\delta\lambda(|\tau|-|\tau_0|) \middle | (\*U^*, \*U) \in E(\gamma_1, \gamma_2)\right\} \\
    &\leq \P_{\*U}\left\{\|\*U\|^2 > \frac{\delta\lambda(|\tau|-|\tau_0|)}{\gamma_2^2 \sigma_0^2 }\right\}\\
    & \leq  \exp\left\{-\frac{n}{2}\log(1-2t_2)-t_2\frac{\delta\lambda(|\tau|-|\tau_0|)}{\gamma_2^2 \sigma_0^2}\right\}. \numberthis \label{eq: bound_I2}
\end{align*}
}

Now, by making of $(1-\gamma^2_1)(1 - \delta-1/6) =  (1-\gamma^2_1)^2(\delta-1/6)^2/4$, we obtain $\delta = \frac{2}{1- \gamma^2_1}(\sqrt{\frac{5}{3} - \frac{2}{3}\gamma_1^2} - 1) + \frac{1}{6} \geq 0.74$.
Further
we make $t_1=t_2= \frac{\gamma_2}{2.04}$, so we have $-\frac{n}{2}\log(1-2t_1)=-\frac{n}{2}\log(1-2t_2)=-\frac{n}{2}\log(1-\frac{\gamma_2}{1.02}) \leq 2n\gamma_2$.
Then, intersect with the event $\{(\*U^*, \*U) \in E(\gamma_1, \gamma_2)\}$, we have
\begin{align*}
    & \P_{(\*U^*, \*U)}\big\{\hat\tau_{\*U^*} \neq \tau_0, \big | (\*U^*, \*U) \in E(\gamma_1, \gamma_2)\big\} \\
    & \qquad \leq  2\sum_{i=1}^{|\tau_0|} \sum_{j=0}^{i} \genfrac(){0pt}{0}{p-|\tau_0|}{j}\genfrac(){0pt}{0}{|\tau_0|}{i} \exp\left\{-\left(\sqrt{\frac{5}{3} - \frac{2}{3}\gamma_1^2} - 1\right)^2\frac{niC_{\min}}{2.04 \gamma_2\sigma_0^2 } + 2n\gamma_2
    \right\} \\
    &   + 2\sum_{i=0}^{|\tau_0|} \sum_{j=i+1}^{p} \genfrac(){0pt}{0}{p-|\tau_0|}{j}\genfrac(){0pt}{0}{|\tau_0|}{i} \exp\left\{-\frac{0.74\lambda(j-i)}{2.04\gamma_2\sigma_0^2 } + 2\gamma_2n \right\}\\
    & \qquad \leq
    2\sum_{i=1}^{|\tau_0|} \sum_{j=0}^{i} \genfrac(){0pt}{0}{p-|\tau_0|}{j}\genfrac(){0pt}{0}{|\tau_0|}{i} \exp\left\{- \frac{niC_{\min}}{26\sigma_0^2} \frac{\left(1 - \gamma_1^2\right)^2}{\gamma_2} + 2n\gamma_2
    \right\}\\
     &   + 2\sum_{i=0}^{|\tau_0|} \sum_{j=i+1}^{p} \genfrac(){0pt}{0}{p-|\tau_0|}{j}\genfrac(){0pt}{0}{|\tau_0|}{i} \exp\left\{-\frac{\lambda(j-i)}{4\gamma_2\sigma_0^2} + 2\gamma_2n \right\}.
     \end{align*}
The last inequality holds since $(\sqrt{\frac{5}{3} - \frac{2}{3}\gamma_1^2} - 1)^2 \geq 1.02 (1 - \gamma_1^2)^2/13$ for $\gamma_1^2 \in (0,1)$.

By similar calculation to that in \eqref{eq:U12a}, the first part of the above can be bounded by
\begin{align*}
   L_p(\gamma_1, \gamma_2)=  3\exp\left\{- \frac{n}{26\sigma_0^2}\left[ \frac{\left(1 - \gamma_1^2\right)^2}{\gamma_2} C_{\min} - 52\left( \frac{  \log(p/2)}{n}+ \gamma_2\right) \sigma_0^2\right]
    \right\}.
\end{align*}
{
As for the second part,
\begin{align*}
   &  2\sum_{i=0}^{|\tau_0|} \sum_{j=i+1}^{p} \genfrac(){0pt}{0}{p-|\tau_0|}{j}\genfrac(){0pt}{0}{|\tau_0|}{i} \exp\left\{-\frac{\lambda(j-i)}{4\gamma_2\sigma_0^2} + 2\gamma_2n \right\} \\
    & \leq  2\sum_{i=0}^{|\tau_0|} |\tau_0|^i \exp\left\{\gamma_2 n+ \frac{\lambda i}{4\gamma_2 \sigma_0^2} \right\} \sum_{j=i+1}^{p} \exp \left[-j\left\{\frac{\lambda}{4\gamma_2 \sigma_0^2} - \log(p-|\tau_0|)\right\}\right]\\
&\leq \frac{2\sum_{i=0}^{|\tau_0|} \exp\left[-\frac{\lambda}{4\gamma_2 \sigma_0^2} +\gamma_2n + \log(p-|\tau_0|)+ i \{\log(|\tau_0|) + \log (p-|\tau_0|)\}\right]}{1-\exp \left\{\frac{\lambda}{4\gamma_2 \sigma_0^2} - \log(p-|\tau_0|)\right\} }\\
&\leq \frac{2\exp\left[-\frac{\lambda}{4\gamma_2 \sigma_0^2} + \gamma_2n + (|\tau_0|+1) \{\log |\tau_0| + \log(p-|\tau_0|\}\right]}{1-\exp \left\{\frac{\lambda}{4\gamma_2 \sigma_0^2} - \log(p-|\tau_0|)\right\}}\\
&\leq 3\exp\left(-\frac{1}{4\gamma_2\sigma_0^2} nt  \right), \numberthis \label{eq:bound_penalty}
\end{align*}
if
\begin{align}
 \label{eq:lambda_0}
 \frac{\lambda}{n} \in \left\{\frac{\lambda^{(1)}_0}{n}+t, \frac{1}{6}(1-\gamma_1^2)C_{\min}\right\},
\end{align}
where $\lambda^{(1)}_0=4\gamma_2\sigma_0^2\left[\gamma_2 n +  (|\tau_0|+1) \{\log |\tau_0| + \log(p-|\tau_0|\}\right].$

{
It then follows from \eqref{eq:ABC_indequality} and  Lemma~\ref{lem::angle}  that
\begin{align*}
         & \P_{(\*U^*, \*U | \cdot)}(\hat \tau_{\*U^*} \neq \tau_0| \rho(\*U^*, \*U) > 1- \gamma_2^2) \nonumber\\
     &\leq  \P_{(\*U^*, \*U | \cdot)}\{\hat\tau_{\*U^*} \neq \tau_0 | (\*U^*, \*U) \in E(\gamma_1, \gamma_2)\} + \P\left(\max_{\tau \neq \tau_0, |\tau|\leq |\tau_0|} \rho_{\tau^\bot}(\*U, \*X_{\tau_0} \bbeta_0) \geq \gamma_1^2\right) \nonumber\\
     & \leq L_p(\gamma_1, \gamma_2) + 3\exp\left(-\frac{1}{4\gamma_2\sigma_0^2} nt  \right) +  4(\arccos \gamma_1)^{n-|\tau_0|-1} p^{|\tau_0|}.  \numberthis
     \label{eq:bound_penalty_single}
\end{align*}
}

}

{
We then make $\gamma_1 = \sqrt{1-\gamma_2^{1/4}} \geq 1-1.6\gamma_2^{1/3} > 0$ for $\gamma_2 \in [0,0.24].$ Therefore $\arccos\gamma_1 \leq \arccos (1-1.6\gamma_2^{1/3})
\leq 2\gamma_2^{1/6} <1. $  In addition, we make $t=\sqrt{\gamma_2}\sigma_0^2.$ Hence the above probability bound reduces to

\begin{align*}
     & \P_{(\*U^*, \*U|\cdot)}(\hat \tau_{\*U^*} \neq \tau_0| \rho(\*U^*, \*U) > 1- \gamma_2^2) \\
     & \leq   3\exp\left\{- \frac{n}{26\sigma_0^2}\left[ \frac{C_{\min} }{\sqrt{\gamma_2}} - 52\left( \frac{  \log(p/2)}{n}+ \gamma_2\right) \sigma_0^2\right]
    \right\} + 3\exp\left(-\frac{n}{4\gamma_2^{1/2}}   \right) \\
     & \hspace{4cm} +  4(64\gamma_2)^{\frac{n-|\tau_0|-1}{6}} p^{|\tau_0|},
\end{align*} }
where $\gamma_2 < 1/64$ since $\gamma_2^{1/4} < 0.35. $

{

Finally, we will show that the range required for the tuning parameter $\lambda$ in Lemma~\ref{lem::finite_penalize} satisfies \eqref{eq:lambda_0}  and is nonempty. Make $t = \sqrt{\gamma_2}\sigma_0^2,$ it then follows from \eqref{eq:lambda_0} and the fact that $\log(|\tau_0|)
 + \log(p- |\tau_0|) \leq 2\log(p/2)$ when $\gamma^{1/4}_2 <  \frac{ C_{\min}}{24 \{2 + 2(|\tau_0|+1)\log(p/2)/n\}\sigma_0^2}$ that
\begin{align*}
    \frac{\lambda_0^{(1)}}{n} + t & \leq 4 \gamma_2^{1/2}\left\{\gamma_2^{3/2} + 1 + 2\gamma_2^{1/2}(|\tau_0|+1)\frac{\log(p/2)}{n} \right\}\sigma_0^2\\
    & < 4 \gamma_2^{1/2}\left\{2 + 2(|\tau_0|+1)\frac{\log(p/2)}{n} \right\}\sigma^2_0\\
    & < \frac{1}{6}\gamma_2^{1/4}C_{\min} =  \frac{1}{6}(1- \gamma_1^2)C_{\min}.   \numberthis \label{eq:lambda_range_proof}
\end{align*}
The second to last inequality shows that the range for $\lambda$ specified in Lemma~\ref{lem::finite_penalize}, $ \lambda \in \left[4n\gamma_2^{1/2}\left\{2 + 2(|\tau_0|+1)\frac{\log(p/2)}{n} \right\}\sigma_0^2,
     \frac{n\gamma_2^{1/4}}{6} C_{\min}\right],$ always exists, and it satisfies \eqref{eq:lambda_0}.
    \label{par:proof_lambda}

}
\end{proof}

\subsection{Proof of Lemma~\ref{lem::finite_complete} and  Lemma~\ref{lem::finite_penalize_complete} }
\label{sec:proof_lem35}
In order to prove Lemma~\ref{lem::finite_complete} and  Lemma~\ref{lem::finite_penalize_complete}, we first introduce the following techinical lemma.
\begin{lemma}\label{lem:angle_ud}
Suppose $\*U^*_1, \ldots, \*U^*_d$ are $d$ i.i.d. copies of $\*U^* \sim N(0, \*I),$ then
\begin{align*}
\P\left(\bigcap_{b=1}^d\{\rho(\*U^*_b, \*U) \leq  1- \gamma_2^2\} \right) \leq \left(1-\frac{\gamma_2^{n-1}}{n-1}\right)^d.
\end{align*}
\end{lemma}
\begin{proof}[Proof of Lemma~\ref{lem:angle_ud}]
By \eqref{eq:u_epsi-1}, $\rho(\*U_b^*, \*U)$ and $\*U$ are independent. It then follows from Lemma~\ref{lem::angle} and the fact $\arcsin(\gamma_2) > \gamma_2$ that
\begin{align*}
    & \P\left(\bigcap_{b=1}^d\{\rho(\*U_b^*, \*U) \leq  1- \gamma_2^2\} \right) \\
    & = E\left\{\P\left(\bigcap_{b=1}^d\{\rho(\*U_b^*, \*U) \leq  1- \gamma_2^2\} \middle | \*U \right)\right\}\\
    & = E\left\{\left(1- \P(\rho(\*U^*, \*U) >  1- \gamma_2^2| \*U )\right)^d\right\}\\
    & = \left(1- \P(\rho(\*U^*, \*U) >  1- \gamma_2^2)\right)^d \numberthis \label{eq:bound_Ub}\\
    & \leq \left(1-\frac{\gamma_2^{n-2}\arcsin(\gamma_2)}{n-1}\right)^d \leq \left(1-\frac{\gamma_2^{n-1}}{n-1}\right)^d.
\end{align*}
\end{proof}
\begin{proof}[Proof of Lemma~\ref{lem::finite_complete}]
We can decompose the probability $\tau_0 \notin S^{(d)}$ into
\begin{align}
\label{eq:prob_decompose}
     & \P(\tau_0 \notin S^{(d)}) \nonumber \\
     & = \P\left(\tau_0 \notin S^{(d)},\bigcup_{b=1}^d\{\rho(\*U_b^*, \*U) >  1- \gamma_2^2\} \right) + \P\left(\tau_0 \notin S^{(d)},\bigcap_{b=1}^d\{\rho(\*U_b^*, \*U) \leq  1- \gamma_2^2\} \right)\nonumber\\
     & \leq \P\left(\tau_0 \notin S^{(d)},\bigcup_{b=1}^d\{\rho(\*U_b^*, \*U) >  1- \gamma_2^2\} \right) + \P\left(\bigcap_{b=1}^d\{\rho(\*U_b^*, \*U) \leq  1- \gamma_2^2\} \right).
\end{align}
To bound the first term of \eqref{eq:prob_decompose}, let $D_b = \{\rho(\*U_b^*, \*U) >  1- \gamma_2^2\}\bigcap_{b' < b} \{\rho(\*U_{b'}^*, \*U) \leq  1- \gamma_2^2 \}$. By the fact that $\bigcup_{b=1}^d\{\rho(\*U_b^*, \*U) >  1- \gamma_2^2\} = \bigcup_{b=1}^d D_b,$ and $D_1, \dots, D_d$ are mutually exclusive, we have
\begin{align*}
     &\P\left(\tau_0 \notin S^{(d)},\bigcup_{b=1}^d\{\rho(\*U_b^*, \*U) >  1- \gamma_2^2\} \right)
      = \sum_{b=1}^d \P\left(\tau_0 \notin S^{(d)},D_b \right)  = \sum_{b=1}^d \P \left(\tau_0 \notin S^{(d)}\middle | D_b \right) \P(D_b).
\end{align*}
Then by Lemma~\ref{lem::angle}, $\*U$ and the event $D_b$ are independent, therefore
\begin{align*}
    \P \left(\tau_0 \notin S^{(d)}\middle | D_b \right)  \leq \P (\hat{\tau}_{\*U_b^*} \neq \tau_0 | D_b) = E_{\*U}\{\P (\hat{\tau}_{\*U_b^*} \neq \tau_0 | D_b, \*U) \}.
\end{align*}
Then because given $\*U,$ both $\hat{\tau}_{\*U_b^*}$ and  $\rho(\*U^*_b, \*U)$ are independent of $\{\rho(\*U^*_{b'}, \*U), b' \neq b \},$
\begin{align*}
    \P (\hat{\tau}_{\*U_b^*} \neq \tau_0 | D_b, \*U)  = \P (\hat{\tau}_{\*U_b^*} \neq \tau_0 | \rho(\*U_b^*, \*U) \leq  1- \gamma_2^2, \*U)  = \P (\hat{\tau}_{\*U^*} \neq \tau_0 | \rho(\*U^*, \*U) \leq  1- \gamma_2^2, \*U),
\end{align*}
from which it follows
\begin{align*}
   & \P\left(\tau_0 \notin S^{(d)},\bigcup_{b=1}^d\{\rho(\*U_b^*, \*U) >  1- \gamma_2^2\} \right) \\
   & \leq E_{\*U}\left\{\P (\hat{\tau}_{\*U^*} \neq \tau_0 | \rho(\*U^*, \*U) \leq  1- \gamma_2^2, \*U) \right\}\sum_{b=1}^d \P(D_b) \\
   & \leq \P (\hat{\tau}_{\*U^*} \neq \tau_0 | \rho(\*U^*, \*U) \leq  1- \gamma_2^2) \P \left(\bigcup_{b=1}^d\{\rho(\*U_b^*, \*U) >  1- \gamma_2^2\}\right)\\
   & \leq \P (\hat{\tau}_{\*U^*} \neq \tau_0 | \rho(\*U^*, \*U) \leq  1- \gamma_2^2).
\end{align*}
It then follows that \eqref{eq:prob_decompose} reduces to
\begin{align}
    \label{eq:inequality_final}
    \P(\tau_0 \notin S^{(d)})\leq \P\left(\hat\tau_{\*U^*} \neq \tau_0\middle |\rho(\*U^*, \*U) >  1- \gamma_2^2 \right) + \P\left(\bigcap_{b=1}^d\{\rho(\*U_b^*, \*U) \leq  1- \gamma_2^2\} \right).
\end{align}

Then the probability bound in \eqref{eq:prob_bound_constraint} follows immediately from Lemma~\ref{lem::finite} and Lemma~\ref{lem:angle_ud}.
\end{proof}
\begin{proof}[Proof of Lemma~\ref{lem::finite_penalize_complete}]
By \eqref{eq:inequality_final}, the probability bound in \eqref{eq:prob_bound_penalize_finite} follows immediately from Lemma~\ref{lem::finite_penalize} and Lemma~\ref{lem:angle_ud}.
\end{proof}

 \subsection{Proof of  Lemma~\ref{lem::asymptotic_constraint} and Lemma~\ref{lem::symptotic_bound_penalize0}}
\label{sec:proof_lem6}

 \begin{proof}[Proof of Lemma~\ref{lem::asymptotic_constraint}]
 By \eqref{eq:dist}, for any $\delta \in (0,1)$, for any $\*u^*$ such that $$ \max_{\tau \neq \tau_0, |\tau|\leq |\tau_0|} \rho_{\tau^\bot}(\*u^*, \*X_{\tau_0}\bbeta_0) < \gamma_1^2,$$
we have
\begin{align*}
   & \P_{\*U}\left\{ D(\tau,\*u^*) -D(\tau_0,\*u^*)< 0\right\}\\
    & \leq \P_{\*U}\bigg\{(1- \gamma^2_1) \|(\*I-\*H_\tau)\*X_{\tau_0}\bbeta_0\|^2 -\sigma_0^2\*U^\top( \*H_{\tau,\*u^*} - \*H_{\tau_0,\*u^*})\*U \\
    & \quad \quad  \hspace{3cm}+ 2\sigma_0\*U^\top(\*I - \*H_{\tau,\*u^*})\*X_{\tau_0}\bbeta_0 <0\bigg\}\\
    & \leq \P_{\*U}\bigg\{(1-\gamma^2_1)(1-\delta)\|(\*I-\*H_\tau)\*X_{\tau_0}\bbeta_0\|^2 \nonumber- \sigma_0^2\*U^\top( \*H_{\tau,\*u^*} - \*H_{\tau_0,\*u^*})\*U <0\bigg\} \\
    & \qquad + \P_{\*U}\bigg\{(1-\gamma^2_1)\delta\|(\*I-\*H_\tau)\*X_{\tau_0}\bbeta_0\|^2 + 2\sigma_0 \*U^\top(\*I - \*H_{\tau,\*u^*})\*X_{\tau_0}\bbeta_0 < 0\bigg\} \\ & = (I_1) + (I_2).
\end{align*}
By Lemma 4 of \citep{shen_constrained_2013},  we bound the log of the moment generating function $M(t)$ of $\*U^\top( \*H_{\tau,\*u^*} - \*H_{\tau_0,\*u^*})\*U$
\begin{align}
\label{eq:log_mgf}
\log\{M(t)\} &= \sum_{r=1}^{\infty} \frac{2^{r-1}t^r}{r} \tr\{(\*H_{\tau,\*u^*} - \*H_{\tau_0,\*u^*})^r\} \nonumber\\
&\leq t(|\tau| - |\tau_0|) + \frac{t^2}{1-2t}\tr\{(\*H_{\tau,\*u^*} - \*H_{\tau_0,\*u^*})^2\}\leq 2t |\tau \setminus \tau_0| \leq 2t|\tau_0 \setminus \tau|,
\end{align}
for any $0<t<1/2.$ Therefore by Markov Inequality
\begin{align*}
    (I_1)  \leq \exp\left\{2t_1|\tau\setminus \tau_0| - \frac{t_1(1-\delta)(1-\gamma^2_1)n |\tau_0\setminus \tau| C_{\min}}{\sigma_0^2}\right\},
\end{align*}
for any $0<t_1<1/2.$
Further because $2\sigma_0\*U^\top(\*I - \*H_{\tau,\*u^*})\*X_{\tau_0}\bbeta_0$ follows $N(0, 4\sigma^2_0\|(\*I - \*H_{\tau,\*u^*})\*X_{\tau_0}\bbeta_0\|^2)$, then by Markov inequality and moment generating function of the normal distribution , we have
\begin{align*}
  (I_2) \leq \exp\left\{\frac{(2t_2^2- \delta t_2) (1-\gamma_1^2) n |\tau_0 \setminus \tau| C_{\min}}{\sigma_0^2}\right\}  \end{align*}
for any $0<t_2<1/2.$
It then follows that
\begin{align*}
    \P_{\*U}(\hat\tau_{\*u^*} \neq \tau_0) &  \leq \sum_{i=0}^{|\tau_0|} \sum_{j=0}^{i} \genfrac(){0pt}{0}{|\tau_0|}{i}\genfrac(){0pt}{0}{p-|\tau_0|}{j}  \Bigg[\exp\left\{2t_1j - \frac{t_1(1-\delta)(1-\gamma^2_1)n i C_{\min}}{\sigma_0^2}\right\}   \\ &  \qquad \qquad +   \exp\left\{\frac{(2t_2^2- \delta t_2) (1-\gamma_1^2) n i C_{\min}}{\sigma_0^2}\right\}\Bigg].\\
\end{align*}
We can make $t_1=t_2=1/3$, $\delta=5/6,$ therefore $t_1(1-\delta) =-(2t_2^2- \delta t_2)=1/18.$ Then by the fact that $\genfrac(){0pt}{0}{a}{b}\leq a^b$, the probability bound above can be simplified as
\begin{align*}
    & \P_{\*U}(\hat\tau_{\*u^*} \neq \tau_0) \leq 2\sum_{i=0}^{|\tau_0|} \sum_{j=0}^{i}(p-|\tau_0|)^j |\tau_0|^i\exp\left\{-\frac{(1-\gamma_1^2)nC_{\min}}{18 \sigma_0^2}i + \frac{2}{3}j\right\}\\
  &    = 2\sum_{i=0}^{|\tau_0|} \exp\left[-i\left\{\frac{(1-\gamma_1^2)nC_{\min}}{18 \sigma_0^2} -\log |\tau_0|\right\}\right] \sum_{j=0}^{i} \exp\left[j\left\{\frac{2}{3} + \log(p-|\tau_0|)\right\} \right].\\
\end{align*}
Then we have
\begin{align*}
  \sum_{j=0}^{i} \exp\left[j\left\{\frac{2}{3} + \log(p-|\tau_0|)\right\} \right] \leq \frac{\exp\left[(i+1)\left\{\frac{2}{3} + \log(p-|\tau_0|)\right\} \right]}{ \exp\left\{\frac{2}{3} + \log(p-|\tau_0|)\right\} - 1 }\\
\leq \frac{\exp\left[i\left\{\frac{2}{3} + \log(p-|\tau_0|)\right\} \right]}{1-e^{-2/3}}.
\end{align*}
It then follows from $\log(p-|\tau_0|) + \log(|\tau_0|) \leq 2\log p -1$ that
\begin{align*}
    & \P_{\*U}(\hat\tau_{\*u^*} \neq \tau_0) \leq \frac{2}{1-e^{-2/3}}\sum_{i=1}^{|\tau_0|} \exp\left[-i\left\{\frac{(1-\gamma_1^2)nC_{\min}}{18 \sigma_0^2} -2\log p\right\}\right]\\
    & \qquad \leq  \frac{2}{1-e^{-2/3}}\frac{\exp\left[-\frac{n}{18 \sigma_0^2}\left\{(1-\gamma_1^2)C_{\min} -36\frac{\log p}{n}\sigma_0^2\right\}\right]}{1- \exp\left[-\frac{n}{18 \sigma_0^2}\left\{(1-\gamma_1^2)C_{\min} -36\frac{\log p}{n}\sigma_0^2\right\}\right] }.
\end{align*}
It then follows that for any $$ \*u^* \in \left\{ \max_{\tau \neq \tau_0, |\tau|\leq |\tau_0|} \rho_{\tau^\bot}(\*u^*, \*X_{\tau_0}\bbeta_0) < \gamma_1^2 \right\}, $$
we have the probability bound
\begin{align}
\label{eq:L_gamma_1}
    & \P_{\*U}(\hat\tau_{\*u^*} \neq \tau_0) \nonumber \\
    &  \leq \left\{\frac{2}{1-e^{-2/3}} +  \P_{\*U}(\hat\tau_{\*u^*}\neq \tau_0) \right\}\exp\left[-\frac{n}{18 \sigma_0^2}\left\{(1-\gamma_1^2)C_{\min} -36\frac{\log p}{n}\sigma_0^2\right\}\right] \nonumber \\
    & \leq \frac{3-e^{-2/3}}{1-e^{-2/3}} \exp\left[-\frac{n}{18 \sigma_0^2}\left\{(1-\gamma_1^2)C_{\min} -36\frac{\log p}{n}\sigma_0^2\right\}\right]\nonumber\\
    & \leq 6\exp\left[-\frac{n}{18 \sigma_0^2}\left\{(1-\gamma_1^2)C_{\min} -36\frac{\log p}{n}\sigma_0^2\right\}\right] =
    L(\gamma_1).
\end{align}
By Lemma \ref{lem::angle}
\begin{align}
\label{eq:prob_bound_gamma1_tau}
     &  \P_{\*U^*}\left\{
     \max_{\tau \neq \tau_0, |\tau| \leq |\tau_0|}\rho_{\tau^\bot}(\*U^*, \*X_{\tau_0}\bbeta_0) \geq \gamma_1^2
     \right\} \nonumber
     \\& \qquad
    \leq  \sum_{\tau \neq \tau_0, |\tau| \leq |\tau_0|} \P_{\*U^*}\left\{\rho_{\tau^\bot}(\*U^*, \*X_{\tau_0}\bbeta_0) \geq \gamma_1^2  \right\} \nonumber
    \\
    & \qquad \leq  \sum_{\{\tau: |\tau| \leq |\tau_0|\}} 2\{\arccos (\gamma_1)\}^{n-|\tau|-1} = \sum_{k = 1}^{|\tau_0|} \genfrac(){0pt}{0}{p}{k} 2\{\arccos (\gamma_1)\}^{n-k-1}
   \nonumber
    \\
     & \qquad
    \leq 2\{\arccos (\gamma_1)\}^{n-|\tau_0|-1 } p^{|\tau_0|}.
    \end{align}
Then let $i_{\min} = \arg\min_{1\leq i \leq d} \max_{\tau \neq \tau_0, |\tau|\leq |\tau_0|} \rho_{\tau^\bot}(\*U^*_i, \*X_{\tau_0}\bbeta_0),$ we have
\begin{align*}
 & \P_{({\cal U}^d, \*Y)} (\tau_0 \notin S^{(d)} )  \leq \P_{({\cal U}^d, \*Y)} \left\{\tau_0 \notin S^{(d)} , \min_{1\leq i \leq d} \max_{\tau \neq \tau_0, |\tau|\leq |\tau_0|} \rho_{\tau^\bot}(\*U^*_i, \*X_{\tau_0}\bbeta_0) < \gamma_1^2\right\} \\
 & \qquad + \P_{({\cal U}^d, \*Y)} \left\{\tau_0 \notin S^{(d)} , \min_{1\leq i \leq d} \max_{\tau \neq \tau_0, |\tau|\leq |\tau_0|} \rho_{\tau^\bot}(\*U^*_i, \*X_{\tau_0}\bbeta_0) \geq \gamma_1^2\right\} \\
 &  \leq  \sum^d_{i=1}\P_{({\cal U}^d, \*Y| \cdot)} \left\{\tau_0 \notin S^{(d)} \middle | i_{\min} =i\right\} \P_{({\cal U}^d)} (i_{\min} =i) \\
 & \qquad + \P_{({\cal U}^d)} \left\{ \min_{1\leq i \leq d} \max_{\tau \neq \tau_0, |\tau|\leq |\tau_0|} \rho_{\tau^\bot}(\*U^*_i, \*X_{\tau_0}\bbeta_0) \geq \gamma_1^2\right\} \\
 & \leq  \sum^d_{i=1}\P_{({\cal U}^d, \*Y|  \cdot)} \left\{\hat\tau_{\*U^*_i} \neq \tau_0 \middle | i_{\min} =i\right\} \P_{({\cal U}^d)} (i_{\min} =i)\\
  & \qquad + \P_{({\cal U}^d, \*Y)} \left\{ \min_{1\leq i \leq d} \max_{\tau \neq \tau_0, |\tau|\leq |\tau_0|} \rho_{\tau^\bot}(\*U^*_i, \*X_{\tau_0}\bbeta_0) \geq \gamma_1^2\right\}.
\end{align*}
Further, because $\*U$, $\*U_1^*, \dots, \*U^*_d$ are independent, it then follows from \eqref{eq:L_gamma_1} that
\begin{align*}
    \P_{({\cal U}^d, \*Y)} \left(\hat\tau_{\*U^*_i} \neq \tau _0\middle | i_{\min} =i\right) & =  \P_{(\*U, \*U_i^*| \cdot)} \left(\hat\tau_{\*U^*_i} \neq \tau _0\middle |  \max_{\tau \neq \tau_0, |\tau|\leq |\tau_0|} \rho_{\tau^\bot}(\*U^*_i, \*X_{\tau_0}\bbeta_0) \geq \gamma_1^2\right) \leq L(\gamma_1),
\end{align*}
From which and \eqref{eq:prob_bound_gamma1_tau} Lemma~\ref{lem::asymptotic_constraint} follows immediately.

 \end{proof}

\begin{proof}[Proof of Lemma~\ref{lem::symptotic_bound_penalize0}]
By Lemma~\ref{lem_proj}, we let
$D(\tau, \*u^*)=  \frac{1}{2}\|(\*I - \*H_{\tau,\*u^*})\*Y\|^2 + \lambda |\tau| =  \frac{1}{2}\|(\*I-\*H_\tau-  O_{ \tau_0^{\bot} \*u})\*y\|^2 + \lambda |\tau|$
for any $\tau \neq \tau_0.$ Then By \eqref{eq:dist_penalize}, for any $\delta \in (0,1)$ an any $\*u^*$ such that $$ \max_{\tau \neq \tau_0, |\tau|\leq |\tau_0|} \rho_{\tau^\bot}(\*u^*, \*X_{\tau_0}\bbeta_0) < \gamma_1^2,$$
we have
\begin{align*}
   & \P_{ \*U}\{ D(\tau, \*u^*) -D(\tau_0,\*u^*)< 0\}\\
    & \leq \P_{\*U}\{(1- \gamma^2_1) \|(\*I-\*H_\tau)\*X_{\tau_0}\bbeta_0\|^2 -\sigma_0^2\*U^\top( H_{\tau}+\*O_{ \tau^{\bot}  \*u^*} -  \*H_{\tau_0}-\*O_{ \tau_0^{\bot} \*u^*} )\*U \\
    & \hspace{5cm} + 2\sigma_0\*U^\top(\*I - \*H_{\tau}-\*O_{ \tau^{\bot}  \*u^*})\*X_{\tau_0}\bbeta_0 +  2\lambda(|\tau|-|\tau_0|) <0 \}\\
    & \leq  \P_{\*U}\{(1-\gamma^2_1)(1-\delta)\|(\*I-\*H_\tau)\*X_{\tau_0}\bbeta_0\|^2 - \sigma_0^2\*U^\top( H_{\tau}+\*O_{ \tau^{\bot}  \*u^*} -  \*H_{\tau_0}-\*O_{ \tau_0^{\bot} \*u^*} )\*U\\
    & \hspace{5cm} + \lambda(|\tau|-|\tau_0|) <0\} \\
    & \quad + \P_{\*U}\{(1-\gamma^2_1)\delta\|(\*I-\*H_\tau)\*X_{\tau_0}\bbeta_0\|^2 + 2\sigma_0\*U^\top(\*I - \*H_{\tau}-\*O_{ \tau^{\bot}  \*u^*})\*X_{\tau_0}\bbeta_0 \\
    & \hspace{5cm} +\lambda(|\tau|-|\tau_0|) ) < 0\} \\
    &= (I_1) + (I_2).
   \end{align*}

Then it follows from \eqref{eq:log_mgf}  and  Markov Inequality
\begin{align*}
    (I_1)  \leq \exp\left\{2t_1|\tau \setminus \tau_0| - \frac{t_1(1-\delta)(1-\gamma^2_1)n |\tau_0\setminus \tau| C_{\min} + t_1 \lambda (|\tau| - |\tau_0|)}{\sigma_0^2} \right\}, \numberthis \label{eq:I1_asymp_pen}
\end{align*}
for any $0<t_1<1/2.$
Further because $2\sigma_0\*U^\top(\*I - \*H_{\tau,\*u^*})\*X_{\tau_0}\bbeta_0$ follows $N(0, 4\sigma^2_0\|(\*I - \*H_{\tau,\*U^*})\*X_{\tau_0}\bbeta_0\|^2)$, then by Markov inequality and moment generating function of the normal distribution, we
\begin{align*}
  (I_2) \leq \exp\left\{\frac{(2t_2^2- \delta t_2) (1-\gamma_1^2) n |\tau_0 \setminus \tau| C_{\min} - t_2 \lambda (|\tau| - |\tau_0|)}{\sigma_0^2}\right\},  \numberthis \label{eq:I2_asymp_pen}
  \end{align*}
for any $0<t_2<1/2.$
It then follows that
\begin{align*}
   &  \P_{\*U}(\hat\tau_{\*u^*} \neq \tau_0)  \\
   & \leq  \sum_{i=1}^{|\tau_0|} \sum_{j=0}^i\genfrac(){0pt}{0}{|\tau_0|}{i} \genfrac(){0pt}{0}{p-|\tau_0|}{j}\Bigg[\exp\bigg\{\frac{(2t^2_1-\delta t_1)(1-\gamma_1^2)niC_{\min} + t_1 \lambda (i-j)}{\sigma_0^2}\bigg\}\\
 & \quad \quad \quad    + \exp\bigg\{\frac{-(1-\delta)t_2(1-\gamma^2_1)niC_{\min}+ t_2 \lambda(i-j)}{\sigma_0^2} + 2t_2j\bigg\}\Bigg]\\
  &  + \sum_{i=0}^{|\tau_0|} \sum_{j=i+1}^{p} \genfrac(){0pt}{0}{|\tau_0|}{i}\genfrac(){0pt}{0}{p-|\tau_0|}{j}\left[\exp\left\{\frac{t_1 \lambda (i-j)}{\sigma_0^2}\right\} + \exp\left\{ \frac{t_2 \lambda (i-j)}{\sigma_0^2} + 2t_2j\right\} \right].
\end{align*}

We can make $\delta=1/2, t_1=t_2=1/3,$ then by \eqref{eq:L_gamma_1},
if $$\frac{\lambda}{n} \in \left[\frac{3\sigma_0^2(|\tau_0|+1)\{\log(p-|\tau_0|)  +\log(|\tau_0|)+ \frac{2}{3}\}}{n}+t, \frac{(1-\gamma_1^2)C_{\min}}{6},\right]$$
we have
\begin{align*}
  &  \P_{\*U}(\hat\tau_{\*u^*} \neq \tau_0) \\
  &\leq 2\sum_{i=1}^{|\tau_0|} \sum_{j=0}^i (p-|\tau_0|)^j |\tau_0|^i \exp \left\{- \frac{(1-\gamma_1^2)niC_{\min}}{18\sigma_0^2}  + \frac{\lambda(i-j)}{3\sigma_0^2} + \frac{2}{3}j\right\}\\
  & \quad \quad \quad  + 2\sum_{i=0}^{|\tau_0|} \sum_{j=i+1}^{p} (p-|\tau_0|)^j |\tau_0|^i \exp \left\{ -\frac{\lambda(j-i)}{3\sigma_0^2} + \frac{2}{3}j\right\}\\
 & \leq L(\gamma_1) + 2 \frac{\sum_{i=0}^{|\tau_0|} \exp \left[-\frac{\lambda}{3\sigma_0^2} + \log(p-|\tau_0|)+ \frac{2}{3}+ i\left\{\log(|\tau_0|) + \log(p-|\tau_0|)+ \frac{2}{3}\right\}\right]}{1- \exp\left\{-\frac{\lambda}{3\sigma_0^2} + \frac{2}{3} + \log(p-|\tau_0|)\right\}}\\
   & \leq L(\gamma_1) + 2\frac{\exp\left[-\frac{\lambda}{3\sigma_0^2} + (|\tau_0|+1)\{\log(p-|\tau_0|)  +\log(|\tau_0|)+ \frac{2}{3}\} \right]}{1- \exp\left\{-\frac{\lambda}{3\sigma_0^2} + \frac{2}{3} + \log(p-|\tau_0|)\right\}}\\
   &\leq L(\gamma_1) + 3\exp\left(-\frac{nt}{3\sigma_0^2}\right). \numberthis \label{eq:asymp_gamma1_bound}
\end{align*}
 Then let $i_{\min} = \arg\min_{1\leq i \leq d} \max_{\tau \neq \tau_0, |\tau|\leq |\tau_0|} \rho_{\tau^\bot}(\*U^*_i, \*X_{\tau_0}\bbeta_0),$ by \eqref{eq:prob_bound_gamma1_tau} we have
\begin{align*}
 & \P_{({\cal U}^d, \*Y)} (\tau_0 \notin S^{(d)} )  \leq \P_{({\cal U}^d, \*Y)} \left\{\tau_0 \notin S^{(d)} , \min_{1\leq i \leq d} \max_{\tau \neq \tau_0, |\tau|\leq |\tau_0|} \rho_{\tau^\bot}(\*U^*_i, \*X_{\tau_0}\bbeta_0) < \gamma_1^2\right\} \\
 & \qquad + \P_{({\cal U}^d, \*Y)} \left\{\tau_0 \notin S^{(d)} , \min_{1\leq i \leq d} \max_{\tau \neq \tau_0, |\tau|\leq |\tau_0|} \rho_{\tau^\bot}(\*U^*_i, \*X_{\tau_0}\bbeta_0) \geq \gamma_1^2\right\} \\
 & \leq  \P_{({\cal U}^d, \*Y)} \left\{\hat\tau_{\*U^*_{i_{\min}}} \neq \tau_0 ,  \max_{\tau \neq \tau_0, |\tau|\leq |\tau_0|} \rho_{\tau^\bot}(\*U^*_{i_{\min}}, \*X_{\tau_0}\bbeta_0) < \gamma_1^2\right\} \\
  & \qquad + \P_{({\cal U}^d, \*Y)} \left\{ \min_{1\leq i \leq d} \max_{\tau \neq \tau_0, |\tau|\leq |\tau_0|} \rho_{\tau^\bot}(\*U^*_i, \*X_{\tau_0}\bbeta_0) \geq \gamma_1^2\right\} \\
  & \leq  \P_{( \*U, {\cal U}^d | \cdot)} \left\{\hat\tau_{\*U^*_{i_{\min}}} \neq \tau_0 \middle| \max_{\tau \neq \tau_0, |\tau|\leq |\tau_0|} \rho_{\tau^\bot}(\*U^*_{i_{\min}}, \*X_{\tau_0}\bbeta_0) < \gamma_1^2\right\} \\
  & \hspace{5cm} \P_{{\cal U}^d} \left\{\max_{\tau \neq \tau_0, |\tau|\leq |\tau_0|} \rho_{\tau^\bot}(\*U^*_{i_{\min}}, \*X_{\tau_0}\bbeta_0) < \gamma_1^2\right\} \\
  & \qquad + \P_{({\cal U}^d, \*Y)} \left\{ \min_{1\leq i \leq d} \max_{\tau \neq \tau_0, |\tau|\leq |\tau_0|} \rho_{\tau^\bot}(\*U^*_i, \*X_{\tau_0}\bbeta_0) \geq \gamma_1^2\right\} \\
  & \leq \max\left\{ \P_{\*U}(\hat\tau_{\*u^*} \neq \tau_0) : \max_{\tau \neq \tau_0, |\tau|\leq |\tau_0|} \rho_{\tau^\bot}(\*u^*, \*X_{\tau_0}\bbeta_0) < \gamma_1^2 \right\} \\
  & \qquad  + \prod^d_{i=1}  \P_{\*U^*_i} \left\{  \max_{\tau \neq \tau_0, |\tau|\leq |\tau_0|} \rho_{\tau^\bot}(\*U^*_i, \*X_{\tau_0}\bbeta_0) \geq \gamma_1^2\right\}\\ \numberthis \label{eq:aymptotic_bound}
  &  \leq L(\gamma_1) + 3\exp\left(-\frac{nt}{3\sigma_0^2}\right) + \left[2\{\arccos (\gamma_1)\}^{n-|\tau_0|-1 } p^{|\tau_0|}\right]^d.
\end{align*}

\end{proof}

{
\subsection{ Proof of Theorem~\ref{the:size}}
\label{size}

\begin{proof}[Proof of Theorem~\ref{the:size}]

First, with a slight abuse of notation,
 let $\gamma^2_1 = \rho((\*I - \*H_\tau)\*U^*, (\*I - \*H_{\tau})\*X\bbeta_0),$ and $\gamma^2_2 = \rho(\*U, \*U^*),$ be the random quantities that measures the angle between $(\*I - \*H_\tau)\*U^*$ and $(\*I - \*H_{\tau})\*X\bbeta_0$, and between $\*U$ and $\*U^*$ respectively.

By \eqref{eq:I1_asymp_pen} and \eqref{eq:I2_asymp_pen}, we have
\begin{align*}
    \P(\hat\tau_{\*U^*}  = \tau| \gamma_1) \leq \exp\left\{2t_1|\tau \setminus \tau_0| - \frac{t_1(1-\delta)(1-\gamma^2_1)n |\tau_0\setminus \tau| C_{\tau} + t_1 \lambda (|\tau| - |\tau_0|)}{\sigma_0^2} \right\} \\
     + \exp\left\{\frac{(2t_2^2- \delta t_2) (1-\gamma_1^2) n |\tau_0 \setminus \tau| C_{\tau} - t_2 \lambda (|\tau| - |\tau_0|)}{\sigma_0^2}\right\}.
\end{align*}
Then make $\delta = 1/2,$ $t_1= t_2 = 1/3, $ then
\begin{align*}
    \P(\hat\tau_{\*U^*}  = \tau | \gamma_1) \leq 2 \exp\left\{ - \frac{(1 - \gamma_1^2)n |\tau_0 \setminus \tau|}{18 \sigma_0^2}C_{\tau} - \frac{\lambda(|\tau| - |\tau_0|)}{3\sigma_0^2} + \frac{2|\tau \setminus \tau_0|}{3} \right\}.
\end{align*}

To bound $\P(\hat\tau_{\*U^*}  = \tau) = E\P(\hat\tau_{\*U^*}  = \tau | \gamma_1)$  , we first try to obtain the bound of expectation $E\{e^{-k(1-\gamma^2_1)}\},$  where $k>0$ is a constant. By \eqref{eq:density_angle} and $1-\gamma_1^2 = \sin^2(\varphi),$ the density function of $Z = (1- \gamma_1^2)$ is
\begin{align*}
 f(z) \sim  z^{(n- |\tau| -3)/2} (1-z)^{-1/2}, 0 \leq z \leq 1.  \numberthis \label{eq:beta_distribution}
\end{align*}
Therefore $Z \sim Beta(\frac{n - |\tau| -1}{2}, \frac{1}{2}).$
Then by Jensen's inequality, $E\{e^{-k(1-\gamma^2_1)}\} \leq \exp\{- \frac{n-|\tau|-1}{n-|\tau|}k\}.$ It then follows that
\begin{align*}
      \P(\hat\tau_{\*U^*}  = \tau) & \leq 2 \exp\left\{ - \frac{n (n-|\tau| -1) |\tau_0 \setminus \tau|}{18 \sigma_0^2 (n-|\tau|)}C_{\tau} - \frac{\lambda(|\tau| - |\tau_0|)}{3\sigma_0^2} + \frac{2|\tau \setminus \tau_0|}{3} \right\}
\end{align*}
When $|\tau| \leq |\tau_0|,$ and  $n - |\tau_0| \geq 10. $
\begin{align*}
      \P(\hat\tau_{\*U^*}  = \tau)  \leq 2 \exp\left\{ - \frac{n |\tau_0 \setminus \tau|}{20 \sigma_0^2 }C_{\tau} + \frac{\lambda(|\tau_0| - |\tau|)}{3\sigma_0^2} + \frac{2|\tau \setminus \tau_0|}{3} \right\}\\
      \leq  2 \exp\left\{ - \frac{n|\tau_0 \setminus \tau|}{20 \sigma^2} \left\{C_{\tau} -  \frac{20\lambda(|\tau_0| - |\tau|)}{3n|\tau_0\setminus\tau|} - \frac{40|\tau \setminus \tau_0|\sigma_0^2}{3n}\right\} \right\}\\
      \leq  2 \exp\left\{ - \frac{n|\tau_0 \setminus \tau|}{20 \sigma^2} \left\{C_{\tau} -  \frac{7\lambda}{n} - \frac{14|\tau \setminus \tau_0|\sigma_0^2}{n}\right\} \right\}\\
       \leq  2 \exp\left\{ - \frac{n}{20 \sigma^2} \left\{C_{\tau} -  \frac{7\lambda}{n} - \frac{14\sigma_0^2}{n}\right\} \right\}.
\end{align*}

If $$\frac{\lambda}{n} \geq \frac{3\sigma_0^2(|\tau_0|+1)\{\log(p-|\tau_0|)  +\log(|\tau_0|)+ \frac{2}{3}\}}{n}+t,$$ where $t>0,$
we have
\begin{align*}
  &  \sum_{\{\tau: |\tau| > |\tau_0|\}}\P(\hat\tau_{\*u^*} \neq \tau_0) \\
    & \leq 2\sum_{i=0}^{|\tau_0|} \sum_{j=i+1}^{p} (p-|\tau_0|)^j |\tau_0|^i \exp \left\{ -\frac{\lambda(j-i)}{3\sigma_0^2} + \frac{2}{3}j\right\}\\
 & \leq 2 \frac{\sum_{i=0}^{|\tau_0|} \exp \left[-\frac{\lambda}{3\sigma_0^2} + \log(p-|\tau_0|)+ \frac{2}{3}+ i\left\{\log(|\tau_0|) + \log(p-|\tau_0|)+ \frac{2}{3}\right\}\right]}{1- \exp\left\{-\frac{\lambda}{3\sigma_0^2} + \frac{2}{3} + \log(p-|\tau_0|)\right\}}\\
   & \leq 2\frac{\exp\left[-\frac{\lambda}{3\sigma_0^2} + (|\tau_0|+1)\{\log(p-|\tau_0|)  +\log(|\tau_0|)+ \frac{2}{3}\} \right]}{1- \exp\left\{-\frac{\lambda}{3\sigma_0^2} + \frac{2}{3} + \log(p-|\tau_0|)\right\}}\\
   &\leq 3\exp\left(-\frac{nt}{3\sigma_0^2}\right).
\end{align*}

It then follows that
\begin{align*}
    E(|S^{(d)}|)  & = E\{E(|S^{(d)}|| \*U)\} = E\left\{\sum_{\tau}\P(\tau \in S^{(d)}|\*U)\right\}\\
                &  = E\left\{\sum_{\tau} [ 1- \{1 - \P(\hat\tau_{\*U^*_d}   =   \tau| \*U)\}^d]\right\}\\
                & \leq \sum_{\tau} [ 1- \{1 - \P(\hat\tau_{\*U^*}   =   \tau)\}^d] \qquad \mbox{(by Jenesn's inequality)}\\
                & \leq \sum_{\tau} (d \P (\hat\tau_{\*U^*} = \tau) \wedge 1)\\
                & \leq \left|\left\{\tau: C_{\tau}  \leq \frac{7\lambda+ 14\sigma_0^2( 1+ 1.5\log d )}{n}\right\}\right| \\
                & \quad  + \sum_{\left\{\tau: C_{\tau}  \leq \frac{7\lambda+ 14\sigma_0^2( 1+ 1.5\log d )}{n}\right\}} \exp\left\{ - \frac{n}{20 \sigma^2} \left\{C_{\tau} -  \frac{7\lambda+ 14\sigma_0^2( 1+ 1.5\log d )}{n}\right\} \right\}\\
                & \qquad \qquad + 3\exp\left(-\frac{nt}{3\sigma_0^2}\right).
\end{align*}
\end{proof}

 }\subsection{Proof of  Theorems~\ref{thm:cond_rps}- \ref{the:confidence_set_asymptotic}}

\begin{proof}[Proof of Theorem~\ref{thm:cond_rps}]
The proof of Theorem~\ref{thm:cond_rps} is a direct consequence of our repro samples idea. Specifically,
 $\P\left(\tau_0 \in \Gamma^{\tau}_{\alpha}(\*Y)\right)
 \geq \P(T(\*U, \btheta) \in B_\alpha(\tau, \*W(\*U, \btheta))
    =
    \E\{\P(T(\*U, \btheta) $  $ \in B_\alpha(\tau, \*W(\*U, \btheta)) | \*W(\*U, \btheta))\}
    \geq \alpha.
$
\end{proof}

\begin{proof}[Proof of Theorem~\ref{the:confidence_set}]
First, for a given $\tau$, the distribution of $(\*I-\*H_\tau) \*U/\|(\*I-\*H_\tau) \*U\| = (\*I-\*H_\tau) \*Y_{\btheta}/\|(\*I-\*H_\tau) \*Y_{\btheta}\|$ is free of $(\bbeta_\tau, \sigma).$ Therefore $(\*I-\*H_\tau) \*Y_{\btheta}/\|(\*I-\*H_\tau) \*Y_{\btheta}\|$ is ancillary for $(\bbeta_\tau, \sigma).$ Because $\widetilde{\*W}(\*Y_{\btheta}, \tau)$ is minimal sufficient for $(\bbeta_\tau, \sigma),$ then by Basu's theorem $\widetilde{\*W}(\*Y_{\btheta}, \tau)$ is independent of $(\*I-\*H_\tau) \*U/\|(\*I-\*H_\tau) \*U\|.$ Apparently, $\*A_{\btheta}(\*U) = \widetilde{\*A}_{\tau}(\*Y_{\btheta})$ and $b_{\btheta}(\*U) = \tilde b_{\tau}(\*Y_{\btheta})$ are independent. It then follows that  $\widetilde{\*A}_{\tau}(\*Y_{\btheta}),$ $\tilde b_{\tau}(\*Y_{\btheta})$ and $(\*I-\*H_\tau) \*U/\|(\*I-\*H_\tau) \*U\|$ are mutually independent. As a result, we conclude that the conditional distribution
\begin{align*}
    \left\{\*Y_{\btheta} | \widetilde{\*W}(\*Y_{\btheta}, \tau) =(\*a_{obs}, b_{obs})\right\}  \sim \left\{\*a_{obs} + b_{obs} \frac{(\*I-\*H_\tau)\*U}{ \|(\*I-\*H_\tau) \*U\|}\right\} \sim \*Y^*,
\end{align*}
where $\*Y^* =  \left\{\*a_{obs} + b_{obs} \frac{(\*I-\*H_\tau)\*U^*}{ \|(\*I-\*H_\tau) \*U^*\|}\right\}$ and $\*U^* \sim \*U,$ is free of $(\bbeta_\tau, \sigma)$ for any $\*a_{obs}, b_{obs}.$ Then the conditional probability in \eqref{eq:ptilde} is free of $(\bbeta_\tau, \sigma),$ hence the Borel set $B_{\alpha}(\tau, \*w)$ defined \eqref{eq:borel_model_cs} is also free of $(\bbeta_\tau, \sigma).$ Moreover, it follows from \eqref{eq:borel_model_cs} that
\begin{align*}
& {\P}_{\*Y_{\btheta}| \*w}\left\{\tilde T(\*Y_{\theta}, \tau) \in B_{\alpha}(\tau, \*w) \big| \widetilde{\*W}(\*Y_{\btheta}, \tau) = \*w \right\} \\
& = \sum_{\tau^* \in B_\alpha(\tau, \*w) } p_{(\*w, \tau)}(\tau^*) = 1 -  \sum_{\tau^* \not\in B_\alpha(\tau, \*w) } p_{(\*w, \tau)}(\tau^*)
\geq \alpha,
\end{align*}
which proves \eqref{eq:cond_prob_borel}. Then following from \eqref{eq:cond_prob_borel}, \eqref{eq:bar-CS} and Theorem~\ref{thm:cond_rps},
\begin{align}
\label{eq:upper_p_cs}
\P_{({\cal U}^d, \*Y)}\left\{\tau_0 \not\in  {\bar\Gamma}^{\tau}_{\alpha}(\*Y) \right\} \leq \P_{({\cal U}^d, \*Y)}\left\{\tau_0 \not\in  {\Gamma}^{\tau}_{\alpha}(\*Y) \right\}  + \P_{({\cal U}^d, \*Y)}\left\{\tau_0 \not\in  S^{(d)} \right\} \nonumber \\
\leq 1 -\alpha +  \P_{({\cal U}^d, \*Y)}\left\{\tau_0 \not\in  S^{(d)} \right\}.
\end{align}
Then it follows from Theorem~\ref{the:finite_pen} that $\P_{({\cal U}^d, \*Y)}\left\{\tau_0 \not\in  S^{(d)} \right\} = o(e^{-c_1 d}) $ for some $c_1<-\log\left(1- \frac{\gamma_{\delta}^{n-1}}{n-1}\right).$ Therefore $\P_{({\cal U}^d, \*Y)}\left\{\tau_0 \not\in  {\bar\Gamma}^{\tau}_{\alpha}(\*Y) \right\} \leq  1 -\alpha +  o(e^{-c_1 d}). $
Further let $c_\delta = -\log\left(1- \frac{\gamma_{\delta}^{n-1}}{n-1}\right),$ then by Markov Inequality and Theorem~\ref{the:finite_pen}
\begin{align}
\label{eq:bound_candidate_set_exp}
    \P_{{\cal U}^d}\left[\P_{\*Y| {\cal U}^d }\left\{\tau_0 \not\in  S^{(d)} \right\}  - \delta \geq e^{-c_1 d} \right] \leq \frac{\E_{{\cal U}^d}\left[\P_{\*Y| {\cal U}^d }\left\{\tau_0 \not\in  S^{(d)} \right\} -\delta\right] }{e^{-c_1 d}} \nonumber
    \\ = \frac{\P_{ ({\cal U}^d, \*Y) }\left\{\tau_0 \not\in  S^{(d)}  \right\} -\delta}{e^{-c_1 d}} = e^{-(c_\delta-c_1)d} \rightarrow 0,
\end{align}
as $d \rightarrow \infty$. The last part of Theorem~\ref{the:confidence_set} then follows immediately.
\end{proof}

\begin{proof}[proof of Theorem~\ref{the:confidence_set_asymptotic}]
Under the conditions in Theorem~\ref{the:aymp_pen}, let the constant $c_2>0$~and
{\small \begin{align}
\label{eq:c_2}
 c_2 < c_a = \min\left\{\frac{1}{18\sigma_0^2}\left(0.3C_{\min}-  \frac{36 \log (p+1)}{n}\sigma_0^2\right),\frac{t}{3\sigma_0^2}, d\left(0.23- \frac{|\tau_0|\log(p)+2}{n}\right)\right\}.
\end{align}}
Then Theorem~\ref{the:confidence_set_asymptotic} follows from \eqref{eq:upper_p_cs}, Theorem~\ref{the:aymp_pen} and the following
\begin{align}
\label{eq:bound_candidate_exp_asymp}
       \P_{{\cal U}^d}\left[\P_{\*Y| {\cal U}^d }\left\{\tau_0 \not\in  S^{(d)} \right\}  \geq e^{-c_2 n} \right] \leq \frac{\E_{{\cal U}^d}\left[\P_{\*Y| {\cal U}^d }\left\{\tau_0 \not\in  S^{(d)} \right\} \right] }{e^{-c_2 n}} \nonumber
    \\ = \frac{\P_{ ({\cal U}^d, \*Y) }\left\{\tau_0 \not\in  S^{(d)}  \right\} }{e^{-c_a n}} = e^{-(c_a-c_2)n} \rightarrow 0, \quad \hbox{as $n \rightarrow \infty.$}
\end{align}
\end{proof}
Next, we present the proofs of Theorems~\ref{cor:coverage_beta_Lambda} and \ref{cor:coverage_beta_Lambda_asymp}, showing the validity of the inference for any subset of regression coefficients, both in finite samples and asymptotically.
\subsection{Proofs of Theorems~\ref{cor:coverage_beta_Lambda} and \ref{cor:coverage_beta_Lambda_asymp}}

\begin{proof}[Proof of Theorem~\ref{cor:coverage_beta_Lambda}]
We first write
 {\small
\begin{align}
\label{eq:beta_lambda_p_bound}
  \P_{\*Y} \{\bbeta_{0,\Lambda} \not\in \Gamma^{\bbeta_\Lambda}_{\alpha}(\*Y)  \}   & =  \P_{\*Y} \{\bbeta_{0,\Lambda} \not\in \Gamma^{\bbeta_\Lambda}_{\alpha}(\*Y), \tau_0 \in S^{(d)}  \} + \P_{\*Y} \{\beta_{0,\Lambda} \not\in \Gamma^{\beta_\Lambda}_{\alpha}(\*Y), \tau_0 \not\in S^{(d)}  \}.
\end{align}}
Then let $\bm\eta_{0, \Lambda} = (\tau_0,\beta_{0,\Lambda}),$
\begin{align}
\label{eq:beta_lambda_T}
    \P_{\*Y} \{\bbeta_{0,\Lambda} \not\in \Gamma^{\bbeta_\Lambda}_{\alpha}(\*Y), \tau_0 \in S^{(d)}  \} \leq  \P_\*U\{\tilde T(\*Y, \bm \eta_{0,\Lambda}) \not\in B_{\bm \eta_{0,\Lambda}}(\alpha)\}  = 1- \alpha.
\end{align}
Therefore, from the above and Theorem~\ref{the:finite_pen},for some $c_1 <-\log\left(1- \frac{\gamma_{\delta}^{n-1}}{n-1}\right)$
\begin{align}
    \P_{\*Y} \{\beta_{0,\Lambda} \not\in \Gamma^{\beta_\Lambda}_{\alpha}(\*Y), \tau_0 \not\in S^{(d)}  \}\leq \P_{\*Y} \{\tau_0 \not\in S^{(d)} \} = \delta + o(e^{-c_1d}). \nonumber
\end{align}
Theorem~\ref{cor:coverage_beta_Lambda} then follows immediately from the above and \eqref{eq:bound_candidate_set_exp}.
\end{proof}

\begin{proof}[Proof of Theorem~\ref{cor:coverage_beta_Lambda_asymp}]
Define the constant $c_2$ as in \eqref{eq:c_2}. It follows from Theorem~\ref{the:aymp_pen} that
\begin{align}
    \P_{\*Y} \{\beta_{0,\Lambda} \not\in \Gamma^{\beta_\Lambda}_{\alpha}(\*Y), \tau_0 \not\in S^{(d)}  \}\leq \P_{\*Y} \{\tau_0 \not\in S^{(d)} \} =  o(e^{-c_2n}). \nonumber
\end{align}
Hence Theorem~\ref{cor:coverage_beta_Lambda_asymp} follows from \eqref{eq:bound_candidate_exp_asymp}, \eqref{eq:beta_lambda_p_bound}, \eqref{eq:beta_lambda_T} and the above.
\end{proof}
\subsection{Proofs of Theorems~\ref{the:coverage_eta} and \ref{the:coverage_eta_asym}}
\begin{proof}[Proof of Theorems~\ref{the:coverage_eta} and~\ref{the:coverage_eta_asym}]
Since $\big\{\tilde T(\*y_{obs}, \bm \eta_\tau) \in B_{\alpha_2}({\bm \eta_\tau}), \tau \in {\bar \Gamma}_{\alpha_1}^\tau(\*y_{obs}) \big\}^C = \big\{ \tilde T(\*y_{obs}, \bm \eta_\tau) \not\in B_{\alpha_2}({\bm \eta_\tau}) \big\} \cup  \big\{ \tau \not\in {\bar \Gamma}_{\alpha_1}^\tau(\*y_{obs})\big\},$  we have $\P(\bm\eta_0   \not \in \Gamma^{\bm\eta_\tau}_\alpha(\*Y)) \leq P(\tilde T(\*Y, \bm \eta_0) \not\in B_{\alpha_2}({\bm \eta_0})) + \P(\tau_0 \not\in  {\bar \Gamma}_{\alpha_1}^\tau(\*Y)).$ Then, Theorem~\ref{the:coverage_eta} follows from the above inequality and Theorem~\ref{the:confidence_set} and Theorem  \ref{the:coverage_eta_asym} follows from the above inequality and Theorem~\ref{the:confidence_set_asymptotic}.
\end{proof}

{
\section{Theoretical Proofs for non-Gaussian and Sub-Gaussian Errors}
\label{sec:non_normal}

We first introduce the following technical lemma.
\begin{lemma}
    \label{lemma:u_d}
   For any random vector $\*U$ and $\*U^*,$ let $$E(\gamma_1, \gamma_2)= \left\{ \max_{\tau \neq \tau_0, |\tau|\leq |\tau_0|} \rho_{\tau^\bot}(\*U^*, \*X_{\tau_0}\bbeta_{0}) < \gamma_1^2, \rho(\*U^*, \*U) > 1- \gamma_2^2  \right\},$$

for any $0<\gamma_1, \gamma_2 <1$ and  $\rho(\*u, \tau) = \frac{\|\*H_\tau\*u\|^2}{\|\*u\|^2}.$  Then
\begin{align*}
     \left\{\rho(\*U^*, \*U) > 1- \gamma_2^2, \max_{\tau \neq \tau_0, |\tau|\leq |\tau_0|} \rho_{\tau^\bot}(\*U, \*X_{\tau_0}\bbeta_{0}) < \tilde\gamma_1^2,  \max_{\tau \neq \tau_0, |\tau|\leq |\tau_0|} \rho(\*U, \tau) < 1- \gamma_2  \right\}\\
     \subset E(\gamma_1, \gamma_2),
\end{align*}
where  $\tilde\gamma_1=(1-\sqrt{\gamma_2})\gamma_1 - \sqrt{2-2\sqrt{1-\gamma^2_2}}.$
\end{lemma}

\begin{proof}[Proof of Lemma~\ref{lemma:u_d}]

Denote by $g_\tau(\*U)= \frac{\|(\*I-\*H_\tau)\*U\|}{\|\*U\|}$ and $g_{\tau}(\*U^*)= \frac{\|(\*I-\*H_\tau)\*U^*\|}{\|\*U^*\|}$.
For each $\*U^*$, given $\left\{\*U  \in \bigcap_{\{\tau: |\tau|  \leq |\tau_0|\}} A(\tilde\gamma_1^2, \tau)\right\}$, we have
\begin{align*}
     & \frac{1}{\|(\*I-\*H_\tau)\*U^*\|}((\*U^*)^\top(\*I-\*H_\tau)\*X_{\tau_0}\bbeta_0\\
     & =   \frac{1}{\|\*U\|} \*U^\top(\*I-\*H_\tau)\*X_{\tau_0}\bbeta_0
     + \left((\*U^*)^\top/\|\*U^*\|-\*U^\top/\|\*U\|\right)(\*I-\*H_\tau)\*X_{\tau_0}\bbeta_0  \\ & \quad + \left(\frac{1}{\|(\*I-\*H_\tau)\*U^*\|}
      -\frac{1}{\|\*U^*\|}\right)(\*U^*)^\top(\*I-\*H_\tau)\*X_{\tau_0}\bbeta_0\\
     & \leq \frac{\|\*I-\*H_\tau)\*U\|}{\|\*U\|}\frac{1}{\|(\*I-\*H\tau)\*U\|}\*U^\top(\*I-\*H_\tau)\*X_{\tau_0}\bbeta_0  +  \left\|\frac{(\*U^*)^\top}{\|\*U^*\|}-\frac{\*U^\top}{\|\*U\|}\right\|\|(\*I-\*H_\tau)\*X_{\tau_0}\bbeta_0\| \\
     &   \quad  + \frac{\|\*U^*\|-\|(\*I-\*H_\tau)\*U^*\|}{\|\*U^*\|}\frac{1}{\|(\*I-\*H_\tau)\*U^*\|}(\*U^*)^\top(\*I-\*H_\tau)\*X_{\tau_0}\bbeta_0 \\
     &  \leq g_{\tau}(\*U)\tilde\gamma_1\|(\*I-\*H_\tau)\*X_{\tau_0}\bbeta_0\|  +  \sqrt{2-2\frac{(\*U^*)^\top \*U}{\|(\*U^*)^\top\| \|\*U\|}}\|(\*I-\*H_\tau)\*X_{\tau_0}\bbeta_0\| \\
     & \quad  +(1-g_{\tau}(\*U^*))\frac{1}{\|(\*I-\*H_\tau)\*U^*\|}(\*U^*)^\top(\*I-\*H_\tau)\*X_{\tau_0}\bbeta_0\\
     &   \leq g_{\tau}(\*U) \tilde\gamma_1\|(\*I-\*H_\tau)\*X_{\tau_0}\bbeta_0\|  +  \sqrt{2-2\sqrt{1-\gamma^2_2}}\|(\*I-\*H_\tau)\*X_{\tau_0}\bbeta_0\|\\
     & \quad + (1-g_{\tau}(\*U^*))\frac{1}{\|(\*I-\*H_\tau)\*U^*\|}(\*U^*)^\top(\*I-\*H_\tau)\*X_{\tau_0}\bbeta_0.
\end{align*}
It then follows that
\begin{align*}
    & \frac{1}{\|(\*I-\*H_\tau)\*U^*\|}(\*U^*)^\top(\*I-\*H_\tau)\*X_{\tau_0}\bbeta_0\\
    & \leq  \frac{g_{\tau}(\*U)}{g_{\tau}(\*U^*)} \tilde\gamma_1\|(\*I-\*H_\tau)\*X_{\tau_0}\bbeta_0\|  + \frac{1}{g_{\tau}(\*U^*)} \sqrt{2-2\sqrt{1-\gamma^2_2}}\|(\*I-\*H_\tau)\*X_{\tau_0}\bbeta_0\| \\
\end{align*}
Further because $\|(\*I - \*H_\tau)\*U^*\|  \leq \|(\*I - \*H_\tau \P_{\*U})\*U^*\|$, if $(\*U^*, \*U)  \in B(\gamma_2)$ we have
\begin{align*}
    g_{\tau}(\*U^*) & \leq \frac{\|(\*I-\*P_{\*U})\*U^*\|}{\|\*U^*\|} + \frac{\|(\*P_{\*U}-\*H_\tau \*P_{\*U})\*U^*\|}{\|\*U^*\|} \\ &\leq \gamma_2 + \frac{\|(\*I-\*H_\tau)\*P_{\*U}\*U^*\|}{\|\*P_{\*U}\*U^*\|} = \gamma_2 + g_{\tau}(\*U).
\end{align*}
Similarly, we can show that $g_{\tau}(\*U)\leq  g_{\tau}(\*U^*) + \gamma_2.$
Then
\begin{align*}
\frac{g_{\tau}(\*U^*)}{g_{\tau}(\*U)}\geq 1 - \frac{\gamma_2}{g_{\tau}(\*U)}.
\end{align*}
It then follows that a sufficient condition for $\frac{1}{\|(\*I-\*H_\tau)\*U^*\|}(\*U^*)^\top(\*I-\*H_\tau)\*X_{\tau_0}\bbeta_0 \leq \gamma_1 \|(\*I-\*H_\tau)\*X_{\tau_0}\bbeta_0\|$ is
\begin{align*}
\tilde\gamma_1\leq \left(1 - \frac{\gamma_2}{g_{\tau}(\*U)}\right) \gamma_1 - \sqrt{2-2\sqrt{1-\gamma^2_2}}\leq  \frac{g_{\tau}(\*U^*)}{g_{\tau}(\*U)}\gamma_1 - \sqrt{2-2\sqrt{1-\gamma^2_2}}.
\end{align*}
Then it follows from $\rho(\*U, \tau) < 1- \gamma_2$ that $g_{\tau}(\*U) = \sqrt{1-\rho^2(\*U, \tau) } > \sqrt{\gamma_2}$ that the above holds for $\tilde\gamma_1=(1-\sqrt{\gamma_2})\gamma_1 - \sqrt{2-2\sqrt{1-\gamma^2_2}}.$
Therefore
\begin{align*}
  \{\rho(\*U^*, \*U)  > 1- \gamma_2^2\}\hspace{-0.5mm}\bigcap \hspace{-0.5mm}\bigg\{ \max_{\tau \neq \tau_0, |\tau|\leq |\tau_0|} \rho_{\tau^\bot}(\*U, \*X_{\tau_0}\bbeta_0) < \tilde\gamma_1^2, \max_{\tau \neq \tau_0, |\tau|\leq |\tau_0|} \rho(\*U, \tau) < 1- \gamma_2 \bigg\}\\
  \subset  E(\gamma_1, \gamma_2).
\end{align*}
\end{proof}

\subsection{Proof of Theorem~\ref{the:uw}}
To prove the results in Theorem~\ref{the:uw}, we first define some notations, Given $\*\Omega = \bf \omega, $ let $\omega_{\max} = \max\{\omega_i: \omega_i > 0, i =1, \dots, n\}$ be the largest nonzero elements of $\omega,$ and $\omega_{\min} = \min\{\omega_i: \omega_i > 0, i =1, \dots, n\}$ be the smallest non-zero elements of $\omega.$ Further let $\tilde n = |\{\omega_i: \omega_i > 0, i =1, \dots, n\}| $ be the number of non-zero elements in $\omega,$ and $\tilde{\*U} = (U_1 I(\omega_1 >0), \dots, U_n I(\omega_n >0))^\top.$ We assume $\tilde n > |\tau_0|$ as in Theorem~\ref{the:uw}.  Then we introduce the following lemma.

\begin{lemma}
    \label{lem::angle_uw}
Suppose $|\tau| < n$. For any $-1 \leq \gamma_1, \gamma_2 \leq 1,$ if $\*U^* \sim \*U \sim N(0, \*I),$
\begin{align}
\label{eq: angle_ustar_uw}
    \P_{(\*U^*, \*U)}\{\rho(\*U^*, \*U_\omega) > 1-\gamma_2^2\} > \frac{\gamma_2^{n-2}\arcsin (\gamma_2)}{n-1}.
\end{align}
  $\rho_{\tau^\bot}(\*U^*, \*X_{\tau_0}\bbeta_0)$ and $\rho(\*U^*, \*U_\omega)$ are independent, and $(\rho_{\tau^\bot}(\*U^*, \*X_{\tau_0}\bbeta_0) ,\rho(\*U^*, \*U_\omega))$ are independent of $\|\*U_\omega\|.$ Moreover, $\|\*U\|$ is independent of the event $(\*U_{\omega},\*U^* ) \in E(\gamma_1, \gamma_2).$
\end{lemma}

\begin{proof}[Proof of Lemma~\ref{lem::angle_uw}]

To prove \eqref{eq: angle_ustar_uw}, we first derive the conditional distribution of $\rho(\*U_{\omega}, \*U^*)$,  by similar arguments to the proof of Lemma~\ref{lem::angle}, given $\*U_{\omega} = \*w$,
\begin{align}\label{eq:u_epsi_uw}
    & \P_{\*U^*|\*U_{\omega}}\left\{{\|\*U_{\omega}^T \*U^*\|}\big/{(\|\*U_{\omega}\|\| \*U^*\|)} > \sqrt{1 - \gamma^2} \bigg|  \*U_{\omega}=\*w \right\}\nonumber\\
    & \qquad =  \P_{ \psi}\left\{  |\sin(\psi)|  < \gamma  \right\} \nonumber \\
    & \qquad = \frac{2}{c_1}\int_0^{\arcsin (\gamma)} \sin^{n-2}(s) ds  \nonumber \\
    & \qquad >  \frac{2}{c_1}\int_0^{\arcsin (\gamma)} (\frac{s \gamma }{\arcsin \gamma})^{n-2} ds>  \frac{\gamma^{n-2}\arcsin (\gamma)}{n-1},
\end{align}
where the first inequality follows from the fact that $\sin(s)$ is a concave function. Here,  $\psi = \arccos{\sqrt{ \rho( \*w, \*U^*)}}$ is the (positive) angle between $\*U^*$ and $\*w$, whose density function is $\sin^{n-2}(\psi)/c_1$, with a normalizing constant $c_1 = \int_0^{\pi} \sin^{n-2} (\psi) d\psi = 2 \int_0^{\frac \pi 2} \sin^{n-2} (\psi) d\psi \leq 2 \int_0^{\frac \pi 2} \sin(\psi) d\psi = 2 $. This density function is derived using a spherical transformation on $\*U^*$ in $\mathbb{R}^n$ space, with $\psi$ being the first angular coordinate and a Jacobian equal to $r^{n-1} \sin^{n-2}(\psi) \prod_{ d=2}^{n-1} \sin^{n-d-1}(\psi_d)$, where $r$ is the radius and $\psi_2, \dots, \psi_{n-2}$ are the second to $(n-2)$th angular coordinates.
Also, $sin(s) < \frac{s \gamma }{\arcsin \gamma}$ for $s \in (0, \arcsin \gamma)$ and a small $\gamma >0$.

Note that \eqref{eq:u_epsi_uw} does not involve $\*U^*$ and $\*w$. We have
\begin{align*}
    \P_{( \*U^*, \*U_{\omega})}\big\{ \rho(\*U_{\omega},, \*U^*) > 1- \gamma^2 \big\} & =  \E_{\*U_{\omega}}\left[   \P_{\*U^*|\*U_{\omega}}\left\{{\|\*U_{\omega}^T \*U^*\|}\big/{(\|\*U_{\omega}\|\| \*U^*\|)} > \sqrt{1 - \gamma^2} \bigg| \*U_{\omega}\right\}\right]
  \\ & > \frac{\gamma^{n-2}\arcsin \gamma}{n-1},
\end{align*}
from which \eqref{eq: angle_ustar_uw} of the lemma holds.

Furthermore, from the second equation of \eqref{eq:u_epsi},
we see that the conditional distribution of $\rho(\*U_{\omega}, \*U^*)$, given $\*U_{\omega}  = \*w$, does not involve $\*w.$ Thus, $\rho(\*U_{\omega}, \*U^*)$ and $\*U_{\omega}$ (and thus $\*U$) are independent.
Hence, $\rho(\*U^*,\*U_{\omega})$ and   $\rho((\*I - \*H_\tau)\*U_{\omega}, (\*I - \*H_\tau)\*X_{\tau_0} \*\beta_0)$ are also independent.

{
Finally, by the aforementioned spherical transformation, $\|\*U\|$ is independent with its direction $\*U/\|\*U\|.$ It then follows that  $\|\*U\|,$ $\*U/\|\*U\|$ and $\*U^*$ are mutually independent, since $\*U$ and $\*U^*$ are independent. Therefore because $$\frac{\*U_{\omega}}{\|\*U_{\omega}\|} = \frac{\mbox{diag}(\omega)\*U/\|\*U\|}{\left\|\mbox{diag}(\omega)\*U/\|\*U\|\right\|},$$
is a function of $\*U/\|\*U\|,$ $\|\*U\|, \*U_{\omega}/\|\*U_{\omega}\|$ and $\*U^*$ are mutually independent. It then follows immediately that $\|\*U\|$ is independent of the event $(\*U_{\omega},\*U^* ) \in E(\gamma_1, \gamma_2).$ }
\end{proof}

\begin{proof}[Proof of Theorem~\ref{the:uw}]

Similar to the decomposition in \eqref{eq:decom_prob}, for any $\bf\Omega = \bf\omega,$ we have
\begin{align*}
   & \P_{(\*U^*, \*U_\omega| \cdot)}\left\{ D(\tau, \*U^*) -D(\tau_0,\*U^*)< 0|  (\*U^*, \*U_\omega) \in E(\gamma_1, \gamma_2) \right\}\\
    & \leq \P_{(\*U^*, \*U_\omega|\cdot)}\big\{(1- \gamma^2_1) \|(\*I-\*H_\tau)\*X_{\tau_0}\bbeta_0\|^2 - \sigma_0^2\*U_\omega^\top( \*H_{\tau}+ \*O_{ \tau^{\bot}\*U^*}-  \*H_{\tau_0}- \*O_{ \tau_0^{\bot} \*U^*} )\*U_\omega \\
    & \qquad + 2\sigma_0\*U_\omega^\top(\*I - \*H_{\tau}-\*O_{ \tau^{\bot}  \*U^*})\*X_{\tau_0}\bbeta_0 +  2\lambda(|\tau|-|\tau_0|) <0\big| (\*U^*, \*U_\omega) \in E(\gamma_1, \gamma_2) \big\}\\
    & \leq  \P_{(\*U^*, \*U_\omega| \cdot)}\big\{(1-\gamma^2_1)(1-\delta)\|(\*I-\*H_\tau)\*X_{\tau_0}\bbeta_0\|^2 \\
    & \qquad{} - \sigma_0^2 \*U_\omega^\top(\*H_{\tau}+\*O_{ \tau^{\bot}  \*U^*} -  \*H_{\tau_0}-\*O_{ \tau_0^{\bot} \*U^*} )\*U_\omega  + \lambda(|\tau|-|\tau_0|) <0\big| (\*U^*, \*U_\omega) \in E(\gamma_1, \gamma_2)\big\} \\
    & \quad + \P_{(\*U^*, \*U_\omega |\cdot)}\big\{(1-\gamma^2_1)\delta\|(\*I-\*H_\tau)\*X_{\tau_0}\bbeta_0\|^2 \\
    & \qquad{} + 2\sigma_0\*U_\omega^\top(\*I - \*H_{\tau}-\*O_{ \tau^{\bot}  \*U^*})\*X_{\tau_0}\bbeta_0 +\lambda(|\tau|-|\tau_0|) ) < 0\big| (\*U^*, \*U_\omega) \in E(\gamma_1, \gamma_2)\big\} \\
    &= (I_1) + (I_2).
\end{align*}
   Then following the proof of Lemma~\ref{lem::finite_penalize} and \eqref{eq: I_bound},  for any $|\tau| \leq  |\tau_0|$ and
   $\frac{\lambda}{n} < \frac{1}{6}(1-\gamma_1^2)C_{\min},$
by and Lemma~\ref{lem::angle_uw}, we have
\begin{align*}
    & (I_1) \\
    & \leq \P_{(\*U^*, \*U|\cdot)}\left\{\|\*U_\omega\|^2 > \frac{(1-\gamma^2_1)(1-\delta)}{\gamma^2_2}\frac{\|(\*I-\*H_\tau)\*X_{\tau_0}\bbeta_0\|^2 }{\sigma_0^2} + \frac{\lambda(|\tau|-|\tau_0|)}{\gamma_2^2\sigma_0^2} \middle | (\*U^*,
    \*U_\omega) \in E(\gamma_1, \gamma_2)\right\}\\
      & \leq \P_{(\*U^*, \*U|\cdot)}\left\{\omega^2_{\max}\|\*U\|^2 > \frac{(1-\gamma^2_1)(1-\delta)}{\gamma^2_2}\frac{\|(\*I-\*H_\tau)\*X_{\tau_0}\bbeta_0\|^2 }{\sigma_0^2} + \frac{\lambda(|\tau|-|\tau_0|)}{\gamma_2^2\sigma_0^2} \middle | (\*U^*,
    \*U_\omega) \in E(\gamma_1, \gamma_2)\right\}\\
    & =  \P_{\*U}\left\{\omega^2_{\max}\|\*U\|^2 > \frac{(1-\gamma^2_1)(1-\delta)}{\gamma^2_2}\frac{\|(\*I-\*H_\tau)\*X_{\tau_0}\bbeta_0\|^2 }{\sigma_0^2} + \frac{\lambda(|\tau|-|\tau_0|)}{\gamma_2^2\sigma_0^2}\right\}\\
    &\leq  \P\left\{\chi^2_n> \frac{(1-\gamma^2_1)(1-\delta-1/6)}{\gamma^2_2}\frac{niC_{\min}}{\omega^2_{\max}\sigma_0^2}\right\}\\
    &\leq  \exp\left\{-\frac{n}{2}\log(1-2t_1) - t_1\frac{(1-\gamma^2_1)(1-\delta-1/6  )}{\gamma^2_2}\frac{niC_{\min}}{\omega^2_{\max}\sigma_0^2}  \right\},
\end{align*}
for any $0<t_1<1/2,$ where the equality follows from Lemma~\ref{lem::angle_uw}.
For any $|\tau| > |\tau_0|,$ we  have
\begin{align*}
    (I_1) \leq  \exp\left\{-\frac{n}{2}\log(1-2t_1) -  t_1 \frac{\lambda(|\tau|-|\tau_0|)}{\gamma_2^2\omega_{\max}\sigma_0^2} \right\}.
\end{align*}
The derivations of the above inequalities are similar to those in the proof of Lemma~\ref{lem::finite_penalize}.

For $(I_2)$,  if $\frac{\lambda}{n} < \frac{1}{6}(1-\gamma_1^2)C_{\min},$ by Cauchy-Schwartz inequality and Lemma~\ref{lem::angle}, when $|\tau| \leq |\tau_0|,$  we have
\begin{align*}
    & (I_2)\\
    &\leq \P_{(\*U^*, \*U_\omega|\cdot)}\big\{(1- \gamma^2_1) \delta \|(\*I-\*H_\tau)\*X_{\tau_0}\bbeta_0\|^2  <  2\sigma_0\|\*U_\omega^\top(\*I - \*H_{\tau}-\*O_{ \tau^{\bot}  \*U^*})\|\|(\*I-\*H_\tau)\*X_{\tau_0}\bbeta_0\|\\
     & \qquad  \hspace{8cm}  - \lambda(|\tau|-|\tau_0|) \big| (\*U^*, \*U_\omega) \in E(\gamma_1, \gamma_2)\big\} \\
    &= \P_{(\*U^*, \*U_\omega|\cdot)}\big\{2\sigma_0\|\*U_\omega^\top(\*I - \*H_{\tau}-\*O_{ \tau^{\bot}  \*U^*})\| >(1-\gamma_1^2 ) \delta \|(\*I-\*H_\tau)\*X_{\tau_0}\bbeta_0\|\\
     & \qquad \hspace{4cm} + \lambda(|\tau|-|\tau_0|)/ \|(\*I-\*H_\tau)\*X_{\tau_0}\bbeta_0\|\big| (\*U^*, \*U_\omega) \in E(\gamma_1, \gamma_2)\big\} \\
         &\leq \P_{(\*U^*, \*U |\cdot)}\left\{\|\*U_{\omega}\|^2 > \frac{(1-\gamma^2_1)^2(\delta-1/6)^2}{4\gamma^2_2}\frac{\|(\*I-\*H_\tau)\*X_{\tau_0}\bbeta_0\|^2}{\sigma_0^2}\middle | (\*U^*, \*U_\omega) \in E(\gamma_1, \gamma_2)\right\}\\
          &\leq \P_{( \*U |\cdot)}\left\{\omega^2_{\max}\|\*U\|^2 > \frac{(1-\gamma^2_1)^2(\delta-1/6)^2}{4\gamma^2_2}\frac{\|(\*I-\*H_\tau)\*X_{\tau_0}\bbeta_0\|^2}{\sigma_0^2}\middle | (\*U^*, \*U_\omega) \in E(\gamma_1, \gamma_2)\right\}\\
    &\leq \P_{\*U}\left\{\omega_{\max}^2\|\*U\|^2 > \frac{(1-\gamma^2_1)^2(\delta-1/6)^2}{4\gamma^2_2}\frac{\|(\*I-\*H_\tau)\*X_{\tau_0}\bbeta_0\|^2}{\sigma_0^2}\right\}\\
    &\leq \P\left\{\chi^2_n > \frac{(1-\gamma^2_1)^2(\delta-1/6)^2}{4\gamma^2_2}\frac{niC_{\min}}{\omega_{\max}^2\sigma_0^2}\right\} \\
    &\leq \exp\left\{-\frac{n}{2}\log(1-2t_2)- t_2\frac{(1-\gamma^2_1)^2(\delta-1/6)^2}{4\gamma^2_2}\frac{niC_{\min}}{\omega_{\max}^2\sigma_0^2}\right\},
\end{align*}
for any $0<t_2<1/2.$

When $|\tau|  >  |\tau_0|$, by similar arguments to \eqref{eq: bound_I2}, we have
\begin{align*}
   (I_2)
     &\leq \P_{(\*U^*, \*U_{\omega}|\cdot)}\left\{\sigma_0^2\|\*U_{\omega}^\top(\*I - \*H_{\tau}-\*O_{ \tau^{\bot}  \*U^*})\|^2 > \delta\lambda(|\tau|-|\tau_0|) \middle | (\*U^*, \*U_{\omega}) \in E(\gamma_1, \gamma_2)\right\} \\
    &\leq \P_{\*U_{\omega}|\cdot}\left\{\|\*U_\omega\|^2 > \frac{\delta\lambda(|\tau|-|\tau_0|)}{\gamma_2^2 \sigma_0^2}\middle | (\*U^*, \*U_{\omega}) \in E(\gamma_1, \gamma_2)\right\}\\
        &\leq \P_{\*U|\cdot}\left\{\omega^2_{\max}\|\*U\|^2 > \frac{\delta\lambda(|\tau|-|\tau_0|)}{\gamma_2^2 \sigma_0^2}\middle | (\*U^*, \*U_{\omega}) \in E(\gamma_1, \gamma_2)\right\}\\
           & = \P_{\*U|\cdot}\left\{\omega^2_{\max}\|\*U\|^2 > \frac{\delta\lambda(|\tau|-|\tau_0|)}{\gamma_2^2 \sigma_0^2}\right\}\\
    & \leq  \exp\left\{-\frac{n}{2}\log(1-2t_2)-t_2\frac{\delta\lambda(|\tau|-|\tau_0|)}{\gamma_2^2 \omega^2_{\max} \sigma_0^2}\right\}.
\end{align*}

Now, by making of $(1-\gamma^2_1)(1 - \delta-1/6) =  (1-\gamma^2_1)^2(\delta-1/6)^2/4$, we obtain $\delta = \frac{2}{1- \gamma^2_1}(\sqrt{\frac{5}{3} - \frac{2}{3}\gamma_1^2} - 1) + \frac{1}{6} > 0.74$.
Further
we make $t_1=t_2= \frac{\gamma_2}{2.04}$, so we have $-\frac{n}{2}\log(1-2t_1)=-\frac{n}{2}\log(1-2t_2)=-\frac{n}{2}\log(1-\frac{\gamma_2}{1.02}) \leq 2n\gamma_2$.
Then, intersect with the event $\{(\*U^*, \*U_\omega)) \in E(\gamma_1, \gamma_2)\}$, we have
\begin{align*}
    & \P_{(\*U^*, \*U_\omega|\cdot)}\big\{\hat\tau_{\*U^*} \neq \tau_0, \big | (\*U^*, \*U_{\omega}) \in E(\gamma_1, \gamma_2)\big\} \\
    & \qquad \leq  2\sum_{i=1}^{|\tau_0|} \sum_{j=0}^{i} \genfrac(){0pt}{0}{p-|\tau_0|}{j}\genfrac(){0pt}{0}{|\tau_0|}{i} \exp\left\{-\left(\sqrt{\frac{5}{3} - \frac{2}{3}\gamma_1^2} - 1\right)^2\frac{niC_{\min}}{2.04 \gamma_2\omega_{\max}^2\sigma_0^2} + 2n\gamma_2
    \right\} \\
    &   + 2\sum_{i=0}^{|\tau_0|} \sum_{j=i+1}^{p} \genfrac(){0pt}{0}{p-|\tau_0|}{j}\genfrac(){0pt}{0}{|\tau_0|}{i} \exp\left\{-\frac{0.74\lambda(j-i)}{2.04\gamma_2\omega_{\max}^2\sigma_0^2} + 2\gamma_2n \right\}\\
    & \qquad \leq
    2\sum_{i=1}^{|\tau_0|} \sum_{j=0}^{i} \genfrac(){0pt}{0}{p-|\tau_0|}{j}\genfrac(){0pt}{0}{|\tau_0|}{i} \exp\left\{- \frac{niC_{\min}}{26\omega_{\max}^2\sigma_0^2} \frac{\left(1 - \gamma_1^2\right)^2}{\gamma_2} + 2n\gamma_2
    \right\}\\
     &   + 2\sum_{i=0}^{|\tau_0|} \sum_{j=i+1}^{p} \genfrac(){0pt}{0}{p-|\tau_0|}{j}\genfrac(){0pt}{0}{|\tau_0|}{i} \exp\left\{-\frac{\lambda(j-i)}{4\gamma_2\omega_{\max}^2\sigma_0^2} + 2\gamma_2n \right\}.
     \end{align*}
The last inequality holds since $(\sqrt{\frac{5}{3} - \frac{2}{3}\gamma_1^2} - 1)^2 \geq 1.02 (1 - \gamma_1^2)^2/13$ for $\gamma_1^2 \in (0,1)$.

By similar calculation to that in \eqref{eq:U12a}, the first part of the above can be bounded by
\begin{align*}
   L_p(\gamma_1, \gamma_2, \omega_{\max})=  3\exp\left\{- \frac{n}{26\sigma_0^2}\left[ \frac{\left(1 - \gamma_1^2\right)^2}{\gamma_2} C_{\min} - 52\left( \frac{  \log(p/2)}{n}+ \gamma_2\right) \sigma_0^2\omega_{\max}^2\right]
    \right\}\\
    \leq L_p(\gamma_1, \gamma_2, \bar\omega).
    \end{align*}
As for the second part,
\begin{align*}
   &  2\sum_{i=0}^{|\tau_0|} \sum_{j=i+1}^{p} \genfrac(){0pt}{0}{p-|\tau_0|}{j}\genfrac(){0pt}{0}{|\tau_0|}{i} \exp\left\{-\frac{\lambda(j-i)}{4\gamma_2\omega_{\max}^2\sigma_0^2} + 2\gamma_2n \right\} \\
    & \leq  2\sum_{i=0}^{|\tau_0|} |\tau_0|^i \exp\left\{\gamma_2 n+ \frac{\lambda i}{4\gamma_2 \omega_{\max}^2\sigma_0^2} \right\} \sum_{j=i+1}^{p} \exp \left[-j\left\{\frac{\lambda}{4\gamma_2 \omega_{\max}^2\sigma_0^2} - \log(p-|\tau_0|)\right\}\right]\\
&\leq \frac{2\sum_{i=0}^{|\tau_0|} \exp\left[-\frac{\lambda}{4\gamma_2\omega_{\max}^2 \sigma_0^2} +\gamma_2n + \log(p-|\tau_0|)+ i \{\log(|\tau_0|) + \log (p-|\tau_0|)\}\right]}{1-\exp \left\{\frac{\lambda}{4\gamma_2 \omega_{\max}^2\sigma_0^2} - \log(p-|\tau_0|)\right\} }\\
&\leq \frac{2\exp\left[-\frac{\lambda}{4\gamma_2\omega_{\max}^2 \sigma_0^2} + \gamma_2n + (|\tau_0|+1) \{\log |\tau_0| + \log(p-|\tau_0|\}\right]}{1-\exp \left\{\frac{\lambda}{4\gamma_2\omega_{\max}^2 \sigma_0^2} - \log(p-|\tau_0|)\right\}}\\
&\leq 3\exp\left(-\frac{1}{4\gamma_2\omega_{\max}^2\sigma_0^2} nt  \right) \leq 3\exp\left(-\frac{1}{4\gamma_2\bar\omega^2\sigma_0^2} nt  \right),
\end{align*}
For any $\omega_{\max} \leq \bar\omega$.
Then if
\begin{align}
\label{eq:tuning_range}
 \frac{\lambda}{n} \in \left[\frac{\lambda^{(1)}_0}{n}+t, \frac{1}{6}(1-\gamma_1^2)C_{\min}\right],
\end{align}
where
$\lambda^{(1)}_0={4\gamma_2\sigma_0^2\omega_{\max}^2}\left[\gamma_2 n +  (|\tau_0|+1) \{\log |\tau_0| + \log(p-|\tau_0|\}\right].$

By Lemma~\ref{lem::angle}, $\*U_\omega$ is independent of $\rho(\*U^*, \*U_\omega).$ It then follows from \eqref{eq:ABC_indequality} and Lemma~\ref{lemma:u_d}  that
\begin{align*}
     & \P_{(\*U^*, \*U_\omega |\cdot)}(\hat \tau_{\*U^*} \neq \tau_0 | \rho(\*U^*, \*U_\omega) > 1- \gamma_2^2) \\
     &\leq  \P_{(\*U^*, \*U_\omega |\cdot)}\{\hat\tau_{\*U^*} \neq \tau_0 | (\*U^*, \*U_\omega) \in E(\gamma_1, \gamma_2)\} + \P\left(\max_{\tau \neq \tau_0, |\tau|\leq |\tau_0|} \rho_{\tau^\bot}(\*U_\omega, \*X_{\tau_0}\bbeta_0) \geq \tilde\gamma_1^2\right)\\
     & \hspace{4cm}+ \P\left(\max_{\tau \neq \tau_0, |\tau|\leq |\tau_0|} \rho(\*U_\omega, \tau) \geq 1- \gamma_2 \right)\\
     & \leq L_p(\gamma_1, \gamma_2, \bar\omega) + 3\exp\left(-\frac{1}{4\gamma_2\bar\omega\sigma_0^2} nt  \right) +  \P_{\*U_\omega}\left(\max_{\tau \neq \tau_0, |\tau|\leq |\tau_0|} \rho_{\tau^\bot}(\*U_\omega, \*X_{\tau_0}\bbeta_0) \geq \tilde\gamma_1^2\right)\\
     & \hspace{4cm}+ \P_{\*U_\omega}\left(\max_{\tau \neq \tau_0, |\tau|\leq |\tau_0|} \rho(\*U_\omega, \tau) \geq 1- \gamma_2 \right). \numberthis \label{:eq:total_bound_w}
\end{align*}

We then make $\gamma_1 = \sqrt{1-\gamma_2^{1/4}},$ from which we have
\begin{align*}
    & L_p(\gamma_1,\gamma_2, \bar \omega) = L_P\left(\sqrt{1-\gamma_2^{1/4}}, \gamma_2, \bar\omega\right) \\
    & =  3\exp\left\{- \frac{n}{26\sigma_0^2}\left[ \frac{C_{\min} }{\sqrt{\gamma_2}} - 52\left( \frac{  \log(p/2)}{n}+ \gamma_2\right) \sigma_0^2\right]
    \right\} \rightarrow 0, \numberthis \label{eq:bound_Lp}
\end{align*}
as $\gamma_2 \rightarrow 0.$ Moreover $\tilde\gamma_1=(1-\sqrt{\gamma_2})\sqrt{1-\gamma_2^{1/4}}- \sqrt{2-2\sqrt{1-\gamma^2_2}} \geq 1-1.6\gamma_2^{1/3} > 0$ for $\gamma_2 \in [0,0.24].$
Then
$$
\P_{\*U_\omega}\left(\max_{\tau \neq \tau_0, |\tau|\leq |\tau_0|} \rho_{\tau^\bot}(\*U_\omega, \*X_{\tau_0}\bbeta_0) \geq \tilde\gamma_1^2\right) \leq \P_{\*U_\omega}\left(\max_{\tau \neq \tau_0, |\tau|\leq |\tau_0|} \rho_{\tau^\bot}(\*U_\omega, \*X_{\tau_0}\bbeta_0) \geq 1-1.6\gamma_2^{1/3}\right).
$$

Let $\*a = (\*I - \*H_{\tau}) \*X_{\tau_0} \bbeta_0$, we then apply the eigen decomposition on the matrix $diag(\Omega)(\*I - \*H_{\*a})diag(\Omega) = \sum_{j  = 1}^{n - 1} \Lambda_j \*L_j \*L^T_j,$ where $\Lambda_j, j = 1 , \dots, n - 1$ are the non-zeroeigenvalues, and $\*L_j$ are the corresponding eigen vectors. Then we have
\begin{align*}
    \|((\*I - \*H_{\*a})\*U_{\omega}\|^2 = \sum^{n-1}_{j= 1}  \Lambda_j\*U^\top \*L_j \*L^T_j \*U^\top \geq \omega^2_{\min} \sum_{j=1}^{n-1} (\*U^\top \*L_j)^2.
\end{align*}
Therefore,
\begin{align*}
     1- \rho_{\tau^\bot}^2(\*U_\omega, \*X_{\tau_0}\bbeta_0)  = \frac{\|((\*I - \*H_{\*a})\*U_{\omega}\|^2}{\|\*U_{\omega}\|^2} \geq \frac{\omega^2_{\min} \sum_{j=1}^{n-1} (\*U^\top \*L_j)^2}{\omega^2_{\max} \|\*U\|^2}\\ = \frac{\omega^2_{\min}}{\omega_{\max}^2} \sum_{j=1}^{n-1} \rho^2(\*U,\*L_j) \geq \frac{\omega^2_{\min}}{\omega_{\max}^2} \rho^2(\*U,\*L_1),
\end{align*}
since $\|\*L_j\| = 1. $
Then
\begin{align*}
    \P_{\*U_\omega}\left(\rho_{\tau^\bot}(\*U_\omega, \*X_{\tau_0}\bbeta_0) \geq 1-1.6\gamma_2^{1/3}\right)  \leq  \*P_{\*U} (\rho^2(\*U,\*L_1) \leq \frac{\omega^2_{\max}}{\omega^2_{\min}} 1.96 \gamma_2^{2/3}) \\
    \leq \*P_{\*U} (\rho^2(\*U,\*L_1) \leq 1.96 \bar r_{\omega} \gamma_2^{2/3})=  p_{1, \tau}(\bar r_{\omega}, \gamma_2)
\rightarrow 0,
\end{align*}
as $\gamma_2 \rightarrow 0, $ for any fixed $\bar r_{\omega} < \infty.$ It then follows
\begin{align}
\label{eq:bound_p1}
    \P_{\*U_\omega}\left(\max_{\tau \neq \tau_0, |\tau|\leq |\tau_0|} \rho_{\tau^\bot}(\*U_\omega, \*X_{\tau_0}\bbeta_0) \geq \tilde\gamma_1^2\right) \leq \P_{\*U_\omega}\left(\max_{\tau \neq \tau_0, |\tau|\leq |\tau_0|} \rho_{\tau^\bot}(\*U_\omega, \*X_{\tau_0}\bbeta_0) \geq 1-1.6\gamma_2^{1/3}\right) \rightarrow 0,
\end{align}
as $\gamma_2 \rightarrow 0$

We then try to bound $\P_{\*U_\omega}\left(\max_{\tau \neq \tau_0, |\tau|\leq |\tau_0|} \rho(\*U_\omega, \tau) \geq 1- \gamma_2 \right).$
Because
\begin{align*}
\frac{\|\*H_\tau \*U_\omega\|^2}{\|\*U_\omega\|^2} =  \frac{\|\*H_\tau \*U_\omega\|^2}{\|\*H_\tau\*U_\omega\|^2 + \|(\*I- \*H_\tau)\*U_\omega\|^2},
\end{align*}
then it follows that \begin{align*}
   & \P_{\*U}(\rho(\*U_{\omega}, \tau) \geq 1- \gamma_2 )=\P_{\*U}(g_{\tau}(\*U_\omega) \leq \sqrt{\gamma_2})= \P_{\*U}\left(\frac{\|(\*I-\*H_\tau)\*U_\omega\|^2}{\|\*H_{\tau}\*U_\omega\|^2} \leq \frac{\gamma_2}{1-\gamma_2}\right).
\end{align*}
Further because the non zero eigenvalues of $diag(\Omega)\*H_\tau diag(\Omega)$ and $diag(\Omega)(\*I - \*H_\tau diag(\Omega))$ are bounded below by $\omega_{\min}^2$ and above $\omega^2_{\max},$ it then follows from the eigen decompositions of $diag(\Omega)\*H_\tau diag(\Omega)$ and $diag(\Omega)(\*I - \*H_\tau) diag(\Omega))$  that $\|(\*I-\*H_\tau)\*U_\omega\|^2$ is bounded below by $\omega^2_{\min} \chi^2_{\tilde n - |\tau|}$ and $\|\*H_\tau \*U_\omega\|^2$ is bounded above by $\omega^2_{\max} \chi^2_{|\tau|},$ therefore
\begin{align*}
    \P_{\*U}\left(\frac{\|(\*I-\*H_\tau)\*U_\omega\|^2}{\|\*H_{\tau}\*U_\omega\|^2} \leq \frac{\gamma_2}{1-\gamma_2}\right) \leq F_{\tilde n-|\tau|, |\tau|}\left(\frac{\omega_{\max}^2\gamma_2/(\tilde n-|\tau|)}{\omega_{\min}^2(1-\gamma_2)/|\tau|}\right) \leq F_{\tilde n-|\tau|, |\tau|}\left(\bar r^2_{\omega}\frac{\gamma_2/(\tilde n-|\tau|)}{(1-\gamma_2)/|\tau|}\right) \\ = p_{2, \tau}(\bar r^2_{\omega}, \gamma_2).
\end{align*}
Apparently, for any fixed $r^2_{\omega},$ $p_{2, \tau}(\bar r^2_{\omega}, \gamma_2) \rightarrow 0$ as $\gamma_2 \rightarrow 0. $ It then follows immediately that for any fixed $\bar r_{\omega} < \infty,$ we have
\begin{align*}
\P_{\*U_\omega}\left(\max_{\tau \neq \tau_0, |\tau|\leq |\tau_0|} \rho(\*U_\omega, \tau) \geq 1- \gamma_2 \right) \rightarrow 0. \numberthis \label{eq:bound_Uw_tau}
\end{align*}
as $\gamma_2 \rightarrow 0.$

 We then make $t = \sqrt{\gamma_2}\sigma_0^2\bar\omega^2,$ it then follows from \eqref{eq:lambda_0} and the fact that $\log(|\tau_0|)
 + \log(p- |\tau_0|) \leq 2\log(p/2)$ when $\gamma^{1/4}_2 <  \frac{ C_{\min}}{24 \{2 + 2(|\tau_0|+1)\log(p/2)/n\}\sigma_0^2\bar\omega^2}$ that
\begin{align*}
    \frac{\lambda_0^{(1)}}{n} + t & \leq 4 \gamma_2^{1/2}\left\{\gamma_2^{3/2} + 1 + 2\gamma_2^{1/2}(|\tau_0|+1)\frac{\log(p/2)}{n} \right\}\sigma_0^2 \bar\omega^2\\
    & < 4 \gamma_2^{1/2}\left\{2 + 2(|\tau_0|+1)\frac{\log(p/2)}{n} \right\}\sigma^2_0 \bar\omega\\
    & < \frac{1}{6}\gamma_2^{1/4}C_{\min}.
\end{align*}
Then the range for $\lambda$ in \eqref{eq:tuning_range} reduces to $$ \lambda \in \left[4n\gamma_2^{1/2}\left\{2 + 2(|\tau_0|+1)\frac{\log(p/2)}{n} \right\}\sigma_0^2\bar\omega^2,
     \frac{n\gamma_2^{1/4}}{6}C_{\min}\right],$$ which is always non empty by the above.

Moreover when $\gamma_1 = \sqrt{1-\gamma_2^{1/4}}$ and $t= \sqrt{\gamma_2}\sigma_0^2 \bar\omega^2,$
\begin{align*}
     3\exp\left(-\frac{1}{4\gamma_2\bar\omega^2\sigma_0^2} nt  \right) =  3\exp\left(-\frac{n}{4\gamma_2^{1/2}}  \right) \rightarrow 0,
\end{align*}
as $\gamma_2 \rightarrow 0.$ Then by the above, \eqref{:eq:total_bound_w}, \eqref{eq:bound_Lp}, \eqref{eq:bound_p1}, and \eqref{eq:bound_Uw_tau},
\begin{align*}
    & \P_{(\*U^*, \*U_\omega |\cdot)}(\hat \tau_{\*U^*} \neq \tau_0 | \rho(\*U^*, \*U_\omega) > 1- \gamma_2^2)  \rightarrow 0,
\end{align*}
as $\gamma_2 \rightarrow 0,$ for any $\omega_{\max} \leq \bar\omega <\infty,$ and $\omega_{\max}/\omega_{\min} \leq \bar r_{\omega} < \infty$

Then by the proof of Lemma~\ref{lem::finite_complete} and \eqref{eq:inequality_final},
\begin{align*}
    \P(\tau_0 \notin S^{(d)})\leq \P\left(\hat\tau_{\*U^*} \neq \tau_0\middle |\rho(\*U^*, \*U_{\omega}) >  1- \gamma_2^2 \right) + \P\left(\bigcap_{b=1}^d\{\rho(\*U_b^*, \*U_\omega) \leq  1- \gamma_2^2\} \right).
\end{align*}
To bound the second term of the above, it follows from \eqref{eq: angle_ustar_uw} and  and  Lemma~\ref{lem:angle_ud} that $\P\left(\bigcap_{b=1}^d\{\rho(\*U_b^*, \*U_\omega) \leq  1- \gamma_2^2\} \right) \leq \left(1-\frac{\gamma_2^{n-1}}{n-1}\right)^d \rightarrow 0,$ as $d \rightarrow \infty. $ Then there exists a $\gamma_{\delta}$, such that when $\omega_{\max} < \bar \omega,$ $\omega_{\max}/\omega_{\min} < \bar r_{\omega},$
\begin{align*}
     \P(\tau_0 \notin S^{(d)}) < \left(1-\frac{\gamma_\delta^{n-1}}{n-1}\right)^d + \delta/2,
\end{align*}
Theorem~\ref{the:uw} then follows from the fact that $P(\Omega_{\max} < \bar \omega)  \rightarrow 0$ as $\bar\omega \rightarrow \infty,$ and $P(\Omega_{\max}/\Omega_{\min} < \bar  r_\omega) \rightarrow 0$ as $\bar r_\omega \rightarrow \infty.$

\end{proof}

\subsection{Proof of Theorem~\ref{the:sub_finite_pen}, Corollary~\ref{cor:second_moment_u}, and Theorem~\ref{the:sub_aymp_pen}}

To prove Theorem~\ref{the:sub_finite_pen}, we first introduce a technical lemma on sub-Gaussian vector.
\begin{lemma}
\label{lem:sub_gaussian}
If $\*U$ is a sub-Gaussian vector, then for any $\epsilon > 0$, there exists a constant $\kappa,$ such that $\P_{\*U}(\*U/\|\*U\| \in S_{\kappa}) < \epsilon,$
where $S_{\kappa} = \{\*v: \P(\|\*U^2\| > C|\*U/\|\*U\| = v) > \kappa \P(\chi^2_n > C^2) \mbox { for some } C> 0\}$
\end{lemma}

\begin{proof}[Proof of Lemma~\ref{lem:sub_gaussian}.]
If $\*U$ is a sub-gaussian, then by definition, there exists a constant $\kappa'$ such that  $\P(\|\*U\| > C)$ for any $C > 0,$ $\P(\|\*U\| > C) \leq \kappa'\P(\chi_n^2 > C^2). $

Then if  Lemma~\ref{lem:sub_gaussian} does not hold, there exists an $\epsilon > 0$ such that $\P_{\*U}(\*U/\|\*U\| \in S_{\kappa}) \geq \epsilon $ for any $\kappa > 0,$ then we make $\kappa = \kappa'/\epsilon,$ which leads to
\begin{align*}
    \P(\|\*U\| > C) >  \kappa'/\epsilon \P(\chi^2_{n} > {C}^2) P(\*U/\|\*U\| \in S_{\kappa'/\epsilon}) \geq \kappa' \P(\chi_n^2 \geq C^2).
\end{align*}
The above contradicts the fact that $\*U$ is sub-Gaussian. Lemma~\ref{lem:sub_gaussian}  then follows.

\end{proof}

\begin{proof}[Proof of Theorem~\ref{the:sub_finite_pen}]

First we let $F(\gamma_1, \gamma_2, \kappa) = E(\gamma_1, \gamma_2) \cap S^C_\kappa.$
We then follow the proof of Lemma~\ref{lem::finite_penalize}, but make slight changes to prove the theorem.

Similar to \eqref{eq:decom_prob}, we  decompose the conditional probability  of $D(\tau, \*U^*) -D(\tau_0,\*U^*)< $ given $  (\*U^*, \*U) \in F(\gamma_1, \gamma_2, \kappa) $ as follows
\begin{align*}
   & \P_{(\*U^*, \*U| \cdot)}\left\{ D(\tau, \*U^*) -D(\tau_0,\*U^*)< 0|  (\*U^*, \*U) \in F(\gamma_1, \gamma_2, \kappa) \right\}\\
    & \leq \P_{(\*U^*, \*U|\cdot)}\big\{(1- \gamma^2_1) \|(\*I-\*H_\tau)\*X_{\tau_0}\bbeta_0\|^2 - \sigma_0^2\*U^\top( \*H_{\tau}+ \*O_{ \tau^{\bot}\*U^*}-  \*H_{\tau_0}- \*O_{ \tau_0^{\bot} \*U^*} )\*U \\
    & \qquad + 2\sigma_0\*U^\top(\*I - \*H_{\tau}-\*O_{ \tau^{\bot}  \*U^*})\*X_{\tau_0}\bbeta_0 +  2\lambda(|\tau|-|\tau_0|) <0\big| (\*U^*, \*U) \in F(\gamma_1, \gamma_2, \kappa) \big\}\\
    & \leq  \P_{(\*U^*, \*U| \cdot)}\big\{(1-\gamma^2_1)(1-\delta)\|(\*I-\*H_\tau)\*X_{\tau_0}\bbeta_0\|^2 \\
    & \qquad{} - \sigma_0^2 \*U^\top(\*H_{\tau}+\*O_{ \tau^{\bot}  \*U^*} -  \*H_{\tau_0}-\*O_{ \tau_0^{\bot} \*U^*} )\*U  + \lambda(|\tau|-|\tau_0|) <0\big| (\*U^*, \*U) \in F(\gamma_1, \gamma_2, \kappa)\big\} \\
    & \quad + \P_{(\*U^*, \*U |\cdot)}\big\{(1-\gamma^2_1)\delta\|(\*I-\*H_\tau)\*X_{\tau_0}\bbeta_0\|^2 \\
    & \qquad{} + 2\sigma_0\*U^\top(\*I - \*H_{\tau}-\*O_{ \tau^{\bot}  \*U^*})\*X_{\tau_0}\bbeta_0 +\lambda(|\tau|-|\tau_0|) ) < 0\big| (\*U^*, \*U) \in F(\gamma_1, \gamma_2, \kappa)\big\} \\
    &= (I_1) + (I_2),
\end{align*}
for any $\delta \in (0,1).$

It hen follows from the above that when $|\tau| \leq  |\tau_0|$ and $\frac{\lambda}{n} < \frac{1}{6}(1-\gamma_1^2)C_{\min},$
\begin{align*}
    (I_1) & \leq \P_{(\*U^*, \*U|\cdot)}\left\{\|\*U\|^2 > \frac{(1-\gamma^2_1)(1-\delta)}{\gamma^2_2}\frac{\|(\*I-\*H_\tau)\*X_{\tau_0}\bbeta_0\|^2 }{\sigma_0^2} + \frac{\lambda(|\tau|-|\tau_0|)}{\gamma_2^2\sigma_0^2} \middle | (\*U^*,
    \*U) \in F(\gamma_1, \gamma_2, \kappa)\right\}.
\end{align*}
By Lemma~\ref{lem:sub_gaussian}, for an arbitrarily small $\epsilon > 0,$ there exists $\kappa > 0, $ such that for conditioning on any $\*U^* = \*u^*,$ we have
\begin{align*}
 &    P_{(\*U|\cdot)}\left\{\|\*U\|^2 > \frac{(1-\gamma^2_1)(1-\delta)}{\gamma^2_2}\frac{\|(\*I-\*H_\tau)\*X_{\tau_0}\bbeta_0\|^2 }{\sigma_0^2} + \frac{\lambda(|\tau|-|\tau_0|)}{\gamma_2^2\sigma_0^2} \middle | (\*u^*,
    \*U) \in F(\gamma_1, \gamma_2, \kappa)\right\} \\
   &\leq  \kappa \P\left\{\chi^2_n> \frac{(1-\gamma^2_1)(1-\delta-1/6)}{\gamma^2_2}\frac{niC_{\min}}{\sigma_0^2}\right\},
\end{align*}
from which we have
\begin{align*}
    (I_1) \leq  \kappa \exp\left\{-\frac{n}{2}\log(1-2t_1) - t_1\frac{(1-\gamma^2_1)(1-\delta-1/6  )}{\gamma^2_2}\frac{niC_{\min}}{\sigma_0^2}  \right\}, \numberthis \label{eq:I1_bound_sub_gaussian}
\end{align*}
for any $0 < t_1 <1/2.$ When $|\tau| > |\tau_0|,$ by Lemma~\ref{lem:sub_gaussian}, Markov inequality and moment generating function of chi-square distribution we have
\begin{align*}
    (I_1) \leq  \kappa\exp\left\{-\frac{n}{2}\log(1-2t_1) -  t_1 \frac{\lambda(|\tau|-|\tau_0|)}{\gamma_2^2\sigma_0^2} \right\}.
\end{align*}

For $(I_2)$,  if $\frac{\lambda}{n} < \frac{1}{6}(1-\gamma_1^2)C_{\min},$ by Cauchy-Schwartz inequality and Lemma~\ref{lem:sub_gaussian} and similar argument as \eqref{eq:I1_bound_sub_gaussian}, when $|\tau| \leq |\tau_0|,$  we have
\begin{align*}
    & (I_2)\\
    &\leq \P_{(\*U^*, \*U|\cdot)}\big\{(1- \gamma^2_1) \delta \|(\*I-\*H_\tau)\*X_{\tau_0}\bbeta_0\|^2  <  2\sigma_0\|\*U^\top(\*I - \*H_{\tau}-\*O_{ \tau^{\bot}  \*U^*})\|\|(\*I-\*H_\tau)\*X_{\tau_0}\bbeta_0\|\\
     & \qquad  \hspace{6cm}  - \lambda(|\tau|-|\tau_0|) \big| (\*U^*, \*U) \in F(\gamma_1, \gamma_2, \kappa)\big\} \\
    &= \P_{(\*U^*, \*U|\cdot)}\big\{2\sigma_0\|\*U^\top(\*I - \*H_{\tau}-\*O_{ \tau^{\bot}  \*U^*})\| >(1-\gamma_1^2 ) \delta \|(\*I-\*H_\tau)\*X_{\tau_0}\bbeta_0\|\\
     & \qquad \hspace{3cm} + \lambda(|\tau|-|\tau_0|)/\|(\*I-\*H_\tau)\*X_{\tau_0}\bbeta_0\|\big| (\*U^*, \*U) \in F(\gamma_1, \gamma_2, \kappa)\big\} \\
         &\leq \P_{(\*U^*, \*U |\cdot)}\left\{\|\*U\|^2 > \frac{(1-\gamma^2_1)^2(\delta-1/6)^2}{4\gamma^2_2}\frac{\|(\*I-\*H_\tau)\*X_{\tau_0}\bbeta_0\|^2}{\sigma_0^2}\middle | (\*U^*, \*U) \in F(\gamma_1, \gamma_2, \kappa)\right\}\\
    &\leq \kappa \P\left\{\chi^2_n > \frac{(1-\gamma^2_1)^2(\delta-1/6)^2}{4\gamma^2_2}\frac{niC_{\min}}{\sigma_0^2}\right\} \\
    &\leq \kappa \exp\left\{-\frac{n}{2}\log(1-2t_2)- t_2\frac{(1-\gamma^2_1)^2(\delta-1/6)^2}{4\gamma^2_2}\frac{niC_{\min}}{\sigma_0^2}\right\},
\end{align*}
for any $0<t_2<1/2.$

When $|\tau|  >  |\tau_0|$, by similar argument to \eqref{eq: bound_I2},  we have
\begin{align*}
   (I_2)
     &\leq \P_{(\*U^*, \*U|\cdot)}\left\{\sigma_0^2\|\*U^\top(\*I - \*H_{\tau}-\*O_{ \tau^{\bot}  \*U^*})\|^2 >\delta\lambda(|\tau|-|\tau_0|) \middle | (\*U^*, \*U) \in F(\gamma_1, \gamma_2, \kappa)\right\} \\
    &\leq \kappa \P\left\{\chi^2_n > \frac{\delta\lambda(|\tau|-|\tau_0|)}{\gamma_2^2 \sigma_0^2}\right\}\\
    & \leq  \kappa \exp\left\{-\frac{n}{2}\log(1-2t_2)-t_2\frac{\delta\lambda(|\tau|-|\tau_0|)}{\gamma_2^2 \sigma_0^2}\right\}.
\end{align*}

Now, by making of $(1-\gamma^2_1)(1 - \delta-1/6) =  (1-\gamma^2_1)^2(\delta-1/6)^2/4$, we obtain $\delta = \frac{2}{1- \gamma^2_1}(\sqrt{\frac{5}{3} - \frac{2}{3}\gamma_1^2} - 1) + \frac{1}{6}$.
Further
we make $t_1=t_2= \frac{\gamma_2}{2.04}$, so we have $-\frac{n}{2}\log(1-2t_1)=-\frac{n}{2}\log(1-2t_2)=-\frac{n}{2}\log(1-\frac{\gamma_2}{1.02}) \leq 2n\gamma_2$.
Then, conditional on the event $\{(\*U^*, \*U) \in F(\gamma_1, \gamma_2,\kappa)\}$, we have
\begin{align*}
    & \P_{(\*U^*, \*U)}\big\{\hat\tau_{\*U^*} \neq \tau_0, \big | (\*U^*, \*U) \in F(\gamma_1, \gamma_2, \kappa)\big\} \\
    & \qquad \leq  2\kappa\sum_{i=1}^{|\tau_0|} \sum_{j=0}^{i} \genfrac(){0pt}{0}{p-|\tau_0|}{j}\genfrac(){0pt}{0}{|\tau_0|}{i} \exp\left\{-\left(\sqrt{\frac{5}{3} - \frac{2}{3}\gamma_1^2} - 1\right)^2\frac{niC_{\min}}{2.04 \gamma_2\sigma_0^2} + 2n\gamma_2
    \right\} \\
    &   + 2\kappa \sum_{i=0}^{|\tau_0|} \sum_{j=i+1}^{p} \genfrac(){0pt}{0}{p-|\tau_0|}{j}\genfrac(){0pt}{0}{|\tau_0|}{i} \exp\left\{-\frac{0.74\lambda(j-i)}{2.04\gamma_2\sigma_0^2} + 2\gamma_2n \right\}\\
    & \qquad \leq
    2\kappa \sum_{i=1}^{|\tau_0|} \sum_{j=0}^{i} \genfrac(){0pt}{0}{p-|\tau_0|}{j}\genfrac(){0pt}{0}{|\tau_0|}{i} \exp\left\{- \frac{niC_{\min}}{26\sigma_0^2} \frac{\left(1 - \gamma_1^2\right)^2}{\gamma_2} + 2n\gamma_2
    \right\}\\
     &   + 2\kappa\sum_{i=0}^{|\tau_0|} \sum_{j=i+1}^{p} \genfrac(){0pt}{0}{p-|\tau_0|}{j}\genfrac(){0pt}{0}{|\tau_0|}{i} \exp\left\{-\frac{\lambda(j-i)}{4\gamma_2\sigma_0^2} + 2\gamma_2n \right\}. \numberthis \label{eq:I_and_II_subG}
     \end{align*}
The last inequality holds since $(\sqrt{\frac{5}{3} - \frac{2}{3}\gamma_1^2} - 1)^2 \geq 1.02 (1 - \gamma_1^2)^2/13$ for $\gamma_1^2 \in (0,1)$.

By similar calculation to that in \eqref{eq:U12a}, the first part of the above can be bounded by
\begin{align*}
   \kappa L_p(\gamma_1, \gamma_2)=  3\kappa \exp\left\{- \frac{n}{26\sigma_0^2}\left[ \frac{\left(1 - \gamma_1^2\right)^2}{\gamma_2} C_{\min} - 52\left( \frac{  \log(p/2)}{n}+ \gamma_2\right) \sigma_0^2\right]
    \right\}.
\end{align*}
As for the second part, it follows from the derivation in \eqref{eq:bound_penalty} that
\begin{align*}
   &  2\kappa\sum_{i=0}^{|\tau_0|} \sum_{j=i+1}^{p} \genfrac(){0pt}{0}{p-|\tau_0|}{j}\genfrac(){0pt}{0}{|\tau_0|}{i} \exp\left\{-\frac{(1-\gamma_1^2)\lambda(j-i)}{4\gamma_2\sigma_0^2} + 2\gamma_2n \right\} \\
&\leq 3\kappa \exp\left(-\frac{1}{4\gamma_2\sigma_0^2} nt  \right),
\end{align*}
if $$\frac{\lambda}{n} \in \left[\frac{\lambda^{(1)}_0}{n}+t, \frac{1}{6}(1-\gamma_1^2)C_{\min}\right),$$ where $\lambda^{(1)}_0=4\gamma_2\sigma_0^2\left[\gamma_2 n +  (|\tau_0|+1) \{\log |\tau_0| + \log(p-|\tau_0|\}\right].$

It then follows from \eqref{eq:ABC_indequality}, Lemma~\ref{lem::angle} and  Lemma~\ref{lemma:u_d} that
\begin{align*}
     & \P_{(\*U^*, \*U |\cdot)}(\hat \tau_{\*U^*} \neq \tau_0 | \rho(\*U^*, \*U) > 1- \gamma_2^2) \\
     &\leq  \P_{(\*U^*, \*U |\cdot)}\{\hat\tau_{\*U^*} \neq \tau_0 | (\*U^*, \*U) \in F(\gamma_1, \gamma_2, \kappa)\} + \P\left(\max_{\tau \neq \tau_0, |\tau|\leq |\tau_0|} \rho_{\tau^\bot}(\*U, \*X_{\tau_0}\bbeta_0) \geq \tilde\gamma_1^2\right) \\
     & \hspace{4cm}+ \P\left(\max_{\tau \neq \tau_0, |\tau|\leq |\tau_0|} \rho(\*U, \tau) \geq 1- \gamma_2 \right) + \P(\*U  \in S_\kappa ).
\end{align*}
We then make $\gamma_1 = \sqrt{1-\gamma_2^{1/4}},$ from which we have $\tilde\gamma_1=(1-\sqrt{\gamma_2})\sqrt{1-\gamma_2^{1/4}}- \sqrt{2-2\sqrt{1-\gamma^2_2}} \geq 1-1.6\gamma_2^{1/3} > 0$ for $\gamma_2 \in [0,0.24].$ Therefore $\arccos\tilde\gamma_1 \leq \arccos (1-1.6\gamma_2^{1/3})
\leq 2\gamma_2^{1/6} <1. $  In addition, we make $t=\sqrt{\gamma_2}\sigma_0^2.$ Hence the above probability bound reduces to
\begin{align*}
     & \P_{(\*U^*, \*U|\cdot)}(\hat \tau_{\*U^*} \neq \tau_0| \rho(\*U^*, \*U) > 1- \gamma_2^2) \\
     & \leq   3\kappa\exp\left\{- \frac{n}{26\sigma_0^2}\left[ \frac{C_{\min} }{\sqrt{\gamma_2}} - 52\left( \frac{  \log(p/2)}{n}+ \gamma_2\right) \sigma_0^2\right]
    \right\}
     + \P\left(\max_{\tau \neq \tau_0, |\tau|\leq |\tau_0|} \rho_{\tau^\bot}(\*U, \*X_{\tau_0}\bbeta_0) \geq 1-1.6\gamma_2^{1/3}  \right) \\
     & \hspace{4cm}+ \P\left(\max_{\tau \neq \tau_0, |\tau|\leq |\tau_0|} \rho(\*U, \tau) \geq 1- \gamma_2 \right)
     + 3\exp\left(-\frac{n}{4\gamma_2^{1/2}}   \right) + \epsilon,
\end{align*}
where $\gamma_2 < 1/64$ since $\gamma_2^{1/4} < 0.35. $
Then by \eqref{eq:inequality_final}, Lemma~\ref{lem:angle_ud} and \eqref{eq:bound_Ub},
\begin{align*}
      & \P(\tau_0 \notin S^{(d)})\leq \P\left(\hat\tau_{\*U^*} \neq \tau_0\middle |\rho(\*U^*, \*U) >  1- \gamma_2^2 \right) + \P\left(\bigcap_{b=1}^d\{\rho(\*U_b^*, \*U) \leq  1- \gamma_2^2\} \right)\\
      & \leq   3\kappa\exp\left\{- \frac{n}{26\sigma_0^2}\left[ \frac{C_{\min} }{\sqrt{\gamma_2}} - 52\left( \frac{  \log(p/2)}{n}+ \gamma_2\right) \sigma_0^2\right]
    \right\}
     + \P\left(\max_{\tau \neq \tau_0, |\tau|\leq |\tau_0|} \rho_{\tau^\bot}(\*U, \*X_{\tau_0}\bbeta_0) \geq 1-1.6\gamma_2^{1/3}  \right) \\
     & + \P\left(\max_{\tau \neq \tau_0, |\tau|\leq |\tau_0|} \rho(\*U, \tau) \geq 1- \gamma_2 \right) + 3\exp\left(-\frac{n}{4\gamma_2^{1/2}}   \right)+ \epsilon + \left(1- \P(\rho(\*U^*, \*U) >  1- \gamma_2^2)\right)^d.
\end{align*}
Because  $\epsilon$ is arbitrarily small and  the first three terms of the above converges to 0 as $\gamma_2 \rightarrow 0$,  there exists a $\gamma_2 > 0,$ such that
\begin{align*}
      & \P(\tau_0 \notin S^{(d)})\leq \delta +  \left(1- \P(\rho(\*U^*, \*U) >  1- \gamma_2^2)\right)^d.
\end{align*}
By \eqref{eq:lambda_range_proof}, the range for $\lambda$ in Theorem~\ref{the:sub_finite_pen} is never empty. Then Theorem~\ref{the:sub_finite_pen} follows immediately by making $\gamma_\delta = \gamma_2$ and $\zeta_\delta = -\log\left(1- \P(\rho(\*U^*, \*U) >  1- \gamma_2^2)\right) > 0. $

Finally, the range for $\lambda$, which is always nonempty, follows from the same arguments as \eqref{eq:lambda_range_proof}.

\end{proof}

{
\begin{proof}[Proof of Corollary~\ref{cor:second_moment_u}]

The proof follows similar steps as the proof of Theorem~\ref{the:sub_finite_pen}.
First for any $\epsilon, $ let  $\mathcal D_{K} = \left\{\*v: \P(\|\*U\|^2 > K \middle | \*U/\|\*U\| = \*v) > \epsilon \right\}.$ Let $\*V = \*U/\|\*U\|$, then by Markov Inequality
\begin{align}
\label{eq:bound_DC}
    \P(\*V \in \mathcal D_K) \leq \frac{\E_{\*V} \{\P(\|\*U\|^2 > K |\*V)\} }{\epsilon} = \frac{\P(\|\*U\|^2 > K)}{\epsilon} \leq \frac{\E (\|\*U\|^2)}{K\epsilon}.
\end{align}
Therefore for any arbitrarily small $\epsilon >0,$ there is a large enough $K,$ such that $\P(\*V \in \mathcal D_K)$ is arbitrarily small.

With a slight abuse of notation, let $F(\gamma_1, \gamma_2, K) = E(\gamma_1, \gamma_2) \cap \mathcal D_K^C$. Then by similar steps leading to \eqref{eq:I_and_II_subG} in the proof of Theorem~\ref{the:sub_finite_pen},
\begin{align*}
    & \P_{(\*U^*, \*U)}\big\{\hat\tau_{\*U^*} \neq \tau_0, \big | (\*U^*, \*U) \in F(\gamma_1, \gamma_2, K)\big\} \\
    &  \leq
    2 \sum_{i=1}^{|\tau_0|} \sum_{j=0}^{i} \genfrac(){0pt}{0}{p-|\tau_0|}{j}\genfrac(){0pt}{0}{|\tau_0|}{i} \bar P \left( \frac{niC_{\min}}{26\sigma_0^2} \frac{\left(1 - \gamma_1^2\right)^2}{\gamma_2}
    \right)\\
     & \qquad   + 2\sum_{i=0}^{|\tau_0|} \sum_{j=i+1}^{p} \genfrac(){0pt}{0}{p-|\tau_0|}{j}\genfrac(){0pt}{0}{|\tau_0|}{i} \bar P\left(\frac{\lambda(j-i)}{4\gamma_2\sigma_0^2}  \right),
     \end{align*}
    where $\bar P(C) =  \P_{(\*U^*, \*U|\cdot)}\left\{\|\*U\|^2 > C \middle | (\*U^*,
    \*U) \in F(\gamma_1, \gamma_2, K)\right\} $

Now make $\gamma_1  = \sqrt{1 - \gamma_2^{1/4}},$ and $\frac{\lambda}{n}  \in [\gamma_2^{3/4}, \frac{1}{6} \gamma_2^{1/4} C_{\min}),$ then the above reduces to
\begin{align*}
    & \P_{(\*U^*, \*U)}\Big\{\hat\tau_{\*U^*} \neq \tau_0, \Big | (\*U^*, \*U) \in F(\gamma_1, \gamma_2, K)\Big\} \\
    &  \leq
    2 \sum_{i=1}^{|\tau_0|} \sum_{j=0}^{i} \genfrac(){0pt}{0}{p-|\tau_0|}{j}\genfrac(){0pt}{0}{|\tau_0|}{i} \bar P \left( \frac{niC_{\min}}{26\sigma_0^2 \gamma_2^{1/2}}
    \right)\\
     & \qquad   + 2\sum_{i=0}^{|\tau_0|} \sum_{j=i+1}^{p} \genfrac(){0pt}{0}{p-|\tau_0|}{j}\genfrac(){0pt}{0}{|\tau_0|}{i} \bar P\left(\frac{(j-i)n}{4\gamma^{1/4}_2\sigma_0^2}  \right),
     \end{align*}
Since $\epsilon$ is arbitrarily small, for finite $n,p,$  $\P_{(\*U^*, \*U)}\Big\{\hat\tau_{\*U^*} \neq \tau_0, \Big | (\*U^*, \*U) \in F(\gamma_1, \gamma_2, K)\Big\} < \delta$ for any $\delta >0$ when $\gamma_2$ is small enough.

Moreover, by \eqref{eq:bound_DC}, for any $\epsilon >0$, $\P(\*V \in \mathcal D_K)$ is also arbitrarily small for a small enough $\gamma_2.$ Then the rest of the proof follows the same steps as the proof of Theorem~\ref{the:sub_finite_pen}.
\end{proof}
}

\begin{proof}[Proof of Theorem~\ref{the:sub_aymp_pen}]
The proof is similar to the proof of of Lemma~\ref{lem::symptotic_bound_penalize0} and Theorem~\ref{the:aymp_pen}. By Lemma~\ref{lem_proj}, we let
$D(\tau, \*u^*)=  \frac{1}{2}\|(\*I - \*H_{\tau,\*u^*})\*Y\|^2 + \lambda |\tau| =  \frac{1}{2}\|(\*I-\*H_\tau-  O_{ \tau_0^{\bot} \*u})\*y\|^2 + \lambda |\tau|$
for any $\tau \neq \tau_0.$ Then By \eqref{eq:dist_penalize}, for any $\delta \in (0,1)$ an any $\*u^*$ such that $$ \max_{\tau \neq \tau_0, |\tau|\leq |\tau_0|} \rho_{\tau^\bot}(\*u^*, \*X_{\tau_0}\bbeta_0) < \gamma_1^2,$$
we have
\begin{align*}
   & \P_{ \*U}\{ D(\tau, \*u^*) -D(\tau_0,\*u^*)< 0\}\\
    & \leq \P_{\*U}\{(1- \gamma^2_1) \|(\*I-\*H_\tau)\*X_{\tau_0}\bbeta_0\|^2 -\sigma_0^2\*U^\top( H_{\tau}+\*O_{ \tau^{\bot}  \*u^*} -  \*H_{\tau_0}-\*O_{ \tau_0^{\bot} \*u^*} )\*U \\
    & \hspace{5cm} + 2\sigma_0\*U^\top(\*I - \*H_{\tau}-\*O_{ \tau^{\bot}  \*u^*})\*X_{\tau_0}\bbeta_0 +  2\lambda(|\tau|-|\tau_0|) <0 \}\\
    & \leq  \P_{\*U}\{(1-\gamma^2_1)(1-\delta)\|(\*I-\*H_\tau)\*X_{\tau_0}\bbeta_0\|^2 - \sigma_0^2\*U^\top( H_{\tau}+\*O_{ \tau^{\bot}  \*u^*} -  \*H_{\tau_0}-\*O_{ \tau_0^{\bot} \*u^*} )\*U\\
    & \hspace{5cm} + \lambda(|\tau|-|\tau_0|) <0\} \\
    & \quad + \P_{\*U}\{(1-\gamma^2_1)\delta\|(\*I-\*H_\tau)\*X_{\tau_0}\bbeta_0\|^2 + 2\sigma_0\*U^\top(\*I - \*H_{\tau}-\*O_{ \tau^{\bot}  \*u^*})\*X_{\tau_0}\bbeta_0 \\
    & \hspace{5cm} +\lambda(|\tau|-|\tau_0|) ) < 0\} \\
    &= (I_1) + (I_2).
   \end{align*}

By Remark 2 of \cite{hsu2012tail},  we bound the log of the moment generating function $M(t)$ of $\*U^\top( \*H_{\tau,\*u^*} - \*H_{\tau_0,\*u^*})\*U$ as
\begin{align}
\log\{M(t)\}  \leq t(|\tau| - |\tau_0|) +
\frac{t^2}{1-2t}\tr\{(\*H_{\tau,\*u^*} - \*H_{\tau_0,\*u^*})^2\}\leq 2t |\tau \setminus \tau_0| \leq 2t|\tau_0 \setminus \tau|, \nonumber
\end{align}
for any $0<t<1/2.$ Therefore it follows from the above  and  Markov Inequality
\begin{align*}
    (I_1)  \leq \exp\left\{2t_1|\tau \setminus \tau_0| - \frac{t_1(1-\delta)(1-\gamma^2_1)n |\tau_0\setminus \tau| C_{\min} + t_1 \lambda (|\tau| - |\tau_0|)}{\sigma_0^2} \right\},
\end{align*}
for any $0<t_1<1/2.$
Further by the definition of sub-Gaussian vector, the moment generating function of  $2\sigma_0\*U^\top(\*I - \*H_{\tau,\*u^*})\*X_{\tau_0}\bbeta_0$ is bounded by $$\E\exp\{2t_2\sigma_0\*U^\top(\*I - \*H_{\tau,\*u^*})\*X_{\tau_0}\bbeta_0\} \leq \exp\{2t_2^2\sigma_0^2 (\*U^\top(\*I - \*H_{\tau,\*u^*})\*X_{\tau_0}\bbeta_0)^2\}, $$
then Markov inequality, the exact same bound in \eqref{eq:I2_asymp_pen} follows
\begin{align*}
  (I_2) \leq \exp\left\{\frac{(2t_2^2- \delta t_2) (1-\gamma_1^2) n |\tau_0 \setminus \tau| C_{\min} - t_2 \lambda (|\tau| - |\tau_0|)}{\sigma_0^2}\right\},
  \end{align*}
for any $0<t_2<1/2.$

The rest of the proof follows identical steps as those in the proofs of Lemma~\ref{lem::symptotic_bound_penalize0} and Theorem~\ref{the:aymp_pen}.
We make $\delta=1/2, t_1=t_2=1/3,$ then by \eqref{eq:asymp_gamma1_bound}, \eqref{eq:aymptotic_bound} and the bounds for $(I_1), (I_2)$ above, it follows that
\begin{align*}
  \P_{({\cal U}^d, \*Y)} (\tau_0 \notin S^{(d)} )
  &  \leq L(\gamma_1) + 3\exp\left(-\frac{nt}{3\sigma_0^2}\right) + \left[2\{\arccos (\gamma_1)\}^{n-|\tau_0|-1 } p^{|\tau_0|}\right]^d.
\end{align*}
  Theorem~\ref{the:sub_aymp_pen} then follows by making $\gamma^2_1=0.7$ and simplifying the  lower bound for $\frac{\lambda}{n}$ by the inequality $\log(|\tau_0|)+ \log(p-|\tau_0|)\leq 2\log(p/2).$
\end{proof}

\subsection{Proof of Corollary~\ref{cor:nonGaussian}}
\begin{proof}[Proof of Corollary~\ref{cor:nonGaussian}]
The probability that the confidence set $\widetilde \Gamma^{\bbeta_{\Lambda}}(\*y_{obs})$ does not cover  $\bm\beta_{0, \Lambda }$ is bounded by
\begin{align*}
    \P(\bbeta_{0,\Lambda} \not\in \widetilde \Gamma^{\bbeta_{\Lambda}}(\*Y)) \leq \P(\tau_0 \not\in S^{(d)}) +  \P\big(\bm\beta_{0, \Lambda } \not\in \widetilde \Gamma^{\bm\beta_\Lambda|\tau_0}(\*Y)\big) = 1 - \tilde\alpha +  \P(\tau_0 \not\in S^{(d)}),
\end{align*}
from which  Corollary~\ref{cor:nonGaussian} follows immediately.
\end{proof}

\section{on the number of the repro samples for the model candidate set in Algorithm~1}
\label{sec:size_d}

In this section, we explore the number of the repro samples in Algorithm~1, \( d \), sufficient to ensure a high probability that the model candidate set , \( S^{(d)} \), includes the true model, \( \tau_0 \). Specifically, we have derived a theorem that establishes the scale of \( d \) sufficient to theoretically ensure that the probability \( P(\tau_0 \notin S^{(d)}) \) is 
small.

\begin{theorem}
\label{the:size_d}
  Suppose $|\tau_0| < n$ and $|\tau_0|\log(p)/n = O(1),$ then 
   \begin{itemize}
       \item[(a)] If  $C_{\min} \geq O(\log(p)/n)$, when $d \geq O(1),$ $P(\tau_0 \neq S^{(d)}) \leq e^{-O(n)}.$ 
       \item[(b)] Otherwise, for any $\epsilon > 0 $, when $d \geq \log(1/\epsilon) O(n^{1/2} r_n^{-n+1}),$ where $r_n = e^{-O(\log(p)/n)} \wedge O(\frac{C_{\min}}{\log(p)/n}),$ $P(\tau_0 \neq S^{(d)}) \leq \epsilon.$ 
   \end{itemize}
   
\end{theorem}

\noindent
The above theorem indicates, if other models are separated from the true model $\tau_0$ with $C_{\min} > O(\log(p)/n),$ 
and $\log(p)/n$ is not too large,  
we only need a limited number of repro samples $d$ to achieve a high inclusion probability of the true model in the model candidate set. Furthermore, when the separation $C_{\min}$ is smaller than the scale of $\log(p)/n,$ we would need a repro samples size $d$ of the order $n^{1/2}r_n^{-n+1}$ to achieve a large inclusion probability; this includes the cases where the signals are very small.
Moreover, if $p$ is extremely large, we would also need a large number of repro samples to include 
the true model in the candidate set.

\begin{proof}[Proof of Theorem~\ref{the:size_d}]

First, Theorem~\ref{the:size_d} (a) follows immediately from Theorem~2.

In the following, we will focus on the proof of Theorem~\ref{the:size_d} (b).
First by \eqref{eq:beta_distribution}, $1-\rho_{\tau^\bot}(\*U^*, \*X_0\bbeta_0) \sim Beta(\frac{n-|\tau| - 1}{2}, \frac{1}{2}),$ therefore By \cite{segura_sharp_2016} and Gautschi's inequality, if $1- \gamma_1^2 < \frac{n-|\tau|-1}{n - |\tau|},$ then
\begin{align*}
    \P(\rho_{\tau^\bot}(\*U^*, \*X_0\bbeta_0) \geq \gamma_1^2)  & = \P(1-\rho_{\tau^\bot}(\*U^*, \*X_0\bbeta_0) \leq 1- \gamma_1^2) \\ 
& < \frac{(1-\gamma_1^2)^{(n-|\tau| - 1)/2}\gamma_1}{B(\frac{n-|\tau|-1}{2}, \frac{1}{2})\left(\frac{n-|\tau| - 1}{2} - \frac{(n- |\tau|)(1-\gamma_1^2)}{2}\right)} \\
& < \frac{(1-\gamma_1^2)^{(n-|\tau| - 1)/2}\gamma_1}{\frac{n-|\tau|}{2}B(\frac{n-|\tau|-1}{2}, \frac{1}{2})\left(\frac{n-|\tau| - 1}{n -|\tau|} -(1-\gamma_1^2)\right)} \\
& \approx \frac{\sqrt{2\pi}(1- \gamma^2_1)^{(n-|\tau|-1)/2}}{\sqrt{n - |\tau|}\gamma_1}. \numberthis \label{eq: gamma_1_bound}
\end{align*}

Then
\begin{align*}
     &  \P_{\*U^*}\left\{
     \max_{\tau \neq \tau_0, |\tau| \leq |\tau_0|}\rho_{\tau^\bot}(\*U^*, \*X_0\bbeta_0) \geq \gamma_1^2
     \right\} \nonumber
     \\& \qquad 
    \leq  \sum_{\tau \neq \tau_0, |\tau| \leq |\tau_0|} \P_{\*U^*}\left\{\rho_{\tau^\bot}(\*U^*, \*X_0\bbeta_0) \geq \gamma_1^2  \right\} \nonumber
    \\
    & \qquad \leq \sum_{k = 1}^{|\tau_0|} \genfrac(){0pt}{0}{p}{k}  \frac{\sqrt{2\pi}(1- \gamma^2_1)^{(n-|\tau| - 1)/2}}{\sqrt{n -|\tau|}\gamma_1} \leq  \frac{\sqrt{2\pi}(1- \gamma^2_1)^{(n-|\tau_0|-1)/2}}{\sqrt{n - |\tau_0|}\gamma_1} p^{|\tau_0|}  \numberthis \label{eq:gamma_1_sum}
  %
  %
    \end{align*}

By \eqref{eq: gamma_1_bound} and \eqref{eq:gamma_1_sum}, another bound for $\P\left(\max_{\tau \neq \tau_0, |\tau|\leq |\tau_0|} \rho_{\tau^\bot}(\*U, \*X_0\bbeta_0) \geq \tilde\gamma_1^2\right)$ in \eqref{eq:bound_penalty_single} is 
\begin{align*}
  &\P\left(\max_{\tau \neq \tau_0, |\tau|\leq |\tau_0|} \rho_{\tau^\bot}(\*U, \*X_0\bbeta_0) \geq \tilde\gamma_1^2\right) \\
   & \leq \frac{\sqrt{2\pi}(1- \tilde\gamma^2_1)^{(n-|\tau_0|-1)/2}}{\sqrt{n - |\tau_0|}\tilde\gamma_1} p^{|\tau_0|} \\
   &  = \exp\left\{O\left(\frac{n - |\tau_0| -1}{12} \{\log(\gamma_2) + \frac{12|\tau_0| \log(p)}{n -|\tau_0| -1}\}\right) - \frac{1}{2}\log(n - |\tau_0|)\right\}.
\end{align*}
for a $\gamma_2 < 1.$ Then the last term of \eqref{eq:bound_penalty_single} is 
\begin{align*}
    2 \gamma_2^{\frac{n-|\tau_0|}{2}-1} (\sqrt{n}p)^{|\tau_0|} = \exp\left\{O\left(\frac{n - |\tau_0| -2}{2} \{\log(\gamma_2) + \frac{|\tau_0| \log(p) + 0.5 |\tau_0| \log(n)}{n -|\tau_0| -2}\}\right) \right\}.
\end{align*}
Suppose $n>>|\tau_0|,$ and $\log(n)/n$ is small, then the bound in \eqref{eq:bound_penalty_single} can be simplified as
    \begin{align*}
     & \P_{(\*U^*, \*U|\cdot)}(\hat \tau_{\*U^*} \neq \tau_0| \rho(\*U^*, \*U) > 1- \gamma_2^2) \\
     & \leq   3\exp\left\{- \frac{n}{26\sigma_0^2}\left[ \frac{C_{\min} }{\sqrt{\gamma_2}} - 52\left( \frac{  \log(p/2)}{n}+ \gamma_2\right) \sigma_0^2\right]
    \right\} + 3\exp\left(-\frac{n}{4\gamma_2^{1/2}}   \right) \\
     & \hspace{4cm} + \exp\left\{O\left(n \log(\gamma_2) + 12 \log(p)\right)\right\}. \numberthis \label{eq:new_bound}
\end{align*}
Now to make the first term of \eqref{eq:new_bound} less than or equal to $\epsilon,$ we would need the exponent term $\frac{n}{26\sigma_0^2}\left[ \frac{C_{\min} }{\sqrt{\gamma_2}} - 52\left( \frac{  \log(p/2)}{n}+ \gamma_2\right) \sigma_0^2 \right] \geq M,$ where $M = \log(1/\epsilon).$ And because $\sqrt{\gamma_2} \geq \gamma,$ we would only need $\frac{C_{\min} }{\sqrt{\gamma_2}} - 52\left( \frac{  \log(p/2)}{n}+ \sqrt{\gamma_2}\right) \sigma_0^2  \geq   \frac{26\sigma_0^2}{n}M.$ Then solving this inequality would give us a sufficient condition for the inequality to hold is $\gamma_2 \leq \frac{0.8C_{\min}}{52 \log(p/2)\sigma_0^2/n + M/n},$ by the fact that $\sqrt{1+a} -1 > 0.4a$ for any $0<a<1.$  To make the second term less than $\epsilon,$ we would need $\gamma_2 < O(n^2/M^2),$ and for the last term to be less than $\epsilon,$ we would need $\gamma_2 < \exp\{-O(\frac{\log(p) + M}{n})\}.$ The second can be ignored, since $n$ is generally larger than $M,$ e.g. when $M=20,$ $e^{-20} \leq 2 \times 10^{-9}.$ Moreover, for the same reason, it is also reasonable to assume $M = O(\log(p)).$ Therefore, in order for the bound in \eqref{eq:new_bound} to work, we would need $\gamma_2 \leq \exp\{-O(\log(p)/n)\}$ and $\gamma_2 \leq O(\frac{C_{\min}}{\log(p)/n}).$ 

Then we would only need to bound the probability $\P\left(\bigcap_{b=1}^d\{\rho(\*U^*_b, \*U) \leq  1- \gamma_2^2\} \right).$ To this end, let us first try to find a improved bound for $\P(\rho(\*U, \*U^*) \leq \gamma_2). $ By \eqref{eq:u_epsi-1} and \eqref{eq:u_ustar}, 
\begin{align*}
    \P(\rho(\*U, \*U^*) \geq 1-\gamma^2_2) = \frac{2}{c_1}\int_0^{\arcsin \gamma_2} \sin^{n-2}(s) ds.
\end{align*}
Now, we apply the transformation $v = \sin^2(s),$ then the above reduced to 
\begin{align*}
    \P(\rho(\*U, \*U^*) \geq 1- \gamma^2_2) = \frac{1}{c_1}\int_0^{\gamma^2_2} v^{(n-3)/2} (1-v) ^{-1/2}dv, 
\end{align*}
which is the cdf of the $Beta(\frac{n-1}{2}, \frac{1}{2})$ distribution at $\gamma_2^2.$
By \cite{segura_sharp_2016} and Gautschi's inequality,  the above is bounded
\begin{align*}
    \P(\rho(\*U, \*U^*) \geq 1-\gamma^2_2) & \geq \frac{\gamma_2^{(n-1)}(1-\gamma^2_2)^{1/2}}{\frac{n-1}{2}  B(\frac{n-1}{2}, \frac{1}{2})}\left(1 + \frac{2n}{2n+2}\gamma^2_2\right)\\
    & \gtrsim \frac{\sqrt 2}{n^{1/2}}\gamma_2^{(n-1)}(1-\gamma^2_2)^{1/2}(1 + \gamma_2^2)
\end{align*}
for $\gamma_2<1.$

It then follows  from \eqref{eq:bound_Ub} that 
\begin{align*}
\P\left(\bigcap_{b=1}^d\{\rho(\*U^*_b, \*U) \leq  1- \gamma_2^2\} \right)  & \leq \left(1-\frac{\sqrt 2}{n^{1/2}}\gamma_2^{(n-1)}(1-\gamma^2_2)^{1/2}(1 + \gamma_2^2)\right)^d\\
&\approx \exp\left\{ - d\frac{\sqrt 2}{n^{1/2}}\gamma_2^{(n-1)}(1-\gamma^2_2)^{1/2}(1 + \gamma_2^2)\right\}\\ & 
 =  \exp\left\{ - dO(\frac{1}{n^{1/2}}\gamma_2^{(n-1)})\right\},
\end{align*}
for $\gamma_2 \leq \exp\{-O(\log(p)/n)\}$ and $\gamma_2 \leq O(\frac{C_{\min}}{\log(p)/n}).$  Then part (b) of the theorem is proved by making $r_n \leq \gamma_2. $ 

\end{proof}

\section{{EBIC implementation, additional simulation results and visualizations for the joint confidence set of all regression coefficients }}
\label{sec:joint_beta_full}

{ 
\subsection{Implementation of EBIC for choosing $\lambda$ in Algorithm~1}
In our implementation of Algorithm~1,  to obtain the model candidate set, the following EBIC is used to choose the values of the tuning parameter $\lambda$ in (5),
\begin{align*}\small
    \mbox{EBIC}_{b, \zeta}(\lambda) = n\log\left[\left\|\*y_{obs}-X_{\hat\tau_{b,\lambda }}\hat\beta_{\hat\tau_{b,\lambda }}\right\|^2/n\right]+ \left|\hat\tau_{b,\lambda }\right|\log(n) + 2\zeta \log \genfrac(){0pt}{0}{p}{\left|\hat\tau_{b,\lambda }\right|}.
\end{align*}
Here, $\hat\tau_{b, \lambda}$ is the solution to (5) with the tuning parameter $\lambda,$ $\hat\beta_{\hat\tau_{b,\lambda }}$ is an estimation of $\beta_{\hat\tau_{b,\lambda }}$ and  $0 \leq \zeta \leq 1$ can range between 0 and 1. To increase the efficiency of candidate models search, we pick multiple models for each $\*u^*_{b}.$ Specifically, we pick all $\lambda$'s between $\lambda^0_{b}$ and $\lambda^1_{b}$ i.e. $\Lambda_b=[\lambda^0_{b}, \lambda^1_{b}],$ where $\lambda^\zeta_{b} = \argmin_\lambda \mbox{EBIC}_{b, \zeta}(\lambda).$ This is equivalent to using all $0 \leq \zeta \leq 1$,  because $\lambda^\zeta_{b}$ is monotonically non-decreasing in $\zeta$, and \cite{chen_extended_2008-1} showed that the model selection consistency of EBIC holds for some $0 \leq \zeta \leq 1.$
}

\subsection{Simulation results for the joint confidence set (29)}
Besides getting the model confidence set and the confidence set for single coefficients, our repro samples method also provides a joint inference for $\bbeta_0^{full}$. 
To evaluate the performance of the joint confidence set for $\bbeta_0^{full}$ in (29), we apply (29) on the 200 simulated data sets for models (M1)-(M5), and summarize the results in Table~\ref{tab:simu_tab_joint}. Evidently, the proposed confidence set can achieve the desired coverage rate, since it covers the truth $\bbeta_0^{full}$  94\% -96\% of the times for models all the models. Moreover,  the proposed confidence set, as opposed to those in \citep{zhang_simultaneous_2017, dezeure_high-dimensional_2017}, has a sparse structure in the sense that the vast majority of dimensions of the joint confidence set corresponding to  the zero regression coefficients are shrunk to [0,0], as illustrated by Table~\ref{tab:simu_tab_joint}. This is because if variable $X_i$ is not in any of the models in the model candidate set $S^{(d)},$ then any value of $\bbeta$ with nonzero $\beta_i$ will be excluded from the confidence set, following from the union in (29).  Such sparse confidence sets give researchers two advantages in practice: (1) the size/volume of the confidence set is substantially smaller, and therefore it is more informative; (2) it offers a new tool for confidently and efficiently screening variables. Here the proportions of the confidence set's dimensions shrunk to $[0,0]$ are above 98.5\% for model (M1) and (M3)-(M5) and 91.6\% for model (M2),  demonstrating that the number of variables left after screening is much smaller than $n-1,$ which is suggested for the sure independence screening approach \citep{fan_sure_2008}.

\begin{table}[ht]
\caption{Performances of the Joint Confidence Set for $\bbeta_0^{full}$}
\label{tab:simu_tab_joint}
\begin{tabular}{c|c|c}
\hline
Model& Coverage  Rate &   Proportions of Dimensions Shrunk to [0,0] \\
\hline
M1: $n=50, p=1,000$ & 0.940 (0.016)& 0.997 (0.000) \\
M2: $n=80, p=150$ & 0.945 (0.016) & 0.916 (0.002) \\
M3: $n=100, p=500$ &  0.950 (0.015) &  0.986 (0.001)\\
M4: $n=150, p=200$ &  0.965 (0.013) &  0.967 (0.002)\\
M4: $n=50, p=1000$ &  0.940 (0.016) &  0.996 (0.000)\\
\hline
\end{tabular}
\end{table}

\subsection{Visualization of the joint confidence set in (29)}
We now use a 3-dimensional graph to present a visualization of the joint confidence set  $\Gamma_{\alpha}^{\bbeta}(\*y_{obs})$ for $\bbeta_0^{full}$. 
To do so, 
we consider a particular example of $p = 8$ with the true model $\tau_0 = \{1,2\}$, for which our candidate set contains only three models $S^{(d)} = \big\{\{1,2,3\}, \{1,2\}, \{1,3\}\big\}$, each having three or less covariates. Unlike the confidence set obtained in \citep{zhang_simultaneous_2017, dezeure_high-dimensional_2017}, which, in this example, would typically be a $8$-dimensional shallow disc, our confidence set  $\Gamma_{\alpha}^{\bbeta}(\*y_{obs})$ is a union of three sets, one 3-dimensional ellipsoid and two 2-dimensional ellipsoids, corresponding to models $\{1,2,3\}$, $\{1,2\}$ and $\{1,3\}$, respectively. 
Plotted in Figure~\ref{fig:B2C} are two components:
(a) a confidence curve \citep{xie_confidence_2013} plot of model $\tau_0$ plotted on the candidate model space $S^{(d)}$; and (b) the corresponding confidence regions of the coefficients in the three candidate models. The $y$-axis of plot (a) is the associated confidence level of each model computed via the conditional probability in  (18)
, therefore the plot demonstrates the uncertainty of the models. 
The figure on the right shows the level-$95\%$ confidence sets of $\bbeta$ (the two blue ones) for each of the three models in the candidate set $S^{(d)}$. It demonstrates that our algorithm produces a union of three sets of different dimensions in this example.

\begin{figure}[H]
  \vspace{-6mm}      \includegraphics[scale=0.18]{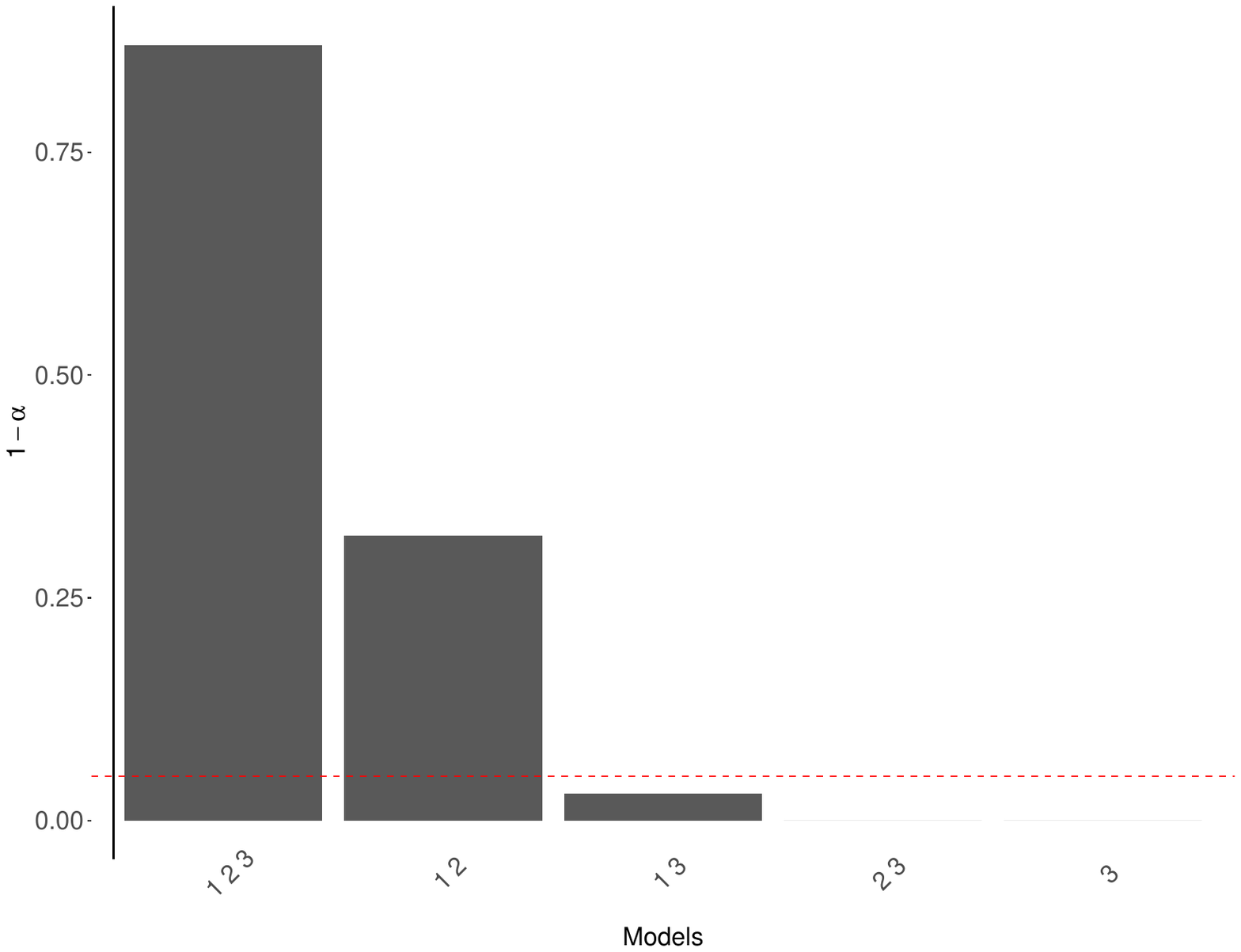}
        \includegraphics[scale=0.23]{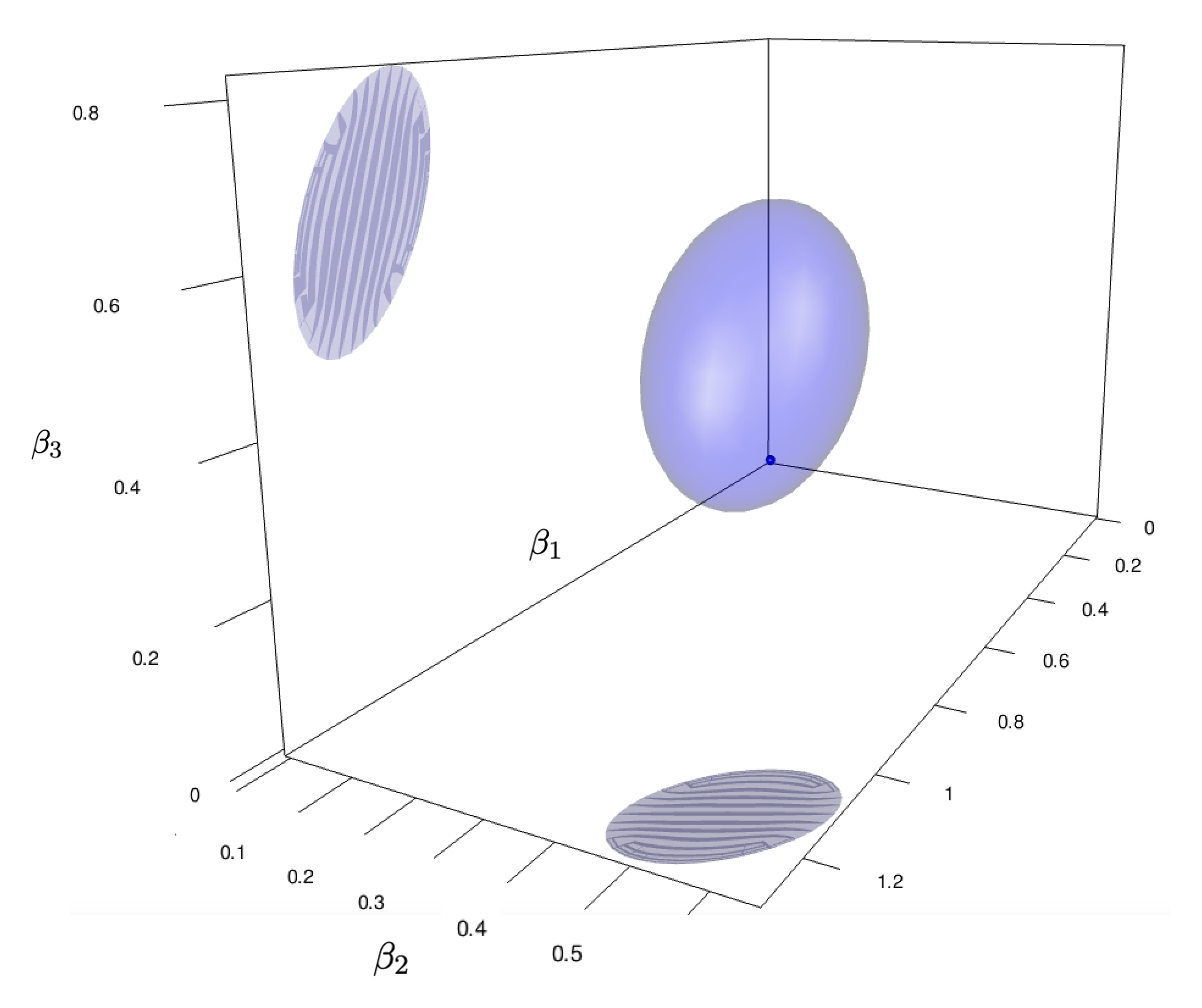} \\ {\qquad \small (a) \qquad\qquad\qquad\qquad \qquad\qquad\qquad\qquad (b) \qquad } 
\caption{\small (a)  Confidence curve \citep{xie_confidence_2013} plot on $S^{(d)}$; (b)  confidence sets of $\bbeta_\tau$ (one 3-dimensional ellipsoid and two 2-dimensional ellipsoids) 
of the three $\tau$ models in candidate set $S^{(d)} = \big\{\{1,2,3\}, \{1,2\}, \{1,3\}\big\}$. In (a), the red line instantiates the case where we aim to construct a level-$0.95$ ($\alpha=0.95$) model confidence set. In this case, our $95\%$ model confidence set for the true $\tau_0$ contains two models; i.e.,   ${\Gamma}^\tau_\alpha(\*y_{obs}) =
\big\{\{1,2,3\}, \{1,2\}\big\}$. In (b), a $95\%$ joint confidence set for $\bbeta_0^{full}$ is the union of these three confidence sets, one 3-dimensional  on the $(\beta_1, \beta_2, \beta_3)$ space and two 2-dimensional ellipsoids on the $(\beta_1, \beta_2)$ and $(\beta_1, \beta_3)$ space, respectively (in each of the cases the remaining $\beta_j$'s are $0$). 
}
\label{fig:B2C}  

\end{figure}

\subsection{Simulation result for an additional setting} \label{sec:add_simu}

Here we conduct a simulation study on Scenario (f) of the simulation conducted by \cite{li_model_2019} to demonstrate the performance of the propose repro samples approach in the low-dimensional setting when $n>p.$ This simulation setting is as follows.
\begin{itemize}
\item[(MA)] (From \cite{li_model_2019}, with many signals) Let the true values $\bbeta_0^{full}=(\underbrace{1, \dots, 1}_{12}, 0, \dots, 0).$ For $j_1,j_2\in[p]$, the correlation between $x_{j_1}$
and $x_{j_2}$ is set to $0.5^{|j_1 - j_2 |}$. We let $n = 300,
p = 200$ and $\sigma = 1$.
\end{itemize}

\begin{table}[ht]
\caption{Comparison of Performance of the Model Candidate Sets}
\label{tab:mod_candidate_low}
\scriptsize
\begin{tabular}{ll|ll}
\hline
 & {Method} & Cardinality of $S^{(d)}$& Inclusion of  $\tau_0$ \\
 \hline
\multirow{4}{*}{\shortstack{Model MA: $n=300, p=200$
}} & Repro samples &  1.010 (0.007)          & 1.000 (0.000)    \\
             &    Bootstrap AIC & 601.645 (12.612)       & 1.000 (0.000)    \\
             &  Bootstrap BIC  &246.995 ( 4.404)       & 1.000 (0.000)  \\
             &  Bootstrap CV  &567.150 (12.446)       & 1.000 (0.000)  \\
                  \hline
\end{tabular}
\end{table}

\begin{table}[ht]
\caption{Repro Confidence Sets for True Model $\tau_0$}
\label{tab:mod_confidence_sets_low}
\scriptsize
\begin{tabular}{ll|ll}
\hline
 & {Method} & Cardinality of $\Gamma^{\tau}_{0.95}$& Coverage of $\Gamma^{\tau}_{0.95}$ \\
 \hline
\multirow{4}{*}{\shortstack{Model MA: $n=300, p=200$
}} & Repro samples &  1.010 (0.007)     & 1.000 (0.000) \\
                  & Bootstrap AIC  & 601.645 (12.612)  & 1.000 (0.000)  \\
                  & Bootstrap BIC & 246.995 ( 4.404)  & 1.000 (0.000)   \\
                  & Bootstrap CV &567.150 (12.446)  & 1.000 (0.000)   \\
                  \hline
\end{tabular}
\end{table}

Table~\ref{tab:mod_candidate_low} and Table~\ref{tab:mod_confidence_sets_low} summarize the results for the candidate set and the 95\% model confidence set respectively from $200$ simulation repetitions. The comparison of the proposed repro samples approach with the bootstrap is similar to those from (M1)-(M5) in Section~\ref{sec:simu_model}. Even in this low-dimensional settings, the bootstrap approach would produce an excessive large amount of models (between 250 and 600), while our repro samples approach only needs about 1 model, smaller than $1.745$ reported for the (M4) setting in Table~\ref{tab:combined_model} as expected (since sample size $n = 300$ is twice of $n = 150$). This performance appears to be notably more efficient than the confidence bound approach reported in \cite{li_model_2019} for this Scenario (f): Figure 3 (f) of \cite{li_model_2019} indicates that the confidence bound needs to encompass at least 50 variables (about 25\% of all $p=200$ variables).
The reliance of their confidence bounds approach on bootstrap techniques,
which generate a large number of models as indicated in Table~\ref{tab:mod_candidate_low} and Table~\ref{tab:mod_confidence_sets_low} may have contributed partly to the excessively broad confidence bounds.

\begin{table}[ht]
\caption{Repro Confidence Sets for (Scalar) Regression Parameter $\beta_{0,i}$ with Comparison to Debiased Lasso}
\label{tab:confidence_interval_low}
\scriptsize
\begin{tabular}{lr|cc|cc|cc}
\hline
 & & \multicolumn{2}{c|}{Repro Samples}    & \multicolumn{2}{c|}{Debiased Lasso (JM) } & \multicolumn{2}{c}{Debiased Lasso (ZZ)}    \\
Model &  $\beta_{0,i}$ & Coverage & Width& Coverage & Width &  Coverage & Width   \\ \hline
\multirow{3}{*}{MA} & All $\beta_{0,i} $ & 0.998(0.000) & 0.018(0.000) & 0.971(0.001) & 0.320(0.000) & 0.964(0.001) & 0.339(0.000) \\
 & $\beta_{0,i} \neq 0$    & 0.960(0.004) & 0.296(0.000) & 0.950(0.004) & 0.319(0.001) & 0.959(0.004) & 0.338(0.000) \\
 & $\beta_{0,i} = 0$      & 1.000(0.000) & 0.000(0.000) & 0.973(0.001) & 0.320(0.000) & 0.965(0.001) & 0.339(0.000)    \\
\hline
\end{tabular}
\end{table}

\begin{table}[ht]
\caption{Revised Comparison of Confidence Sets for Nonzero Regression Coefficient $\beta_{0,i} \neq 0$}
\label{tab:revised_simulation_table_low}
\scriptsize
\begin{tabular}{l|cc|cc|cc}
\hline
 & \multicolumn{2}{c|}{Repro Samples} & \multicolumn{2}{c|}{Debiased Lasso (JM)} & \multicolumn{2}{c}{Debiased Lasso (ZZ)} \\
$\beta_{0,i}$ & Coverage & Width & Coverage & Width & Coverage & Width \\
\hline
$\beta_{0,1}=1$ & 0.940(0.017) & 0.268(0.001) & 0.935(0.017) & 0.294(0.002) & 0.955(0.015) & 0.323(0.001) \\
$\beta_{0,2}=1$ & 0.970(0.012) & 0.300(0.001) & 0.940(0.017) & 0.320(0.002) & 0.945(0.016) & 0.338(0.001) \\
$\beta_{0,3}=1$ & 0.955(0.015) & 0.302(0.001) & 0.930(0.018) & 0.322(0.002) & 0.935(0.017) & 0.340(0.001) \\
$\beta_{0,4}=1$ & 0.960(0.014) & 0.301(0.001) & 0.955(0.015) & 0.320(0.002) & 0.955(0.015) & 0.339(0.001) \\
$\beta_{0,5}=1$ & 0.950(0.015) & 0.303(0.001) & 0.960(0.014) & 0.322(0.002) & 0.970(0.012) & 0.341(0.001) \\
$\beta_{0,6}=1$ & 0.945(0.016) & 0.300(0.001) & 0.960(0.014) & 0.320(0.002) & 0.955(0.015) & 0.339(0.001) \\
$\beta_{0,7}=1$ & 0.965(0.013) & 0.300(0.001) & 0.945(0.016) & 0.320(0.002) & 0.965(0.013) & 0.338(0.001) \\
$\beta_{0,8}=1$ & 0.960(0.014) & 0.300(0.001) & 0.945(0.016) & 0.319(0.002) & 0.960(0.014) & 0.338(0.001) \\
$\beta_{0,9}=1$ & 0.965(0.013) & 0.300(0.001) & 0.960(0.014) & 0.321(0.002) & 0.960(0.014) & 0.339(0.001) \\
$\beta_{0,10}=1$ & 0.980(0.010) & 0.302(0.001) & 0.975(0.011) & 0.322(0.002) & 0.980(0.010) & 0.341(0.001) \\
$\beta_{0,11}=1$ & 0.975(0.011) & 0.301(0.001) & 0.960(0.014) & 0.321(0.002) & 0.970(0.012) & 0.340(0.001) \\
$\beta_{0,12}=1$ & 0.950(0.015) & 0.269(0.001) & 0.940(0.017) & 0.322(0.002) & 0.955(0.015) & 0.340(0.001) \\
\hline
\end{tabular}
\end{table}

We summarized the performance of the repro samples confidence intervals for the regression coefficients along with those of the debiased confidence intervals in Table~\ref{tab:confidence_interval_low} and Table~\ref{tab:revised_simulation_table_low}. In general, both the repro samples approach and the debiased methods  achieve the desired coverage rates. However, proposed repro samples interval are significantly narrower compared to the debiased confidence intervals. In particular, the average length of our repro samples confidence interval is only about 6\% of width of the debiased confidence intervals on average, mainly due to the advantage in the widths of the confidence intervals for the coefficients whose true values are 0's. Even for the signals, we observe from Table~\ref{tab:revised_simulation_table_low} our repro samples approach consistently produces narrower confidence intervals compared to both of the debiased approaches for every single non-zero coefficients, while obtaining the desired coverage rate.

{

\subsection{Simulation results for confidence sets of each nonzero coefficient.}

\begin{table}[ht]
\caption{Repro Samples Method versus Debiased Lasso for Making Inference for Nonzero Regression Coefficient $\beta_{0,i} \neq 0$}
\label{tab:simu_tab_sigals}
\scriptsize
\begin{tabular}{ll|cc|cc|cc}
\hline
 & & \multicolumn{2}{c|}{Repro Samples}    & \multicolumn{2}{c|}{Debiased Lasso (JM)}   & \multicolumn{2}{c}{Debiased Lasso (ZZ)}   \\
Model &  $\beta_{0,i}$ & Coverage & Width& Coverage & Width & Coverage & Width    \\ \hline
\multirow{3}{*}{\shortstack{M1}} & $\beta_{0,1}=3$ & 0.970(0.012)   & 0.714(0.008) & 0.310(0.033)    & 0.244(0.008)  & 0.990(0.007)      & 1.266(0.014)\\
 & $\beta_{0,2}=2$    & 0.960(0.014)   & 0.810(0.010) & 0.440(0.035)    & 0.249(0.009)  & 0.980(0.010)      & 1.346(0.017)  \\
 & $\beta_{0,3}=1.5$  & 0.925(0.019)   & 0.718(0.009) & 0.320(0.033)    & 0.250(0.009)  & 0.975(0.011)      & 1.352(0.016)  \\
\hline
\multirow{6}{*}{\shortstack{M2 }} & $\beta_{0,1}=2$   & 0.990(0.007)   & 0.540(0.006)  & 0.960(0.014)     & 0.498(0.004) & 0.995(0.005)      & 0.942(0.009)  \\
& $\beta_{0,2}=1.5$  & 0.965(0.013)  & 0.544(0.006)& 0.915(0.020)   & 0.498(0.005) & 0.980(0.010)      & 0.941(0.010)  \\
&$\beta_{0,3}=1$   & 0.965(0.013)   & 0.538(0.007)  & 0.935(0.017)    & 0.494(0.005) & 0.995(0.005)      & 0.936(0.010)  \\
& $\beta_{0,4}=0.8$ & 0.980(0.010)   & 0.540(0.006) & 0.930(0.018)     & 0.494(0.004) & 0.990(0.007)      & 0.933(0.009) \\
&$\beta_{0,5}=0.6$ & 0.975(0.011)    & 0.533(0.007)  & 0.950(0.015)     & 0.499(0.004) & 0.985(0.009)      & 0.946(0.009) \\
\hline
\multirow{6}{*}{\shortstack{M3}} & $\beta_{0,1}=3$   &0.980(0.010)   &0.458(0.005) &0.870(0.024)    &0.427(0.003)  & 0.995(0.005)      & 0.942(0.010) \\
& $\beta_{0,2}=2$   &0.950(0.015)   &0.463(0.005) &0.870(0.024)    &0.428(0.003)  & 0.955(0.015)      & 0.941(0.009) \\
&$\beta_{0,3}=1.5$ &0.965(0.013)   &0.461(0.005) &0.865(0.024)    &0.427(0.004)  & 0.980(0.010)      & 0.941(0.010) \\
&$\beta_{0,4}=1$   &0.960(0.014)   &0.465(0.005) &0.850(0.025)    &0.431(0.004) & 0.975(0.011)      & 0.951(0.010) \\
&$\beta_{0,5}=0.8$ &0.945(0.016)   &0.471(0.006) &0.890(0.022)    &0.431(0.003)  & 0.960(0.014)      & 0.950(0.009) \\
&$\beta_{0,6}=0.6$ &0.955(0.015)   &0.436(0.005) &0.865(0.024)    & 0.424(0.003) & 0.995(0.005)      & 0.930(0.009) \\
\hline

\multirow{2}{*}{\shortstack{M4}} & $\beta_{0,1}=1$   & 0.965(0.013)    & 0.399(0.003)  & 0.940(0.017)     &0.416(0.003) & 0.965(0.013)      & 0.498(0.003)    \\
& $\cdots$   &\multicolumn{2}{c|}{$\cdots$} &\multicolumn{2}{c|}{$\cdots$} & \multicolumn{2}{c}{$\cdots$} \\
\hline

\multirow{3}{*}{\shortstack{M5}} & $\beta_{0,1}=3$   &0.970(0.012)   & 0.723(0.009)  & 0.255(0.031)     & 0.243(0.009) & NA & NA \\
& $\beta_{0,2}=2$   &0.965(0.013)   & 0.813(0.010)  & 0.455(0.035)     & 0.249(0.009)  & NA  & NA \\
&$\beta_{0,3}=1.5$ &0.925(0.019)    & 0.728(0.010)  & 0.300(0.032)     & 0.247(0.009)  & NA & NA \\
\hline
\end{tabular}
\end{table}

To further investigate the performance differences between the proposed approach and the two debiased Lasso methods, we compare in Table~\ref{tab:simu_tab_sigals} the coverage rates and widths of the confidence sets for each nonzero regression coefficient. To save space, we only display the results of the confidence sets for $\beta_{0,1} = 1$ of (M4) due to the similarity of the results across the 12 non-zero coefficients. Both the repro samples approach and debiased Lasso (ZZ) achieve the desired coverage regardless of the signal strengths, with the debiased Lasso (ZZ) intervals at least 70\% wider.
Conversely, the debiased Lasso (JM) uniformly undercovers the truths for all signals in all models except for the two coefficients in (M2).
As expected, the under-coverage issue of the debiased Lasso (JM) approach is more serious when $p/n$ is larger, since the second order approximation is more difficult. For {(M1) and (M5)} with $n=50, p=1000$, the coverage rate of the debiased Lasso (JM) is only around 25\%--45\%, and for Model (M3) with $n=100, p=500$ the coverage rate is around 85\%--89\%.
In terms of the width of the confidence sets, for Model (M3) with $n=100, p=500$, the widths of the repro samples confidence sets are less than half of those from the debiased Lasso (ZZ) and comparable to the debiased Lasso (JM), which undercovers all the signals in (M3).  For the other three models, the repro samples confidence sets are also at least 40\% shorter than the debiased Lasso (ZZ) confidence intervals for the signals, providing a more accurate assessment of the uncertainties of the estimation of these regression parameters.
To sum up, the repro samples approach covers all the signals with the desired coverage rate and correctly quantifies the uncertainty of parameter estimation regardless of the dimension of the design matrix and signal strength.

}

}
\end{appendix}

\end{document}